\documentclass[12pt]{article}
\usepackage{amsfonts}
\usepackage{amsmath, amsthm, amssymb, bbm}
\usepackage{xcolor,multirow,graphicx}
\usepackage{enumerate}
\usepackage{enumitem}
\usepackage{mathtools}
\usepackage{geometry}
\usepackage{setspace}
\usepackage{soul}
\usepackage{tikz}
\usepackage{float}
\usepackage{needspace,afterpage}
\usetikzlibrary{positioning}
\usepackage{natbib}
\usepackage[colorlinks,citecolor=blue,urlcolor=blue,linkcolor=blue]{hyperref}
\usepackage{placeins}
\hypersetup{pdftitle={Quantile Methods in Large Panels with Fixed Effects},
pdfauthor={Manuel Arellano, Ayden Higgins, Martin Weidner},
pdfsubject={Panel quantile regression and jackknife bias correction}}

\newcommand{\designfactorwidth}{0.4\textwidth}

\newtheorem*{remarks}{Remarks}

\newtheorem{theorem}{Theorem}
\newtheorem{lemma}{Lemma}

\newtheorem{definition}{Definition}
\newtheorem{example}{Example}
\newtheorem*{example*}{Example}

\newcommand{\plim}{\operatorname*{plim}}

\newcommand{\argmin}{\operatorname*{argmin}}

\newcommand\smallO{%
  \mathchoice
    {\scalebox{.6}{$\mathcal{O}$}}%
    {\scalebox{.6}{$\mathcal{O}$}}%
    {\scalebox{.5}{$\scriptstyle\mathcal{O}$}}%
    {\scalebox{.45}{$\scriptscriptstyle\mathcal{O}$}}%
}

\usepackage{booktabs}
\usepackage{threeparttable}
\usepackage{pdflscape}

\newcommand{\comment}[1]{}  %

\begin{document}

\title{\bf  Quantile Methods in Large Panels \\ with Fixed Effects\footnote{
We are grateful to Carlos Lamarche, Frank Kleibergen, and Iv\'{a}n Fern\'{a}ndez-Val for helpful comments and discussions, and to seminar and conference participants at
Bilkent University,
HU Berlin,
Sciences Po Paris,
CREST Paris,
the University of Nottingham,
NYU,
CEMFI,
Harvard-MIT,
Lund University,
the Amsterdam School of Economics Panel Data Workshop (2015),
the Econometric Study Group Annual Conference in Bristol (2013),
the International Panel Data Conference in Tokyo (2014),
the ERCIM Conference in Pisa (2014),
the IAAE Annual Conference in Thessaloniki (2015),
and the North American Winter Meeting of the Econometric Society in Boston (2015).
}}

\author{\setcounter{footnote}{2}
Manuel Arellano\footnote{
CEMFI, Email: {\tt arellano@cemfi.es}.}
\and
Ayden Higgins\footnote{University of Exeter,
Email {\tt  a.higgins@exeter.ac.uk}
}
\and
Martin Weidner\footnote{University College London,
 Email: {\tt m.weidner@ucl.ac.uk}.
}}

\date{October 2026}

\maketitle

\begin{abstract}\footnotesize\noindent
We study quantile regression (QR) and instrumental variable quantile regression (QIV) in static panel data models with additive individual fixed effects. We derive the asymptotic distribution of the fixed effects estimator as both panel dimensions grow large, showing that it has an asymptotic bias of order $1/T$, where $T$ is the number of observations per individual. The bias analysis relies on a novel characterization of the profiled score as a weighted sum of induced order statistics (concomitants), which yields an exact finite-$T$ expression for the score bias. We propose Jackknife bias corrections and uncover a bias-variance tradeoff that is specific to the non-smooth quantile objective functions. The standard delete-one Jackknife fails to eliminate the leading bias and, even when modified to achieve bias correction, inflates the variance so that only $\sqrt{N}$-consistency is attained, where $N$ is the number of individuals. We propose a modified Jackknife bias correction for general subsample sizes, derive the asymptotic distribution of the resulting estimator, and show that $\sqrt{NT}$-consistency and valid inference are achieved for suitable choices of the subsample size. As an empirical application, we study the effect of age on school children's test scores using school fixed effects.
\end{abstract}

\medskip


\medskip


\newpage

\section{Introduction}
\label{sec:Introduction}

We study panel data quantile regression models with individual specific fixed effects,
covering both exogenous and endogenous regressors.
Fixed effects quantile regression (QR) is a natural method for estimating
effects of covariates on the distribution of outcomes
while controlling for unobserved time-invariant heterogeneity.
If some regressors are endogenous even after controlling for individual specific effects,
we extend the quantile instrumental variable (QIV) framework of
\cite{ChernozhukovHansen2005,ChernozhukovHansen2006}
by allowing for individual specific effects.
Standard QR is a special case of QIV when all regressors are exogenous.

We consider panel data on $N$ units and $T$ periods $\left\{
y_{it},z_{it}\right\} _{t=1}^{T}$ $\left( i=1,...,N\right) $.
In the exogenous case, $z_{it}$ denotes the vector of regressors and, for a given $\tau \in (0,1)$,
we impose the conditional quantile restriction
$$
  Q_{y_{it}}\left(\tau \,\big|\, z_{it}, A_i \right) = z'_{it} \, \theta(\tau) + \alpha_i(\tau),
$$
where $\theta(\tau)$ and $\alpha_i(\tau)$ are unknown parameters
and $A_i$ is a vector of time-invariant individual characteristics that includes $\alpha_i(\tau)$.
In the endogenous case we additionally observe a vector of regressors $x_{it}$
(some components of which may be endogenous) and a vector of instruments $z_{it}$,
and we impose the quantile restriction
$$
  \mathbb{E}\left\{ \mathbbm{1} \big[
   y_{it} \leq  x'_{it} \, \theta(\tau)  + \alpha_i(\tau)
    \big]  \, \Big|  \; z_{it} , \, A_i
                   \right\} =  \tau.
$$
If all regressors are exogenous we set $z_{it} = x_{it}$, reducing to the standard QR model above.

We assume that the data are independently and identically distributed over $t$,
conditional on $A_i$.
This rules out all serial dependence conditional on $A_i$,
including predetermined regressors such as lagged dependent variables,
restricting us to static panel models.
Note, however, that unconditional time-series dependence and persistence are
accommodated through the vector-valued $A_i$, which can be of arbitrary dimension
and can encode rich patterns of individual heterogeneity that induce serial correlation
in the unconditional distribution of the data.
Importantly, $t$ need not index time.
In our empirical application, for example, $i$ indexes schools and $t$ indexes
students within schools, so that the iid assumption across $t$ is natural.

A noteworthy advantage of the QIV model in the panel setting concerns the
rank invariance (or rank similarity) assumption that underlies the
\cite{ChernozhukovHansen2005} framework.
In cross-sectional applications this can be a strong requirement.
In the panel setting with fixed effects, however, rank invariance is only required
to hold conditional on the unobserved heterogeneity $A_i$.
Since $A_i$ absorbs a large part of the unobserved variation across individuals,
the assumption is substantially weakened compared to the purely cross-sectional QIV model.

The application of QR and QIV models to panel data has been hampered by incidental parameter bias.
The quantile model is non-linear so that naive inclusion of individual specific fixed effects
results in a \cite{NeymanScott1948} type incidental parameter problem.
The mainstream approach has addressed this problem in other panel contexts
by seeking alternative estimators that are fixed-$T$ consistent for large $N$.
However, point-identification of the parameters of interest for fixed-$T$ panel data with fixed effects is often impossible.
An alternative approach is therefore to approximate the sampling distribution of the fixed effect estimator
as $N$ and $T$ both grow large. For smooth objective functions this
approach leads to a bias correction that can be easily implemented by
automatic or analytical methods for a large class of models;
see the survey by \cite{ArellanoHahn2007} for details and references.
The current paper extends these large-$T$ bias correction techniques to the QR and QIV setting,
where the non-smoothness of the quantile check function presents novel challenges.
Our main results are as follows.
\begin{enumerate}[label=(\roman*)]
\item The fixed effects estimator has an asymptotic bias of order $1/T$,
which we characterize explicitly, and we derive the asymptotic distribution
of the estimator as $N,T \to \infty$.

\item The bias analysis employs a novel approach based on
induced order statistics (concomitants).
The profile score of the fixed effects objective function can be written as a
weighted sum of induced order statistics, and we derive an exact finite-$T$
characterization of the score bias as an integral involving beta distribution densities.
This exact representation is the key technical result that drives our asymptotic analysis.

\item We propose Jackknife bias corrections and uncover a bias-variance tradeoff
that is specific to the non-smoothness of the quantile objective function.
The standard delete-one Jackknife of \cite{HahnNewey2004} does not eliminate
the leading $1/T$ bias, unlike in smooth nonlinear panel models.
Even when modified to achieve bias correction,
delete-one Jackknife inflates the variance of the estimator from order $1/(NT)$ to order $1/N$,
yielding only $\sqrt{N}$-consistency.
More generally, we consider Jackknife bias correction with subsample size $S$,
including the delete-one choice $S=T-1$ and the half-sample choice $S=T/2$
used by the split-panel Jackknife of \cite{DhaeneJochmans2015}.
We derive the asymptotic distribution of the resulting bias-corrected estimator,
which is $\sqrt{NT}$-consistent for suitable choices of $S$, and characterize the
bias-variance tradeoff.

\item We extend the framework to QIV with fixed effects.
The properties of the panel QIV estimator and its bias correction parallel the exogenous case.

\item In an application, we estimate the effect of child maturity on academic achievement. Using school fixed effects and age relative to the school cutoff as an instrument for child age, we uncover heterogeneous effects, which suggest that maturity and ability are complements in academic achievement. The bias-corrected estimates suggest a somewhat steeper quantile profile than the uncorrected fixed-effects estimates.

\end{enumerate}

We take a moment-based approach. The quantile restriction implies that certain
unconditional moment conditions hold for each individual.
The incidental parameter $\alpha_i(\tau)$ is unknown, but can be estimated as $T$ becomes large.
The profile score, obtained by profiling out $\alpha_i(\tau)$,
cannot be used for finite-$T$ consistent inference in general, but the large-$T$
consistency of the estimated fixed effect implies that the profile score can be
used for large $N$, $T$ consistent inference on $\theta$, under appropriate regularity conditions.

The key technical challenge is that the profile score is non-continuous in $\theta$,
leading to similar complications as in the cross-sectional QR and QIV estimation problems,
with the additional difficulty that both $N$ and $T$ grow to infinity asymptotically.
As the sample size grows, however, the profile objective concentrates around a smooth limit,
and the non-continuity does not affect the final asymptotic distribution of $\widehat\theta$.

\bigskip

\citet{ChiangWu2026} also study the unsmoothed fixed effects QR estimator.
Using a related rank-score representation, they derive its leading bias and establish
the validity of the split-panel Jackknife, while also allowing for serial dependence
and developing inference for individual effects.
Our paper additionally covers endogenous regressors and studies a general class of
Jackknife corrections, including how the subsample size affects bias and variance.

The paper relates to several strands of the literature on panel quantile regressions.
While large-$T$ bias correction in models with smooth objective functions is well understood,
the extension to non-smooth objective functions such as the QR check function had remained elusive.
\cite{GrahamHahnPowell2009} raise the incidental parameter problem in a non-differentiable panel data model.
The most closely related work is
\cite{KatoGalvaoMontes-Rojas2012}, who establish
asymptotic normality of the fixed effects QR estimator
when $N$ and $T$ grow jointly, under conditions requiring $N/T\rightarrow 0$
that eliminate the leading bias;
\cite{GalvaoKato2016}, who derive the large $N$, $T$ asymptotic distribution,
including the leading order $1/T$ bias, for a smoothed FE-QR estimator
and propose an analytical bias correction; and
\cite{GalvaoGuVolgushev2020},
who establish unbiased asymptotic normality for the non-smoothed FE-QR estimator
under improved conditions on the relative growth of $N$ and~$T$,
using a novel analysis of the Bahadur representation.
The present paper complements these contributions in several respects.
It characterizes the $1/T$ bias for the non-smoothed FE-QR estimator under general asymptotics
where $N/T$ may tend to a positive constant,
provides an exact finite-$T$ bias characterization via order statistics,
uncovers the surprising failure of delete-one Jackknife for non-smooth models,
and extends the framework to
IV quantile regression with fixed effects and provides Jackknife-based bias corrections
with accompanying asymptotic theory.
For a survey of large-$T$ bias correction methods in nonlinear panel models
more broadly, see \cite{ArellanoHahn2007} and \cite{FernandezValWeidner2018}.

A number of papers take alternative approaches to panel quantile models.
\cite{Canay2011} shows that when fixed effects are pure
location shifts, the model is identified for $T\geq 2$ and a simple
two-step estimator is available.
\cite{Rosen2012} shows that without additional assumptions the fixed effects model
is only set identified when $T$ is fixed.
\cite{MachadoSantosSilva2019} propose
estimating conditional quantiles through a location-scale model using conditional
means of the outcome and of absolute residuals, which naturally accommodates additive fixed effects
and avoids the non-smoothness of the check function.
\cite{ChernozhukovFernandezValWeidner2024}
construct simultaneous confidence bands for quantile effects in network and panel models
via distribution regression with debiased fixed effects, providing an alternative
to direct quantile regression estimation.
\cite{ChernozhukovFernandezValHahnNewey2013}
provide identification and estimation of average and quantile effects in
fully nonparametric nonseparable panel models.
\cite{GrahamHahnPoirierPowell2018}
extend the correlated random coefficients representation of linear QR to panel data,
obtaining $\sqrt{N}$-consistent estimates when $T$ exceeds the number of regressors.
\cite{Powell2022} proposes a QR estimator for panels with nonadditive fixed effects
that is consistent for small~$T$.
\cite{GuVolgushev2019} develop panel quantile regression with grouped fixed effects,
approximating individual heterogeneity by a finite number of latent groups.
\cite{HardingLamarche2014} extend panel QR to models with interactive (factor) effects,
and \cite{HardingLamarchePesaran2020} combine heterogeneous dynamic panel QR with
the common correlated effects approach to handle cross-sectional dependence.

On the random effects and penalized side,
\cite{Koenker2004} considered the penalized
interpretation of standard random-effects estimation to estimate QRs with
fixed effects and suggested a shrinkage device, and \cite{Lamarche2010}
discussed an optimal choice of the tuning parameter in Koenker's method.
\cite{AbrevayaDahl2008} considered a correlated random effects
model to study the effects of smoking and prenatal care on birthweight.
\cite{ArellanoBonhomme2016} develop a class of
QR estimators for short panels using correlated random effects,
covering static and dynamic models.

On endogenous regressors in panel quantile models,
\cite{HardingLamarche2009}
discusses IV estimation methods with simulation evidence.
\cite{Galvao2011} studies QR for dynamic panel data with
fixed effects using instrumental-variable methods.
\cite{ChetverikovLarsenPalmer2016}
develop IV quantile regression for group-level treatments with group-level unobservables,
applying their method to the distributional effects of trade.

\bigskip

The paper is organized as follows.
Section~\ref{sec:Setup} introduces the model and assumptions for the exogenous case
and states the main asymptotic distribution result for the fixed effects estimator.
Section~\ref{sec:MomentCondition} develops the profiled score analysis using induced order statistics,
characterizing the bias, variance, and Hessian of the score,
and showing that the score has exactly zero mean when $\theta(\tau)$ is constant, implying no incidental parameter bias.
Section~\ref{sec:BiasCorrection} introduces the Jackknife bias corrections
and analyzes the resulting bias-variance tradeoff.
Section~\ref{sec:Endogenous} extends all results to instrumental quantile regression with endogenous regressors.
Section~\ref{sec:Interpretation} provides a detailed interpretation of the leading bias.
Sections~\ref{sec:MC} and~\ref{sec:Empirical} present Monte Carlo simulations
and the empirical application, respectively.
All proofs are contained in the Appendix.

\section{Model and Fixed Effect Estimator}
\label{sec:Setup}
\subsection{Setup and Main Assumptions}
\label{sec:Assumptions}

We consider panel data on $N$ units and $T$ time periods.
For $i=1,\ldots,N$ and $t=1,\ldots,T$ we observe
a scalar outcome variable $y_{it}$ and a column vector of regressors $z_{it}$.\footnote{
We denote the regressors by $z_{it}$ here, because it makes the notational transition to the QIV case easier. For both the QR and QIV cases the exogenous variable is denoted $z_{it}$.
}
We write $y_i = (y_{i1},\ldots,y_{iT})$,
$z_i = (z_{i1},\ldots,z_{iT})$
for the time paths of individual $i$,
and define ${\bf N} = \{1,2,\ldots,N\}$ and ${\bf T}=\{1,2,\ldots,T\}$.

The central modelling assumption is a conditional quantile restriction that allows for
individual specific fixed effects.
We write $Q_{y_{it}}(\tau | z_{it}, A_i)$ for the $\tau$'th conditional quantile of $y_{it}$
given the regressors $z_{it}$ and the unobserved individual characteristics $A_i$.

\newtheorem{QRassumption}{Assumption}
\renewcommand{\theQRassumption}{QR}
\begin{QRassumption}[\bf Quantile Restriction]
     \label{ass:QR}
      For a given $\tau \in (0,1)$ and for all $i \in {\bf N}$, $t \in {\bf T}$,
   $$Q_{y_{it}}(\tau | z_{it}, A_i) = z'_{it} \, \theta(\tau) + \alpha_i(\tau),$$
     where  $\theta(\tau) \in \mathbbm{R}^{\dim z}$ is a
     column vector of unknown
      common parameters,  $\alpha_i(\tau) \in \mathbbm{R}$ is an unknown individual specific effect,
      and $A_i$ is a vector of time-invariant individual characteristics
      of arbitrary dimension that captures all unobserved individual heterogeneity,
      with $\alpha_i(\tau)$ being a function of $A_i$.
\end{QRassumption}

All formal results in this paper are for a single chosen $\tau$, and if that is the only object of interest, then choosing $A_i = \alpha_i(\tau)$ is a possible choice,
though a larger $A_i$ is also useful to satisfy the following conditional independence assumption.
However, the underlying model is naturally coherent across all quantiles simultaneously.
While only a one-dimensional fixed effect $\alpha_i(\tau)$
enters at each quantile, the entire function $\tau \mapsto \alpha_i(\tau)$ can be captured by the vector $A_i$,
which is indeed the main substantive reason for conditioning on $A_i$ as opposed to only $\alpha_i(\tau)$ in Assumption~\ref{ass:QR}.
By the law of iterated expectations, Assumption~\ref{ass:QR} implies $Q_{y_{it}}(\tau | z_{it}, \alpha_i(\tau)) = z'_{it} \, \theta(\tau) + \alpha_i(\tau)$.

\newtheorem{CIassumption}{Assumption}
\renewcommand{\theCIassumption}{CI}
\begin{CIassumption}[\bf Conditional Independence]
     \label{ass:CI}
      For all $i \in {\bf N}$ we assume that
      $\{ y_{it}, z_{it} \}$ is independently and identically distributed over $t$,
      conditional on $A_i$.
\end{CIassumption}

Assumption~\ref{ass:CI} requires that, conditional on $A_i$, the observations
$\{y_{it}, z_{it}\}$ are iid over time.
This rules out all serial dependence conditional on $A_i$,
including predetermined regressors such as lagged dependent variables,
restricting us to static panel models.
Note, however, that unconditional time-series dependence and persistence are
accommodated through the vector-valued $A_i$. Since $A_i$ can be of arbitrary
dimension, it can encode rich patterns of individual heterogeneity that induce
serial correlation in the unconditional distribution of $(y_{it}, z_{it})$.
Under Assumption~\ref{ass:CI}, conditioning on the full regressor path
$z_i = (z_{i1},\ldots,z_{iT})$ instead of $z_{it}$ in Assumption~\ref{ass:QR}
would be equivalent, since $z_{is}$ for $s \neq t$ is independent of
$(y_{it}, z_{it})$ conditional on $A_i$.
Together, Assumptions~\ref{ass:QR} and~\ref{ass:CI}
form the core of our modelling framework, and all results in the main text rely crucially on them.
Appendix~\ref{sec:INID} gives an exact finite-$T$ representation under conditional independence without identical distributions.

\newtheorem{RSassumption}{Assumption}
\renewcommand{\theRSassumption}{RS}
\begin{RSassumption}[\bf Random Sampling]
     \label{ass:RS}
     $\{ y_i, z_i, A_i \}$ is independently and identically distributed across $i \in {\bf N}$.
\end{RSassumption}

Assumption~\ref{ass:RS} imposes iid sampling across individuals, which is a benign assumption for most microeconomic panels.
Allowing for weak cross-sectional correlation could be accommodated without changing any of our insights on the quantile regression incidental parameter bias,
but at the expense of more complicated variance estimation and inference.

Finally, we also require the following set of regularity conditions.
Here $\tau_0$ denotes the quantile of interest and $\theta^0$ its true coefficient.
Note that under Assumption~\ref{ass:CI} the conditional distribution of $y_{it}$ given $z_{it}$ and $A_i$
does not depend on $t$. We therefore denote its cdf by
$F_y(y | z_{it}, A_i) := \Pr( y_{it} \leq y \, | \, z_{it}, A_i )$.

\newtheorem{RCassumption}{Assumption}
\renewcommand{\theRCassumption}{RC}
\begin{RCassumption}[\bf Regularity Condition]~
     \label{ass:RC}
     \begin{itemize}
        \item[(i)] The true parameter $\theta^0$ lies in the interior of a compact set $\Theta \subset \mathbbm{R}^{\dim z}$.

        \item[(ii)]
        $F_y(y | z_{it}, A_i)$ is six times continuously differentiable for all $y \in \mathbbm{R}$,
        and all six derivatives are bounded uniformly over $y \in \mathbbm{R}$, and over all possible realisations of $z_{it}$ and $A_i$.
        In addition, the conditional density $f_y(y | z_{it}, A_i) = \partial_y F_y(y | z_{it}, A_i)$
        is bounded away from zero, uniformly over $y$ in a neighborhood of the conditional
        quantile $z'_{it}\theta(\tau_0) + \alpha_i(\tau_0)$ and over all realisations of $z_{it}$ and $A_i$.

        \item[(iii)] The range of $z_{it}$ is bounded uniformly, and
         $ \mathbb{E}\left[ \left. |y_{it}|^{2+\delta} \, \right| A_i  \right] < c$
        for some $\delta > 0$ and $c<\infty$.
     \end{itemize}
\end{RCassumption}

\noindent
Assumption~\ref{ass:RC}(i) is a standard compactness condition.
The smoothness in~\ref{ass:RC}(ii) ensures that the bias and variance of the profile score
have well-behaved large-$T$ expansions (six derivatives suffice for the
Taylor expansions used in Section~\ref{sec:MomentCondition}).
The boundedness conditions in~\ref{ass:RC}(iii) are used to control remainder terms
uniformly across individuals.

\begin{example}[\bf Nonseparable Panel Model]
\label{ex:DGP}
Consider the data-generating process
$$
y_{it} = z'_{it} \, \theta(U_{it}) + r(A_i, U_{it}),
$$
where $U_{it} \sim U[0,1]$ is independent of $z_{it}$ conditional on $A_i$,
$A_i$ is a vector of unobserved individual characteristics of arbitrary dimension,
$z'_{it}\,\theta(\tau) + r(A_i, \tau)$ is strictly increasing in $\tau$ for all $z_{it}$ and $A_i$,
and both $\theta(\tau)$ and $r(A_i, \tau)$ are sufficiently smooth in $\tau$.
With $\alpha_i(\tau) := r(A_i, \tau)$, the conditional quantile restriction
$Q_{y_{it}}(\tau | z_{it}, A_i) = z'_{it}\,\theta(\tau) + \alpha_i(\tau)$
holds for every $\tau \in (0,1)$, verifying Assumption~\ref{ass:QR}.
Assumption~\ref{ass:CI} is satisfied if the pairs $(z_{it},U_{it})$
are iid over $t$ conditional on $A_i$.
Assumptions~\ref{ass:RS} and~\ref{ass:RC} hold under standard regularity conditions on $z_{it}$ and $r$.
This is a one-factor random coefficient model with the single unobserved factor $U_{it}$.
In Section~\ref{sec:QIVmodel} we generalize this example to accommodate endogenous regressors,
replacing $z_{it}$ with a potentially endogenous $x_{it}$ and introducing instruments.
\end{example}

\subsection{Fixed Effect Estimator}
\label{sec:FEestimator}

In the following we write $\theta^0(\tau)$ and $\alpha^0_i(\tau)$ for the true parameters that
satisfy the assumptions in Section~\ref{sec:Assumptions}, to distinguish them from some generic parameter values
$\theta(\tau)$ and $\alpha_i(\tau)$. We furthermore consider only one quantile of interest
 $\tau_0 \in (0,1)$ in the following sections and we write $\theta^0 = \theta^0(\tau_0)$
 and  $\alpha^0_i = \alpha^0_i(\tau_0)$.
We write $A = (A_1,\ldots,A_N)$ for the collection of all individual characteristics.
All of the analysis in this section is conditional on $A$,
and no assumptions are made on the distribution of $A_i$ or on how $A_i$ is correlated with $z_{it}$.
For our inference procedure we follow a fixed effect approach in the sense that
we ``profile out'' $\alpha^0_i$.
The resulting fixed effect quantile regression estimator for $\theta$ and the
profile objective function read\footnote{%
Here, we write $\in$ instead of $=$ in the definition of $\widehat \theta^{\rm \, FE}$
since the minimizer of the quantile objective function
may not be unique, in which case we assume that $\widehat \theta^{\rm \, FE}$ is somehow
arbitrarily chosen from the set of minimizers.
Additional conditions on the distribution of $z_{it}$ could be imposed
to guarantee that the set
$\argmin_{\theta} {\cal L}_{NT}(\theta)$
is a singleton with probability one, but this is not required for any of the following
results to hold, because even if the set of minimizers of ${\cal L}_{NT}(\theta)$
is not unique it will shrink at a rate that is fast enough to have
no influence on our asymptotic results for $\widehat \theta^{\rm \, FE}$ below.
}
\begin{align*}
    \widehat \theta^{\rm \, FE}  &\in
     \argmin_{\theta \in \mathbbm{R}^{\dim z}} {\cal L}_{NT}(\theta) ,
    &
    {\cal L}_{NT}(\theta)  &= \frac{1}{N} \sum_{i=1}^N L_{i,T}(\theta) ,
    &
    L_{i,T}(\theta) &= \min_{\alpha_{i} \in \mathbbm{R}} \frac{1}{T} \sum_{t=1}^T \ell_{it}(\theta,\alpha_i) ,
\end{align*}
where
$\ell_{it}(\theta,\alpha_i) = \rho_{\tau_0}( y_{it} - z'_{it} \, \theta - \alpha_i )$
and $\rho_{\tau_0}(u)=[\tau_0 - 1(u<0)] u$,
which is the usual objective function used for the quantile regression estimator, see e.g.~\cite{Koenker2005}.

The quantile regression objective function is convex, which implies that also
the profile objective function ${\cal L}_{NT}(\theta)$ is convex.
Using
the results on minimizers of convex objective functions in \cite{HjortPollard1993},
based on the ``convexity lemma'' of \cite{Pollard1991},
one can therefore derive the asymptotic behavior of $\widehat \theta^{\rm \, FE}$ from a pointwise quadratic
approximation of  ${\cal L}_{NT}(\theta)$ around $\theta^0$. One thus obtains that\footnote{
More formally, under the conditions of Theorem~\ref{th:ExAsy}, we have
\begin{align*}
          \sqrt{NT} \left( \widehat \theta^{\rm \, FE} - \theta^0 \right)
            &= - H^{-1} \left( \sqrt{NT}\,
            \frac{d {\cal L}_{NT} ( \theta^0) }{d \theta} \right) + \smallO_P(1) .
\end{align*}
Note that the Hessian matrix $H$ is defined in terms of the expected objective function, because
${\cal L}_{NT}(\theta)$ itself is locally linear. The quadratic approximation of ${\cal L}_{NT}(\theta)$ which
justifies \eqref{ApproxExFE} is therefore not just a standard Taylor expansion, but
instead requires showing that
the remainder $S_{NT}(\theta) := {\cal L}_{NT}(\theta) - {\cal L}_{NT}(\theta^0)
              - \frac{d {\cal L}_{NT}( \theta^0) }{d \theta'} (\theta - \theta^0)$
admits a uniform quadratic approximation $S_{NT}(\theta^0 + (NT)^{-1/2} \phi) = \frac{1}{2NT}\phi'H\phi + \smallO_P((NT)^{-1})$
for bounded $\phi$. When $N/T$ diverges, the proof applies the quadratic approximation
around $\theta^0-H^{-1}B_{NT}/T$, see Appendix~\ref{app:Exog} for details.
}
\begin{align}
   \widehat \theta^{\rm \, FE} - \theta^0  
            &= - H^{-1} \,
            \frac{d {\cal L}_{NT} ( \theta^0) }{d \theta}  + \text{smaller order terms}    ,
       \label{ApproxExFE}
\end{align}
where $H =  \plim_{N,T \rightarrow \infty}
        \left. \frac{d^2}{d \theta d \theta'} \right|_{\theta=\theta^0}\allowbreak
    \mathbb{E}\left[ \left. {\cal L}_{NT}(\theta) \right| A   \right]$.
The asymptotic linearization of the estimator in equation \eqref{ApproxExFE} has exactly the same structure as for
 a cross-sectional quantile regression estimator
without fixed effects, including the subtlety that $H$ is the Hessian
of the expected (rather than sample) objective function.
The only difference is that we work with the profiled objective ${\cal L}_{NT}(\theta)$,
but the linearization step itself is identical to the cross-sectional case.

In Section~\ref{sec:MomentCondition} below we derive large-$T$ approximations
of the bias, variance, and Hessian of the profile score
$\frac{d L_{i,T}(\theta^0)}{d\theta}$.
Specifically, we are going to show that
 \begin{align}
      \mathbb{E}\left[ \left.  \frac{ d L_{i,T}(\theta^0) }{d \theta}  \, \right|  A_i   \right]
           &=   \frac{B_i }{T} +  {\cal O}_P \left( T^{-2} \right) ,
   \nonumber    \\
     {\rm Var} \left[ \left.   \frac{ d L_{i,T}(\theta^0) }{d \theta}  \, \right|  A_i   \right]
         &=  \frac{V_{0,i}}{T}   +  {\cal O}_P \left( T^{-2} \right),
 \nonumber     \\
   \left. \frac{d^2}{d \theta d \theta'} \right|_{\theta=\theta^0}\allowbreak
          \mathbb{E} \left[ \left.  L_{i,T}(\theta)  \right|  A_i
           \right]
        &= H_i     + \smallO_P(1)  .
    \label{LargeTapprox}
 \end{align}
Here $B_i$, $V_{0,i}$, and $H_i$ are quantities that depend on $A_i$ but not on $T$.
Their explicit formulas are derived in Section~\ref{sec:MomentCondition},
see equations~\eqref{DefB}, \eqref{DefV0}, and~\eqref{DefH}.
The subscript zero on $V_{0,i}$ will be useful later to distinguish
it from the variance of alternative score functions.
Combining these approximations with \eqref{ApproxExFE} we obtain the following theorem.

\begin{theorem}
   \label{th:ExAsy}
      Let the Assumptions~\ref{ass:QR}, \ref{ass:CI}, \ref{ass:RS} and \ref{ass:RC} be satisfied.
     Consider  $N,T \rightarrow \infty$ with $N/T^3 \rightarrow 0$.
    Define
    $B_{NT} :=
       \frac{1}{N}
           \sum_{i=1}^N  B_i$,
     and suppose that non-random limits
    $V^{\rm FE} = \plim_{N \rightarrow \infty}
       \frac{1}{N}
           \sum_{i=1}^N  V_{0,i}$, and
    $H = \plim_{N \rightarrow \infty}
       \frac{1}{N}
           \sum_{i=1}^N  H_i$
     exist,\footnote{The quantities $H_i$ and $V_{0,i}$ depend on $A_i$ but not on $T$
(see equations~\eqref{DefH} and~\eqref{DefV0}),
so the probability limits here are over the cross-sectional
randomness in $A_i$ and do not require $T \to \infty$.
The definition of $H$ as $\plim_{N \to \infty} \frac{1}{N}\sum_i H_i$
is equivalent to the one below equation~\eqref{ApproxExFE},
where $H = \plim_{N,T \to \infty} \frac{d^2}{d\theta d\theta'}\mathbb{E}[{\cal L}_{NT}(\theta)|A]\big|_{\theta^0}$,
because the Hessian of the expected objective converges to $\frac{1}{N}\sum_i H_i$
as $T \to \infty$ by Theorem~\ref{th:VarJac}.} and that $H>0$ and $V^{\rm FE} > 0$.
      We then have
   \begin{align*}
       \sqrt{NT} \left(  \widehat \theta^{\rm \, FE} - \theta^0  + \frac{1}{T} H^{-1} B_{NT} \right)
       \; \Rightarrow \;
       {\cal N}\left( 0, \,
       H^{-1} \, V^{\rm FE} \, H^{-1} \right) .
   \end{align*}
\end{theorem}

Consistent estimators of $H$ and $V^{\rm FE}$ needed for inference
are provided in Appendix~\ref{app:VarEstimation}.

The theorem shows that the FE-QR estimator $\widehat \theta^{\, \rm FE}$ has a leading bias of order  $1/T$,
which to our knowledge was not known in the literature previously.
The asymptotic distribution of $\widehat \theta$ obtained here, including the asymptotic bias,
is the same as the one found in \cite{GalvaoKato2016} for the smoothed
FE-QR estimator, if their result is restricted to the case of static panel models discussed here (they also
consider dynamic models). This shows that their smoothing procedure
 (under their bandwidth conditions) does not change the asymptotic bias.
 This, however, was unknown previously, that is, one cannot conclude the asymptotic distribution of
the FE-QR estimator from the known result for the smoothed version.

\section[Profile Score]{Profile Score for $\theta(\tau)$}
\label{sec:MomentCondition}

The linearization~\eqref{ApproxExFE} shows that the asymptotic behavior of the FE estimator
is governed by the profile score 
$$\frac{d {\cal L}_{NT}(\theta^0)}{d\theta}
= \frac{1}{N}\sum_{i=1}^N \frac{d L_{i,T}(\theta^0)}{d\theta} .
$$
By Assumption~\ref{ass:RS} we have independence across $i$,
so understanding the asymptotic distribution of
$\widehat\theta^{\rm FE}$ reduces to characterizing the bias and variance
of the individual profile score $d L_{i,T}(\theta^0) / d\theta$.
This is the focus of the present section.
In the following,
we write $\lfloor \tau T \rfloor$ for the integer part of $\tau T$,
$\langle \tau T \rangle =  \tau T - \lfloor  \tau T \rfloor \in [0,1)$ for its fractional part,
and $\lceil \tau T \rceil$ for the smallest integer greater than or equal to $\tau T$.

Recall from Section~\ref{sec:FEestimator} that
$L_{i,T}(\theta) = \min_{\alpha_i} \frac{1}{T}\sum_{t=1}^T \rho_{\tau_0}(u_{it}(\theta) - \alpha_i)$,
where $u_{it}(\theta) = y_{it} - z_{it}' \, \theta$.
For a given $\theta$, the minimizer $\widehat\alpha_i(\theta)$ is the $\tau_0$-sample quantile
of  $\{u_{it}(\theta)\}_{t=1}^T$.
Denoting the order statistics of this sample by
$u_{i(1)}(\theta) \leq \cdots \leq u_{i(T)}(\theta)$,
the minimizer is
$\widehat\alpha_i(\theta) = u_{i(\lfloor\tau_0 T\rfloor+1)}(\theta)$.
By the subgradient calculus of the quantile regression objective,
the profile score is
\begin{align}
    \frac{d L_{i,T}(\theta)}{d\theta}
    &= \frac{1}{T} \sum_{t=1}^T z_{it} \, \Psi_{it}(\theta) ,
    \label{ScoreAsPsi}
\end{align}
where
\begin{align}
    \Psi_{it}(\theta)
        &:= - \tau_0 + \mathbbm{1}[ u_{it}(\theta) < u_{i (\lfloor \tau_0 T \rfloor + 1)}(\theta)  ]
           +  \langle \tau_0 T \rangle
          \,   \mathbbm{1}[u_{it}(\theta) = u_{i (\lfloor \tau_0 T \rfloor + 1)}(\theta)] .
    \label{DefPsi}
\end{align}
To understand this expression, note that for most observations $\Psi_{it}(\theta)$
is simply $\mathbbm{1}(u_{it}(\theta) < \widehat\alpha_i(\theta)) - \tau_0$,
which equals $1-\tau_0$ when $u_{it}$ falls below the estimated
fixed effect and $-\tau_0$ otherwise.
The profile score~\eqref{ScoreAsPsi} is therefore a weighted average of the regressors $z_{it}$,
where the weights depend on whether the corresponding residual $u_{it}(\theta)$
falls above or below the estimated fixed effect.

The representation~\eqref{ScoreAsPsi} can be rewritten in a way
that turns out to be more tractable for analyzing the bias.
Sort the $T$ pairs $[u_{it}(\theta), z_{it}]$ in increasing order of their
first components\footnote{%
For any non-random $\theta$ and continuously distributed $u_{it}(\theta)$
we almost surely have $u_{it}(\theta) \neq u_{is}(\theta)$ for all $t \neq s$,
so ties are not an issue.}
and denote the resulting ordered pairs by
$[u_{i(1)}(\theta), z_{i[1]}(\theta)]$, \ldots, $[u_{i(T)}(\theta), z_{i[T]}(\theta)]$.
The $z_{i[q]}(\theta)$ are the regressors sorted by residual rank,
known as the \emph{induced order statistics} or \emph{concomitants} \citep{DavidNagaraja2003}.
One can then verify that~\eqref{ScoreAsPsi} can be rewritten as in the following lemma.

\begin{lemma}
      \label{lemma:FEweights}
           For all values $\theta$ where $L_{i,T}(\theta)$ is differentiable
           (which includes $\theta=\theta^0$ with probability one),
        \begin{align}
                \frac{ d L_{i,T}(\theta) }{d \theta}
                      &= \frac{1}{T} \sum_{q=1}^T  w^{\rm FE}_{T,q} \, z_{i[q]}(\theta)     ,
            \label{ScoreAsWeightedOS}
         \end{align}
where, assuming $\tau_0 T \geq 1$ and $(1-\tau_0)T \geq 1$,
the fixed effect weight sequence
$w^{\rm FE}_T=(w^{\rm FE}_{T,1},\ldots,w^{\rm FE}_{T,T})$ is given by
     \begin{align*}
          w_{T,q}^{\rm FE} &:= \left\{
          \begin{array}{rl}
               1-\tau_0, & \text{for $q \leq \lfloor \tau_0 T \rfloor$,}
              \\
             \langle \tau_0 T \rangle  -\tau_0 , & \text{for $q = \lfloor \tau_0 T \rfloor + 1$,}
             \\
              -\tau_0 , & \text{for $q \geq \lfloor \tau_0 T \rfloor +2$.}
          \end{array}
          \right.
     \end{align*}
\end{lemma}

\noindent
The weights $w^{\rm FE}_{T,q}$ are positive ($1-\tau_0$) for ranks below the quantile
and negative ($-\tau_0$) for ranks above it, reflecting the structure of $\Psi_{it}$.
The representation~\eqref{ScoreAsWeightedOS} is the key to analyzing the bias,
because it expresses the profile score as a weighted sum of the regressors sorted by residual rank,
making it possible to exploit the distributional properties of order statistics.
We will analyze the bias and variance not just for the FE weights but for
general weight sequences $w_T = (w_{T,1},\ldots,w_{T,T})$ satisfying
$\sum_{q=1}^T w_{T,q} = 0$, because in Section~\ref{sec:BiasCorrection}
we will use alternative weights to eliminate the leading bias.

\subsection[Finite-T Bias of the Profile Score]{Finite $T$ Bias of the Profile Score}
\label{sec:finiteTbias}

We now characterize the expected value of
$\frac{1}{T}\sum_{q=1}^T w_{T,q}\, z_{i[q]}(\theta^0)$
for a general weight sequence $w_T$.
Under Assumption~\ref{ass:CI}, the pair $(u_{it}, z_{it})$ is iid over $t$
conditional on $A_i$.
Denote the marginal cdf and pdf of $u_{it}$
(conditional on $A_i$) by
$F(u | A_i) := \Pr(u_{it} \leq u \,|\, A_i)$ and $f(u | A_i) := F'(u | A_i)$,
the conditional cdf and pdf given $z_{it}$ by
$F(u|z_{it},A_i) := \Pr(u_{it} \leq u | z_{it}, A_i)$
and $f(u|z_{it},A_i) := F'(u|z_{it},A_i)$,
and write $\alpha^0_i = F^{-1}(\tau_0|A_i)$.
It will be useful to reparameterize the conditional cdf
by the quantile level $\tau$ rather than the value $u$.
Define
\begin{align}
    G(\tau | z_{it}, A_i) := F\!\left( F^{-1}(\tau | A_i) \,\big|\, z_{it}, A_i \right) .
    \label{DefG}
\end{align}
The function $G(\tau|z_{it},A_i)$ is the conditional probability
(given $z_{it}$ and $A_i$) that $u_{it}$ falls below the marginal $\tau$-quantile.
If $z_{it}$ is uninformative about $u_{it}$ conditional on $A_i$,
then $G(\tau|z_{it},A_i) = \tau$ for all $\tau$.
Deviations of $G$ from the identity reflect dependence between $z_{it}$ and $u_{it}$.
Assumption~\ref{ass:QR} at $\tau = \tau_0$ implies
$G(\tau_0 | z_{it}, A_i) = \tau_0$
almost surely.
Using $G$, we define the function that governs the bias of the profile score:
\begin{align}
    g_i(\tau) := \mathbb{E}\left\{ \left. z_{it} \, \left[ G\!\left( \tau \,\big|\, z_{it},A_i  \right)
  - \tau_0 \right] \right| A_i  \right\}.
  \label{Defg}
\end{align}
Note that $g_i$ depends on $A_i$ (hence the subscript $i$) but not on $t$ (by iid).
Since $G(\tau_0|z_{it},A_i) = \tau_0$, we have $g_i(\tau_0) = 0$.
Finally, we also introduce the beta distribution ${\cal B}(a,b)$ with integer parameters $a,b > 0$,
whose pdf reads
$f_{{\cal B}}(\tau, a,b)
      = \frac{(a+b-1)!}{(a-1)! (b-1)!}
        \; \tau^{a-1} \; (1-\tau)^{b-1}$.

\begin{theorem}[\bf Finite $T$ Characterization of Profile Score Bias]
   \label{th:finiteTbias}
    Let $i \in {\bf N}$.
     Let Assumption~\ref{ass:CI} hold, assume $\mathbb E[\|z_{it}\|\mid A_i]<\infty$,
    let $F(u|z_{it},A_i)$
    be continuous in $u \in \mathbbm{R}$ for all
    possible values of $z_{it}$,
   and let $ \tau_0    w_{T,1}
             + (1- \tau_0) w_{T,T} = 0$.\footnote{%
     This boundary condition is satisfied by both the FE weights defined earlier and the MJ/JK weights defined below
     (as can be verified from their definitions), and ensures that boundary contributions
     cancel in the integration-by-parts derivation below.}
    Then
    \begin{align*}
        \mathbb{E}\left[  \left. \frac{1}{T} \sum_{q=1}^T w_{T,q}  \; z_{i[q]}(\theta^0)
         \right|    A_i  \right]  &=
               \int_{0}^1  g_i(\tau)   \,  h_T(\tau,w_T) \, d\tau ,
    \end{align*}
    where
    $ h_T(\tau,w_T) := \sum_{q=1}^{T-1}  (w_{T,q} - w_{T,q+1}) f_{{\cal B}}(\tau, q,T-q) $.
\end{theorem}

This result is exact for any finite $T$ and is the foundation of our bias analysis.
The full proof is in Appendix~\ref{app:MomentCondition}.
The key step is computing the expected value of the $q$'th concomitant
$\mathbb{E}[z_{i[q]}(\theta^0)|A_i]$.
Since $z_{i[q]}$ is the regressor paired with the observation whose residual has rank $q$,
its expectation involves averaging over the distribution of that rank.
Define $\xi_{it} := F(u_{it}(\theta^0)|A_i)$, which guarantees $\xi_{it} \sim \text{Uniform}(0,1)$
conditional on $A_i$.
Also define the $q$'th order statistic $\xi_{i(q)}$ of $\xi_{i1},\ldots,\xi_{iT}$, which has
pdf $f_{{\cal B}}(\tau, q, T+1-q)$.
Writing $z_{i[q]} = \sum_{t=1}^T z_{it}\,\mathbbm{1}(\xi_{it} = \xi_{i(q)})$
and using the conditional iid assumption over $t$:

\begin{align*}
    \mathbb{E}\!\left[\left. z_{i[q]}(\theta^0) \right| A_i \right]
    &= T \cdot \mathbb{E}\!\left[\left. z_{it}\,\mathbbm{1}(\xi_{it} = \xi_{i(q)}) \right| A_i \right]
    \\
    &= T \cdot \mathbb{E}\!\Bigg[\left. z_{it}\,
       \mathbb{E}\!\Big[ \mathbbm{1}(\xi_{it} = \xi_{i(q)}) \,\Big|\, \xi_{it}, \, A_i \Big]
       \,\right| A_i \Bigg]
    \\
    &= T \cdot \mathbb{E}\!\Bigg[ z_{it}\,
       \underbrace{{\textstyle\binom{T-1}{q-1}} \xi_{it}^{q-1}(1-\xi_{it})^{T-q}}_{\Pr(\text{$\xi_{it}$ has rank $q$} \,|\, \xi_{it}, \, A_i)}
       \,\Bigg|\, A_i \Bigg] \\
    &= T \int_0^1
       \mathbb{E}\!\left[\left. z_{it} \,\right|\, \xi_{it} = \tau, \, A_i \right]
       \, {\textstyle\binom{T-1}{q-1}} \tau^{q-1}(1-\tau)^{T-q}\, d\tau
    \\
    &= \int_0^1
       \underbrace{\mathbb{E}\!\left[\left. z_{it} \,\right|\, \xi_{it} = \tau, \, A_i \right]}_{
       = \, g_i'(\tau)}
       \; f_{{\cal B}}(\tau, q, T\!+\!1\!-\!q) \, d\tau ,
\end{align*}
where the first line uses the iid assumption over $t$,
the second line uses the law of iterated expectations conditioning on $\xi_{it}$,
the third line uses that, conditional on $\xi_{it}$ and $A_i$,
the event $\{\xi_{it} = \xi_{i(q)}\}$ is the event that $\xi_{it}$ has rank $q$
among $T$ iid $\text{Uniform}(0,1)$ draws, which has probability
$\binom{T-1}{q-1}\xi_{it}^{q-1}(1-\xi_{it})^{T-q}$,
the fourth line uses the law of iterated expectations conditioning on $\xi_{it} = \tau$
with $\xi_{it} \sim \text{Uniform}(0,1)$,
and the last line uses $T\binom{T-1}{q-1}\tau^{q-1}(1-\tau)^{T-q} = f_{{\cal B}}(\tau, q, T\!+\!1\!-\!q)$.

Thus $\mathbb{E}[z_{i[q]}|A_i]$ is simply the expected regressor
$g_i'(\tau) = \mathbb{E}[z_{it}|u_{it} = F^{-1}(\tau|A_i), A_i]$
at the $\tau$-quantile,
averaged over the distribution of the $q$'th uniform order statistic.
The theorem then follows by forming the weighted sum
$\frac{1}{T}\sum_q w_{T,q}\,\mathbb{E}[z_{i[q]}|A_i]$
and using integration by parts to convert $g_i'$ to $g_i$.\footnote{Specifically, $\frac{1}{T}\sum_q w_{T,q} \int_0^1 g_i'(\tau)\, f_{{\cal B}}(\tau,q,T\!+\!1\!-\!q)\, d\tau
= \frac{1}{T}\sum_q w_{T,q} \int_0^1 g_i(\tau)\, [-f_{{\cal B}}'(\tau,q,T\!+\!1\!-\!q)]\, d\tau
+ \text{boundary terms}$,
where the boundary terms vanish under the condition $\tau_0 w_{T,1} + (1-\tau_0) w_{T,T} = 0$,
and the derivatives $-f_{{\cal B}}'(\tau,q,T\!+\!1\!-\!q)$ combine into the kernel $h_T(\tau,w_T)$
involving $f_{{\cal B}}(\tau,q,T\!-\!q)$.}
An equivalent leave-one-out proof is given in Appendix~\ref{app:MomentCondition}.

\Needspace{5\baselineskip}
\begin{remarks}~
\begin{itemize}

\item[(i)]
The kernel $h_T(\tau,w_T) = \sum_{q=1}^{T-1} (w_{T,q} - w_{T,q+1}) f_{{\cal B}}(\tau,q,T-q)$
acts as a weighting function on $[0,1]$.
For the fixed effect weights, $h_T(\tau,w^{\rm FE}_T)$ is a weighted combination of
$f_{{\cal B}}(\tau,\lfloor\tau_0 T\rfloor, T-\lfloor\tau_0 T\rfloor)$ and
$f_{{\cal B}}(\tau, \lfloor\tau_0 T\rfloor+1, T-\lfloor\tau_0 T\rfloor-1)$,
both of which are beta densities with modes near $\tau_0$.
As $T$ grows, these beta densities become increasingly concentrated around $\tau_0$,
so that $h_T(\tau,w^{\rm FE}_T)$ behaves like a kernel that places all its mass at $\tau_0$ in the limit.

\item[(ii)]
The function $g_i(\tau)$ measures how $G(\tau|z_{it},A_i)$ deviates from $\tau_0$,
weighted by $z_{it}$.
Since $g_i(\tau_0)=0$ and $h_T(\tau,w_T)$ concentrates around $\tau_0$ as $T$ grows
(see Remark~(i)),
the bias $\int_0^1 g_i(\tau)\, h_T(\tau,w_T)\,d\tau$ is small for large $T$,
converging to $g_i(\tau_0) = 0$.

\item[(iii)]
   Theorem~\ref{th:finiteTbias} uses conditional identical distributions to obtain the beta representation.
   Under conditional independence without identical distributions, the expected score instead has the rank-probability representation in Appendix~\ref{sec:INID}.
\end{itemize}
\end{remarks}

\subsection[Large-T Bias of the Profile Score]{Large $T$ Bias of the Profile Score}
\label{sec:largeTbias}

Since $g_i(\tau_0) = 0$ and $h_T$ concentrates around $\tau_0$ as $T$ grows,
the exact bias $\int_0^1 g_i(\tau)\, h_T(\tau,w_T)\,d\tau$ from Theorem~\ref{th:finiteTbias}
can be approximated for large $T$ by Taylor-expanding $g_i$ around $\tau_0$.
Using $g_i(\tau_0) = 0$, the expansion reads
\begin{align*}
\int_0^1 g_i(\tau)\, h_T(\tau,w_T)\,d\tau
&= g_i'(\tau_0) \int_0^1 (\tau-\tau_0)\, h_T(\tau,w_T)\, d\tau
\\ & \quad
+ \frac{g_i''(\tau_0)}{2} \int_0^1 (\tau-\tau_0)^2 h_T(\tau,w_T)\, d\tau
\; + \; \cdots
\end{align*}
The integrals $\int_0^1 (\tau-\tau_0)^r h_T\, d\tau$ are moments of $h_T$ centered at $\tau_0$.
Using $\int_0^1 \tau\, f_{{\cal B}}(\tau,q,T-q)\,d\tau = q/T$
(the mean of the ${\cal B}(q,T-q)$ distribution),
the first moment $\int_0^1 (\tau-\tau_0)\, h_T\, d\tau
= \sum_{q=1}^{T-1}(w_{T,q}-w_{T,q+1})(q/T - \tau_0)$
vanishes for the FE weights
(the two nonzero first differences at $q = \lfloor\tau_0 T\rfloor$ and $q = \lfloor\tau_0 T\rfloor + 1$ contribute terms that cancel exactly),
and more generally equals $M_1(w_T,\tau_0)/\sqrt{T}$
where $M_1$ is defined below.
Under the conditions of Theorem~\ref{th:largeTbias}, this term does not contribute to the leading bias.
Similarly, the variance of ${\cal B}(q,T-q)$ is $q(T-q)/[T^2(T+1)]$,
which is of order $1/T$ for $q$ near $\tau_0 T$.
The second moment of $h_T$ is therefore of order $1/T$,
making the $g_i''(\tau_0)$ term of order $1/T$.
This is the leading contribution to the bias.
We write $f_{i,T}={\cal O}_{\rm u}(g_T)$ if $|f_{i,T}|\leq M|g_T|$ for all $T\geq T_0$, uniformly over $i$, and $f_{i,T}=\smallO_{\rm u}(g_T)$ if $\sup_i|f_{i,T}|/|g_T|\to0$.
For $r \geq 1$ define the moments of the first-differenced weights:
\begin{align*}
    M_r(w_T,\tau_0) :=   \sum_{q=1}^{T-1}
         (w_{T,q} - w_{T,q+1})  \left[ \sqrt{T}  \left(q/T - \tau_0 \right) \right]^r .
\end{align*}

\begin{theorem}[\bf Large $T$ Characterization of Profile Score Bias]
   \label{th:largeTbias}
      Let the Assumptions~\ref{ass:QR}, \ref{ass:CI}, and \ref{ass:RC} hold,
      and consider $T \rightarrow \infty$.
    Assume that $w_{T,1}=  1- \tau_0$,  $w_{T,T} = -\tau_0$,
    $ \sum_{q=1}^{T} w_{T,q} = 0$,
   $\frac{1}{T} \sum_{q=1}^{T-1}    |w_{T,q} - w_{T,q+1}|  = {\cal O}(1)$,
   and
$\frac{1}{T^k} \sum_{q=1}^{T-1}    |w_{T,q} - w_{T,q+1}|  (q - \tau_0 T)^{2k} = {\cal O}(1)$
for $k \in \{1,2,3\}$.
 Then we have
\begin{align*}
     \mathbb{E}\left[ \left.  \frac{1}{T} \sum_{q=1}^T w_{T,q}  \; z_{i[q]}(\theta^0)  \, \right|  A_i
           \right]
        &=
           \frac{(1-\tau_0) \tau_0 + M_2(w_T,\tau_0)}{2T} \;
             g''_i(\tau_0)
       \\ & \quad
         +  \frac{M_3(w_T,\tau_0)}{6 T^{3/2} } \;
             g'''_i(\tau_0)
         +  {\cal O}_{\rm u}( T^{-2} )    ,
\end{align*}

\end{theorem}

The proof is given in Appendix~\ref{app:MomentCondition}.
Applying the theorem with the fixed effect weights
($M_2(w^{\rm FE}_T,\tau_0) = {\cal O}(T^{-1})$,
$M_3(w^{\rm FE}_T,\tau_0) = {\cal O}(T^{-3/2})$) gives
\begin{align}
     \mathbb{E}\left[ \left.  \frac{d L_{i,T}(\theta^0)}{d\theta}  \, \right|  A_i
           \right]
           = \frac{B_i}{T} + {\cal O}_{\rm u}(T^{-2}) ,
    \label{BiasResult}
\end{align}
where
\begin{align}
    B_i := \frac{(1-\tau_0)\tau_0}{2} \, g''_i(\tau_0) ,
    \label{DefB}
\end{align}
verifying the display~\eqref{LargeTapprox}.
The leading bias depends on the slope of the conditional expectation
$g_i'(\tau) = \mathbb{E}[z_{it}|u_{it}=F^{-1}(\tau|A_i),A_i]$
at $\tau = \tau_0$.
When $z_{it}$ is independent of $u_{it}$
conditional on $A_i$ (as happens when $\theta(\tau)$ does not depend on $\tau$),
$g_i'(\tau)$ is constant in $\tau$ and $B_i = 0$.

\subsection{Variance and Jacobian of the Profile Score}
\label{sec:VarJac}

To complete the characterization needed for the asymptotic distribution
in Theorem~\ref{th:ExAsy}, we also need large-$T$ approximations
of the variance and Hessian of the profile score.

To state the results we introduce
\begin{align*}
   \lambda_i &:= \mathbb{E}\!\left[\left.
   z_{it} \frac{f(\alpha^0_i|z_{it},A_i)}{f(\alpha^0_i|A_i)}
   \right| A_i\right]
   = \mathbb{E}\left[ z_{it} \,\big|\, u_{it} = \alpha^0_i, \, A_i \right] ,
\end{align*}
where the second equality follows from Bayes' theorem.
The vector $\lambda_i$ is the conditional expectation of the regressors
given that the error $u_{it}$ equals its $\tau_0$-quantile.
Using $\lambda_i$, define the leading-order variance and Hessian of the profile score as
\begin{align}
    V_{0,i}  &:=    \tau_0 (1-\tau_0) \,
          \mathbb{E}\!\left[\left. (z_{it} - \lambda_i)(z_{it} - \lambda_i)'
               \,\right|\, A_i \right]    ,
    \label{DefV0}
    \\
     H_i
        & :=
         \mathbb{E}\!\left[\left.
       f(\alpha^0_i|z_{it},A_i) \, z_{it} (z_{it} - \lambda_i)'
       \right| A_i\right]  .
    \label{DefH}
\end{align}
Both $V_{0,i}$ and $H_i$ involve de-meaning of $z_{it}$ by $\lambda_i$
rather than by the unconditional mean $\mathbb{E}[z_{it}|A_i]$,
reflecting the profiling out of the fixed effect.
In addition, define
\begin{align*}
    V_{1,i} &:= {\rm Var}\left( z_{it} \,\big|\, u_{it} = \alpha^0_i, \, A_i \right)
    = \mathbb{E}\!\left[\left.
        (z_{it} - \lambda_i)(z_{it} - \lambda_i)' \frac{f(\alpha^0_i|z_{it},A_i)}{f(\alpha^0_i|A_i)}
        \right| A_i \right] ,
\end{align*}
which is the conditional variance of $z_{it}$ given that $u_{it}$ equals its $\tau_0$-quantile.

\begin{theorem}[\bf Large $T$ Characterization of Variance and Jacobian]
   \label{th:VarJac}
    Let the Assumptions of Theorem~\ref{th:largeTbias} hold.
  Then, as $T \rightarrow \infty$,
\begin{align*}
     {\rm Var} \left[ \left.  \frac{1}{T} \sum_{q=1}^T w_{T,q}  \; z_{i[q]}(\theta^0)  \, \right| A_i
           \right]      &= \frac{V_{0,i}}{T} +
            \frac{V_{1,i} }{T^2}   \sum_{q=1}^T \left( w_{T,q} - w^{\rm FE}_{T,q} \right)^2
          \\ &\quad +   \smallO_{\rm u}\!\left( T^{-1}+T^{-2}\sum_{q=1}^T(w_{T,q}-w^{\rm FE}_{T,q})^2 \right) ,
  \end{align*}
    and
    \begin{align*}
         \left. \frac{d}{d \theta} \right|_{\theta=\theta^0}\allowbreak
          \mathbb{E} \left[ \left.  \frac{1}{T} \sum_{q=1}^T w_{T,q}  \; z_{i[q]}(\theta)  \, \right|  A_i
           \right]
        &= H_i   + \smallO_{\rm u}(1)  .
    \end{align*}
\end{theorem}

The proof is given in Appendix~\ref{app:MomentCondition}.
For the fixed effect weights, $w_{T,q}=w^{\rm FE}_{T,q}$, so the $V_{1,i}$ term vanishes.
In this case,
${\rm Var}[\frac{dL_{i,T}(\theta^0)}{d\theta}|A_i]
=V_{0,i}/T+{\cal O}_{\rm u}(T^{-2})$.
This is what we anticipated in display~\eqref{LargeTapprox}.
The $V_{1,i}$ term in the variance formula of Theorem~\ref{th:VarJac} plays a central role in the bias-variance tradeoff analyzed in Section~\ref{sec:BiasCorrection}. For the FE weights, $w_{T,q} = w^{\rm FE}_{T,q}$, the $V_{1,i}$ term does not contribute, because $(w_{T,q} - w^{\rm FE}_{T,q})^2 = 0$. However, for alternative weight sequences that achieve bias correction, the deviation $w_{T,q} - w^{\rm FE}_{T,q}$ can be substantial, and the $V_{1,i}$ term can dominate the variance, as we will see for the delete-one Jackknife below.

\subsection[Finite-T Consistency for Constant Coefficients]{Finite $T$ consistency for $\theta(\tau) = \theta$ constant}
\label{sec:ThetaConstant}

We now consider the important special case where $\theta(\tau)$ does not depend on $\tau$.
Consider the model
$y_{it} = z'_{it} \, \theta^0 + \alpha_i(U_{it})$
with $U_{it} \sim U[0,1]$ independent of $z_{it}$ conditional on $A_i$,
and $\alpha_i(\tau)$ strictly increasing and sufficiently differentiable with $\alpha'_i(\tau_0) > 0$.
Here, $\theta(\tau) = \theta^0$ is constant and
$u_{it}(\theta^0) = \alpha_i(U_{it})$ is independent of $z_{it}$ conditional on $A_i$,
so $g_i'(\tau) = \mathbb{E}[z_{it}|A_i]$ is constant in $\tau$ and
$g_i(\tau) = (\tau - \tau_0)\,\mathbb{E}[z_{it}|A_i]$.
In this case the expected value of the profile score is exactly zero for any finite $T$,
\begin{align}
    \mathbb{E}\left[\left. \frac{d L_{i,T}(\theta^0)}{d\theta} \right| A_i\right] = 0 .
    \label{ScoreZeroMean}
\end{align}
To see this, note that by Theorem~\ref{th:finiteTbias} the expected score equals
$\int_0^1 g_i(\tau)\, h_T(\tau, w^{\rm FE}_T)\, d\tau$,
which with $g_i(\tau) = (\tau - \tau_0)\,\mathbb{E}[z_{it}|A_i]$ reduces to
$\mathbb{E}[z_{it}|A_i] \int_0^1 (\tau - \tau_0)\, h_T(\tau, w^{\rm FE}_T)\, d\tau$.
The integral vanishes because
$\int_0^1 (\tau - \tau_0)\, h_T(\tau, w^{\rm FE}_T)\, d\tau
= \sum_{q=1}^{T-1}(w^{\rm FE}_{T,q} - w^{\rm FE}_{T,q+1})(q/T - \tau_0) = 0$,
where the last equality follows from the explicit form of the FE weights
(the two nonzero first differences at $q = \lfloor\tau_0 T\rfloor$
and $q = \lfloor\tau_0 T\rfloor + 1$ contribute terms that cancel exactly).

Since the score has exactly zero mean for any finite $T$, the FE estimator is consistent even for fixed $T$ under the usual identification and regularity conditions.
For any fixed $T$, the profile objective ${\cal L}_{NT}(\theta) = \frac{1}{N}\sum_i L_{i,T}(\theta)$ is a sample average of iid convex terms across $i$, and the convexity-based linearization applies as $N \to \infty$ alone. Specifically, writing
$H_T :=
        \plim_{N \to \infty} \left. \frac{d^2}{d \theta d \theta'} \right|_{\theta=\theta^0}\allowbreak
    \mathbb{E}\left[ \left. {\cal L}_{NT}(\theta) \right| A   \right]$
for the fixed-$T$ Hessian,
\begin{align}
   \sqrt{N}\left(\widehat \theta^{\rm \, FE} - \theta^0\right)
            &= - H_T^{-1} \left(\sqrt{N}\,
            \frac{d {\cal L}_{NT} ( \theta^0) }{d \theta}\right)  + \smallO_P(1)
       \label{ApproxExFEfixedT}
\end{align}
holds for fixed $T$ as $N \to \infty$, analogous to~\eqref{ApproxExFE}.
Since the score $\frac{d {\cal L}_{NT}(\theta^0)}{d\theta} = \frac{1}{N}\sum_i \frac{d L_{i,T}(\theta^0)}{d\theta}$ has exactly zero mean by~\eqref{ScoreZeroMean},
the central limit theorem gives $\sqrt{N}\,\frac{d {\cal L}_{NT}(\theta^0)}{d\theta} \Rightarrow {\cal N}(0, V_{0,T}/T)$
where
\[
V_{0,T} := \left[\tau_0(1-\tau_0)-\frac{\langle\tau_0T\rangle(1-\langle\tau_0T\rangle)}{T}\right]
\mathbb{E}\!\left[{\rm Var}(z_{it}\mid A_i)\right],
\]
and therefore $\widehat\theta^{\rm FE}$ is $\sqrt{N}$-consistent for $\theta^0$ even when $T$ is fixed. In this special case there is no incidental parameter bias at all.

\section{Jackknife Bias Correction}
\label{sec:BiasCorrection}

We have shown in the preceding sections that the naive quantile fixed effects estimator has a leading bias of order $1/T$.
One approach to bias correction would be to estimate the bias $B_i$ analytically
and subtract it.
However, the formula $B_i = \frac{(1-\tau_0)\tau_0}{2}\,g''_i(\tau_0)$ involves
the second derivative of $g_i(\tau) = \mathbb{E}[z_{it}(G(\tau|z_{it},A_i) - \tau_0)|A_i]$,
which depends on the conditional distribution of $u_{it}$ given $z_{it}$ and $A_i$
in a complicated way.
Estimating $g''_i(\tau_0)$ nonparametrically for each individual $i$
would require estimating conditional densities and their derivatives,
which is difficult with only $T$ observations per individual.
This motivates considering the Jackknife,
which eliminates the leading bias without requiring explicit estimation of $B_i$.
In the large-$T$ panel literature on smooth nonlinear models,
the Jackknife is known to eliminate the leading bias while preserving $\sqrt{NT}$-consistency
\citep{HahnNewey2004, DhaeneJochmans2015}.
In the quantile regression setting, however, the non-smoothness of the objective function
leads to several surprising results that are not expected from the well-known results on smooth objective functions.
Before presenting the technical details, we summarize the main findings of this section.
\begin{itemize}
\item[(i)]
The standard delete-one Jackknife does \emph{not} eliminate the leading $1/T$ bias
in the quantile regression setting, unlike in smooth nonlinear panel models.

\item[(ii)]
The asymptotic variance of the estimator that results from standard delete-one Jackknife correction
can be of order $1/N$ instead of $1/(NT)$.
Thus, the naive delete-one Jackknife not only fails to achieve leading-order bias correction,
but can also increase the asymptotic variance significantly compared to the naive quantile fixed effects estimator.
We also present a modified version of the delete-one Jackknife that restores first-order bias correction,
but that minimal modification still gives only a $\sqrt{N}$-consistent (not $\sqrt{NT}$-consistent) estimator.

\item[(iii)]
The Jackknife averaging over all subsamples of size $S=T/2$ achieves $1/T$ bias correction
without asymptotic variance inflation.
However, given that we are considering a static model, there is no reason to restrict
to subpanels of size $S = T/2$ as in the standard split-panel approach.
Instead, any subpanel size $S \in \{2, 3, \ldots, T-1\}$ can be considered
(subject to $\tau_0 S \geq 1$ and $(1-\tau_0) S \geq 1$).
When doing so, we discover a bias-variance tradeoff
depending on the relative magnitude of $T-S$ and $T$.
When $T-S$ is of order $T$ (as for $S = T/2$),
there is no leading-order variance inflation.
When $T-S$ is small (as in the delete-one case $S = T-1$),
the variance inflation is substantial.
Our formal asymptotic result for this class of estimators
(Theorem~\ref{th:BCAsy}) is honest about this variance inflation,
which matters for finite-sample inference.
\end{itemize}

\subsection{Jackknife and Modified Jackknife Weights}
\label{sec:JKweights}

We now develop these results in detail.
The idea is to apply the Jackknife not to the estimator directly, but to the
profile score $\frac{d L_{i,T}(\theta)}{d\theta}$,
or equivalently to the weighted moment function
$\frac{1}{T}\sum_{q=1}^T w_{T,q}\, z_{i[q]}(\theta)$
from Lemma~\ref{lemma:FEweights}.
Recall from Section~\ref{sec:MomentCondition} that the FE score
$m^{\{1,\ldots,T\}}_{i,T} := \frac{1}{T}\sum_{q=1}^T w^{\rm FE}_{T,q}\, z_{i[q]}(\theta)$
has a bias of order $1/T$.
For a subsample size $S$ with $\tau_0 S \geq 1$ and $(1-\tau_0) S \geq 1$,
and a subset $\{t_1,\ldots,t_S\} \subset {\bf T}$ of time periods,
we can compute the same FE score using only those $S$ periods,
which we denote by $m^{\{t_1,\ldots,t_S\}}_{i,S}$.
This subsample score has a bias of order $1/S$.
The standard Jackknife with subsample size $S$ combines the full-sample score
with an average of all $\binom{T}{S}$ subsample scores, with coefficients
chosen so that a bias of the form $B/T$ (full sample) and $B/S$ (subsample)
would cancel exactly, that is,
\begin{align}
    m^{\rm JK}_{i}(S)
    := \frac{T}{T-S} m^{\{1,\ldots,T\}}_{i,T}
    - \frac{S}{T-S} \binom{T}{S}^{-1}
    \sum_{\{t_1,\ldots,t_S\} \in \binom{{\bf T}}{S}} m^{\{t_1,\ldots,t_S\}}_{i,S} .
    \label{DefJKweights}
\end{align}
The special case $S = T-1$ gives the delete-one Jackknife of \cite{HahnNewey2004},
while $S = \lfloor T/2 \rfloor$ uses the subsample size of the split-panel Jackknife of \cite{DhaeneJochmans2015}.

A key observation is that $m^{\rm JK}_i(S)$ can be written as
a weighted sum of induced order statistics,
$m^{\rm JK}_i(S) = \frac{1}{T}\sum_{q=1}^T w^{\rm JK}_{T,q}(S)\, z_{i[q]}(\theta^0)$,
with Jackknife weights $w^{\rm JK}_{T,q}(S)$ given by
\begin{align}
    w^{\rm JK}_{T,q}(S)
    := \frac{T}{T-S}
    \left[ w^{\rm FE}_{T,q}
    - \binom{T}{S}^{-1}
    \sum_{k=\max(1,q-T+S)}^{\min(q,S)}
    \binom{q-1}{k-1}\binom{T-q}{S-k} w^{\rm FE}_{S,k} \right] .
    \label{HowToComputeWJK}
\end{align}
This formula (derived by combinatorial arguments) expresses the Jackknife score
in the same weighted-order-statistic form as the FE score,
which allows us to apply the bias and variance results of
Theorems~\ref{th:largeTbias} and~\ref{th:VarJac} directly.

\begin{example*}[\bf $S=T-1$, Delete-One Jackknife Weight Sequence]
    Assume that  $\tau_0 T \geq 2$ and $(1-\tau_0) T \geq 2$.
    For $\langle \tau_0 T \rangle < \tau_0$ we define
     \begin{align*}
          w_{T,q}(a,b) &:= \left\{
          \begin{array}{rl}
               1-\tau_0, & \text{for $q \leq \lfloor \tau_0 T \rfloor-1$,}
              \\[3pt]
               1-\tau_0 + a, & \text{for $q = \lfloor \tau_0 T \rfloor$,}
              \\[3pt]
             \langle \tau_0 T \rangle  -\tau_0 - a , & \text{for $q = \lfloor \tau_0 T \rfloor + 1$,}
              \\[3pt]
              -\tau_0 , & \text{for $q \geq \lfloor \tau_0 T \rfloor +2$,}
          \end{array}
          \right.
     \end{align*}
     and  for $\langle \tau_0 T \rangle \geq \tau_0$ we define
     \begin{align*}
          w_{T,q}(a,b) &:= \left\{
          \begin{array}{rl}
               1-\tau_0, & \text{for $q \leq \lfloor \tau_0 T \rfloor$,}
              \\[3pt]
             \langle \tau_0 T \rangle  -\tau_0 + b , & \text{for $q = \lfloor \tau_0 T \rfloor + 1$,}
              \\[3pt]
              -\tau_0 - b, & \text{for $q = \lfloor \tau_0 T \rfloor +2$,}
              \\[3pt]
              -\tau_0 , & \text{for $q \geq \lfloor \tau_0 T \rfloor +3$.}
          \end{array}
          \right.
     \end{align*}
     Then, applying the delete-one Jackknife to the QR profile score (equation \eqref{DefJKweights} with $S=T-1$) gives
     $m^{\rm JK}_i(T-1) = \frac{1}{T}\sum_{q=1}^T w^{\rm JK}_{T,q}(T-1)\, z_{i[q]}(\theta^0)$, where the
      delete-one Jackknife weights
     are given by $w^{\rm JK}_{T,q}(T-1) = w_{T,q}(a,b)$
     with $a=(\tau_0- \langle \tau_0 T \rangle) [ T (1-\tau_0) +    \langle \tau_0 T \rangle]$
     and $b = (\langle \tau_0 T \rangle - \tau_0) [ T \tau_0 + 1 -    \langle \tau_0 T \rangle]$.
     This is simply the special case of \eqref{HowToComputeWJK} with $S=T-1$, but it is useful to discuss this case explicitly.
     Crucially, these delete-one weights do not satisfy $M_2(w_T,\tau_0) = - (1-\tau_0) \tau_0 + {\cal O}(T^{-1})$,
     and therefore are not first-order bias correcting according to Theorem~\ref{th:largeTbias}.
     The squared correction satisfies
     $\frac{1}{T^2}\sum_q(w^{\rm JK}_{T,q}(T-1)-w^{\rm FE}_{T,q})^2
       =2a^2/T^2$ in the first case and $2b^2/T^2$ in the second.
     This quantity depends on the fractional part $\langle\tau_0T\rangle$
     and can remain of order one.
     Thus the standard delete-one Jackknife can inflate the score variance
     to order one and need not preserve the $\sqrt{NT}$ convergence rate.
     When $\langle\tau_0T\rangle=\tau_0$, however, the correction vanishes
     and the standard delete-one weights equal the FE weights.
     This possible variance inflation does not occur in smooth nonlinear
     panel models \citep{HahnNewey2004}.
\end{example*}

\paragraph{Modified Jackknife weights.}
As illustrated by the delete-one example, the standard Jackknife weights
$w^{\rm JK}_{T,q}(S)$ do not in general satisfy the bias correction condition
$M_2(w_T,\tau_0) = -(1-\tau_0)\tau_0$.
We therefore introduce a modification, denoted $w^{\rm MJ}_{T,q}(S)$,
that is designed to satisfy this condition exactly.
For $S<T-1$ or $\tau_0S\notin\mathbb Z$, the modified Jackknife weights are defined by
\begin{align*}
    w^{\rm MJ}_{T}(S) :=  w^{\rm FE}_{T}  + \kappa(\tau_0,T,S) \, \left[ w^{\rm JK}_{T}(S) - w^{\rm FE}_{T} \right] ,
\end{align*}
where 
\begin{align}
\kappa(\tau_0,T,S) := \frac{ -\tau_0(1-\tau_0) - M_2(w_T^{\textup{FE}}, \tau_0) }{ M_2(w_T^{\textup{JK}}(S), \tau_0) - M_2(w_T^{\textup{FE}}, \tau_0) }
   \label{DefKappa}
\end{align}
is chosen such that $M_2(w^{\rm MJ}_T(S),\tau_0) = - \tau_0 (1-\tau_0)$ exactly.\footnote{
Recall the definition
$M_2(w_T,\tau_0) := T  \sum_{q=1}^{T-1}(w_{T,q}-w_{T,q+1}) (q/T-\tau_0)^2$,
which is linear in $w_T$. The bias-correction condition
$M_2(w^{\rm MJ}_T(S),\tau_0) = -\tau_0(1-\tau_0)$
is therefore a linear equation in $\kappa$. Under the stated restriction, its unique solution is \eqref{DefKappa}.
The denominator vanishes exactly when $S=T-1$ and $\tau_0S\in\mathbb Z$;
at these points the standard Jackknife weights equal the FE weights.
See Appendix~\ref{app:WeightRegularity} for the proof.
Using the closed-form expressions
$M_2(w^{\rm FE}_T,\tau_0) = \langle\tau_0 T\rangle\bigl(1-\langle\tau_0 T\rangle\bigr)/T$
and
\[
M_2(w^{\rm JK}_T(S),\tau_0)
= -\frac{S\,\tau_0(1-\tau_0)}{S+1}
+ \frac{\langle\tau_0 T\rangle\bigl(1-\langle\tau_0 T\rangle\bigr)}{T-S}
- \frac{(T+1)\,\langle\tau_0 S\rangle\bigl(1-\langle\tau_0 S\rangle\bigr)}{(T-S)(S+1)},
\]
this can be more explicitly written as
\[
\kappa(\tau_0,T,S)
= \frac{(T-S)(S+1)\,\bigl[\,T\,\tau_0(1-\tau_0) + \langle\tau_0 T\rangle\bigl(1-\langle\tau_0 T\rangle\bigr)\,\bigr]}{T(T-S)S\,\tau_0(1-\tau_0) \;-\; S(S+1)\,\langle\tau_0 T\rangle\bigl(1-\langle\tau_0 T\rangle\bigr) \;+\; T(T+1)\,\langle\tau_0 S\rangle\bigl(1-\langle\tau_0 S\rangle\bigr)}.
\]
}
For large values of $S$, the weights $w^{\rm MJ}_{T,q}(S)$ only differ from $w^{\rm JK}_{T,q}(S)$
for a few values of $q$ near $\tau_0 T$.

\begin{example*}[\bf $S=T-1$, Modified Delete-One Jackknife Weight Sequence]
     Assume $\tau_0T\geq2$ and $(1-\tau_0)T\geq2$, as in the preceding example.
     The following explicit definition also covers the points
     $\tau_0(T-1)\in\mathbb Z$ at which \eqref{DefKappa} is undefined.
     Using the notation $w_{T,q}(a,b)$ from the preceding example,
     the modified delete-one Jackknife weights
     are defined by $w^{\rm MJ}_{T,q}(T-1) := w_{T,q}(a,b)$, with
     $a=b=\frac{T \tau_0 (1-\tau_0)  + (1-\langle \tau_0 T \rangle) \langle \tau_0 T \rangle}{2}$.
     This particular choice of $a$ and $b$ guarantees that
     $M_2(w^{\rm MJ}_T(T-1),\tau_0) = - \tau_0 (1-\tau_0)$. One also finds
     $M_3(w^{\rm MJ}_T(T-1),\tau_0) = {\cal O}(T^{-1/2})$, and applying Theorem~\ref{th:largeTbias} gives
       $\mathbb{E}\left[ \left.  \frac{1}{T} \sum_{q=1}^T w^{\rm MJ}_{T,q}(T-1)  \; z_{i[q]}(\theta^0) \right|  A_i
           \right] = {\cal O}_{\rm u}(T^{-2})$, i.e.\ the modified
           delete-one Jackknife weight sequence is first-order bias correcting.
     However, the modified weights also inflate the variance,
     so the resulting estimator is still only $\sqrt{N}$-consistent.
\end{example*}

\begin{figure}[tb]
     \begin{minipage}{.3\textwidth}
        \includegraphics[width=\linewidth]{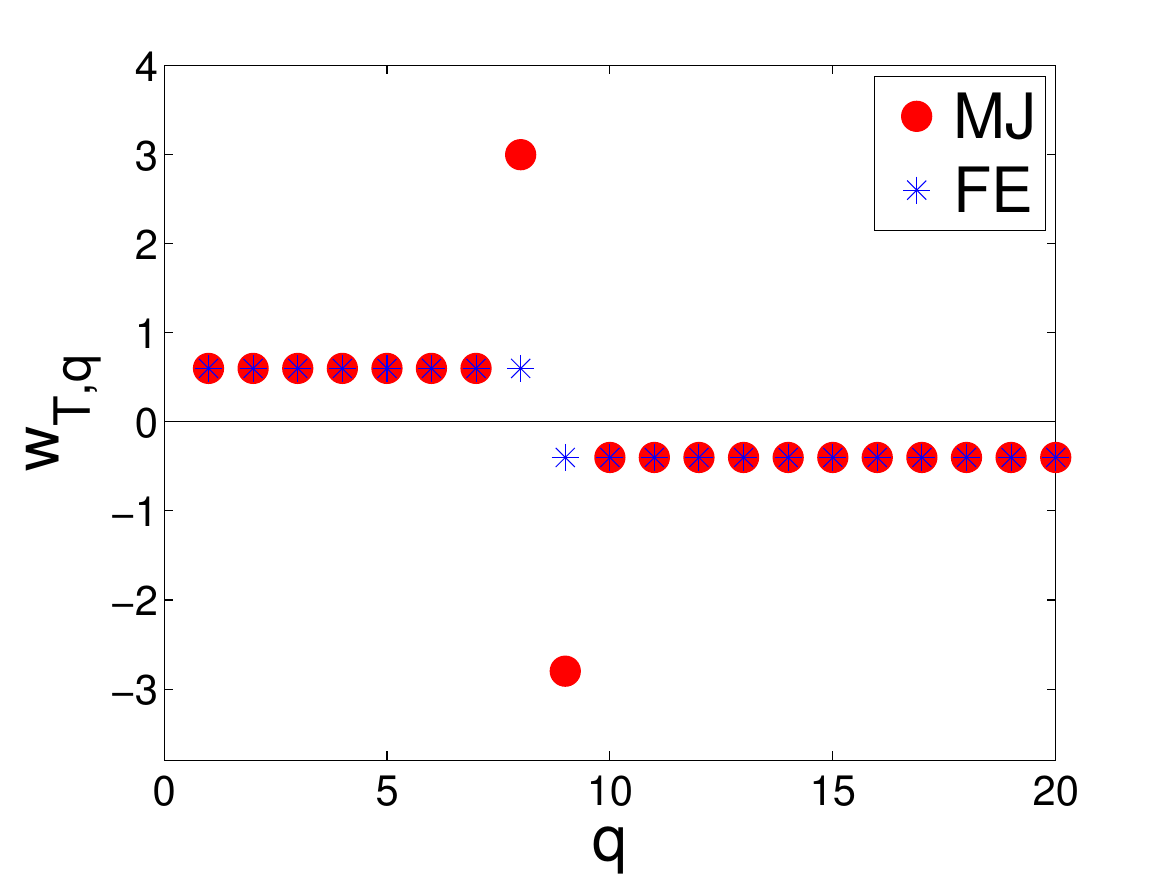}
     \end{minipage}
     \hspace{.02\textwidth}
     \begin{minipage}{.3\textwidth}
        \includegraphics[width=\linewidth]{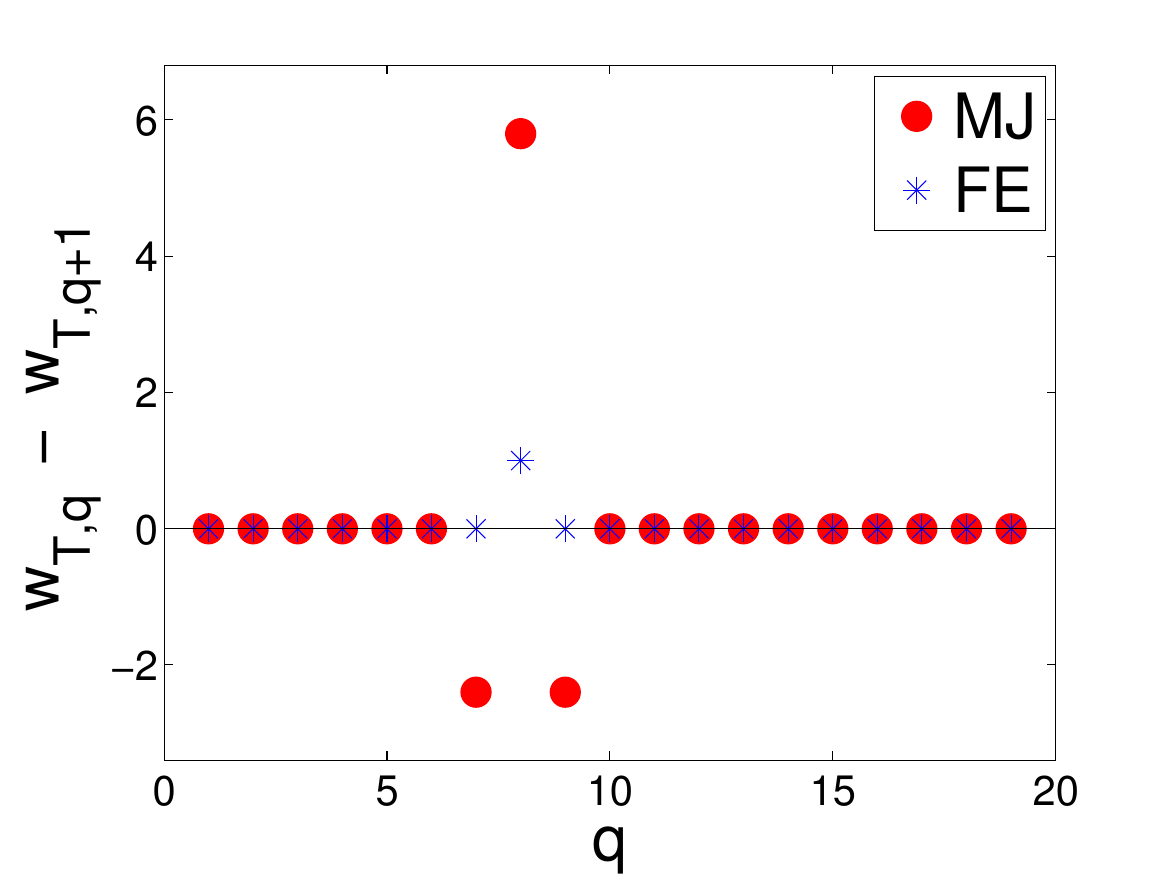}
     \end{minipage}
     \hspace{.02\textwidth}
     \begin{minipage}{.3\textwidth}
     \includegraphics[width=\linewidth]{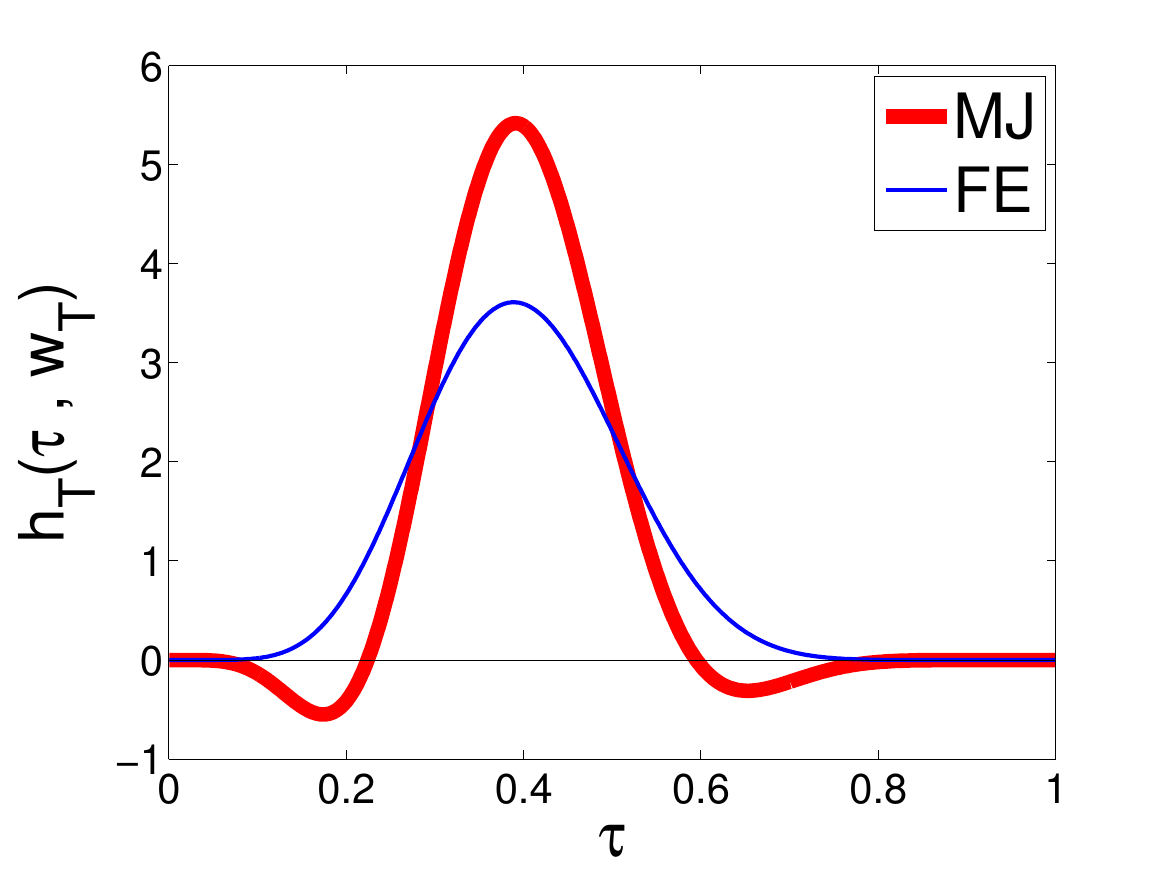}
     \end{minipage}
 \caption{ \label{fig:FEMJweights}
   \footnotesize
   For $T=20$ and $\tau_0=0.4$
   the weights $w_{T,q}=w^{\rm MJ}_{T,q}(T-1)$, their first differences $w_{T,q} - w_{T,q+1}$
      and the function $h_T(\tau,w_T)$ are plotted in red.
   For comparison the corresponding graphs for the fixed effects weights $w^{\rm FE}_{T,q}$ are plotted in blue.
   }
\end{figure}

\Needspace{5\baselineskip}
\begin{remarks}~
      \begin{itemize}

      \item[(i)]
         The weights $w^{\rm MJ}_{T,q}(T-1)$, their first differences,
      and the corresponding function $h_T(\tau,w^{\rm MJ}_T(T-1))$ are plotted in Figure~\ref{fig:FEMJweights}
      for specific values of $T$ and $\tau_0$.
       Note that $h_T(\tau,w^{\rm MJ}_T(T-1))$ has the shape of a higher-order
      kernel function (cf.\ the use of higher-order kernels for bias reduction
      in nonparametric density and regression estimation).
      The analogy is instructive but imperfect: in a kernel regression one can choose the kernel function
       $h_T(\tau,w_T)$ directly, while here we can only choose the weight vector $w_T$.

   \item[(ii)]
   The Jackknife weights are in general well-defined for  $\tau_0 S \geq 1$ and $(1-\tau_0) S \geq 1$,
   but in the above definition of the modified Jackknife weights with $S=T-1$
    we impose the stronger condition
   $\tau_0 T \geq 2$ and $(1-\tau_0) T \geq 2$, in order to guarantee that those weights are equal to
   $1-\tau_0$ for $q=1$ and equal to $-\tau_0$ for $q=T$.
   Asymptotically this condition is always satisfied, since we consider a limit where $\tau_0$ is fixed as $T \rightarrow \infty$,
   but at finite sample we cannot perform bias correction for quantiles $\tau_0$ that are too close to
   zero or one.

           \item[(iii)]
     \cite{HahnNewey2004} show that delete-one Jackknife
      bias correction can be used in nonlinear panel data models with
      smooth objective function to eliminate the leading $1/T$ bias term. It might therefore be surprising that
      the delete-one Jackknife procedure is not first order bias correcting
      in our quantile model: the moment function
      $ \frac{1}{T} \sum_{q=1}^T w^{\rm JK}_{T,q}(T-1)  \; z_{i[q]}(\theta^0)$ still has a leading order bias of
      order $1/T$.
 The reason is that
the second order bias (of order $1/T^2$) has non-standard structure in the quantile regression case.
The coefficient of the leading $1/T$ bias is constant,
but the coefficient of the $1/T^2$ term depends on the fractional part
$\langle \tau_0 T \rangle$. The delete-one difference can turn these
rounding effects into a remaining $1/T$ bias.

   \item[(iv)]
   We found above that the variance of the moment function at $\theta^0$
   was of order $1/T$ for the fixed effects weights $w^{\rm FE}_{T,q}$.
   In contrast, for the modified delete-one Jackknife weights,
   the $V_{1,i}$-term dominates. Using the explicit weight formulas from the
   modified delete-one example above, Appendix~\ref{app:WeightRegularity} shows that
   $\frac{1}{T^2}\sum_q(w_{T,q} - w^{\rm FE}_{T,q})^2 \to \frac{\tau_0^2(1-\tau_0)^2}{2}$,
   so that
   ${\rm Var} \left[ \left.  \frac{1}{T} \sum_{q=1}^T  w_{T,q} \; z_{i[q]}(\theta^0)  \, \right| A_i
           \right]   =  \frac{\tau_0^2 (1-\tau_0)^2}{2} V_{1,i} + {\cal O}_{\rm u}(1/T)$,
    for $w_{T,q} = w^{\rm MJ}_{T,q}(T-1)$,
   i.e.\ the variance of the moment function is   of order one.
   Thus, for large $T$ there is a very significant variance inflation compared to the
   fixed effect weights.
   When using the modified delete-one Jackknife weights for bias correction,
    the bias corrected estimator for $\theta$ is
    only $\sqrt{N}$ (as opposed to $\sqrt{NT}$) consistent
    as $N,T \rightarrow \infty$, and the remaining bias of order $1/T^2$ can be neglected
    as long as $N/ T^4 \rightarrow 0$ asymptotically.
    Given this variance inflation, we do not recommend  using $w^{\rm MJ}_{T,q}(T-1)$ in practice.
    Such a strong variance inflation  is specific to the quantile regression setting, and does not occur
    in \cite{HahnNewey2004} when the delete-one Jackknife is applied to
    a non-linear panel model with smooth objective function (there, the resulting bias corrected estimator
    remains $\sqrt{NT}$ consistent).

      \end{itemize}
\end{remarks}

We now discuss Jackknife bias correction for smaller values of $S$.
If $S=T-c$ for a fixed positive integer $c$, then the asymptotic properties of the moment function
$\frac{1}{T} \sum_{q=1}^T w^{\rm MJ}_{T,q}(T-c) z_{i[q]}(\theta^0)$ are similar to those discussed above
for $S=T-1$: the moment function is first order bias correcting (for the MJ weights),
but the asymptotic variance is of order one, resulting in an estimator for $\theta$
that is only $\sqrt{N}$ consistent.

\begin{figure}[tb]
     \begin{minipage}{.3\textwidth}
        \includegraphics[width=\linewidth]{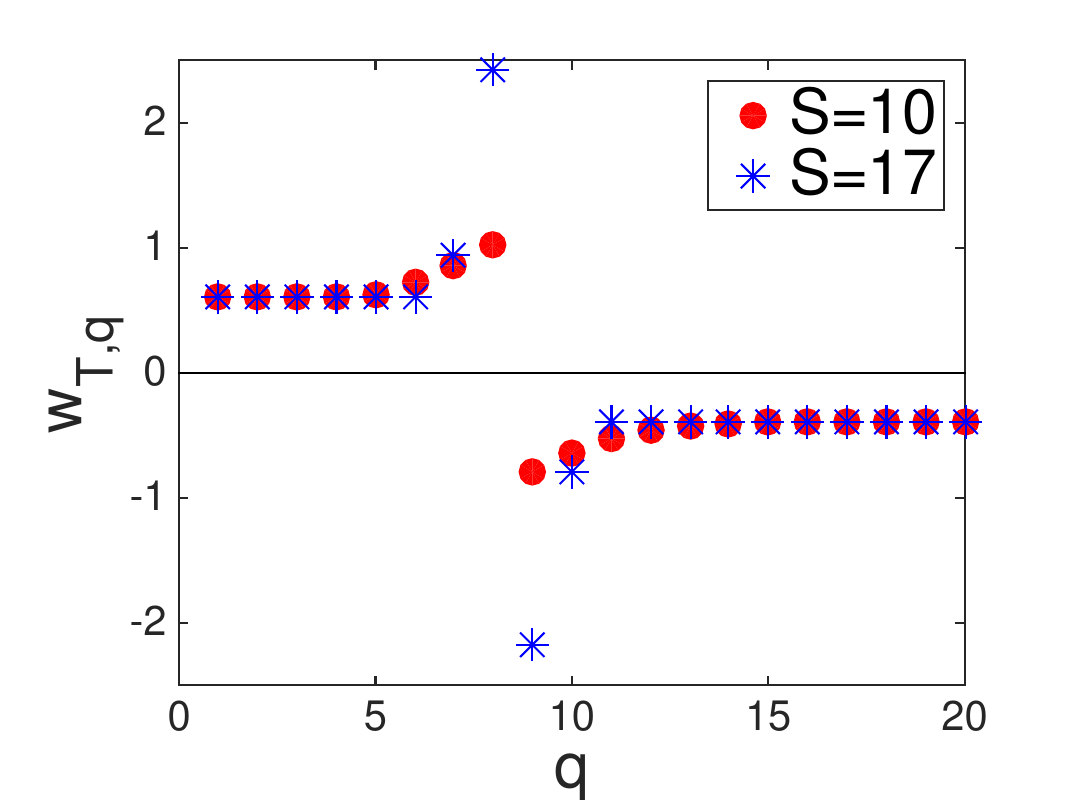}
     \end{minipage}
     \hspace{.02\textwidth}
     \begin{minipage}{.3\textwidth}
        \includegraphics[width=\linewidth]{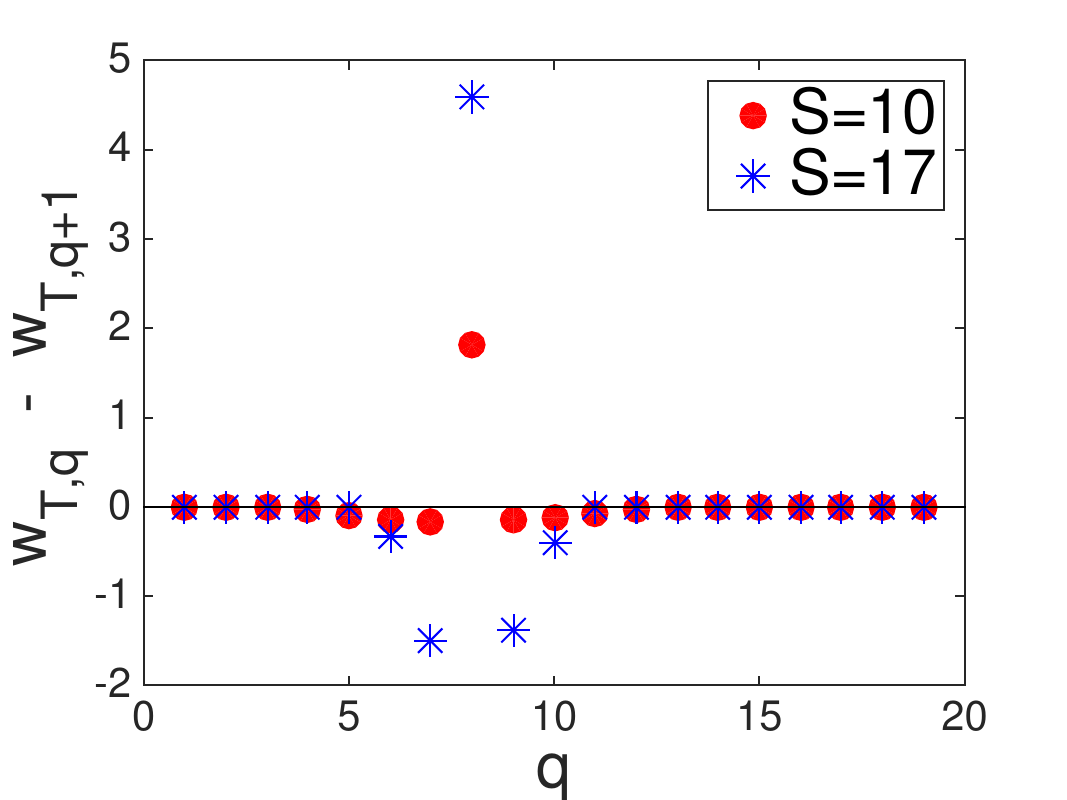}
     \end{minipage}
     \hspace{.02\textwidth}
     \begin{minipage}{.3\textwidth}
     \includegraphics[width=\linewidth]{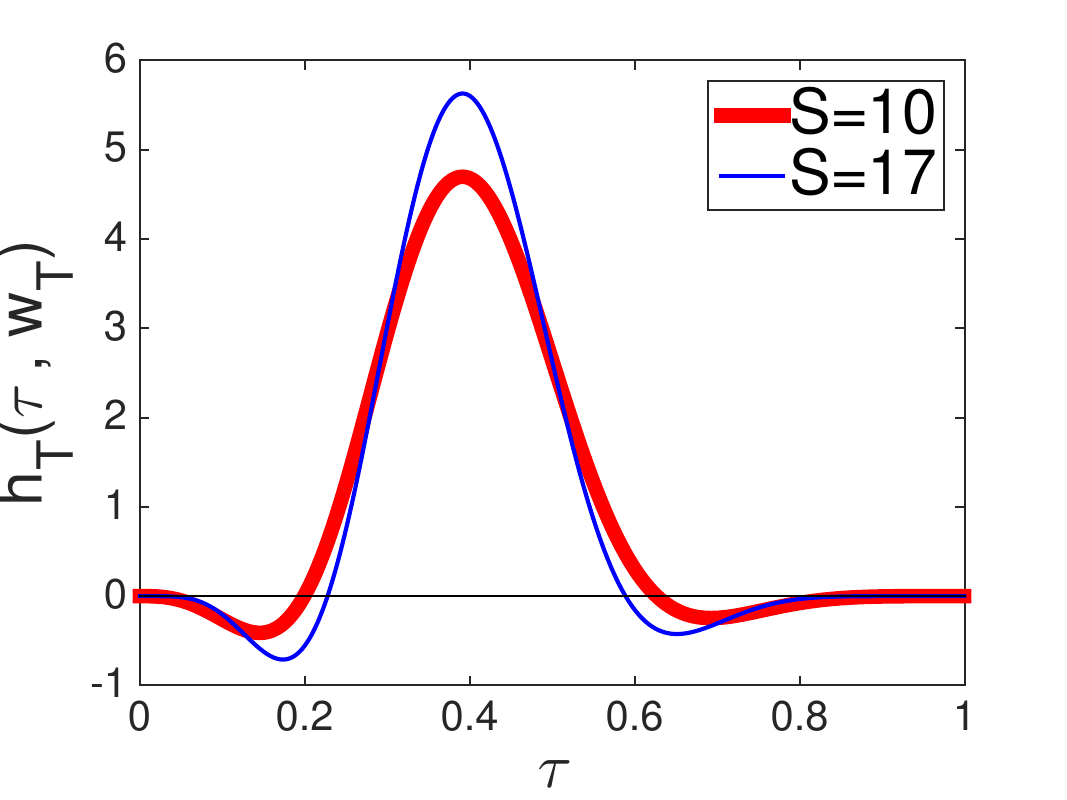}
     \end{minipage}
 \caption{ \label{fig:JKweights}
   \footnotesize
   For $T=20$ and $\tau_0=0.4$
   the Jackknife weights $w_{T,q}=w^{\rm JK}_{T,q}(S)$ are plotted for $S=10$ and $S=17$.
The first differences $w_{T,q} - w_{T,q+1}$
      and the function $h_T(\tau,w_T)$ are plotted for those weights.
   }
\end{figure}

Alternatively, one can consider an asymptotic sequence where $T-S$ grows at the same rate as $T$.
The leading example uses $S= \lfloor T/2 \rfloor$, the subsample size of the split-panel Jackknife of \cite{DhaeneJochmans2015}. In our setting it is natural to consider the more general
choice $S =  \lceil (1-\varphi) T \rceil$, where $\varphi \in (0,1)$.
For fixed $\varphi$ and $T \rightarrow \infty$ we have
\begin{align*}
   \mathbb{E}\left[ \left.  \frac{1}{T} \sum_{q=1}^T w^{\rm JK}_{T,q}(T -  \lfloor \varphi T \rfloor) \; z_{i[q]}(\theta^0)  \, \right|  A_i
           \right]     &=  {\cal O}_{\rm u}( 1/ T^2 ) ,
       \\[5pt]
   {\rm Var} \left[ \left.  \frac{1}{T} \sum_{q=1}^T w^{\rm JK}_{T,q}(T -  \lfloor \varphi T \rfloor)  \; z_{i[q]}(\theta^0)  \, \right| A_i
           \right]      &= \frac{V_{0,i}}{T} + \smallO_{\rm u}(1/T) ,
\end{align*}
i.e.\ the Jackknife moment function is bias correcting and has the same leading-order
asymptotic variance as the fixed effect moment function.
The difference between
$w^{\rm MJ}_{T,q}$ and $w^{\rm JK}_{T,q}$
vanishes as $T$ becomes large with fixed $\varphi$, so the modification is not necessary in that limit.
The advantage of the modified Jackknife weights is that they deliver first order bias correction in both regimes just discussed,
so we expect them to deliver more robust bias correction at finite sample.
Figure~\ref{fig:JKweights} illustrates the Jackknife weights and the corresponding
$h_T$ function for $S=10$ and $S=17$ with $T=20$.

\subsection{Bias-Corrected Estimator}
\label{sec:BCestimator}

We now discuss how to construct a bias corrected estimator.
The idea is to add a linear correction term to the FE objective
that shifts the effective gradient from the FE score
to the bias-correcting MJ score $\frac{1}{T}\sum_q w^{\rm MJ}_{T,q} z_{i[q]}(\theta^0)$.
Specifically, define
$$
    \widehat \theta^*(S)  \in
     \argmin_{\theta \in \Theta} {\cal L}^*_{NT}(\theta,S) ,
$$
where
$$
    {\cal L}^*_{NT}(\theta,S)
      =  {\cal L}_{NT}(\theta) +
    \theta' \, \frac{1}{N} \sum_{i=1}^N \frac{1}{T} \sum_{q=1}^T \left[
   w^{\rm MJ}_{T,q}(S) - w_{T,q}^{\rm FE} \right]    \, z_{i[q]}( \widehat \theta^{\rm \, FE}) .
$$
Computing $\widehat\theta^*(S)$ amounts to solving a convex optimization problem. The FE objective ${\cal L}_{NT}(\theta)$ is convex in $\theta$, and the correction term is linear in $\theta$, so the modified objective ${\cal L}^*_{NT}(\theta,S)$ is convex. Standard convex optimization algorithms can therefore be applied. It is important that the induced order statistics $z_{i[q]}(\widehat\theta^{\rm FE})$ in the correction term are evaluated at the fixed point $\widehat\theta^{\rm FE}$ and are not updated as $\theta$ varies. Ties are ordered by observation index. We do not recommend to iterate between updating $\theta$ and recomputing $z_{i[q]}(\theta)$, because such a procedure would destroy convexity and it is unclear what the iterative scheme would converge to.

Analogous to the proof of Theorem~\ref{th:ExAsy}, one can show that\footnote{As in the linearization~\eqref{ApproxExFE} in Section~\ref{sec:FEestimator}, the display with $\sqrt{NT}$ scaling is only valid under the rate conditions of Theorem~\ref{th:BCAsy}, in particular $N/T^3 \to 0$ and $S/T \to c_1 \in (0,1]$.}
\begin{align*}
          \sqrt{NT}\left( \widehat \theta^*(S) - \theta^0 \right)
            =  - H^{-1}
     \sqrt{NT} \, \frac{1}{N} \sum_{i=1}^N \frac{1}{T} \sum_{q=1}^T
         w^{\rm MJ}_{T,q}(S) z_{i[q]}( \theta^0)  + \smallO_P(1)  ,
\end{align*}
where the key step (Lemma~\ref{lemma:BiasCorrection} in the Appendix) shows that
the induced order statistics evaluated at
$\widehat\theta^{\rm FE}$ can be replaced by those at $\theta^0$
at the cost of an $\smallO_P(1)$ error.
The asymptotic distribution then follows from the profile score results
(Theorems~\ref{th:largeTbias} and~\ref{th:VarJac}) applied to the MJ weights.

The following theorem formalizes the preceding discussion. It shows that the
bias corrected estimator $\widehat\theta^*(S)$ is $\sqrt{NT}$-consistent and
asymptotically normal with zero asymptotic bias, provided the subsample size $S$
satisfies appropriate rate conditions.

\begin{theorem}
   \label{th:BCAsy}
      Let the Assumptions~\ref{ass:QR}, \ref{ass:CI}, \ref{ass:RS} and \ref{ass:RC} be satisfied.
      Consider $N,T,S \rightarrow \infty$ with $N/T^3 \rightarrow 0$,
      $S/T \rightarrow c_1 \in (0,1]$,
      and $T^{2/3}/(T-S) \rightarrow c_2 \in [0,\infty)$.
      Also assume $T^2\log^2(NT)/\{N(T-S)^2\}\to0$.
      Suppose that the limits $H$ and $V^{\rm FE}$ defined in
      Theorem~\ref{th:ExAsy} exist, and that $H > 0$ and $V^{\rm FE} > 0$.
      Also define
      $b_T := \frac{1}{T} \sum_{q=1}^T (w^{\rm MJ}_{T,q}(S) - w^{\rm FE}_{T,q})^2$.
      Then
   \begin{align*}
       \sqrt{NT} \left(  \widehat \theta^*(S) - \theta^0 \right)
       \; \Rightarrow \;
       {\cal N}\left(  0, \,
        H^{-1} \, V^* \, H^{-1} \right) ,
   \end{align*}
   where
   $V^* = \plim_{N,T \rightarrow \infty}
       \frac{1}{N}
           \sum_{i=1}^N  V^*_i$ with $V^*_i := V_{0,i} + b_T \, V_{1,i}$.
\end{theorem}

\Needspace{5\baselineskip}
\begin{remarks}~
\begin{itemize}
\item[(i)]
The condition $T^{2/3}/(T-S) \rightarrow c_2 \in [0,\infty)$ requires $T - S$ to grow at least as fast as $T^{2/3}$.
When $S/T \to c_1 \in (0,1)$ (i.e.\ $T - S$ grows proportionally to $T$), we have $c_2 = 0$ and $b_T = {\cal O}(T^{-1/2})$,
so the asymptotic variance $V^*$ equals $V^{\rm FE}$.
The theorem also allows $S/T \to 1$, provided $T - S$ grows at least as fast as $T^{2/3}$.
In that case $b_T$ has a finite limit, positive when $c_2>0$, and the $V_{1,i}$ term can inflate the asymptotic variance.
The delete-one case $S = T-1$, for which $b_T = {\cal O}(T)$, is excluded because $T^{2/3}/(T-S) = T^{2/3} \to \infty$.
The additional condition $T^2\log^2(NT)/\{N(T-S)^2\}\to0$ controls evaluation of the correction at the preliminary estimator.
\item[(ii)]
The key distinction from Theorem~\ref{th:ExAsy}
is that no bias term appears in the asymptotic distribution,
because the MJ weights eliminate the leading $1/T$ bias.
\end{itemize}
\end{remarks}

\noindent
Consistent estimators of $H$, $V^{\rm FE}$, and $V_{1,i}$ needed for inference
based on Theorems~\ref{th:ExAsy} and~\ref{th:BCAsy}
are provided in Appendix~\ref{app:VarEstimation}.

\subsection{Practical Choice of Subsample Size}
\label{sec:ChoiceOfS}

The modified Jackknife weights are constructed so that $M_2(w^{\rm MJ}_T(S), \tau_0) = -\tau_0(1-\tau_0)$ holds exactly. The leading-order bias term in Theorem~\ref{th:largeTbias},
\[
\frac{(1-\tau_0)\tau_0 + M_2(w_T,\tau_0)}{2T}\,g''_i(\tau_0),
\]
therefore vanishes exactly for every choice of $S$.
For $S$ growing proportionally to $T$, the leading-order bias is zero, so we focus in our practical choice of $S$ on the question of variance.\footnote{One could choose weights that also annihilate $M_3$, $M_4$, and successive higher moments, thereby also eliminating the corresponding higher-order contributions to the bias expansion in Theorem~\ref{th:largeTbias}. There is, however, no natural stopping point in such an iteration, and each additional moment constraint likely inflates the asymptotic variance. We therefore confine the construction to eliminating the leading-order bias.}
The asymptotic variance from Theorem~\ref{th:BCAsy} is $V^*_i = V_{0,i} + b_T\,V_{1,i}$, where
\[
b_T \,:=\, \frac{1}{T}\sum_{q=1}^T \left(w^{\rm MJ}_{T,q}(S) - w^{\rm FE}_{T,q} \right)^2
\]
measures the deviation of the modified Jackknife weights from the FE weights. Direct numerical evaluation shows that $b_T$ is generally increasing in $S$ in the designs considered. For $S$ proportional to $T$, however, $b_T$ alone does not capture the full higher-order variance change.

This tendency for variance to increase with $S$ is the opposite of what holds in smooth nonlinear panel models. \cite{HahnHughesKuersteinerNewey2024} show that for smooth objectives the delete-one Jackknife ($S = T-1$) dominates the split-panel Jackknife ($S = T/2$) in both higher-order bias and higher-order variance. Large $S$ is therefore the natural choice in that setting. In our quantile regression simulations, smaller values of $S$ generally deliver lower variance. We already saw one extreme of this in Section~\ref{sec:JKweights}, where the modified delete-one case $S = T-1$ inflates the variance enough to drop the convergence rate to $\sqrt{N}$.

The smallest admissible $S$ is set by the endpoint conditions $w^{\rm FE}_{S,1}=1-\tau_0$ and $w^{\rm FE}_{S,S}=-\tau_0$. These require $\tau_0 S \geq 1$ and $(1-\tau_0) S \geq 1$, i.e.
\begin{align*}
S \,\geq\, S_{\min}(\tau_0) \,:=\, \left\lceil \frac{1}{\min(\tau_0,\,1-\tau_0)} \right\rceil .
\end{align*}
For example, $S_{\min}(0.5) = 2$, $S_{\min}(0.4) = 3$, $S_{\min}(0.25) = 4$, and $S_{\min}(0.1) = 10$. Quantiles further from the median require larger minimum subsamples.

The smallest admissible choice, $S_{\min}(\tau_0)$, is however a fixed constant that does not grow with $T$, and for quantiles near the median it is very small (e.g.\ $S_{\min}(0.5) = 2$). This sits uneasily with the asymptotic theory of Theorem~\ref{th:BCAsy}, which is derived under $S \to \infty$ with $S/T \to c_1 \in (0,1]$, that is, for subsample sizes that grow proportionally to $T$. A fixed, small $S$ is not covered by these asymptotics and can leave score bias of order $1/T$. In our endogenous simulations it can leave appreciable bias. We therefore recommend choosing
\begin{align*}
S \,=\, \max\!\left( S_{\min}(\tau_0), \, \lfloor T/2 \rfloor \right) .
\end{align*}
For $T\geq S_{\min}(\tau_0)+2$, this choice has $S\leq T-2$ and a nonzero denominator in \eqref{DefKappa}.
This keeps $S$ proportional to $T$, as the theory requires, while staying well below the high-variance region near $S = T-1$. For quantiles in the central part of the distribution the lower bound $\lfloor T/2 \rfloor$ binds, and the subsample size coincides with that used by the split-panel Jackknife of \cite{DhaeneJochmans2015}; only for quantiles far enough into the tails that $S_{\min}(\tau_0) > T/2$ does the admissibility constraint $S_{\min}(\tau_0)$ take over, and even then only for moderate $T$, since $\lfloor T/2 \rfloor$ eventually dominates as $T$ grows. The threshold $T/2$ is to some extent arbitrary, but $T/2$ is the well-established split-panel choice, and the Monte Carlo simulations in Section~\ref{sec:MC} show that it performs well across the designs we consider.

\section{Extension: Instrumental Quantile Regression}
\label{sec:Endogenous}

In this section we extend the framework of Sections~\ref{sec:Setup}--\ref{sec:BiasCorrection}
to the case where (some components of) the regressors are endogenous.
We now distinguish between a column vector of regressors $x_{it}$ and a column vector
of instruments $z_{it}$, where $z_{it}$ satisfies the quantile restriction
in Assumption~\ref{ass:QR} but $x_{it}$ may not.
The data for individual $i$ consist of $\{ y_{it}, x_{it}, z_{it} \}_{t=1}^T$.
Not all components of $x_{it}$ need be endogenous: if $x'_{it} = (d'_{it}, v'_{it})$,
with $d_{it}$ endogenous and $v_{it}$ exogenous,
then we include the exogenous regressors in the instrument vector,
$z'_{it} = (v'_{it}, z^{{\rm rest} \prime}_{it})$.
If all regressors are exogenous, we set $z_{it} = x_{it}$
and the model reduces to the framework of the preceding sections.
The assumptions from Section~\ref{sec:Assumptions} are extended as follows.

\newtheorem{QIVassumption}{Assumption}
\renewcommand{\theQIVassumption}{QIV}
\begin{QIVassumption}
     \label{ass:End}
     \phantom{a}
     \begin{itemize}[leftmargin=2.5em]
        \item[\textup{(QR)}]
        For a given $\tau \in (0,1)$ and for all $i \in {\bf N}$, $t \in {\bf T}$,
        $$\mathbb{E}\left\{ \mathbbm{1} \big[
           y_{it} \leq  x'_{it} \theta(\tau) + \alpha_i(\tau)
            \big]  \Big|  \, z_{it} , \, A_i
                       \right\} =  \tau,$$
        where  $\theta(\tau) \in \mathbbm{R}^{\dim x}$ is a
        column vector of unknown common parameters,
        $\alpha_i(\tau) \in \mathbbm{R}$ is an unknown individual specific effect,
        and $A_i$ is as before.
        The quantile restriction conditions on the instruments $z_{it}$ and $A_i$,
        so that the regressors $x_{it}$ may be endogenous.

        \item[\textup{(CI)}]
        For all $i \in {\bf N}$,
        $\{ y_{it}, x_{it}, z_{it} \}$ is independently and identically distributed over $t$,
        conditional on $A_i$.

        \item[\textup{(RS)}]
        $\{ y_i, x_i, z_i, A_i \}$ is independently and identically distributed across $i \in {\bf N}$.

        \item[\textup{(RC)}]
        The true parameter $\theta^0$ lies in the interior of a compact set $\Theta \subset \mathbbm{R}^{\dim x}$.
        The conditional cdf
        $F_y(y | x_{it}, z_{it}, A_i)$ is six times continuously differentiable for all $y \in \mathbbm{R}$,
        with all six derivatives bounded uniformly over $y \in \mathbbm{R}$ and over all possible realisations of $x_{it}$, $z_{it}$, and $A_i$.
        The conditional density $f_y(y | x_{it}, z_{it}, A_i) = \partial_y F_y(y | x_{it}, z_{it}, A_i)$
        is bounded away from zero, uniformly over $y$ in a neighborhood of $x'_{it}\theta(\tau_0) + \alpha_i(\tau_0)$
        and over all realisations of $x_{it}$, $z_{it}$, and $A_i$.
        The ranges of $x_{it}$ and $z_{it}$ are bounded uniformly, and
        $ \mathbb{E}\left[ \left. |y_{it}|^{2+\delta} \, \right| A_i  \right] < c$
        for some $\delta > 0$ and $c<\infty$.
     \end{itemize}
\end{QIVassumption}

\subsection{The Quantile IV Model}
\label{sec:QIVmodel}

This subsection generalizes Example~\ref{ex:DGP} to accommodate endogenous regressors.
An important model that motivates the endogenous extension is the quantile IV model
of \cite{ChernozhukovHansen2005},
extended to allow for additive unobserved heterogeneity,
     \begin{align}
          y_{it} = x'_{it} \, \theta(U_{it}) + r(A_i, U_{it}) ,
          \label{QIVmodel}
     \end{align}
where $A_i$ is a vector of unobserved individual specific characteristics of arbitrary dimension,
$x'_{it}\theta(\tau) + r(A_i, \tau)$ is strictly increasing in $\tau \in (0,1)$,
and $U_{it} \sim U[0,1]$ is independent of $z_{it}$ conditional on $A_i$.
As in Example~\ref{ex:DGP}, this is a one-factor random coefficient model
with $\alpha_i(\tau) = r(A_i, \tau)$,
so the quantile restriction in Assumption~\ref{ass:End} holds for every $\tau \in (0,1)$.
The key difference from the exogenous case is that $x_{it}$ may be correlated with $U_{it}$
conditional on $A_i$, so that instruments $z_{it}$ are needed.
A noteworthy advantage of the QIV model in the panel setting is that the rank invariance
(or rank similarity) assumption underlying the \cite{ChernozhukovHansen2005} framework
is only required to hold conditional on the unobserved heterogeneity $A_i$.
Since $A_i$ absorbs a large part of the unobserved variation across individuals,
this assumption is substantially weakened compared to the purely cross-sectional QIV model.
A potential outcome framework with $U_{it}(x)$ varying across potential outcomes of $x_{it}$ could also be considered, following \cite{ChernozhukovHansen2005}. In that case, the ``rank invariance'' assumption can be weakened to ``rank-similarity,'' and in our panel framework this is further weakened since rank-similarity only needs to hold conditional on $A_i$.

\subsection{Fixed Effect Quantile IV Estimator}
\label{sec:FEestimatorEND}
Under Assumption~\ref{ass:End}, the regressors $x_{it}$ may be endogenous, and we need to generalize the estimation method from Section~\ref{sec:FEestimator} accordingly.
We define $ {\cal L}_{NT}(\theta,\gamma)  := \frac{1}{N} \sum_{i=1}^N L_{i,T}(\theta,\gamma)$,
with
$L_{i,T}(\theta,\gamma) := \min_{\alpha_{i} \in \mathbbm{R}} \frac{1}{T} \sum_{t=1}^T \ell_{it}(\theta,\gamma,\alpha_i)$,
and $\ell_{it}(\theta,\gamma,\alpha_i) := \rho_{\tau_0}( y_{it} - x'_{it} \theta - z_{it}' \gamma - \alpha_i )$.
In this section, $u_{it}(\theta)=y_{it}-x_{it}'\theta$ and
$u_{it}=u_{it}(\theta^0)$.
We follow \cite{ChernozhukovHansen2005,ChernozhukovHansen2006}
and consider a nested estimation problem of the form
\begin{align*}
    \widehat \gamma(\theta)  &\in
     \argmin_{\gamma \in \mathbb{R}^{\dim z}}  {\cal L}_{NT}(\theta,\gamma)  ,
  &
    \widehat \theta^{\, \text{\rm FE-IV}}     &\in   \argmin_{\theta \in \Theta}
     \;  \widehat \gamma(\theta)'  \, {\cal W}_{NT} \,  \widehat \gamma(\theta) ,
\end{align*}
where $\Theta$ is the compact parameter space in Assumption~\ref{ass:End} (RC), and
${\cal W}_{NT}$ is a symmetric positive definite
$\dim z \times \dim z$ weight matrix, which is assumed to converge to a non-random positive definite
limit ${\cal W}= \plim_{N,T \rightarrow \infty} {\cal W}_{NT}$.
Here, for given $\theta$, the estimator $\widehat \gamma(\theta)$ is obtained by running a fixed effect quantile regression
of $y_{it} - x_{it}' \theta$ on $z_{it}$,
exactly analogous to the calculation of $\widehat \theta^{\rm \, FE}$ in the exogenous case. The instruments
$z_{it}$ feature as auxiliary regressors, whose coefficient under the quantile restriction at $\theta^0$ is $\gamma^0=\vec 0$
(the zero vector in $\mathbbm{R}^{\dim z}$).
The estimator $ \widehat \theta^{\, \text{\rm FE-IV}} $ is then obtained by minimizing the distance between
 $\widehat \gamma(\theta)$ and its true value $\vec 0$, with distance measure specified by the matrix ${\cal W}_{NT}$.

For the case of exogenous regressors Theorem~\ref{th:ExAsy} shows that our assumptions imply
consistency of $\widehat \theta^{\, \text{\rm FE}}$. Here, the situation is more complicated,
because
we can only expect $\widehat \theta^{\, \text{\rm FE-IV}}$ to be consistent if $z_{it}$ is a sufficiently relevant
instrument.
\cite{ChernozhukovHansen2005} provide conditions for identification of
a quantile IV model in a cross-sectional treatment effect context, where the endogenous variable is binary.
This  identification discussion is applicable to our panel model
if one considers a single individual~$i$ and aims to identify $\theta$
  from the sample of observations for that individual over
time $t \in {\bf T}$, as $T \rightarrow \infty$. There is no incidental parameter problem in that identification discussion,
because the observations for individual $i$ are only affected by the finite vector of parameters
$(\alpha_i, \theta')$.
Having data from many individuals available only helps identification of $\theta$.
In the following we therefore just assume consistency of
$\widehat \theta^{\, \text{\rm FE-IV}}$, admitting that the goal of our paper is not to derive new identification results,
but to discuss the incidental parameter bias
and corresponding bias corrections for $   \widehat \theta^{\, \text{\rm FE-IV}} $, estimated
from an $N \times T$  panel with both $N$ and $T$ large.

Once consistency is established, then
the appropriate generalization of equation \eqref{ApproxExFE} to the endogenous case reads
\begin{align}
  \widehat \theta^{\, \text{\rm FE-IV}} - \theta^0
    &= - \left( J W  J' \right)^{-1}  J W
  \,  \frac{\partial {\cal L}_{NT}( \theta^0, \vec 0) }{\partial \gamma}   + \text{smaller order terms}   ,
     \label{ApproxThetaEnd}
\end{align}
where $W:= H^{-1} {\cal W} H^{-1}$, with
\[
H :=  \plim_{N,T \rightarrow \infty}
        \left. \frac{\partial^2}{\partial \gamma \partial \gamma'} \right|_{(\theta,\gamma)=(\theta^0,\vec 0)}
    \mathbb{E}\left[ \left. {\cal L}_{NT}(\theta,\gamma) \right| A   \right],
\]
which gives the same Hessian matrix $H$ as defined for the exogenous case,  and
\[
J :=  \plim_{N,T \rightarrow \infty} \allowbreak
        \left. \frac{\partial^2}{\partial \theta \partial \gamma'} \right|_{(\theta,\gamma)=(\theta^0,\vec 0)}
    \mathbb{E}\left[ \left. {\cal L}_{NT}(\theta,\gamma) \right| A   \right]
\]
is the limiting Jacobian.
See Appendix~\ref{app:Endog} for a formalization of \eqref{ApproxThetaEnd}.

\begin{definition}
     \label{DefJac}
     The definitions of $B_i$, $V_{0,i}$ and $ H_i$
     are unchanged, see equations~\eqref{DefB}, \eqref{DefV0}, and~\eqref{DefH}. In addition, we define
     the Jacobian
     \begin{align*}
        J_i
         &:=
          \mathbb{E}\!\left[\left.
        f(\alpha^0_i|z_{it},x_{it},A_i) \, x_{it} (z_{it} - \lambda_i)'
        \right| A_i\right]  ,
     \end{align*}
     where $f(\alpha^0_i|z_{it},x_{it},A_i)$ is the conditional pdf of $u_{it}$ given $(z_{it},x_{it},A_i)$.
     The Jacobian satisfies $J_i = \lim_{T \to \infty}
     \frac{\partial^2}{\partial \theta \partial \gamma'}\mathbb{E}[L_{i,T}(\theta,\gamma)|A_i]\big|_{(\theta^0,\vec 0)}$
     (see the Jacobian argument in Appendix~\ref{app:MomentCondition}).
     In the exogenous case $z_{it} = x_{it}$, we have $J_i = H_i$.
\end{definition}

Analogous to the results in display~\eqref{LargeTapprox} we have
 \begin{align}
      \mathbb{E}\left[ \left.     \frac{\partial L_{i,T}( \theta^0, \vec 0) }{\partial \gamma}  \, \right|  A_i   \right]
           &=   \frac{B_i }{T} +  {\cal O}_P \left( T^{-2} \right) ,
   \nonumber    \\
     {\rm Var} \left[ \left.      \frac{\partial L_{i,T}( \theta^0, \vec 0) }{\partial \gamma}  \, \right|  A_i   \right]
         &=  \frac{V_{0,i}}{T}   +  {\cal O}_P \left( T^{-2} \right),
 \nonumber     \\
   \left. \frac{\partial^2}{\partial \gamma \partial \gamma'} \right|_{(\theta,\gamma)=(\theta^0,\vec 0)}
          \mathbb{E} \left[ \left.  L_{i,T}( \theta, \gamma)   \right|  A_i
           \right]
        &= H_i     + \smallO_P(1)  ,
 \nonumber     \\
    \left. \frac{\partial^2}{\partial \theta \partial \gamma'} \right|_{(\theta,\gamma)=(\theta^0,\vec 0)}
          \mathbb{E} \left[ \left.  L_{i,T}( \theta, \gamma)   \right|  A_i
           \right]
        &= J_i     + \smallO_P(1)  .
    \label{LargeTapproxEND}
 \end{align}
 In fact, the first three equations in \eqref{LargeTapproxEND} are exactly identical to the results in \eqref{LargeTapprox},
 because  $L_{i,T}( \theta^0, \gamma)$ as a function of $\gamma$ is identical to $L_{i,T}( \theta)$ as a function of
 $\theta$ in the exogenous case (with $z_{it}$ as regressor), only the last result on the Jacobian matrix is new.
 The derivations follow from the profile score results in Section~\ref{sec:MomentCondition}.
 In addition, the profile score with respect to $\gamma$ at $\gamma = \vec 0$ satisfies
         \begin{align*}
            \frac{ \partial L_{i,T}(\theta,\vec 0)   }{\partial \gamma}
                   &= \frac{1}{T} \sum_{t=1}^T z_{it}   \Psi_{it}(\theta)
                   = \frac{1}{T} \sum_{q=1}^T w^{\rm FE}_{T,q}  \; z_{i[q]}(\theta) ,
           \end{align*}
for all values $\theta$ where $L_{i,T}(\theta,\gamma)$ is differentiable with respect to $\gamma$
at $\gamma = \vec 0$ (which includes $\theta=\theta^0$ with probability one).
The Jacobian of the expected
moment function equals $J_i'+\smallO_{\rm u}(1)$, which reduces to $H_i$ in the exogenous case $z_{it}=x_{it}$.

Applying those results together with \eqref{ApproxThetaEnd} we obtain the following theorem.

\begin{theorem}
     \label{th:EndAsy}
      Let Assumption~\ref{ass:End} hold.
      Consider $N,T \rightarrow \infty$ with $N\log^2(NT)/T^3 \rightarrow 0$.
Define
    $B_{NT} :=
       \frac{1}{N}
           \sum_{i=1}^N  B_i$,
     and suppose that non-random limits
     $H = \plim_{N,T \rightarrow \infty}
       \frac{1}{N}
           \sum_{i=1}^N  H_i$,
     $J = \plim_{N,T \rightarrow \infty}
       \frac{1}{N}
           \sum_{i=1}^N  J_i$,
    $V^{\rm FE} = \plim_{N,T \rightarrow \infty}
       \frac{1}{N}
           \sum_{i=1}^N  V_{0,i}$,
      $G = \plim_{N,T \rightarrow \infty}  \left. \frac{\partial^2}{\partial \theta \partial \theta'}  \right|_{\theta=\theta^0}\allowbreak
    \mathbb{E}\left[ \left. {\cal L}_{NT}(\theta,\vec 0) \right| A  \right]$,\footnote{%
    The matrix $G$ does not appear in the theorem's conclusion but enters through the joint Hessian
    ${\cal H} = \left(\begin{smallmatrix} G & J \\ J' & H \end{smallmatrix}\right)$
    in the proof (Appendix~\ref{app:Endog}). Assuming its existence is a regularity condition
    for the joint quadratic expansion.}
    and ${\cal W} = \plim_{N,T \rightarrow \infty}  {\cal W}_{NT}$
            exist.
     Also define $W = H^{-1} {\cal W} H^{-1}$ and $W_{NT}=H^{-1}{\cal W}_{NT}H^{-1}$,
      and assume that  $H > 0$, $V^{\rm FE} > 0$, and ${\rm rank}(J)  \geq \dim x$.
     If $\widehat \theta^{\, \text{\rm FE-IV}}$ is any sequence of
     minimizers of $\widehat \gamma(\theta)' {\cal W}_{NT} \widehat \gamma(\theta)$
     satisfying  $\widehat \theta^{\, \text{\rm FE-IV}} = \theta^0 + \smallO_P(1)$,
      then
\begin{align*}
      &\sqrt{NT} \left( \widehat \theta^{\, \text{\rm FE-IV}}    - \theta^0
      + \frac{1}{T} (J W_{NT} J')^{-1} J W_{NT} B_{NT}   \right)
       \\ &\quad \Rightarrow \;
       {\cal N}\left(  0, \,
       (J W J')^{-1} \, J W V^{\rm FE} W J' \, (J W J')^{-1} \right) .
\end{align*}
\end{theorem}%
Consistent estimators of the covariance matrix in Theorem~\ref{th:EndAsy}
are given in Appendix~\ref{app:VarEstimation:End}.
The fixed-$T$ consistency result from Section~\ref{sec:ThetaConstant} for the case $\theta(\tau) = \theta^0$ constant also applies to the QIV setting. When $\theta(\tau)$ does not depend on $\tau$, the score with respect to $\gamma$ at $(\theta^0,\vec 0)$ has exactly zero finite-$T$ mean, and the QIV estimator $\widehat\theta^{\,\rm FE\text{-}IV}$ is $\sqrt{N}$-consistent for fixed $T$ under the usual identification and regularity conditions.

\subsection{Bias Corrected Quantile IV Estimator}

For the endogenous case, define the bias corrected estimator by the nested problem
\begin{align*}
    \widehat \gamma^*(\theta,S)  &\in
     \argmin_{\gamma \in \mathbbm{R}^{\dim z}}  {\cal L}^*_{NT}(\theta,\gamma,S)  ,
  &
     \widehat \theta^*(S)    &\in   \argmin_{\theta \in \Theta}
     \;  \widehat \gamma^*(\theta,S)'  \, {\cal W}_{NT} \,  \widehat \gamma^*(\theta,S) ,
\end{align*}
where
\begin{align*}
   {\cal L}^*_{NT}(\theta,\gamma,S)
     &=  {\cal L}_{NT}(\theta,\gamma)  +
      \gamma' \; \frac{1}{N} \sum_{i=1}^N \frac{1}{T} \sum_{q=1}^T
      \left[ w^{\rm MJ}_{T,q}(S) - w_{T,q}^{\rm FE} \right]  \, z_{i[q]}( \widehat \theta^{\rm \, FE\text{-}IV}) .
\end{align*}
By a similar argument, using Lemma~\ref{lemma:BiasCorrection} and the results
from Section~\ref{sec:MomentCondition}, we obtain the following asymptotic distribution
for the bias corrected endogenous estimator.

\begin{theorem}
   \label{th:BCAsy:End}
      Let Assumption~\ref{ass:End} and the matrix conditions of Theorem~\ref{th:EndAsy} hold, and let the preliminary FE-IV estimator be consistent.
      Consider $N,T,S \rightarrow \infty$ with $N/T^3 \rightarrow 0$,
      $S/T \rightarrow c_1 \in (0,1]$,
      and $T^{2/3}/(T-S) \rightarrow c_2 \in [0,\infty)$.
      Also assume $T^2\log^2(NT)/\{N(T-S)^2\}\to0$.
      Suppose that $H > 0$ and ${\rm rank}(J) \geq \dim x$.
      Also define $W = H^{-1} {\cal W} H^{-1}$ and
      $b_T := \frac{1}{T} \sum_{q=1}^T (w^{\rm MJ}_{T,q}(S) - w^{\rm FE}_{T,q})^2$.
       If
      $\widehat \theta^*(S)$
      is any sequence of minimizers of
      $\widehat \gamma^*(\theta,S)' \, {\cal W}_{NT} \, \widehat \gamma^*(\theta,S)$
      satisfying $\widehat \theta^*(S) = \theta^0 + \smallO_P(1)$,
      then
   \begin{align*}
       \sqrt{NT} \left(  \widehat \theta^*(S) - \theta^0 \right)
       \; \Rightarrow \;
       {\cal N}\left(  0, \,
        (J W J')^{-1} \, J W V^* W J' \, (J W J')^{-1} \right) ,
   \end{align*}
   where
   $V^* = \plim_{N,T \rightarrow \infty}
       \frac{1}{N}
           \sum_{i=1}^N  V^*_i$ with $V^*_i := V_{0,i} + b_T \, V_{1,i}$.
\end{theorem}

\noindent
The quantity $b_T$ depends on $S$ and $T$; for the subsample size $S = \lfloor T/2 \rfloor$,
$b_T$ is ${\cal O}(T^{-1/2})$, so the $V_{1,i}$ term does not contribute to the asymptotic variance.
For the delete-one case $S = T-1$, $b_T = {\cal O}(T)$ and $V_{1,i}$ dominates.
When $x_{it}=z_{it}$, we have $J=H$, and the limiting covariance
reduces to $H^{-1}V^*H^{-1}$, as in Theorem~\ref{th:BCAsy}.
A consistent covariance estimator for Theorem~\ref{th:BCAsy:End}
is given in Appendix~\ref{app:VarEstimation:End}.

\section{Interpretation of the Bias}
\label{sec:Interpretation}

The fixed effect estimator replaces $\alpha_i$ by the sample $\tau_0$-quantile of the
residuals $y_{it} - z_{it}' \theta$. With $T$ observations per unit, this sample quantile is
a noisy estimate of $\alpha_i(\tau_0)$. Its rank in the residual distribution fluctuates
around $\tau_0$ with standard deviation $\sqrt{\tau_0(1-\tau_0)/T}$, the sampling variability
of a sample quantile.

This fluctuation drives the bias. The coefficient $\theta(\tau)$ changes with the quantile
level, so estimating it while the reference quantile fluctuates amounts to reading $\theta$ at
a perturbed quantile. Averaging over the fluctuation, the estimator measures a local average
of $\theta$ around $\tau_0$ rather than $\theta(\tau_0)$ itself. For a scalar coefficient, this gives the heuristic approximation
\begin{align}
   \mathbb{E}\left[ \widehat\theta^{\, \text{\rm FE}} \right]
   \approx \int_0^1 \theta(\tau)\, \phi_{T,\tau_0}(\tau)\, d\tau ,
   \label{SmoothingInterpretation}
\end{align}
where $\phi_{T,\tau_0}$ is the density of a normal distribution with variance
$\tau_0(1-\tau_0)/T$ and mean $\tau_0 + \delta$. The shift $\delta$ depends on the design and
is of order $1/T$. It is far smaller than the bandwidth, so the kernel is centered near
$\tau_0$.

The right-hand side of~\eqref{SmoothingInterpretation} smooths $\theta(\cdot)$ with a normal
kernel of bandwidth\linebreak[1] $\sqrt{\tau_0(1-\tau_0)/T}$. Expanding the integral,
\begin{align*}
   \int_0^1 \theta(\tau)\, \phi_{T,\tau_0}(\tau)\, d\tau
   \approx \theta(\tau_0) + \delta\, \theta'(\tau_0)
   + \frac{\tau_0(1-\tau_0)}{2T}\, \theta''(\tau_0) .
\end{align*}
The last two terms give the bias in this approximation. The kernel variance equals the
sampling variance of the estimated quantile's rank. It is determined by $T$ and $\tau_0$, not
chosen by the analyst. Appendix~\ref{app:Interpretation}
derives~\eqref{SmoothingInterpretation} and the shift $\delta$.

A kernel average has a resolution limit. The estimator cannot separate $\theta(\tau_0)$ from
the average of $\theta$ over a window of width $\sqrt{\tau_0(1-\tau_0)/T}$ around $\tau_0$.
When $\theta$ is constant the average equals $\theta(\tau_0)$ and there is no bias, as in
Section~\ref{sec:ThetaConstant}. When $\theta$ varies, the average blurs that variation. The
size and direction of the resulting distortion depend on the shape of $\theta(\tau)$ and on
the design, through the kernel variance and the shift $\delta$. The empirical application of
Section~\ref{sec:Empirical} illustrates one case. The age effect rises across quantiles, the
fixed effect estimator without bias correction reports a flatter profile, and the bias
correction gives the steeper slope.

In the endogenous case of Section~\ref{sec:Endogenous} the same picture holds.
Appendix~\ref{app:Interpretation} gives the details.

\section{Monte Carlo Simulations}\label{sec:MC}

In this section we support the theoretical developments in this paper through the use of Monte Carlo simulations. We consider data generated according to
\begin{align}
	y_{it} = 100 \times ( \alpha_i + x_{it} + ( 1 + 0.5 x_{it} ) \varepsilon_{it} ),
	\label{MC-DGP}
\end{align}
where $x_{it} = 0.3 \alpha_i + z_{it} + \rho \varepsilon_{it}$, $z_{it} = 0.3 \alpha_i + \eta_{it}$, $\eta_{it} \sim \textup{iid} \chi^2_3$, $\alpha_i \sim \textup{iid} U[0,1]$, and $\varepsilon_{it} \sim \textup{iid} {\cal N}(0,1)$. This largely coincides with the Monte Carlo design in \cite{GalvaoKato2016}, except that the error $\varepsilon_{it}$ enters the data generating process for $x_{it}$, and the fixed effects enter the data generating process for $z_{it}$. The extent to which the regressor $x_{it}$ is endogenous is governed by the parameter $\rho$, with it being strictly exogenous for $\rho = 0$, and endogenous for $\rho \neq 0$. In addition to this departure, we also rescale $y_{it}$ by a factor of 100 for convenience, and focus on a static model, while \cite{GalvaoKato2016} also allow for $\varepsilon_{it}$ to be serially correlated. Subsection~\ref{subsec:MC-exog} below considers a design in which $x_{it}$ is exogenous, while Subsection~\ref{subsec:MC-endog} considers a design in which $x_{it}$ is endogenous and is instrumented by $z_{it}$. The nominal slope coefficient is $\theta(\tau_0)=100+50\Phi^{-1}(\tau_0)$.\footnote{The simulations extend beyond our regularity conditions because the regressors are unbounded. For $\rho=0.5$, $\varepsilon_{it}$ is determined by $(x_{it},z_{it},\alpha_i)$, so the conditional density required by Assumption~\ref{ass:End}(RC) does not exist. For $\rho=0.5$, the scale $1+0.5x_{it}$ can also be negative. At the nominal coefficients, the conditional moment in Assumption~\ref{ass:End}(QR) equals $\tau_0-\Phi(-4-0.6\alpha_i-2z_{it})$ for the simulated quantiles. The discrepancy is less than $\Phi(-4)$.}

\subsection{Exogenous Regressor}\label{subsec:MC-exog}

Table~\ref{tab:MC1} reports empirical bias, standard deviation, and root mean squared error of the fixed effects quantile regression estimator (FE-QR) and three bias-corrected estimators: the (standard)  Jackknife bias-corrected estimator (JK), the modified  Jackknife bias-corrected estimator (MJ), and the  split-panel Jackknife bias-corrected estimator (SPJ). We display results for $\tau_0 = \{ 0.2, 0.5, 0.7 \}$, $N = 1000$, and $T = \{ 10, 20, 40, 80 \}$.\footnote{It might seem more natural to consider  $\tau_0 = 0.8$ rather than  $\tau_0 = 0.7$ (or, conversely,  $\tau_0 = 0.3$ instead of  $\tau_0 = 0.2$). However, as discussed below, the data generating process exhibits certain symmetries around $\tau_0 = 0.5$ making it more informative to examine asymmetric quantiles.} For  the JK and MJ bias-corrected estimators we display results for $S = T - 1$, $S = T / 2$, and $S = \max\{ \lceil 1 / \tau_0 \rceil, \lceil 1 / ( 1 - \tau_0 ) \rceil \} \eqqcolon S_{\min}$, the smallest subsample size ensuring the endpoint conditions on the weights.\footnote{Additional figures in Appendices \ref{exog:bias} and \ref{exog:std} plot empirical bias and standard deviation across all admissible $S$ for each $( T, \tau_0 )$ combination.} We use $1000$ Monte Carlo replications throughout.

\afterpage{\clearpage
\begin{landscape}
	\begin{table}[p]
		\vspace*{\fill}
		\centering
		\setlength{\tabcolsep}{3pt}
		\renewcommand{\arraystretch}{1.05}
		\caption{Monte Carlo results for FE-QR and Jackknife bias-corrected estimators, $\rho = 0$}
		\label{tab:MC1}
		\medskip
		\footnotesize
		\begin{threeparttable}
			\begin{tabular}{c|l c|rrr|rrr|rrr|rrr}
				\toprule
				& & & \multicolumn{3}{c|}{$T=10$}
				& \multicolumn{3}{c|}{$T=20$}
				& \multicolumn{3}{c|}{$T=40$}
				& \multicolumn{3}{c}{$T=80$} \\
				\midrule[1.1pt]
				& & $S$
				& \multicolumn{1}{c}{bias} & \multicolumn{1}{c}{std} & \multicolumn{1}{c|}{rmse}
				& \multicolumn{1}{c}{bias} & \multicolumn{1}{c}{std} & \multicolumn{1}{c|}{rmse}
				& \multicolumn{1}{c}{bias} & \multicolumn{1}{c}{std} & \multicolumn{1}{c|}{rmse}
				& \multicolumn{1}{c}{bias} & \multicolumn{1}{c}{std} & \multicolumn{1}{c}{rmse} \\
				\midrule[1pt]
				& FE-QR  & ---
				&  \phantom{-}3.565 & \phantom{-}2.379 & 4.286
				&  \phantom{-}1.785 & \phantom{-}1.706 & 2.469
				&  \phantom{-}0.874 & \phantom{-}1.148 & 1.442
				&  \phantom{-}0.473 & \phantom{-}0.793 & \phantom{-}0.923 \\
				\cmidrule(l){2-15}
				& JK     & $T-1$
				& -6.761 & \phantom{-}4.265 & 7.994
				& -2.320 & \phantom{-}3.710 & 4.375
				& -1.066 & \phantom{-}3.402 & 3.565
				& -0.403 & \phantom{-}3.046 & \phantom{-}3.072 \\
				& JK     & $T/2$
				&  \phantom{-}0.583 & \phantom{-}2.666 & 2.729
				&  \phantom{-}0.144 & \phantom{-}1.852 & 1.857
				&  \phantom{-}0.020 & \phantom{-}1.236 & 1.236
				&  \phantom{-}0.052 & \phantom{-}0.844 & \phantom{-}0.845 \\
				\multirow{2}{*}{\ \, $\tau_0 = 0.2$\ \,}
				& JK     & $S_{\min}$
				&  \phantom{-}0.583 & \phantom{-}2.666 & 2.729
				&  \phantom{-}0.320 & \phantom{-}1.769 & 1.798
				&  \phantom{-}0.151 & \phantom{-}1.166 & 1.175
				&  \phantom{-}0.116 & \phantom{-}0.797 & \phantom{-}0.805 \\
				\cmidrule(l){2-15}
				& MJ     & $T-1$
				& -1.531 & \phantom{-}3.162 & 3.513
				& -0.234 & \phantom{-}2.457 & 2.468
				& -0.076 & \phantom{-}2.036 & 2.037
				&  \phantom{-}0.045 & \phantom{-}1.705 & \phantom{-}1.706 \\
				& MJ     & $T/2$
				& -0.033 & \phantom{-}2.740 & 2.740
				& -0.020 & \phantom{-}1.874 & 1.874
				& -0.023 & \phantom{-}1.242 & 1.242
				&  \phantom{-}0.041 & \phantom{-}0.846 & \phantom{-}0.847 \\
				& MJ     & $S_{\min}$
				& -0.033 & \phantom{-}2.740 & 2.740
				&  \phantom{-}0.023 & \phantom{-}1.785 & 1.785
				&  \phantom{-}0.010 & \phantom{-}1.171 & 1.171
				&  \phantom{-}0.045 & \phantom{-}0.797 & \phantom{-}0.799 \\
				\cmidrule(l){2-15}
				& SPJ    & ---
				&  \phantom{-}0.242 & \phantom{-}3.101 & 3.110
				&  \phantom{-}0.027 & \phantom{-}2.021 & 2.021
				& -0.006 & \phantom{-}1.317 & 1.317
				&  \phantom{-}0.047 & \phantom{-}0.879 & \phantom{-}0.880 \\
				\midrule[1pt]
				& FE-QR  & ---
				&  \phantom{-}0.037 & \phantom{-}2.109 & 2.109
				&  \phantom{-}0.045 & \phantom{-}1.469 & 1.469
				& -0.048 & \phantom{-}1.010 & 1.012
				&  \phantom{-}0.031 & \phantom{-}0.668 & \phantom{-}0.669 \\
				\cmidrule(l){2-15}
				& JK     & $T-1$
				&  \phantom{-}0.040 & \phantom{-}4.262 & 4.262
				&  \phantom{-}0.050 & \phantom{-}3.794 & 3.795
				& -0.084 & \phantom{-}3.626 & 3.627
				&  \phantom{-}0.118 & \phantom{-}3.282 & \phantom{-}3.284 \\
				& JK     & $T/2$
				&  \phantom{-}0.038 & \phantom{-}2.334 & 2.334
				&  \phantom{-}0.054 & \phantom{-}1.591 & 1.592
				& -0.048 & \phantom{-}1.090 & 1.091
				&  \phantom{-}0.031 & \phantom{-}0.701 & \phantom{-}0.702 \\
				\multirow{2}{*}{\ \, {$\tau_0 = 0.5$}\ \,}
				& JK     & $S_{\min}$
				&  \phantom{-}0.037 & \phantom{-}2.159 & 2.159
				&  \phantom{-}0.048 & \phantom{-}1.483 & 1.484
				& -0.048 & \phantom{-}1.015 & 1.016
				&  \phantom{-}0.031 & \phantom{-}0.669 & \phantom{-}0.670 \\
				\cmidrule(l){2-15}
				& MJ     & $T-1$
				&  \phantom{-}0.027 & \phantom{-}2.962 & 2.962
				&  \phantom{-}0.046 & \phantom{-}2.402 & 2.402
				& -0.061 & \phantom{-}2.077 & 2.078
				&  \phantom{-}0.075 & \phantom{-}1.749 & \phantom{-}1.750 \\
				& MJ     & $T/2$
				&  \phantom{-}0.038 & \phantom{-}2.288 & 2.288
				&  \phantom{-}0.057 & \phantom{-}1.606 & 1.607
				& -0.048 & \phantom{-}1.094 & 1.095
				&  \phantom{-}0.031 & \phantom{-}0.703 & \phantom{-}0.703 \\
				& MJ     & $S_{\min}$
				&  \phantom{-}0.038 & \phantom{-}2.184 & 2.184
				&  \phantom{-}0.050 & \phantom{-}1.491 & 1.492
				& -0.049 & \phantom{-}1.018 & 1.019
				&  \phantom{-}0.031 & \phantom{-}0.670 & \phantom{-}0.671 \\
				\cmidrule(l){2-15}
				& SPJ    & ---
				&  \phantom{-}0.028 & \phantom{-}2.538 & 2.539
				&  \phantom{-}0.073 & \phantom{-}1.705 & 1.707
				& -0.067 & \phantom{-}1.143 & 1.144
				&  \phantom{-}0.034 & \phantom{-}0.723 & \phantom{-}0.724 \\
				\midrule[1pt]
				& FE-QR  & ---
				& -1.785 & \phantom{-}2.155 & 2.798
				& -0.958 & \phantom{-}1.472 & 1.756
				& -0.464 & \phantom{-}1.052 & 1.150
				& -0.209 & \phantom{-}0.727 & \phantom{-}0.756 \\
				\cmidrule(l){2-15}
				& JK     & $T-1$
				&  \phantom{-}2.955 & \phantom{-}4.345 & 5.255
				&  \phantom{-}1.133 & \phantom{-}3.760 & 3.927
				&  \phantom{-}0.526 & \phantom{-}3.573 & 3.611
				&  \phantom{-}0.202 & \phantom{-}3.297 & \phantom{-}3.303 \\
				& JK     & $T/2$
				&  \phantom{-}1.050 & \phantom{-}2.485 & 2.698
				& -0.075 & \phantom{-}1.605 & 1.607
				& -0.013 & \phantom{-}1.135 & 1.135
				&  \phantom{-}0.008 & \phantom{-}0.766 & \phantom{-}0.766 \\
				\multirow{2}{*}{\ \, $\tau_0 = 0.7$\ \,}
				& JK     & $S_{\min}$
				&  \phantom{-}0.143 & \phantom{-}2.373 & 2.377
				& -0.055 & \phantom{-}1.500 & 1.501
				& -0.026 & \phantom{-}1.065 & 1.065
				&  \phantom{-}0.005 & \phantom{-}0.731 & \phantom{-}0.731 \\
				\cmidrule(l){2-15}
				& MJ     & $T-1$
				&  \phantom{-}0.574 & \phantom{-}3.046 & 3.099
				&  \phantom{-}0.083 & \phantom{-}2.378 & 2.380
				&  \phantom{-}0.026 & \phantom{-}2.065 & 2.065
				& -0.009 & \phantom{-}1.757 & \phantom{-}1.757 \\
				& MJ     & $T/2$
				&  \phantom{-}0.435 & \phantom{-}2.408 & 2.447
				&  \phantom{-}0.012 & \phantom{-}1.622 & 1.622
				&  \phantom{-}0.010 & \phantom{-}1.140 & 1.140
				&  \phantom{-}0.013 & \phantom{-}0.768 & \phantom{-}0.768 \\
				& MJ     & $S_{\min}$
				&  \phantom{-}0.003 & \phantom{-}2.359 & 2.359
				& -0.055 & \phantom{-}1.500 & 1.501
				& -0.015 & \phantom{-}1.065 & 1.065
				&  \phantom{-}0.013 & \phantom{-}0.731 & \phantom{-}0.731 \\
				\cmidrule(l){2-15}
				& SPJ    & ---
				&  \phantom{-}1.487 & \phantom{-}2.754 & 3.130
				& -0.006 & \phantom{-}1.728 & 1.728
				&  \phantom{-}0.026 & \phantom{-}1.203 & 1.203
				&  \phantom{-}0.023 & \phantom{-}0.792 & \phantom{-}0.792 \\
				\bottomrule
			\end{tabular}
			\smallskip
			\begin{tablenotes}[para,flushleft]
				\footnotesize
				Notes: Each panel reports the empirical bias, standard deviation, and root mean squared error of the FE-QR estimator and of Jackknife bias-corrected estimators.
			\end{tablenotes}
		\end{threeparttable}
		\vspace*{\fill}
	\end{table}
\end{landscape}
}

\paragraph{FE-QR Estimator.}
The FE-QR estimator displays bias which is positive at $\tau_0 = 0.2$, close to zero at $\tau_0 = 0.5$, and negative at $\tau_0 = 0.7$. The bias roughly halves with each doubling of $T$, consistent with it originating primarily in a term of order $1/T$. The sign and variation across quantiles follow from Theorem~\ref{th:largeTbias} and the estimator expansion in Theorem~\ref{th:ExAsy}. For this design, the leading FE-QR bias is proportional to
\begin{align}
	\frac{\tau_0(1-\tau_0)}{T} \left( -\frac{q(\tau_0)}{\phi(q(\tau_0))^{2}} \right),
	\label{eq:FE-bias}
\end{align}
where $\phi( \cdot )$ denotes the standard normal density and $q(\cdot) = \Phi^{-1}(\cdot)$. The proportionality factor is positive and does not depend on $\tau_0$ or $T$.

\begin{figure}[!htbp]
	\centering
	\vspace{0.25cm}
	\begin{tikzpicture}
		\node[inner sep=0pt] (img) {\includegraphics[width=\designfactorwidth]{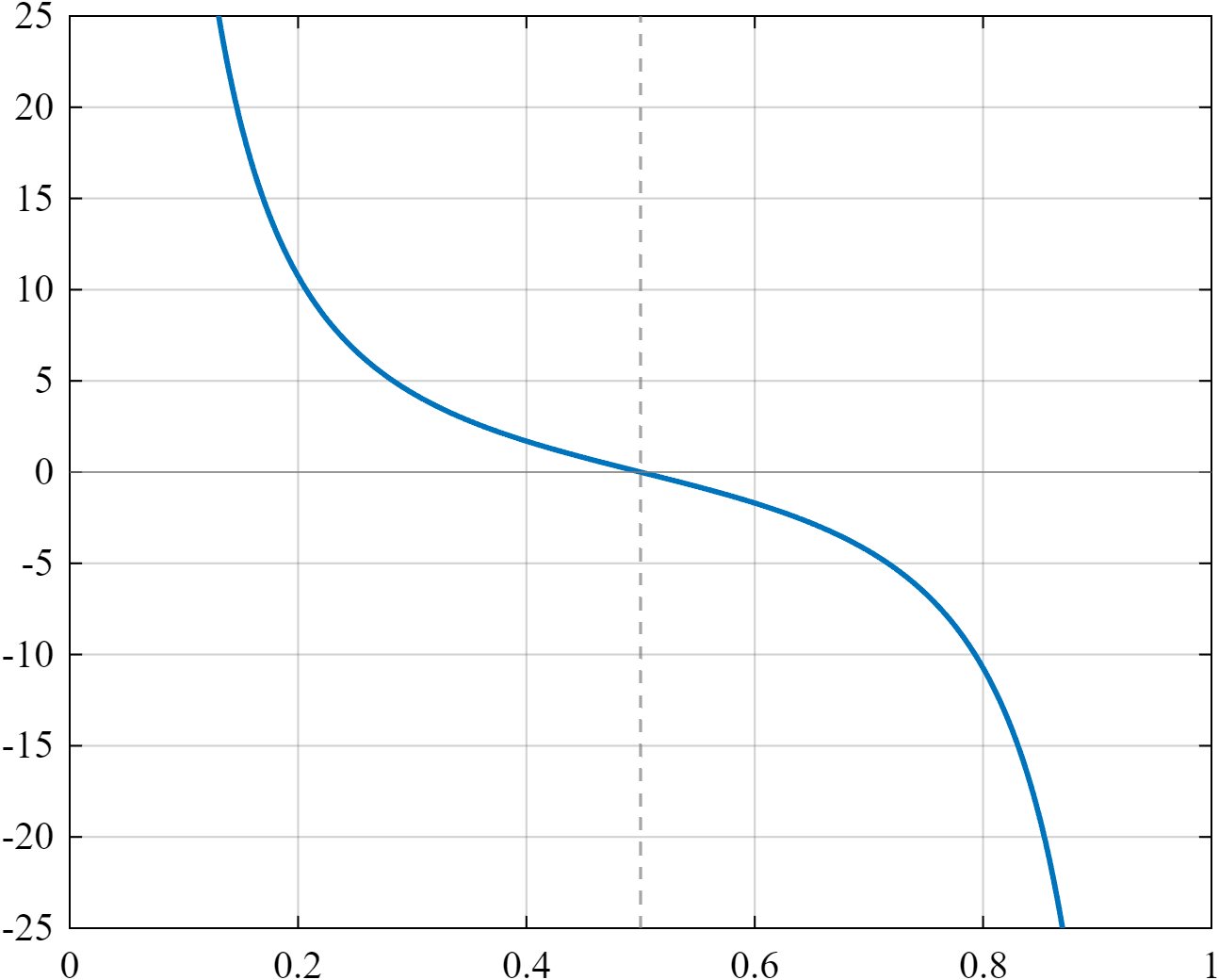}};
		\node[below=2pt of img] {$\tau_0$};
		\node[left=2pt of img, rotate=90, anchor=south] {$-q(\tau_0)/\phi(q(\tau_0))^{2}$};
	\end{tikzpicture}
	\caption{$-q(\tau_0)/\phi(q(\tau_0))^{2}$ as a function of $\tau_0$.}
	\label{fig:design-factor}
\end{figure}

The expression comprises two components; $\tau_0(1-\tau_0)$ and $-q(\tau_0)/\phi(q(\tau_0))^{2}$. The latter, plotted in Figure~\ref{fig:design-factor}, is anti-symmetric about $\tau_0 = 0.5$, where it equals zero, and increases sharply in absolute value as $\tau_0$ approaches either $0$ or $1$. This explains why the bias is positive and largest in magnitude at $\tau_0 = 0.2$, negative and smaller in magnitude at $\tau_0 = 0.7$, and near to zero at $\tau_0 = 0.5$. 

\begin{figure}[H]
	\centering			
	\vspace{0.25cm}
	\begin{tabular}{ccc}
		\begin{tikzpicture}
			\node[inner sep=0pt] (img) {\includegraphics[width=0.28\textwidth]{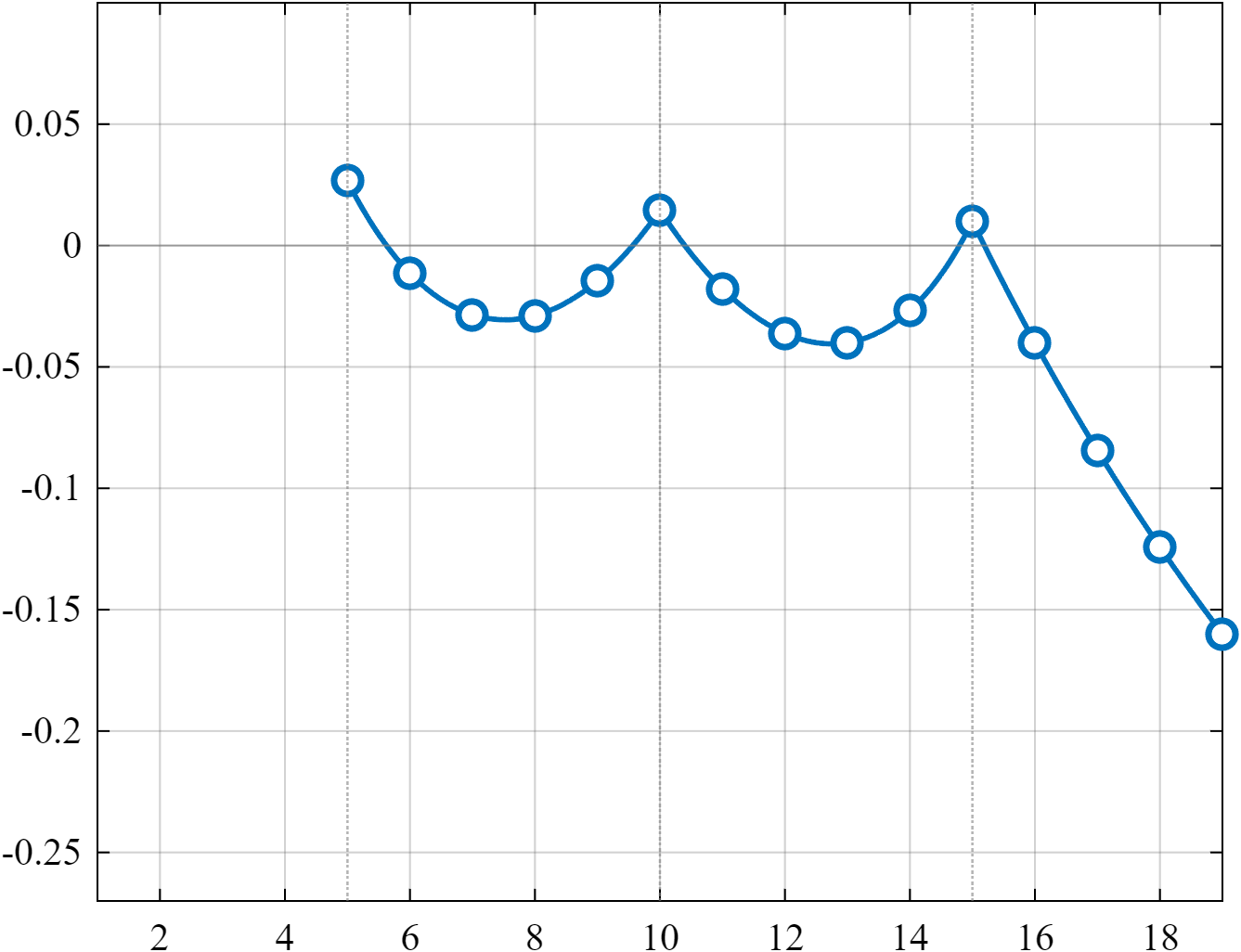}};
			\node[below=2pt of img] {\footnotesize $S$};
			\node[left=2pt of img, rotate=90, anchor=south] {\footnotesize $\tau_0(1-\tau_0) + M_2(w_T^{\textup{JK}}(S), \tau_0)$};
		\end{tikzpicture} &
		\begin{tikzpicture}
			\node[inner sep=0pt] (img) {\includegraphics[width=0.28\textwidth]{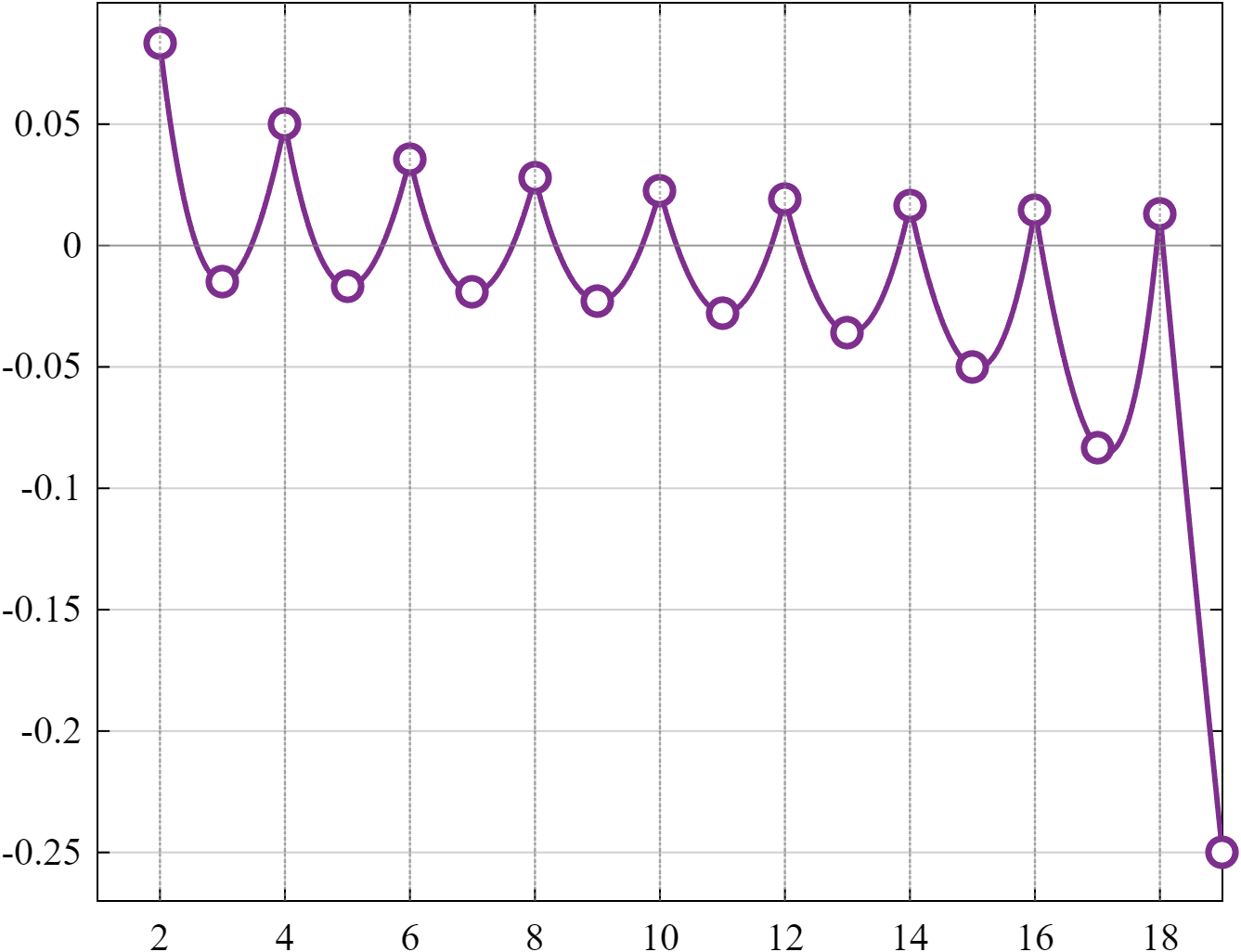}};
			\node[below=2pt of img] {\footnotesize $S$};
		\end{tikzpicture} &
		\begin{tikzpicture}
			\node[inner sep=0pt] (img) {\includegraphics[width=0.28\textwidth]{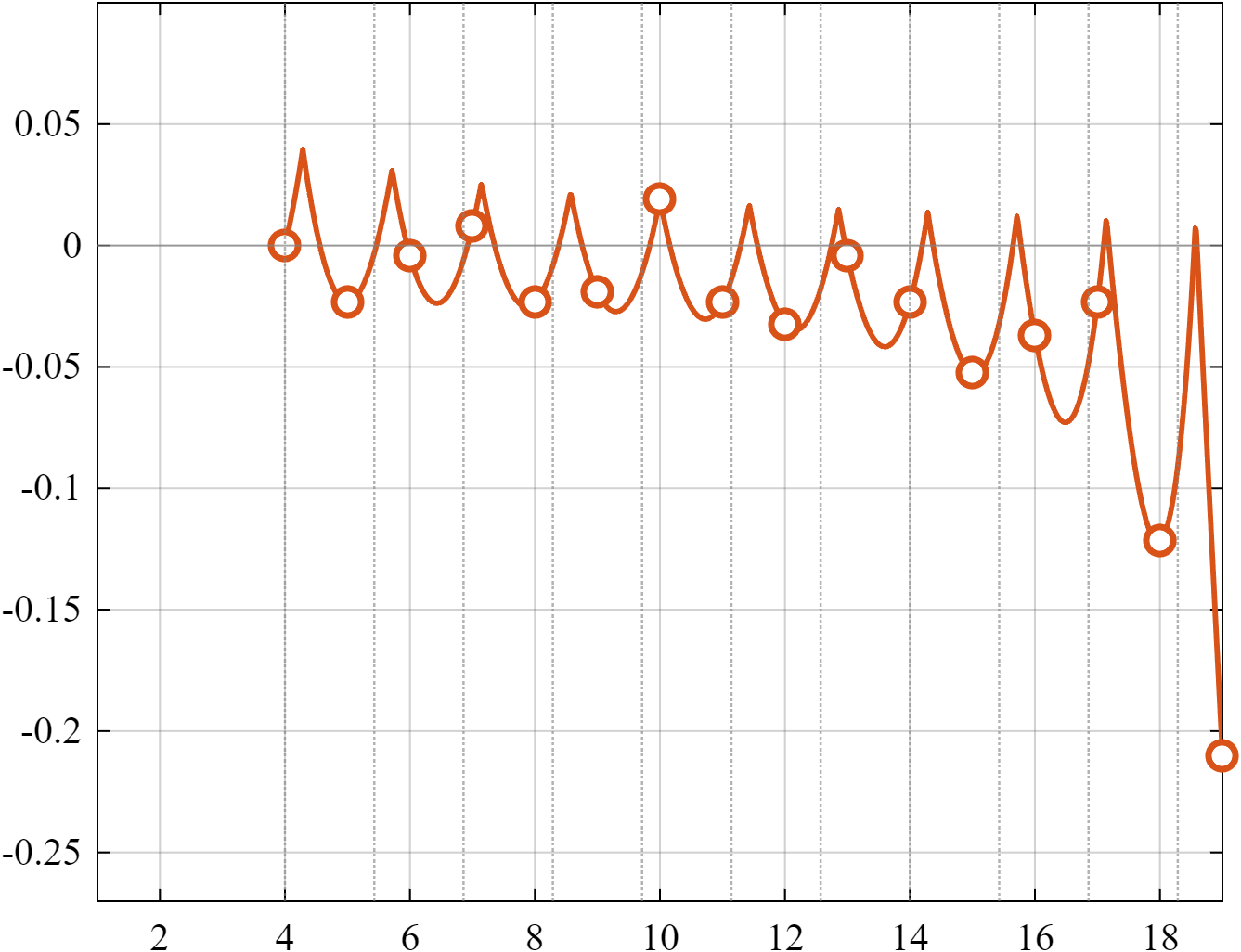}};
			\node[below=2pt of img] {\footnotesize $S$};
		\end{tikzpicture} \\[2pt]
		$\tau_0 = 0.2$ & $\tau_0 = 0.5$ & $\tau_0 = 0.7$
	\end{tabular}	
	\caption{$\tau_0(1-\tau_0) + M_2(w_T^{\textup{JK}}(S), \tau_0)$ at $T = 20$ plotted against real $S$. Markers indicate integer values of $S$.}		
	\label{fig:master-curve}
\end{figure}

\paragraph{JK Bias-corrected Estimator.}
The bias of the JK estimator varies substantially with $S$. It is largest at $S = T-1$, indeed, often larger in magnitude than the FE-QR bias, and is generally smaller away from the delete-one choice. The corresponding quadratic approximation to the JK bias is proportional to
\begin{align}
	\frac{\tau_0(1-\tau_0) + M_2(w_T^{\textup{JK}}(S), \tau_0)}{T}
	\left( -\frac{q(\tau_0)}{\phi(q(\tau_0))^{2}} \right),
	\label{eq:JK-bias}
\end{align}
with the same positive proportionality factor as in \eqref{eq:FE-bias}. The dependence on $S$ enters only through the factor $\tau_0(1-\tau_0) + M_2(w_T^{\textup{JK}}(S), \tau_0)$, which is plotted as a function of $S$ at $T = 20$ in Figure~\ref{fig:master-curve}.

\begin{figure}[H]
	\centering
	\vspace{0.25cm}
	\begin{tabular}{ccc}
		\begin{tikzpicture}
			\node[inner sep=0pt] (img) {\includegraphics[width=0.28\textwidth]{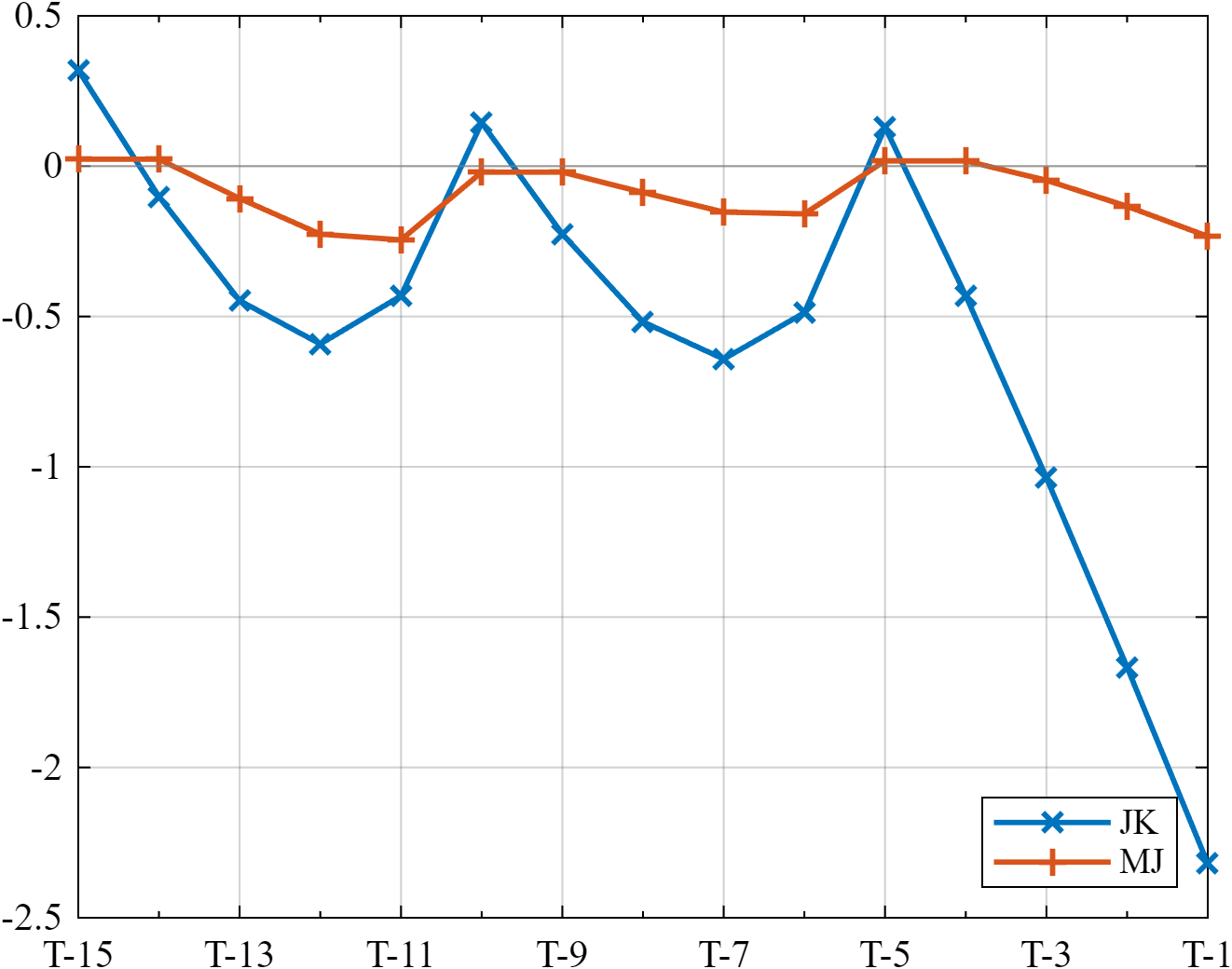}};
			\node[below=2pt of img] {\footnotesize $S$};
			\node[left=2pt of img, rotate=90, anchor=south] {\footnotesize Empirical Bias};
		\end{tikzpicture} &
		\begin{tikzpicture}
			\node[inner sep=0pt] (img) {\includegraphics[width=0.28\textwidth]{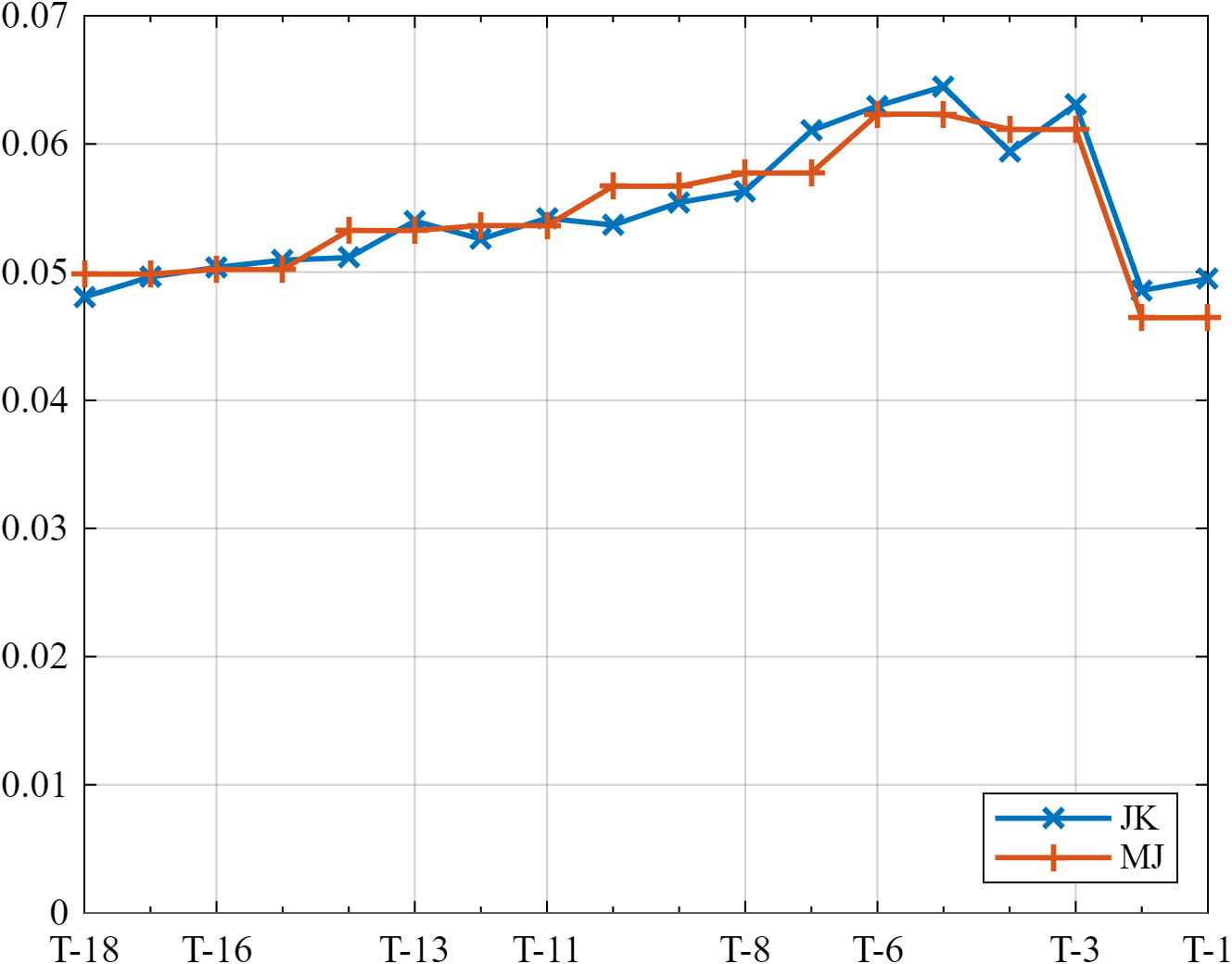}};
			\node[below=2pt of img] {\footnotesize $S$};
		\end{tikzpicture} &
		\begin{tikzpicture}
			\node[inner sep=0pt] (img) {\includegraphics[width=0.28\textwidth]{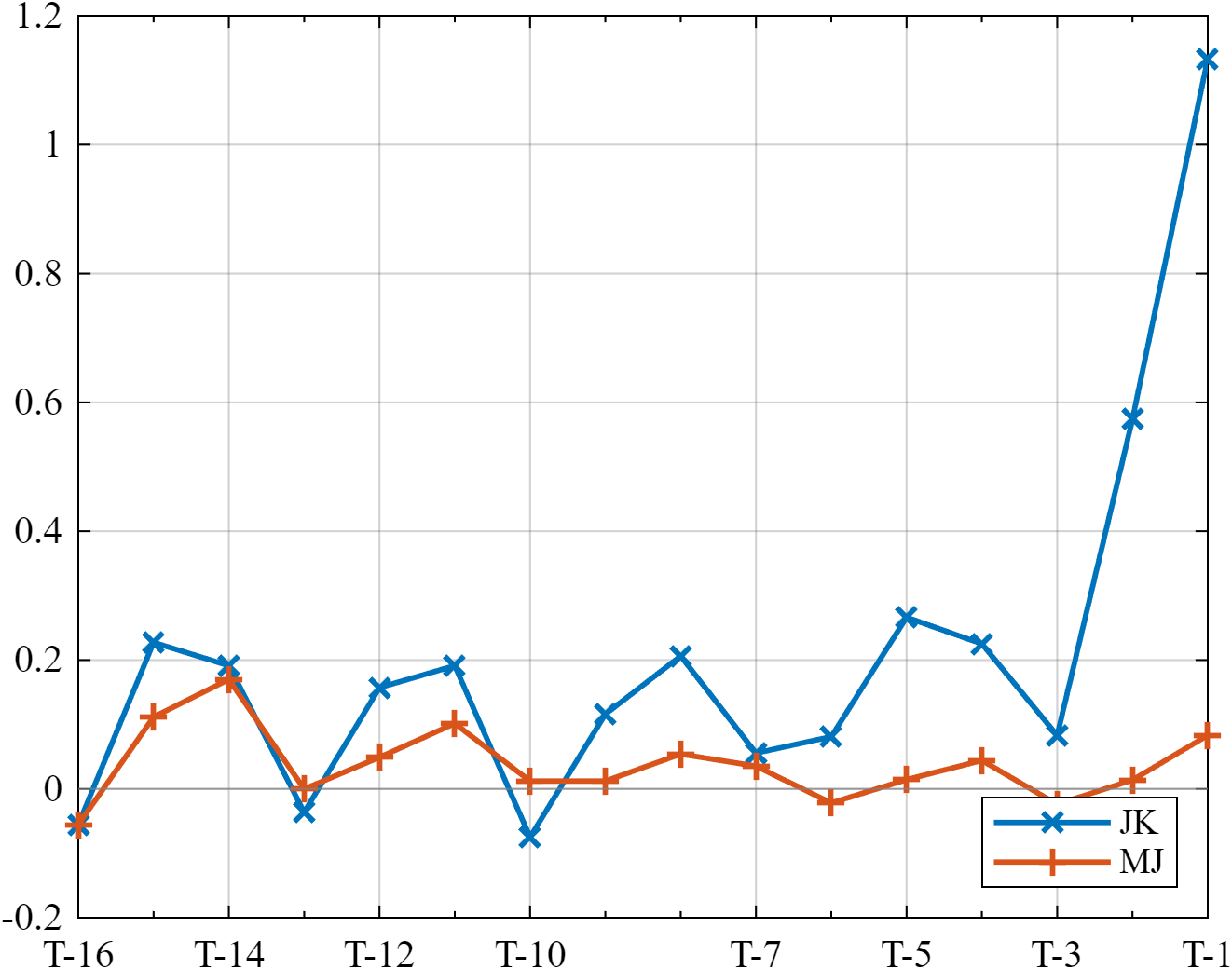}};
			\node[below=2pt of img] {\footnotesize $S$};
		\end{tikzpicture} \\[2pt]
		$\tau_0 = 0.2$ & $\tau_0 = 0.5$ & $\tau_0 = 0.7$
	\end{tabular}
	\caption{Empirical bias of the JK (blue) and MJ (red) estimators at $T = 20$ plotted against $S$.}
	\label{fig:master-curve-emp}
\end{figure}

We observe that the sign of the function varies with $S$, and the function falls sharply near $S = T-1$. Figure~\ref{fig:master-curve-emp} plots empirical bias of the JK estimator at $T = 20$ across different values of $S$. This matches closely what is observed in Figure \ref{fig:master-curve} for $\tau_0 = 0.2$ and $\tau_0 = 0.7$, noting that in the latter case, the empirical bias has the opposite sign because $q( 0.7 ) > 0$. For $\tau_0 = 0.5$ the empirical bias is close to zero since $q(0.5) = 0$. 

The variance inflation discussed in Section~\ref{sec:JKweights} is clearly visible in Table~\ref{tab:MC1}. In particular, the ratio of the JK standard deviations at $S = T-1$ and $S = T/2$ grows with $T$. This can be further seen by inspection of the figures provided in Appendix \ref{exog:std}, which plot empirical standard deviation against $S$ for each $(T, \tau_0)$ combination. For the JK estimator, the standard deviation generally increases with $S$, doing so very rapidly as $S$ approaches $T-1$.

\paragraph{MJ Bias-corrected Estimator.}
For $S$ growing proportionally to $T$, the MJ estimator is constructed to entirely eliminate the leading $1/T$ bias of the FE-QR estimator. This is reflected in Table~\ref{tab:MC1}, where the MJ estimator generally displays substantially smaller bias than both the FE-QR and JK estimators. The same pattern is visible in Figure~\ref{fig:master-curve-emp}, where the MJ bias profile is much flatter than that of  its JK counterpart across the range of $S$. The difference is particularly pronounced at $S=T-1$ where the MJ correction significantly reduces bias relative to the corresponding JK correction. As with the JK estimator, inflation in the standard deviation is observed when $S=T-1$, but the increase is considerably smaller for the MJ estimator. For $S=T/2$ and $S=S_{\min}$, the MJ estimator delivers very small biases across all $(T,\tau_0)$ combinations, while maintaining standard deviations close to those of the FE-QR estimator.

\paragraph{Comparison.}
Overall the results in Table~\ref{tab:MC1} match the theoretical predictions closely. The JK estimator at $S = T-1$ exhibits large biases and inflated standard deviations away from the median. Moving $S$ away from $T-1$ toward $S_{\min}$ generally reduces both bias and standard deviation, with performance of both the JK and MJ estimators becoming roughly similar. The effect of the MJ correction is most notable close to $S = T-1$, where it substantially reduces bias and standard deviation compared to the JK correction. The SPJ estimator is broadly comparable in performance to the MJ estimator at $S = T/2$, both in terms of its bias and its standard deviation. Nonetheless, their biases can differ at small $T$.

		\subsection{Endogenous Regressor}\label{subsec:MC-endog}

		Setting $\rho = 0.5$ in \eqref{MC-DGP} renders $x_{it}$ endogenous, in which case we use $z_{it}$ as an instrumental variable. Table~\ref{tab:MC_IV} reports empirical bias, standard deviation, and root mean squared error of the fixed effects quantile regression estimator (FE-QR), the instrumental variables fixed effects quantile regression estimator (FE-QIV), and three bias-corrected estimators: the (standard) Jackknife bias-corrected estimator (JK), the modified Jackknife bias-corrected estimator (MJ), and the split-panel Jackknife bias-corrected estimator (SPJ). The values of $\tau_0$, $N$, $T$, and $S$ match those in Subsection~\ref{subsec:MC-exog}. Additional figures plotting the empirical bias and standard deviation of the JK and MJ estimators against $S$ for each $(T, \tau_0)$ combination are collected in Appendices \ref{endog:bias} and \ref{endog:std}.

	\afterpage{\clearpage
\begin{landscape}
		\begin{table}[p]
			\vspace*{\fill}
			\centering
			\setlength{\tabcolsep}{3pt}
			\renewcommand{\arraystretch}{1.05}
			\caption{Monte Carlo results for FE-QR, FE-QIV and Jackknife bias-corrected estimators, $\rho = 0.5$}
			\label{tab:MC_IV}
			\medskip
			\footnotesize
			\resizebox{\textwidth}{!}{%
				\begin{threeparttable}
					\begin{tabular}{c|l c|rrr|rrr|rrr|rrr}
						\toprule
						& & & \multicolumn{3}{c|}{$T=10$}
						& \multicolumn{3}{c|}{$T=20$}
						& \multicolumn{3}{c|}{$T=40$}
						& \multicolumn{3}{c}{$T=80$} \\
						\midrule[1.1pt]
						& & $S$ & \multicolumn{1}{c}{bias} & \multicolumn{1}{c}{std} & \multicolumn{1}{c|}{rmse}
						& \multicolumn{1}{c}{bias} & \multicolumn{1}{c}{std} & \multicolumn{1}{c|}{rmse}
						& \multicolumn{1}{c}{bias} & \multicolumn{1}{c}{std} & \multicolumn{1}{c|}{rmse}
						& \multicolumn{1}{c}{bias} & \multicolumn{1}{c}{std} & \multicolumn{1}{c}{rmse} \\
						\midrule[1pt]
						& FE-QR    & ---
						& \phantom{-}27.872 & \phantom{-}2.171 & 27.956
						& \phantom{-}26.103 & \phantom{-}1.443 & 26.143
						& \phantom{-}25.399 & \phantom{-}0.980 & 25.418
						& \phantom{-}25.108 & \phantom{-}0.673 & \phantom{-}25.117 \\
						\cmidrule(l){2-15}
						& FE-QIV   & ---
						& \phantom{-}2.049 & \phantom{-}2.334 & 3.106
						& \phantom{-}0.739 & \phantom{-}1.542 & 1.710
						& \phantom{-}0.295 & \phantom{-}1.083 & 1.123
						& \phantom{-}0.167 & \phantom{-}0.761 & \phantom{-}0.779 \\
						\cmidrule(l){2-15}
						& JK       & $T-1$
						& -6.541 & \phantom{-}4.139 & 7.740
						& -1.896 & \phantom{-}3.383 & 3.878
						& -0.555 & \phantom{-}3.040 & 3.091
						& -0.294 & \phantom{-}2.796 & \phantom{-}2.811 \\
						& JK       & $T/2$
						& -0.133 & \phantom{-}2.617 & 2.620
						& -0.297 & \phantom{-}1.686 & 1.712
						& -0.121 & \phantom{-}1.164 & 1.171
						& -0.007 & \phantom{-}0.805 & \phantom{-}0.805 \\
						\multirow{2}{*}{\ \,$\tau_0 = 0.2$\ \,}
						& JK       & $S_{\min}$
						& -0.133 & \phantom{-}2.617 & 2.620
						& -0.283 & \phantom{-}1.591 & 1.616
						& -0.171 & \phantom{-}1.100 & 1.113
						& -0.060 & \phantom{-}0.766 & \phantom{-}0.768 \\
						\cmidrule(l){2-15}
						& MJ       & $T-1$
						& -2.160 & \phantom{-}3.034 & 3.724
						& -0.568 & \phantom{-}2.246 & 2.317
						& -0.120 & \phantom{-}1.841 & 1.845
						& -0.051 & \phantom{-}1.577 & \phantom{-}1.578 \\
						& MJ       & $T/2$
						& -0.573 & \phantom{-}2.689 & 2.750
						& -0.402 & \phantom{-}1.706 & 1.753
						& -0.141 & \phantom{-}1.169 & 1.178
						& -0.012 & \phantom{-}0.806 & \phantom{-}0.807 \\
						& MJ       & $S_{\min}$
						& -0.573 & \phantom{-}2.689 & 2.750
						& -0.478 & \phantom{-}1.603 & 1.673
						& -0.265 & \phantom{-}1.103 & 1.134
						& -0.105 & \phantom{-}0.767 & \phantom{-}0.774 \\
						\cmidrule(l){2-15}
						& SPJ      & ---
						& -0.484 & \phantom{-}3.009 & 3.048
						& -0.406 & \phantom{-}1.837 & 1.881
						& -0.129 & \phantom{-}1.261 & 1.268
						& -0.017 & \phantom{-}0.844 & \phantom{-}0.844 \\
						\midrule[1pt]
						& FE-QR    & ---
						& \phantom{-}26.540 & \phantom{-}1.897 & 26.608
						& \phantom{-}27.023 & \phantom{-}1.268 & 27.053
						& \phantom{-}27.528 & \phantom{-}0.881 & 27.542
						& \phantom{-}27.787 & \phantom{-}0.664 & \phantom{-}27.795 \\
						\cmidrule(l){2-15}
						& FE-QIV   & ---
						& -1.185 & \phantom{-}2.042 & 2.360
						& -0.694 & \phantom{-}1.406 & 1.568
						& -0.417 & \phantom{-}1.021 & 1.103
						& -0.198 & \phantom{-}0.710 & \phantom{-}0.737 \\
						\cmidrule(l){2-15}
						& JK       & $T-1$
						& \phantom{-}0.950 & \phantom{-}4.081 & 4.190
						& \phantom{-}0.580 & \phantom{-}3.753 & 3.798
						& \phantom{-}0.213 & \phantom{-}3.487 & 3.494
						& \phantom{-}0.235 & \phantom{-}3.349 & \phantom{-}3.357 \\
						& JK       & $T/2$
						& -0.120 & \phantom{-}2.280 & 2.283
						& -0.140 & \phantom{-}1.527 & 1.534
						& -0.090 & \phantom{-}1.091 & 1.094
						& -0.023 & \phantom{-}0.746 & \phantom{-}0.746 \\
						\multirow{2}{*}{\ \,$\tau_0 = 0.5$\ \,}
						& JK       & $S_{\min}$
						& -0.757 & \phantom{-}2.106 & 2.238
						& -0.466 & \phantom{-}1.420 & 1.494
						& -0.295 & \phantom{-}1.025 & 1.066
						& -0.137 & \phantom{-}0.713 & \phantom{-}0.726 \\
						\cmidrule(l){2-15}
						& MJ       & $T-1$
						& -0.108 & \phantom{-}2.856 & 2.858
						& -0.062 & \phantom{-}2.356 & 2.356
						& -0.096 & \phantom{-}1.992 & 1.994
						& \phantom{-}0.020 & \phantom{-}1.787 & \phantom{-}1.787 \\
						& MJ       & $T/2$
						& -0.298 & \phantom{-}2.238 & 2.258
						& -0.085 & \phantom{-}1.542 & 1.545
						& -0.074 & \phantom{-}1.095 & 1.098
						& -0.019 & \phantom{-}0.747 & \phantom{-}0.747 \\
						& MJ       & $S_{\min}$
						& -0.553 & \phantom{-}2.135 & 2.205
						& -0.351 & \phantom{-}1.423 & 1.465
						& -0.235 & \phantom{-}1.026 & 1.052
						& -0.105 & \phantom{-}0.713 & \phantom{-}0.721 \\
						\cmidrule(l){2-15}
						& SPJ      & ---
						& \phantom{-}0.029 & \phantom{-}2.464 & 2.464
						& -0.094 & \phantom{-}1.653 & 1.655
						& -0.078 & \phantom{-}1.157 & 1.160
						& -0.023 & \phantom{-}0.783 & \phantom{-}0.783 \\
						\midrule[1pt]
						& FE-QR    & ---
						& \phantom{-}26.215 & \phantom{-}2.098 & 26.299
						& \phantom{-}28.148 & \phantom{-}1.392 & 28.183
						& \phantom{-}29.253 & \phantom{-}1.010 & 29.270
						& \phantom{-}29.919 & \phantom{-}0.706 & \phantom{-}29.928 \\
						\cmidrule(l){2-15}
						& FE-QIV   & ---
						& -3.153 & \phantom{-}2.259 & 3.879
						& -1.732 & \phantom{-}1.537 & 2.316
						& -0.843 & \phantom{-}1.110 & 1.394
						& -0.378 & \phantom{-}0.756 & \phantom{-}0.845 \\
						\cmidrule(l){2-15}
						& JK       & $T-1$
						& \phantom{-}3.836 & \phantom{-}4.452 & 5.876
						& \phantom{-}1.635 & \phantom{-}3.840 & 4.173
						& \phantom{-}0.993 & \phantom{-}3.550 & 3.686
						& \phantom{-}0.555 & \phantom{-}3.291 & \phantom{-}3.337 \\
						& JK       & $T/2$
						& \phantom{-}0.848 & \phantom{-}2.536 & 2.674
						& -0.266 & \phantom{-}1.666 & 1.687
						& -0.052 & \phantom{-}1.194 & 1.195
						& \phantom{-}0.026 & \phantom{-}0.796 & \phantom{-}0.797 \\
						\multirow{2}{*}{\ \,$\tau_0 = 0.7$\ \,}
						& JK       & $S_{\min}$
						& -0.282 & \phantom{-}2.441 & 2.457
						& -0.366 & \phantom{-}1.566 & 1.608
						& -0.165 & \phantom{-}1.122 & 1.134
						& -0.042 & \phantom{-}0.761 & \phantom{-}0.762 \\
						\cmidrule(l){2-15}
						& MJ       & $T-1$
						& \phantom{-}0.332 & \phantom{-}3.139 & 3.157
						& -0.059 & \phantom{-}2.419 & 2.419
						& \phantom{-}0.070 & \phantom{-}2.053 & 2.054
						& \phantom{-}0.085 & \phantom{-}1.768 & \phantom{-}1.770 \\
						& MJ       & $T/2$
						& \phantom{-}0.001 & \phantom{-}2.479 & 2.480
						& -0.122 & \phantom{-}1.681 & 1.686
						& -0.014 & \phantom{-}1.201 & 1.201
						& \phantom{-}0.036 & \phantom{-}0.797 & \phantom{-}0.798 \\
						& MJ       & $S_{\min}$
						& -0.495 & \phantom{-}2.418 & 2.468
						& -0.366 & \phantom{-}1.566 & 1.608
						& -0.146 & \phantom{-}1.123 & 1.132
						& -0.029 & \phantom{-}0.761 & \phantom{-}0.761 \\
						\cmidrule(l){2-15}
						& SPJ      & ---
						& \phantom{-}1.367 & \phantom{-}2.820 & 3.133
						& -0.140 & \phantom{-}1.815 & 1.820
						& -0.041 & \phantom{-}1.278 & 1.278
						& \phantom{-}0.044 & \phantom{-}0.827 & \phantom{-}0.828 \\
						\bottomrule
					\end{tabular}
					\smallskip
					\begin{tablenotes}[para,flushleft]
						\footnotesize
						Notes: Each panel reports the empirical bias, standard deviation, and root mean squared error of the FE-QR estimator, the FE-QIV estimator, and of Jackknife bias-corrected estimators. 
					\end{tablenotes}
				\end{threeparttable}%
			}%
			\vspace*{\fill}
		\end{table}
	\end{landscape}
}

	\paragraph{FE-QR Estimator.}
			With $x_{it}$ endogenous, the FE-QR estimator is no longer consistent. Its empirical bias lies roughly between $25$ and $30$ at every $(T, \tau_0)$ combination, much larger than its standard deviation, and does not diminish as $T$ grows. This pattern contrasts sharply with the exogenous design in Subsection~\ref{subsec:MC-exog}, where the FE-QR bias halves with each doubling of $T$.

		\paragraph{FE-QIV Estimator.}
			Using $z_{it}$ as an instrument for $x_{it}$ substantially reduces the empirical bias. The bias of the FE-QIV estimator roughly halves with each doubling of $T$, consistent with a leading bias of order $1/T$. Similar to the exogenous design, the bias is positive at $\tau_0 = 0.2$ and negative at $\tau_0 = 0.7$. Unlike the exogenous design, it is also appreciably negative at $\tau_0=0.5$. 

\begin{figure}[!htbp]
			\centering
			\vspace{0.25cm}
			\begin{tabular}{ccc}
				\begin{tikzpicture}
					\node[inner sep=0pt] (img) {\includegraphics[width=0.28\textwidth]{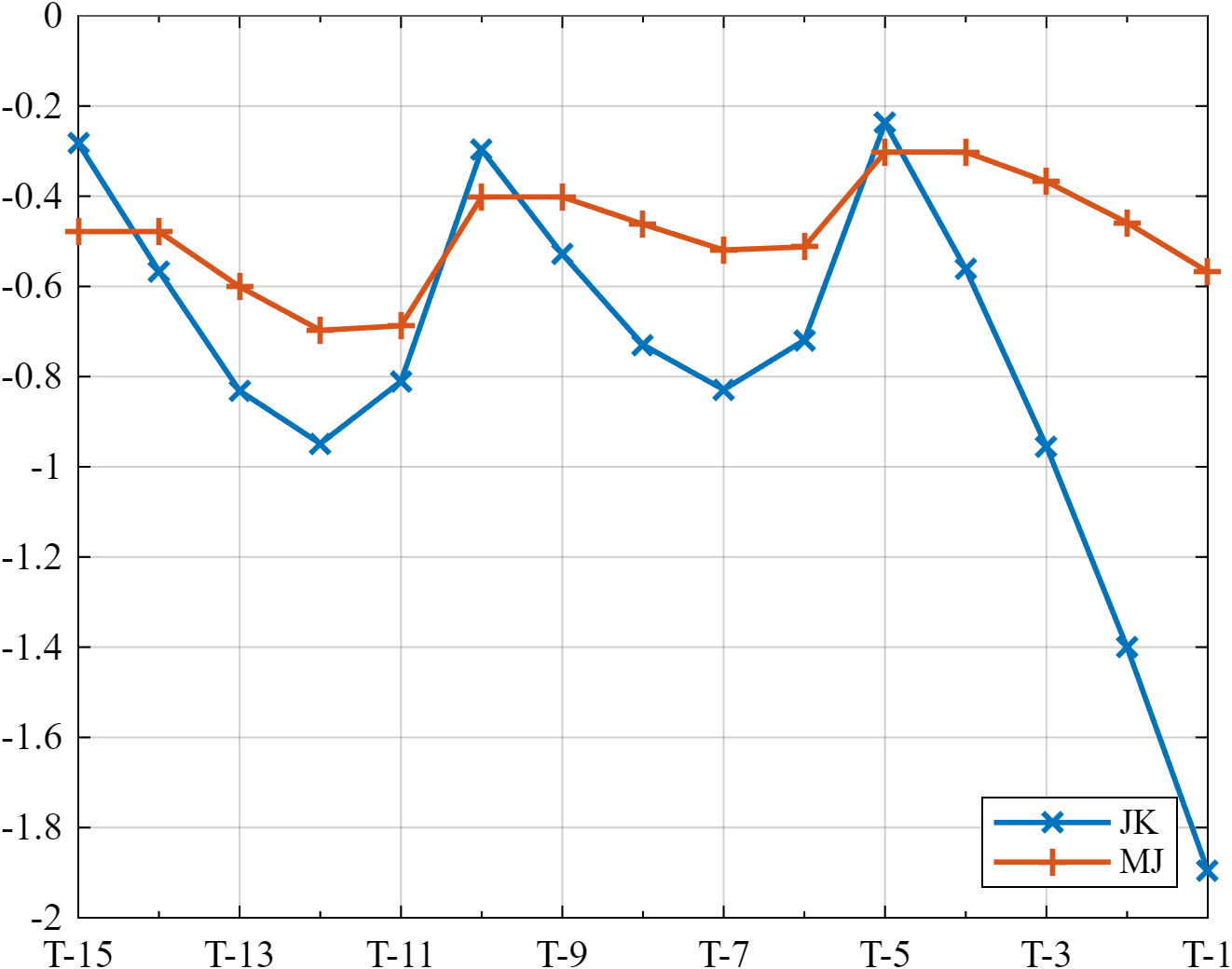}};
					\node[below=2pt of img] {\footnotesize $S$};
					\node[left=2pt of img, rotate=90, anchor=south] {\footnotesize Empirical Bias};
				\end{tikzpicture} &
				\begin{tikzpicture}
					\node[inner sep=0pt] (img) {\includegraphics[width=0.28\textwidth]{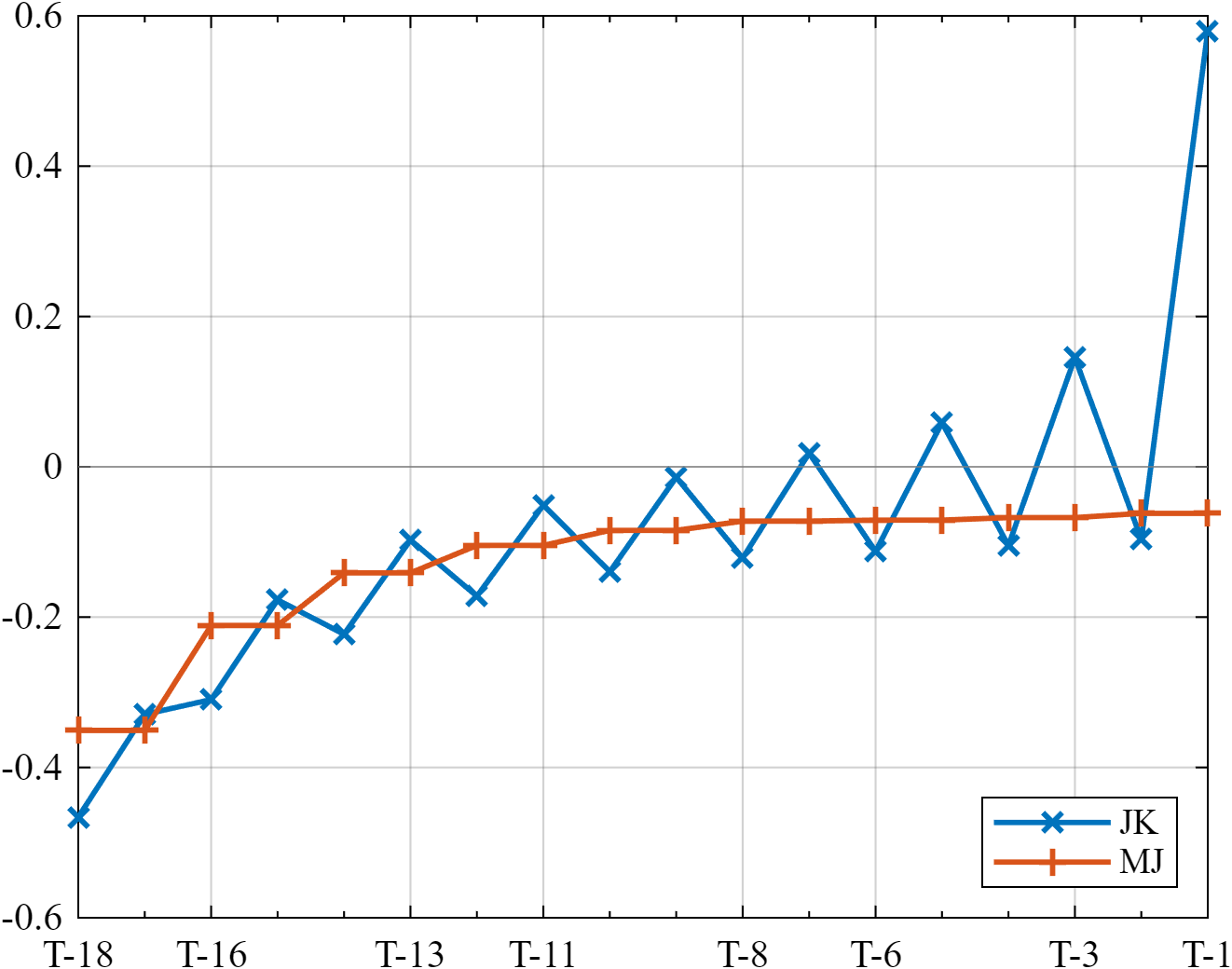}};
					\node[below=2pt of img] {\footnotesize $S$};
				\end{tikzpicture} &
				\begin{tikzpicture}
					\node[inner sep=0pt] (img) {\includegraphics[width=0.28\textwidth]{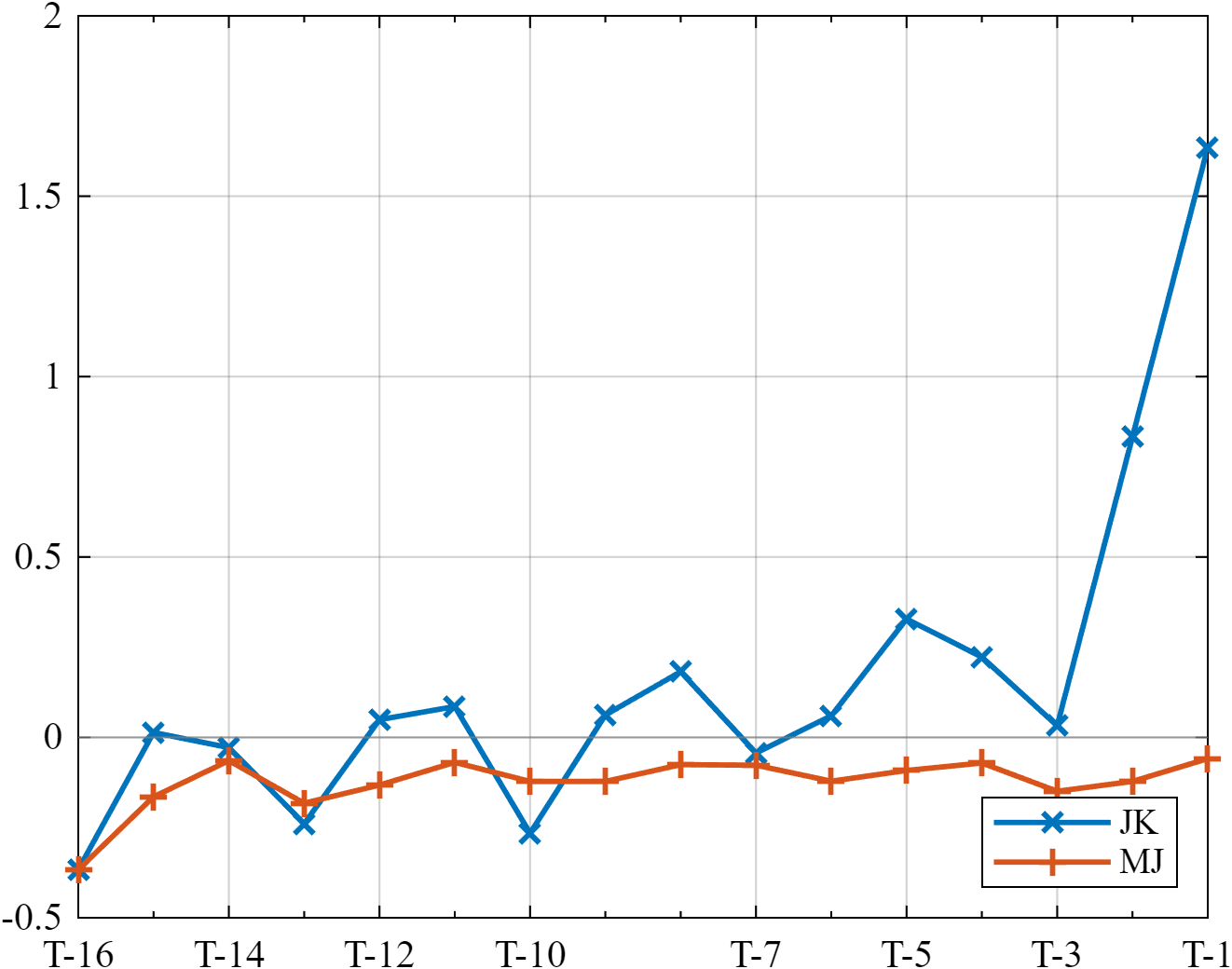}};
					\node[below=2pt of img] {\footnotesize $S$};
				\end{tikzpicture} \\[2pt]
				$\tau_0 = 0.2$ & $\tau_0 = 0.5$ & $\tau_0 = 0.7$
			\end{tabular}
			\caption{Empirical bias of the JK (blue) and MJ (red) estimators at $T = 20$ plotted against $S$.}
			\label{fig:master-curve-emp-endog}
	\end{figure}

	\paragraph{JK Bias-corrected Estimator.}
	The JK bias-corrected estimator behaves similarly to the exogenous case. At $S = T-1$ it exhibits a large bias which has the opposite sign to that of the FE-QIV estimator at every $\tau_0$. The standard deviation at $S=T-1$ is larger than at $S=T/2$, and the ratio grows with $T$. At $S = T/2$ the bias is smaller and the standard deviation is comparable to that of the FE-QIV estimator. Figure~\ref{fig:master-curve-emp-endog} plots the empirical bias of the JK estimator against $S$ at $T=20$. The oscillations in the JK bias resemble those of $\tau_0(1-\tau_0)+M_2(w_T^{\textup{JK}}(S),\tau_0)$. 

	\paragraph{MJ Bias-corrected Estimator.}
		The MJ estimator also behaves similarly to the exogenous case. At $S = T-1$ it exhibits much less bias than the JK estimator and less inflation of the standard deviation. At $S = T/2$ the standard deviation is close to that of the FE-QIV estimator. As in the exogenous case, the profile of the MJ bias is notably flatter than its JK counterpart, although residual bias remains. 

	\paragraph{Comparison.}
		Overall the picture from Table~\ref{tab:MC_IV} is similar to that of Table~\ref{tab:MC1}. The principal difference in the endogenous design is inconsistency of the FE-QR estimator. The JK bias-corrected estimator displays large bias and inflated standard deviations at $S = T - 1$. The MJ bias-corrected estimator is much less affected. At moderate values of $S$ both estimators reduce bias relative to the FE-QIV estimator while maintaining similar standard deviations. The SPJ estimator is broadly comparable in performance to MJ at $S = T/2$, both in terms of its bias and its standard deviation, though, as before, their biases can differ at small $T$.

\section{Empirical Application}
\label{sec:Empirical}

We study the effect of age on academic achievement of school children, following
\citet{BedardDhuey2006}. Those age effects are important, because initial maturity
differences in kindergarten and primary school can have long-lasting effects.
\citet{BedardDhuey2006} consider multiple countries and several different age
groups. We focus instead on New Zealand, and in particular on 3rd and 4th graders (9 year
olds in 1995). The data are from the Trends in International Mathematics and Science Study
(TIMSS), which provides internationally comparable student test scores in mathematics, together with various background characteristics. After omitting students with missing controls, as well as schools with fewer than five observations, the New Zealand panel contains $4{,}307$
students in $N = 148$ schools, with an average of 29 students per school. 
This pseudo-panel is unbalanced, but our techniques generalize to the
unbalanced case.
The outcome variable $y_{it}$ is the math test score of student $t$ in school $i$,
normalized to have mean $50$ and standard deviation $10$ over the whole sample.
The fixed effect $\alpha_i$ is a school-specific effect, and the main regressor of
interest is age, measured in months (mean $121$, std $7.2$). All specifications
include additional controls for gender, grade (3rd or 4th), whether the mother and
father are born in New Zealand, whether both parents are present, whether there is
a calculator, computer, or more than 100 books at home, and
whether the school is in a rural area (the latter only when school fixed effects
are excluded).

Age is potentially endogenous because of grade retention and early or late school
enrolment, the details of which are not observed. Following \citet{BedardDhuey2006}
we use age relative to the school cutoff date to instrument for age. The school
cutoff date in New Zealand is May 1, so we define relative age as $z_{it} = 0$ for
children born in April and $z_{it} = 11$ for children born in May. The exogeneity
of relative age is argued carefully in \citet{BedardDhuey2006}. Since we control
for school fixed effects, we only require exogeneity of relative age
conditional on those school-specific effects.

As a benchmark, we first estimate the model by OLS and 2SLS, with and without
school fixed effects. Results for the coefficient on age are reported in
Table~\ref{tab:nz_linear}. Full estimation results are reported
in Appendix~\ref{app:FullTables}.

\begin{table}[h]
	\centering
	\caption{Estimated Age Effect (standard error).}
    \vspace{6pt}
	\label{tab:nz_linear}
	\begin{tabular}{@{}cccc@{}}
		\toprule
		OLS & 2SLS & OLS + FE & 2SLS + FE \\
		\midrule
		$0.073$   & $0.464$   & $0.038$   & $0.351$   \\
		    $(0.031)$ & $(0.079)$ & $(0.030)$ & $(0.035)$ \\
		\bottomrule
	\end{tabular}
\end{table}

The 2SLS estimate is much larger than the OLS estimate, both with and without
school fixed effects, which is consistent with age being endogenous. Adding school
fixed effects reduces the 2SLS estimate from $0.46$ to $0.35$, which suggests that
relative age is correlated with unobserved school characteristics and motivates the
fixed effects quantile analysis below.

Table~\ref{tab:nz_quantile} presents estimates of the age effect $\theta(\tau)$ at
$\tau \in \{0.1,0.3,0.5,0.7,0.9\}$, obtained from instrumental variable quantile regression, without school fixed
effects (QIV), with school fixed effects (QIV+FE), and with fixed effects and
Jackknife bias correction (QIV+FE+BC).
\begin{table}[h]
	\centering
	\caption{Estimated age effect (standard error).}
    \vspace{6pt}
	\label{tab:nz_quantile}
	\begin{tabular}{@{}l ccccc@{}}
		\toprule
		& $\tau=0.1$ & $\tau=0.3$ & $\tau=0.5$ & $\tau=0.7$ & $\tau=0.9$ \\
		\midrule
		QIV  & $0.39$   & $0.55$   & $0.44$   & $0.39$   & $0.36$   \\
		                & $(0.13)$ & $(0.11)$ & $(0.08)$ & $(0.09)$ & $(0.14)$ \\[3pt]
		QIV + FE     & $0.29$   & $0.43$   & $0.38$   & $0.46$   & $0.42$   \\
		                & $(0.13)$ & $(0.12)$ & $(0.11)$ & $(0.15)$ & $(0.15)$ \\[3pt]
		QIV + FE + BC & $0.28$   & $0.43$   & $0.35$   & $0.41$   & $0.48$   \\
		                & $(0.13)$ & $(0.13)$ & $(0.11)$ & $(0.15)$ & $(0.15)$ \\
		\bottomrule
	\end{tabular}
\end{table}

Without controlling for school fixed effects (first row), the age effect is
largest in the lower-middle of the distribution, and declines thereafter. Taken at
face value, this would suggest that the age effect is larger for students with lower academic achievement. However, since the distribution of relative age may vary across schools, the instrument may itself be correlated with unobserved school characteristics, leading to bias in the QIV estimator. 
Indeed, once we include school fixed effects, but without bias correction (second row), the decline disappears and the effect is
instead slightly increasing in $\tau$. Finally, after applying our Jackknife bias
correction (third row), the estimates increase more clearly in $\tau$. Thus, contrary to the original QIV estimates, once we control for school fixed effects, the age effect is larger for students with higher academic achievement.

The results for New Zealand are particularly interesting when compared to other
countries in the TIMSS data set. Table~\ref{tab:slopes} reports the slope of the
estimated quantile profile $\theta(\tau)$, obtained by an unweighted least-squares
regression of the estimates at the nine deciles $\tau\in\{0.1,0.2,\ldots,0.9\}$
on $\tau$. The five estimates shown in Table~\ref{tab:nz_quantile} are a subset
of the nine used for this calculation. For every other country the bias-corrected slope is positive and changes
little across the three estimators. In contrast, for New Zealand, the
slope is negative ($-0.19$), becomes positive once school fixed effects are included $(0.11)$, and rises further with
bias correction ($0.19$).

\begin{table}[h]
	\centering
	\caption{Slope of $\theta(\tau)$.}
    \vspace{6pt}
	\label{tab:slopes}
	\begin{tabular}{@{}l ccc@{}}
		\toprule
		Country & \phantom{-}QIV & \phantom{-}QIV + FE & QIV + FE + BC \\
		\midrule
		Canada      & $\phantom{-}0.09$  & $\phantom{-}0.07$  & $0.08$ \\
		Czech Rep.  & $\phantom{-}0.04$  & $-0.04$ & $0.07$ \\
		England     & $\phantom{-}0.21$  & $\phantom{-}0.17$  & $0.17$ \\
		New Zealand & $-0.19$ & $\phantom{-}0.11$  & $0.19$ \\
		Norway      & $\phantom{-}0.11$  & $\phantom{-}0.04$  & $0.03$ \\
		Portugal    & $\phantom{-}0.25$  & $\phantom{-}0.27$  & $0.26$ \\
		\bottomrule
	\end{tabular}
\end{table}

Overall, these results illustrate the importance of controlling for fixed effects when studying quantile heterogeneity in panel models. In this case, correlation between the instrument and unobserved school characteristics may produce bias in the QIV estimates, to the extent that the effect of age on academic achievement is found to be greatest amongst students with the lowest academic performance. Controlling for school fixed effects points to the opposite conclusion, that the effect of age on academic achievement is instead greatest amongst students with the highest academic performance, a result which is supported by international comparison.

\section{Conclusions}
\label{sec:conclusions}

This paper develops inference methods for quantile regression models with
individual specific fixed effects in panels where both $N$ and $T$ are large.
The fixed effects approach causes an incidental parameter bias of order $1/T$,
which we characterize explicitly using induced order statistics and the beta distribution.
We propose bias-corrected estimators based on modified Jackknife reweighting
of the profiled moment conditions, establish their $\sqrt{NT}$-consistency
and asymptotic normality, and show that the choice of Jackknife subsample size~$S$
controls a bias-variance tradeoff that is specific to the quantile regression setting.
The framework covers both exogenous and endogenous regressors,
extending the quantile IV approach of \cite{ChernozhukovHansen2005,ChernozhukovHansen2006}
to panels with fixed effects.

Our empirical application to the effect of age on school children's test scores
illustrates the practical relevance of the bias correction.
Without controlling for school fixed effects, or when controlling for them
without bias correction, one may obtain a misleading picture of how the age effect
varies across the test score distribution. After bias correction, the age effect
increases more clearly across quantiles.

Several extensions are natural.
Our results are pointwise in the quantile $\tau_0$, but in applications
one may wish to do inference on the functional $\theta(\tau)$ across all $\tau$.
The bias characterization in Sections~\ref{sec:MomentCondition} and~\ref{sec:Interpretation} extends
directly to this setting, but joint inference requires more work.
Our conditional iid assumption rules out dynamic models with lagged dependent variables
or more general predetermined regressors.
Extending the bias correction framework to such settings would be valuable,
as dynamic quantile panel models are of considerable applied interest.
On the instrumental variables side,
extending the analysis to settings with potentially weak instruments,
building on the work of \cite{ChernozhukovHansen2008},
is another promising direction.
Finally, incorporating common time effects alongside the individual fixed effects
would broaden the applicability of the method.
We leave these extensions for future work.

\section*{Acknowledgment}
This paper benefited from the use of generative AI tools to assist with language editing, \LaTeX{} formatting, and checking proofs. All output was carefully reviewed by the authors. All substantive content, results, and any remaining errors are the authors' responsibility.


%

\appendix

\section{Additional Results and Implementation Details}
\label{app:AdditionalResults}

\subsection{Extension to Non-Identically Distributed Errors}
\label{sec:INID}

Fix an individual $i$ and condition on $(z_i,A_i)$. Suppose that the
residuals $u_{it}(\theta^0)$ are independent under this conditioning and
have continuous distribution functions
\[
 F_{it}(u)=\Pr\{u_{it}(\theta^0)\leq u\mid z_i,A_i\}.
\]
Identical distributions over $t$ are not required. The corresponding
conditional quantile restriction is $F_{it}(\alpha_i^0)=\tau_0$ for every
$t$. The finite-sample formula below holds even without this quantile
restriction.

For $j=0,\ldots,T-1$, define
\begin{equation}
 P_{i,-t,j}(u)
 =\sum_{\substack{B\subseteq{\bf T}\setminus\{t\}\\ |B|=j}}
       \prod_{s\in B}F_{is}(u)
       \prod_{s\in{\bf T}\setminus(B\cup\{t\})}\{1-F_{is}(u)\}.
 \label{eq:app-inid-rank-probability}
\end{equation}
This is the probability that exactly $j$ of the other residuals are below
$u$.
Conditioning on $u_{it}=u$ gives the exact identity
\begin{equation}
 \mathbb E\left[\left.\frac1T\sum_{q=1}^T
          w_{T,q}z_{i[q]}(\theta^0)\right|z_i,A_i\right]
 =\frac1T\sum_{t=1}^Tz_{it}
       \sum_{q=1}^Tw_{T,q}
             \int_{\mathbb R}P_{i,-t,q-1}(u)\,dF_{it}(u).
 \label{eq:app-inid-exact-bias}
\end{equation}
Indeed, observation $t$ has rank $q$ if and only if $q-1$ other residuals
are below it. Continuity excludes ties, and conditional independence gives
the product probabilities in \eqref{eq:app-inid-rank-probability}.

An equivalent expression uses
\[
 d_q=w_{T,q}-w_{T,q+1},\qquad
 P_{i,-t,\leq q-1}(u)=\sum_{j=0}^{q-1}P_{i,-t,j}(u).
\]
If $\tau_0w_{T,1}+(1-\tau_0)w_{T,T}=0$, then
\begin{equation}
 \mathbb E\left[\left.\frac1T\sum_{q=1}^T
          w_{T,q}z_{i[q]}(\theta^0)\right|z_i,A_i\right]
 =\frac1T\sum_{t=1}^Tz_{it}\sum_{q=1}^{T-1}d_q
       \left\{\int_{\mathbb R}P_{i,-t,\leq q-1}(u)\,dF_{it}(u)
                    -\tau_0\right\}.
 \label{eq:app-inid-summation-by-parts}
\end{equation}
This follows from $w_{T,r}=w_{T,T}+\sum_{q=r}^{T-1}d_q$ and
$\sum_{j=0}^{T-1}P_{i,-t,j}(u)=1$.

Under Assumption~\ref{ass:CI}, averaging over $z_i$ conditional on $A_i$
gives the beta representation in Theorem~\ref{th:finiteTbias}. The
leave-one-out residuals then have a common marginal distribution given
$A_i$, and the probability transforms of their order statistics have the
corresponding beta distributions. Under conditional independence alone,
the rank probabilities instead retain the products in
\eqref{eq:app-inid-rank-probability}.

Conditioning on $\alpha_i^0$ instead of $A_i$ requires a separate
independence assumption. Independence conditional on $A_i$ need not survive
averaging over components of $A_i$ that are not determined by $\alpha_i^0$.
Large-$T$ approximations under non-identical distributions also require
uniform restrictions on these distributions. The i.i.d.\ beta-moment expansions
in Theorems~\ref{th:largeTbias} and~\ref{th:VarJac} are not asserted under
conditional independence alone.

\subsection{Estimation of the Asymptotic Variance}
\label{app:VarEstimation}

We estimate the asymptotic covariance matrices in
Theorems~\ref{th:ExAsy}, \ref{th:BCAsy}, \ref{th:EndAsy}
and~\ref{th:BCAsy:End}.

\subsubsection{Exogenous case}

Define $\hat\varepsilon_{it}=y_{it}-z_{it}'\widehat\theta^{\rm FE}-\widehat\alpha_i$,
where $\widehat\alpha_i$ is the sample $\tau_0$-quantile of
$\{y_{it}-z_{it}'\widehat\theta^{\rm FE}\}_{t=1}^T$.
Let $K$ be a bounded, nonnegative, symmetric, Lipschitz continuous function
with compact support and $\int K(v)\,dv=1$.
Write $K_h(u)=K(u/h)/h$. We use a bandwidth $h=h_{NT}$ satisfying
\begin{align}
 h\longrightarrow0,\qquad
 \frac{\sqrt{\log(N+1)/T}+T^{-1}+(NT)^{-1/2}}{h^2}
 \longrightarrow0.
 \label{eq:variance-bandwidth}
\end{align}
Under $N/T^3\to0$, the choice $h=T^{-1/8}$ satisfies them.
Define
\begin{align*}
 \hat f_i&=\frac1T\sum_{t=1}^T K_h(\hat\varepsilon_{it}),
 &\hat\lambda_i&=
 \frac{\sum_{t=1}^T K_h(\hat\varepsilon_{it})z_{it}}
 {T\hat f_i}.
\end{align*}
The denominators are bounded away from zero with probability tending to one.
On the complementary event, set ratios with zero denominator equal to zero.
The estimators of $H$, $V^{\rm FE}$, and $V_{1,i}$ are
\begin{align*}
 \widehat H&=\frac1{NT}\sum_{i=1}^N\sum_{t=1}^T
 K_h(\hat\varepsilon_{it})z_{it}(z_{it}-\hat\lambda_i)',\\
 \widehat V^{\rm FE}&=\frac{\tau_0(1-\tau_0)}{NT}
 \sum_{i=1}^N\sum_{t=1}^T
 (z_{it}-\hat\lambda_i)(z_{it}-\hat\lambda_i)',\\
 \widehat V_{1,i}&=\frac1{T\hat f_i}\sum_{t=1}^T
 K_h(\hat\varepsilon_{it})(z_{it}-\hat\lambda_i)(z_{it}-\hat\lambda_i)'.
\end{align*}
For the bias-corrected estimator, use
$\widehat V^*=\widehat V^{\rm FE}+b_TN^{-1}\sum_i\widehat V_{1,i}$.
The corresponding covariance estimators are
$\widehat H^{-1}\widehat V^{\rm FE}\widehat H^{-1}$ and
$\widehat H^{-1}\widehat V^*\widehat H^{-1}$.

\subsubsection{Endogenous case}
\label{app:VarEstimation:End}

Define $\hat\varepsilon_{it}=y_{it}-x_{it}'\widehat\theta^{\,\rm FE\text{-}IV}
-\widehat\alpha_i$, with $\widehat\alpha_i$ the sample quantile of
$\{y_{it}-x_{it}'\widehat\theta^{\,\rm FE\text{-}IV}\}_{t=1}^T$.
Use these residuals in the definitions of $\hat f_i$, $\hat\lambda_i$,
$\widehat H$, $\widehat V^{\rm FE}$, and $\widehat V_{1,i}$.
The Jacobian estimator is
\[
 \widehat J=\frac1{NT}\sum_{i=1}^N\sum_{t=1}^T
 K_h(\hat\varepsilon_{it})x_{it}(z_{it}-\hat\lambda_i)'.
\]
Set $\widehat W=\widehat H^{-1}{\cal W}_{NT}\widehat H^{-1}$ and
$\widehat A=(\widehat J\widehat W\widehat J')^{-1}\widehat J\widehat W$.
The covariance estimators are $\widehat A\widehat V^{\rm FE}\widehat A'$
and $\widehat A\widehat V^*\widehat A'$.
When $x_{it}=z_{it}$, these formulas reduce to the exogenous formulas.

\subsubsection{Consistency}

Under the conditions of Theorem~\ref{th:ExAsy}
and~\eqref{eq:variance-bandwidth},
\[
 \widehat H^{-1}\widehat V^{\rm FE}\widehat H^{-1}
       \longrightarrow_P H^{-1}V^{\rm FE}H^{-1}.
\]
Under the conditions of Theorem~\ref{th:BCAsy}
and~\eqref{eq:variance-bandwidth}, the same conclusion
holds with $\widehat V^*$ and $V^*$ in place of
$\widehat V^{\rm FE}$ and $V^{\rm FE}$.
For the endogenous case, under the conditions of
Theorem~\ref{th:EndAsy} and~\eqref{eq:variance-bandwidth},
\[
 \widehat A\widehat V^{\rm FE}\widehat A'
 \longrightarrow_P
 (JWJ')^{-1}JW V^{\rm FE}WJ'(JWJ')^{-1}.
\]
Under the conditions of Theorem~\ref{th:BCAsy:End}
and~\eqref{eq:variance-bandwidth}, this conclusion
holds with $\widehat V^*$ and $V^*$ instead. In this last case,
the logarithmic growth condition of Theorem~\ref{th:EndAsy} is not needed.
The proof is in Appendix~\ref{app:VarianceConsistency}.

\subsection{Details for the Interpretation of the Bias}
\label{app:Interpretation}

Consider the scalar model
$y_{it}=z_{it}\theta(U_{it})+\alpha_i(U_{it})$, where $U_{it}$ is uniform
and independent of $z_{it}$ conditional on $A_i$. We first fix $i$ and
suppress its index and the conditioning on $A_i$.
To calculate the first-order effect of a varying slope, put
\[
 \theta_\varepsilon(v)=\theta^0+\varepsilon b(v),
 \qquad b(p)=0,\qquad p=\tau_0.
\]
Assume that $\alpha$ is strictly increasing and that $b$ is bounded. On a
fixed neighborhood of $p$, suppose that $b$ and $\alpha$ have four bounded
derivatives and that $\alpha'$ is bounded away from zero. The regressor is
bounded and has positive variance. For small $\varepsilon$, the expansions
below and their first two derivatives in $v$ hold uniformly on a smaller
neighborhood of $p$.

Write $c_\varepsilon(v)$ for the marginal $v$-quantile of the residual
$\alpha(U)+\varepsilon z b(U)$, and let $\bar z=\mathbb E[z]$.
Implicit differentiation of its distribution gives
\[
 c_\varepsilon(v)=\alpha(v)+\varepsilon\bar z b(v)+O(\varepsilon^2).
\]
If $v^*_{\varepsilon,z}$ solves
$\alpha(v^*_{\varepsilon,z})+\varepsilon z b(v^*_{\varepsilon,z})
=c_\varepsilon(v)$, substitution gives
\[
 G_\varepsilon(v|z)
 =v^*_{\varepsilon,z}
 =v-\varepsilon\frac{(z-\bar z)b(v)}{\alpha'(v)}
       +O_{C^2}(\varepsilon^2).
\]
The notation $O_{C^2}(\varepsilon^2)$ means that the remainder and its
first two derivatives are uniformly bounded by a constant times
$\varepsilon^2$. The stated derivative bounds and the positive lower bound
for $\alpha'$ justify this conclusion by differentiating the implicit
equation.

It follows that
\[
 g_\varepsilon(v)=\bar z(v-p)
    -\operatorname{Var}(z)\frac{\theta_\varepsilon(v)-\theta^0}{\alpha'(v)}
    +O_{C^2}(\varepsilon^2).
\]
Differentiating twice at $p$ gives
\begin{align}
 g_\varepsilon''(p)
 &=\operatorname{Var}(z)
    \left[\frac{2\alpha''(p)\theta_\varepsilon'(p)}{\alpha'(p)^2}
           -\frac{\theta_\varepsilon''(p)}{\alpha'(p)}\right]
       +O(\varepsilon^2).
 \label{gpp}
\end{align}
Since $H_\varepsilon=\operatorname{Var}(z)/\alpha'(p)+O(\varepsilon)$,
the leading asymptotic bias coefficient gives
\begin{align}
 -\frac1T H_\varepsilon^{-1}B_\varepsilon
 &=\frac{p(1-p)}{2T}\theta_\varepsilon''(p)
       +\delta\theta_\varepsilon'(p)+O(\varepsilon^2/T),
       \label{est-bias}\\
 \delta&=-\frac{p(1-p)}T\frac{\alpha''(p)}{\alpha'(p)}.\nonumber
\end{align}

Let $\phi_{T,p}$ denote the normal density with mean $p+\delta$ and
variance $p(1-p)/T$. Taylor expansion and the normal moments give
\[
 \int_0^1\theta_\varepsilon(v)\phi_{T,p}(v)\,dv
 =\theta^0+\delta\theta_\varepsilon'(p)
       +\frac{p(1-p)}{2T}\theta_\varepsilon''(p)+O(T^{-2}).
\]
For fixed interior $p$, the normal probability outside a fixed
neighborhood of $p$ is exponentially small. The normal integral therefore
reproduces the terms in \eqref{est-bias} that are first order in
$\varepsilon$ and leading order in $T^{-1}$, giving the informal smoothing
interpretation in \eqref{SmoothingInterpretation}.
The beta laws of the residual order statistics have the same leading
variance $p(1-p)/T$. Their finite-sample means need not equal $p+\delta$.

In the scalar endogenous model, assume $U$ is independent of the instrument
$z$ conditional on $A$, while $x$ may depend on $U$.
Write
\[
 C(v)=\operatorname{Cov}(z,x|A,U=v),\qquad
 h(v)=\frac{C(v)}{\alpha'(v)}.
\]
Suppose $h(p)\ne0$. Sufficient conditions for the local expansion are a
representation $x=X(U,z,V)$ with $U$ independent of $(z,V)$ conditional
on $A$, bounded $X$ and $z$, and four uniformly bounded continuous
derivatives of $X(u,z,V)$ in $u$ near $p$.
The conditions on $\alpha$ and $b$ are unchanged.
Conditional on $(z,V)$, the residual is
$\alpha(u)+\varepsilon X(u,z,V)b(u)$. Its local inverse has the bounded
derivatives used in the exogenous calculation. Averaging over $V$ given
$z$ preserves the $C^2$ remainder bound.
For the residual $\alpha(U)+\varepsilon x b(U)$, differentiation of its
conditional distribution at $\varepsilon=0$ gives
\[
 \left.\partial_\varepsilon F_\varepsilon(\alpha(v)|z,A)
       \right|_{\varepsilon=0}
 =-\frac{b(v)}{\alpha'(v)}\mathbb E[x|z,A,U=v].
\]
The corresponding marginal quantile derivative is
$b(v)\mathbb E[x|A,U=v]$.
Multiplication by $z$ and averaging therefore gives
\[
 g_\varepsilon(v)=\bar z(v-p)
       -h(v)\{\theta_\varepsilon(v)-\theta^0\}
       +O_{C^2}(\varepsilon^2).
\]
Thus $g_\varepsilon''(p)=-h(p)\theta_\varepsilon''(p)
-2h'(p)\theta_\varepsilon'(p)+O(\varepsilon^2)$ and
$J_\varepsilon=h(p)+O(\varepsilon)$.
The just-identified bias is
\[
 -\frac1T J_\varepsilon^{-1}B_\varepsilon
 =\frac{p(1-p)}{2T}\theta_\varepsilon''(p)
    +\frac{p(1-p)}T
       \left\{\frac{C'(p)}{C(p)}-\frac{\alpha''(p)}{\alpha'(p)}\right\}
          \theta_\varepsilon'(p)+O(\varepsilon^2/T).
\]
This covariance is conditional on $U=p$. It cannot generally be replaced
by the unconditional covariance of $z$ and $x$.

For heterogeneous individuals, let $h_i(v)=C_i(v)/\alpha_i'(v)$ and
write an overbar for the limiting cross-sectional average. The scalar,
just-identified displacement is
$p(1-p)\overline{h_i'(p)}/\{T\overline{h_i(p)}\}$. With several instruments
and one endogenous regressor, it is
$p(1-p)(JWJ')^{-1}JW\overline{h_i'(p)}/T$, where
$J'=\overline{h_i(p)}$. For a vector-valued coefficient the corresponding
correction is a matrix multiplying $\theta'(p)$.

\subsection{Full Estimation Results for the Empirical Application}
\label{app:FullTables}

This subsection reports the full estimation results for all covariates
in the empirical application of Section~\ref{sec:Empirical}.
Table~\ref{tab:APP-1} presents results for the linear model.
Tables~\ref{tab:APP-2}, \ref{tab:APP-3}, and~\ref{tab:APP-4}
present quantile IV results without fixed effects,
with fixed effects but no bias correction,
and with fixed effects and bias correction, respectively.

\begin{table}[H]
	\centering
	\caption{OLS and 2SLS estimates.}
    \vspace{6pt}
	\label{tab:APP-1}
	\begin{tabular}{@{}lcccc@{}}
		\toprule
		& OLS & 2SLS & OLS + FE & 2SLS + FE \\
		\midrule
		age           & $0.073$  & $0.464$   & $0.038$ & $0.351$ \\
		              & $(0.031)$ & $(0.079)$ & $(0.030)$ & $(0.035)$ \\[3pt]
		female        & $0.504$  & $0.656$   & $0.217$ & $0.356$ \\
		              & $(0.265)$ & $(0.272)$ & $(0.240)$ & $(0.257)$ \\[3pt]
		grade         & $5.130$  & $0.472$   & $5.727$ & $0.239$ \\
		              & $(0.447)$ & $(0.976)$ & $(0.432)$ & $(0.985)$ \\[3pt]
		mother-native & $0.145$  & $0.145$   & $0.627$ & $-0.276$ \\
		              & $(0.351)$ & $(0.361)$ & $(0.323)$ & $(0.356)$ \\[3pt]
		father-native & $-0.078$ & $-0.092$  & $0.173$ & $-0.323$ \\
		              & $(0.323)$ & $(0.331)$ & $(0.302)$ & $(0.321)$ \\[3pt]
		both-parents  & $2.269$  & $2.348$   & $1.853$ & $2.010$ \\
		              & $(0.307)$ & $(0.314)$ & $(0.282)$ & $(0.307)$ \\[3pt]
		calculator    & $4.485$  & $4.399$   & $3.293$ & $3.547$ \\
		              & $(0.431)$ & $(0.442)$ & $(0.408)$ & $(0.513)$ \\[3pt]
		computer      & $2.208$  & $2.137$   & $1.153$ & $2.030$ \\
		              & $(0.270)$ & $(0.274)$ & $(0.248)$ & $(0.260)$ \\[3pt]
		books100      & $3.525$  & $3.624$   & $2.575$ & $3.487$ \\
		              & $(0.278)$ & $(0.283)$ & $(0.259)$ & $(0.268)$ \\[3pt]
		rural         & $-0.396$ & $-0.371$  &         &         \\
		              & $(0.356)$ & $(0.362)$ &         &         \\[3pt]
		constant      & $14.731$ & $-16.236$ &         &         \\
		              & $(2.697)$ & $(6.404)$ &         &         \\
		\bottomrule
	\end{tabular}
\end{table}

\begin{table}[H]
	\centering
	\caption{QIV without fixed effects.}
    \vspace{6pt}
	\label{tab:APP-2}
	\begin{tabular}{@{}lccccc@{}}
		\toprule
		& $\tau=0.1$ & $\tau=0.3$ & $\tau=0.5$ & $\tau=0.7$ & $\tau=0.9$ \\
		\midrule
		age           & $0.39$ & $0.55$ & $0.44$ & $0.39$ & $0.36$ \\
		              & $(0.13)$ & $(0.11)$ & $(0.08)$ & $(0.09)$ & $(0.14)$ \\[3pt]
		female        & $1.74$ & $1.35$ & $0.49$ & $0.36$ & $-0.44$ \\
		              & $(0.48)$ & $(0.36)$ & $(0.32)$ & $(0.35)$ & $(0.49)$ \\[3pt]
		grade         & $0.49$ & $-1.12$ & $0.75$ & $1.73$ & $2.20$ \\
		              & $(1.76)$ & $(1.34)$ & $(1.00)$ & $(1.10)$ & $(1.79)$ \\[3pt]
		mother-native & $1.24$ & $0.42$ & $-0.23$ & $-0.11$ & $-0.67$ \\
		              & $(0.63)$ & $(0.58)$ & $(0.40)$ & $(0.52)$ & $(0.55)$ \\[3pt]
		father-native & $-0.35$ & $-0.24$ & $-0.00$ & $0.14$ & $-0.09$ \\
		              & $(0.61)$ & $(0.51)$ & $(0.36)$ & $(0.44)$ & $(0.59)$ \\[3pt]
		both-parents  & $1.93$ & $2.27$ & $2.57$ & $2.55$ & $2.32$ \\
		              & $(0.54)$ & $(0.40)$ & $(0.37)$ & $(0.40)$ & $(0.59)$ \\[3pt]
		calculator    & $4.32$ & $4.53$ & $3.81$ & $5.24$ & $3.56$ \\
		              & $(0.67)$ & $(0.60)$ & $(0.50)$ & $(0.58)$ & $(0.82)$ \\[3pt]
		computer      & $2.40$ & $2.16$ & $2.00$ & $1.99$ & $3.00$ \\
		              & $(0.52)$ & $(0.36)$ & $(0.33)$ & $(0.36)$ & $(0.51)$ \\[3pt]
		books100      & $3.50$ & $3.69$ & $3.74$ & $3.44$ & $2.97$ \\
		              & $(0.49)$ & $(0.36)$ & $(0.34)$ & $(0.37)$ & $(0.51)$ \\[3pt]
		rural         & $-1.16$ & $-0.05$ & $-1.03$ & $-0.32$ & $0.42$ \\
		              & $(0.77)$ & $(0.42)$ & $(0.43)$ & $(0.50)$ & $(0.72)$ \\[3pt]
		constant      & $-18.84$ & $-25.79$ & $-14.07$ & $-7.27$ & $2.82$ \\
		              & $(10.88)$ & $(9.54)$ & $(6.37)$ & $(7.42)$ & $(10.48)$ \\
		\bottomrule
	\end{tabular}
\end{table}

\begin{table}[H]
	\centering
	\caption{QIV with fixed effects.}
        \vspace{6pt}
	\label{tab:APP-3}
	\begin{tabular}{@{}lccccc@{}}
		\toprule
		& $\tau=0.1$ & $\tau=0.3$ & $\tau=0.5$ & $\tau=0.7$ & $\tau=0.9$ \\
		\midrule
		age           & $0.29$ & $0.43$ & $0.38$ & $0.46$ & $0.42$ \\
		              & $(0.13)$ & $(0.12)$ & $(0.11)$ & $(0.15)$ & $(0.15)$ \\[3pt]
		female        & $1.33$ & $0.86$ & $0.65$ & $0.27$ & $0.12$ \\
		              & $(0.58)$ & $(0.41)$ & $(0.40)$ & $(0.49)$ & $(0.60)$ \\[3pt]
		grade         & $2.56$ & $0.97$ & $1.84$ & $1.08$ & $1.52$ \\
		              & $(1.71)$ & $(1.44)$ & $(1.40)$ & $(1.79)$ & $(2.13)$ \\[3pt]
		mother-native & $0.87$ & $0.77$ & $0.88$ & $0.70$ & $0.03$ \\
		              & $(0.82)$ & $(0.62)$ & $(0.48)$ & $(0.62)$ & $(0.94)$ \\[3pt]
		father-native & $-0.13$ & $0.23$ & $0.02$ & $-0.22$ & $0.41$ \\
		              & $(0.67)$ & $(0.51)$ & $(0.45)$ & $(0.53)$ & $(0.75)$ \\[3pt]
		both-parents  & $1.73$ & $2.12$ & $1.82$ & $1.39$ & $1.53$ \\
		              & $(0.70)$ & $(0.57)$ & $(0.48)$ & $(0.53)$ & $(0.66)$ \\[3pt]
		calculator    & $1.95$ & $2.42$ & $2.88$ & $3.31$ & $3.06$ \\
		              & $(1.04)$ & $(0.78)$ & $(0.79)$ & $(0.68)$ & $(1.04)$ \\[3pt]
		computer      & $0.51$ & $1.18$ & $1.23$ & $1.28$ & $1.63$ \\
		              & $(0.57)$ & $(0.46)$ & $(0.41)$ & $(0.42)$ & $(0.60)$ \\[3pt]
		books100      & $2.66$ & $2.66$ & $2.63$ & $2.59$ & $3.11$ \\
		              & $(0.50)$ & $(0.47)$ & $(0.40)$ & $(0.50)$ & $(0.68)$ \\
		\bottomrule
	\end{tabular}
\end{table}

\begin{table}[!htbp]
	\centering
	\caption{QIV with fixed effects and bias correction.}
        \vspace{6pt}
	\label{tab:APP-4}
	\begin{tabular}{@{}lccccc@{}}
		\toprule
		& $\tau=0.1$ & $\tau=0.3$ & $\tau=0.5$ & $\tau=0.7$ & $\tau=0.9$ \\
		\midrule
		age           & $0.28$ & $0.43$ & $0.35$ & $0.41$ & $0.48$ \\
		              & $(0.13)$ & $(0.13)$ & $(0.11)$ & $(0.15)$ & $(0.15)$ \\[3pt]
		female        & $1.58$ & $0.86$ & $0.73$ & $0.31$ & $0.26$ \\
		              & $(0.59)$ & $(0.42)$ & $(0.41)$ & $(0.53)$ & $(0.60)$ \\[3pt]
		grade         & $2.93$ & $0.99$ & $2.29$ & $1.90$ & $0.96$ \\
		              & $(1.80)$ & $(1.54)$ & $(1.36)$ & $(1.75)$ & $(1.88)$ \\[3pt]
		mother-native & $0.96$ & $0.97$ & $1.03$ & $0.59$ & $-0.42$ \\
		              & $(0.84)$ & $(0.65)$ & $(0.47)$ & $(0.67)$ & $(1.04)$ \\[3pt]
		father-native & $0.11$ & $0.11$ & $-0.11$ & $-0.17$ & $-0.02$ \\
		              & $(0.76)$ & $(0.54)$ & $(0.46)$ & $(0.55)$ & $(0.78)$ \\[3pt]
		both-parents  & $1.72$ & $2.15$ & $1.67$ & $1.45$ & $1.03$ \\
		              & $(0.72)$ & $(0.59)$ & $(0.48)$ & $(0.55)$ & $(0.74)$ \\[3pt]
		calculator    & $2.11$ & $2.70$ & $3.04$ & $3.46$ & $3.30$ \\
		              & $(1.07)$ & $(0.80)$ & $(0.76)$ & $(0.73)$ & $(1.30)$ \\[3pt]
		computer      & $0.23$ & $1.22$ & $1.32$ & $1.20$ & $1.86$ \\
		              & $(0.56)$ & $(0.47)$ & $(0.42)$ & $(0.46)$ & $(0.64)$ \\[3pt]
		books100      & $2.88$ & $2.65$ & $2.69$ & $2.78$ & $3.12$ \\
		              & $(0.53)$ & $(0.49)$ & $(0.41)$ & $(0.52)$ & $(0.74)$ \\
		\bottomrule
	\end{tabular}
\end{table}

\FloatBarrier
\FloatBarrier

\section{Proofs}
\label{app:MomentCondition}
\label{app:Proofs}

\subsection{Proof of Theorem~\ref{th:ExAsy}}
\label{app:Exog}

We condition on $A=(A_1,\ldots,A_N)$ and write $\mathbb E_A$,
$\operatorname{Var}_A$ and $\mathbb P_A$ for conditional expectations,
variances and probabilities. Let $r_{NT}=(NT)^{-1/2}$ and
$s_{i,T}(\theta)=dL_{i,T}(\theta)/d\theta$.
We use $f_y(\cdot\mid v,A_i)$ for the conditional density of the outcome
$y$ given a regressor vector $v$ and $A_i$, to distinguish it from the
residual densities below.

\subsubsection{Properties of the profile objective}

\begin{lemma}[Properties of the profile objective]
\label{lemma:profile}
Under Assumptions~\ref{ass:CI} and~\ref{ass:RC}, the profile objective is
convex and integrable. At each fixed $\theta$ it is differentiable with
probability one, with
\begin{align}
 s_{i,T}(\theta)
   =\frac1T\sum_{q=1}^T w^{\rm FE}_{T,q}z_{i[q]}(\theta)
   =\frac1T\sum_{t=1}^Tz_{it}\Psi_{it}(\theta).
 \label{GradientFormula}
\end{align}
Its conditional expectation is twice continuously differentiable near
$\theta^0$, and differentiation and conditional expectation commute.
\end{lemma}

\begin{proof}
The check loss is jointly convex in $(\theta,\alpha_i)$. Minimizing over
$\alpha_i$ preserves convexity. Also,
$0\leq L_{i,T}(\theta)\leq T^{-1}\sum_t|y_{it}-z_{it}'\theta|$,
which is conditionally integrable by Assumption~\ref{ass:RC}(iii).

The gradient formula is proved in Appendix~\ref{app:FEweights}. It is
uniformly bounded. Dominated
convergence therefore gives
\begin{align*}
 \frac{d\mathbb E_A L_{i,T}(\theta)}{d\theta}
       =\mathbb E_A s_{i,T}(\theta).
\end{align*}
The second derivative and its continuity follow from the formula in
Lemma~\ref{lemma:hessian} below.
\end{proof}

\subsubsection{Hessian of the expected objective}

\begin{lemma}[Hessian smoothness]
\label{lemma:hessian}
Under Assumptions~\ref{ass:CI} and~\ref{ass:RC}, uniformly over $i$,
\begin{align*}
 \left.\frac{d^2\mathbb E_A L_{i,T}(\theta)}{d\theta d\theta'}
 \right|_{\theta=\theta^0}=H_i+O(T^{-1}).
\end{align*}
On a fixed neighborhood of $\theta^0$, these Hessians are continuously
differentiable with uniformly bounded first derivatives. They are
therefore Lipschitz continuous with a constant independent of $i$ and $T$.
The same conclusions apply to the joint profile objective with bounded
regressor vector $v_{it}=(x_{it}',z_{it}')'$ under Assumption~\ref{ass:End}.
\end{lemma}

\begin{proof}
For a bounded regressor vector $v$, write the residual as $y-v'\beta$.
Conditional on $A_i$, denote its distribution and density by
$F_{i,\beta}$ and $f_{i,\beta}$, and its $p$-quantile by
$c_{i,\beta}(p)$. Define
\begin{align*}
 \lambda_{i,\beta}(p)
   &=\mathbb E[v\mid y-v'\beta=c_{i,\beta}(p),A_i],\\
 K_{i,\beta}(p)
   &=f_{i,\beta}(c_{i,\beta}(p))
      \operatorname{Var}[v\mid y-v'\beta=c_{i,\beta}(p),A_i].
\end{align*}
Let $k=\lfloor\tau_0T\rfloor$, $a=\langle\tau_0T\rangle$, and
$f_{\cal B}(p,j,l)=p^{j-1}(1-p)^{l-1}/\mathrm B(j,l)$.
For all sufficiently large $T$, the expected profile Hessian is
\begin{align}
 \mathcal H_{i,T}(\beta)
  &=\int_0^1 K_{i,\beta}(p)
       \big\{(1-a)f_{\cal B}(p,k,T-k)+a f_{\cal B}(p,k+1,T-k-1)\big\}\,dp.
 \label{eq:profile-hessian-beta}
\end{align}

To obtain this formula, consider a pair of residuals that cross at $u$,
with $q-1$ of the remaining $T-2$ residuals below $u$. The change in the
gradient across the crossing is determined by
$d_q=w^{\rm FE}_{T,q}-w^{\rm FE}_{T,q+1}$. Integrating over the crossing
and summing over the $T(T-1)/2$ unordered pairs gives
\begin{align*}
 \mathcal H_{i,T}(\beta)
  =\frac{T-1}{2}\sum_{q=1}^{T-1}d_q\int_{\mathbb R}
  &\mathbb E\!\left[(v-\widetilde v)(v-\widetilde v)'
     f_y(u+v'\beta\mid v,A_i)
     f_y(u+\widetilde v'\beta\mid\widetilde v,A_i)\mid A_i\right]\\[-2mm]
  &\qquad\times \binom{T-2}{q-1}
       F_{i,\beta}(u)^{q-1}
       [1-F_{i,\beta}(u)]^{T-q-1}\,du,
\end{align*}
where $\widetilde v$ is an independent copy of $v$ conditional on $A_i$.
The expectation in this integral equals
$2f_{i,\beta}(u)^2\operatorname{Var}(v\mid y-v'\beta=u,A_i)$.
Under the substitution $p=F_{i,\beta}(u)$, we have
$f_{i,\beta}(u)\,du=dp$ and
$(T-1)\binom{T-2}{q-1}=1/\mathrm B(q,T-q)$.
The remaining factor is $K_{i,\beta}(p)$.
Using $d_k=1-a$, $d_{k+1}=a$ gives~\eqref{eq:profile-hessian-beta}.
The bounded conditional densities
justify differentiation under the crossing integrals. The formula is
continuous in $\beta$.

The mixture in~\eqref{eq:profile-hessian-beta} has mean exactly
$(1-a)k/T+a(k+1)/T=\tau_0$, variance $O(T^{-1})$, and exponentially
small mass outside any fixed neighborhood of $\tau_0$.
The density is bounded away from zero near the target quantile, uniformly
for $\beta$ near its true value. The first two derivatives of
$K_{i,\beta}(p)$ with respect to $p$ are consequently bounded there.
Also $K_{i,\beta}$ is bounded on $(0,1)$, because $v$ and the density are
bounded. Taylor expansion on the neighborhood and the exponential bound
outside it give
\begin{align*}
 \mathcal H_{i,T}(\beta)=K_{i,\beta}(\tau_0)+O(T^{-1}).
\end{align*}
At the true value, the $z,z$ block is $H_i$, the $x,z$ block is $J_i$,
and the $x,x$ block gives the corresponding joint Hessian.

For the Lipschitz bound, suppress the indices and let
$s=c_\beta(p)+v'\beta$. Differentiating
$F_\beta(c_\beta(p))=p$ gives
$\partial c_\beta(p)/\partial\beta=-\lambda_\beta(p)$.
Write $K=M_2-M_1M_1'/M_0$, where
$M_j=\mathbb E[v^{\otimes j}f_y(s\mid v,A_i)\mid A_i]$.
The derivatives of the conditional mean cancel, leaving
\begin{align*}
 \frac{\partial K_\beta(p)}{\partial\beta_l}
 =\mathbb E\!\left[(v-\lambda_\beta(p))(v-\lambda_\beta(p))'
       (v_l-\lambda_{\beta,l}(p))
       f_y'(s\mid v,A_i)\mid A_i\right].
\end{align*}
This is uniformly bounded. The quantile map is Lipschitz in $\beta$,
since every residual changes by at most $C\|\beta'-\beta\|$.
The derivative formula holds almost everywhere along parameter segments.
At density-zero points extend $K$ continuously by zero. Integrating the bound
in~\eqref{eq:profile-hessian-beta} proves the Lipschitz bound.
The derivative integrand is continuous for almost every $p$, since
$f_{i,\beta}(c_{i,\beta}(p))>0$ for almost every $p$. Dominated
convergence also gives continuity of the Hessian derivative.
\end{proof}

\subsubsection{Probability and moment bounds}
\label{app:measurability}

The suprema used below and in Appendices~\ref{app:Endog}
and~\ref{app:BiasCorrection} are measurable.
For each sample, the residual-ordering hyperplanes divide the parameter
space into finitely many faces, including their lower-dimensional
boundaries. The selected rank scores are constant on each face.
Their conditional means are continuous in the parameters by dominated
convergence, since ties have probability zero at each fixed parameter.
On each face intersected with a parameter ball, choose a countable dense
set by measurable selections. Such selections can be constructed by
projecting rational points onto closed subsets with strict inequality
slacks at least $1/n$, for $n=1,2,\ldots$.
Continuity within a face makes each centered supremum a countable
supremum of measurable evaluations. The same argument applies to pairs
of parameters and to the independent samples used in symmetrization.
For continuous objective remainders, an ordinary countable dense set
suffices.

\begin{lemma}[Probability bounds for indicator changes]
\label{lemma:indicator}
Let $\bar\theta$ be fixed conditional on $A$, in a fixed neighborhood of
$\theta^0$, and let $\|h\|=d$. Define
$\bar\Psi_{it}=\int_0^1\Psi_{it}(\bar\theta+vh)\,dv$.
Uniformly over the indices and such $\bar\theta,h$,
\begin{align*}
 \mathbb P_A(\bar\Psi_{it}\ne\Psi_{it}(\bar\theta))
   &\leq C(d+T^{-1}),\\
 \mathbb P_A\big((\bar\Psi_{it}-\Psi_{it}(\bar\theta))
       (\bar\Psi_{is}-\Psi_{is}(\bar\theta))\ne0\big)
   &\leq C(d^2+d/T+T^{-2})
\end{align*}
for $s\ne t$. The same bounds hold with $\bar\Psi_{it}$ replaced by
$\Psi_{it}(\bar\theta+h)$.
\end{lemma}

\begin{proof}
Each residual and each residual order statistic changes by at most $C d$
along the segment. A change of rank weight is therefore possible only
for a residual within $2Cd$ of the original target order statistic.
The full-sample target lies between the adjacent leave-one-out order
statistics. Conditional on the remaining observations, the probability
of the enlarged interval is bounded by $Cd$ plus its probability
spacing. The expected probability spacing is $O(T^{-1})$.

For two observations, the full-sample target lies between order
statistics $q_T-2$ and $q_T$ of the leave-two-out sample. Both residuals
must lie in this interval enlarged by $2Cd$. Conditional independence
bounds their joint probability by the square of the interval probability.
The first two moments of its probability spacing are $O(T^{-1})$ and
$O(T^{-2})$. Expanding the square gives the second bound. This argument
uses probability spacings, so it does not require a density lower bound
outside the target neighborhood.
\end{proof}

\begin{lemma}[Score increments]
\label{lemma:scoreincrement}
For fixed parameters $\beta,\beta'$ in the parameter neighborhood, put
$d=\|\beta'-\beta\|$. If $k/T\in[\epsilon,1-\epsilon]$ for a fixed
$\epsilon>0$, and $P_{i,k}(\beta)$ is the sum of the bounded regressors
attached to the $k$ smallest residuals, then
\begin{align}
 \mathbb E_A\|P_{i,k}(\beta')-P_{i,k}(\beta)\|^2
       \leq C\{Td+(Td)^2\}.
 \label{eq:central-cutoff-count-bound}
\end{align}
The constant is uniform over these cutoffs. In particular, the FE score
and each component of the joint profile score satisfy
\begin{align}
 \mathbb E_A\|s_{i,T}(\beta')-s_{i,T}(\beta)\|^2
       \leq C(d/T+d^2).
 \label{eq:FE-score-increment}
\end{align}
\end{lemma}

\begin{proof}
Every residual and its $k$th order statistic moves by at most $Cd$.
An observation that changes membership in the lowest $k$ ranks must
therefore have its original residual within $2Cd$ of the original
$k$th order statistic. Let $J$ count the observations in that interval
other than the $k$th observation itself. If the $k$th observation changes
membership, at least one other observation must change membership too.
The total number of changed memberships is at most $2J$.

Let $\xi=F_{i,\beta}(u_{i(k)}(\beta))$. Conditional on $\xi$, the
observations below and above the cutoff are independent truncated
samples. Their interval probabilities are bounded by $Cd/\xi$ and
$Cd/(1-\xi)$, respectively. Their counts therefore satisfy
\begin{align*}
 \mathbb E_A[J^2\mid\xi]
 \leq C\left[Td\{\xi^{-1}+(1-\xi)^{-1}\}
       +T^2d^2\{\xi^{-2}+(1-\xi)^{-2}\}\right].
\end{align*}
The inverse moments of central uniform order statistics are uniformly
bounded. Together with bounded regressors, this proves
\eqref{eq:central-cutoff-count-bound}.
The FE weights are the sum of a constant and two adjacent cutoff
indicators with nonnegative coefficients summing to one. Dividing the
cutoff bound by $T^2$ gives~\eqref{eq:FE-score-increment}.
\end{proof}

Replacing one observation in~\eqref{GradientFormula} changes
$s_{i,T}(\theta)$ by at most $C/T$. The other observations move by at
most one rank, and the FE weights have total variation one. The
bounded-difference inequality, applied coordinatewise, consequently gives
\begin{align}
 \sup_{i,T,\theta}
 \mathbb E_A\big\|\sqrt T
       (s_{i,T}(\theta)-\mathbb E_A s_{i,T}(\theta))\big\|^4<C.
 \label{eq:profile-score-fourth-moment}
\end{align}
The supremum is over the fixed parameter neighborhood. The
tail bound is $2\exp(-cTu^2)$ for each coordinate of the centered score,
and integrating this bound proves~\eqref{eq:profile-score-fourth-moment}.

\subsubsection{Quadratic expansion and linearization}

\begin{lemma}[Quadratic expansion]
\label{lemma:expansion}
Let $\bar\theta$ be measurable with respect to $A$ and converge to
$\theta^0$. Suppose that the conditional expected Hessian at
$\bar\theta$ converges to $H$. For every fixed $M<\infty$,
\begin{align*}
 \sup_{\|h\|\leq M r_{NT}}
 \left|\mathcal L_{NT}(\bar\theta+h)-\mathcal L_{NT}(\bar\theta)
       -s_{NT}(\bar\theta)'h-\tfrac12h'Hh\right|
   =o_P(r_{NT}^2),
\end{align*}
where $s_{NT}=N^{-1}\sum_i s_{i,T}$.
The result also holds for the joint parameter $(\theta',\gamma')'$,
without requiring its full Hessian to be positive definite, and is
unchanged by adding an affine function to the objective.
\end{lemma}

\begin{proof}
Let
$R_{i,T}(h)=L_{i,T}(\bar\theta+h)-L_{i,T}(\bar\theta)
-s_{i,T}(\bar\theta)'h$.
Absolute continuity along the segment gives
\begin{align*}
 R_{i,T}(h)
   =\frac1T\sum_t z_{it}'h
       \int_0^1[\Psi_{it}(\bar\theta+vh)
                        -\Psi_{it}(\bar\theta)]\,dv.
\end{align*}
By Lemma~\ref{lemma:indicator} and independence across individuals,
\begin{align}
 \operatorname{Var}_A\left[\frac1N\sum_iR_{i,T}(h)\right]
   \leq \frac C N\left(d^4+\frac{d^3}{T}+\frac{d^2}{T^2}\right),
 \qquad d=\|h\|.
 \label{eq:profile-remainder-variance}
\end{align}
At $d=O(r_{NT})$, division by $r_{NT}^4$ leaves terms of orders
$N^{-1}$, $r_{NT}$ and $T^{-1}$. They converge to zero.
Lemma~\ref{lemma:hessian} and the integral Taylor formula give
$N^{-1}\sum_i\mathbb E_A R_{i,T}(h)=\tfrac12h'Hh+o_P(r_{NT}^2)$.
Subtracting a linear function preserves convexity. The convexity lemma
of \cite{Pollard1991} therefore makes this pointwise expansion uniform
on bounded sets of $h/r_{NT}$.
\end{proof}

\begin{lemma}[Linearization]
\label{lemma:linearize}
Under the conditions of Lemma~\ref{lemma:expansion}, suppose that
$H>0$ and $s_{NT}(\bar\theta)=O_P(r_{NT})$. Every minimizer of the
convex objective satisfies
\begin{align*}
 \widehat\theta-\bar\theta
    =-H^{-1}s_{NT}(\bar\theta)+o_P(r_{NT}).
\end{align*}
\end{lemma}

\begin{proof}
For a sufficiently large fixed $M$, the quadratic term in the uniform
expansion exceeds the linear term on $\|h\|=Mr_{NT}$ with arbitrarily
high probability. The objective attains a minimum inside this ball.
Convexity makes it a global minimum and excludes other minimizers
outside the ball. Completing the square and comparing the objective at
a minimizer and at $\bar\theta-H^{-1}s_{NT}(\bar\theta)$ gives
\begin{align*}
 \tfrac12\big(\widehat\theta-\bar\theta
       +H^{-1}s_{NT}(\bar\theta)\big)'H
       \big(\widehat\theta-\bar\theta
       +H^{-1}s_{NT}(\bar\theta)\big)=o_P(r_{NT}^2).
\end{align*}
Positive definiteness gives the conclusion.
\end{proof}

\begin{proof}[Proof of Theorem~\ref{th:ExAsy}]
The moment results give
\begin{align*}
 \mathbb E_A s_{NT}(\theta^0)&=B_{NT}/T+O(T^{-2}),\\
 \frac1N\sum_i T\operatorname{Var}_A(s_{i,T}(\theta^0))
    &\longrightarrow_P V^{\rm FE}.
\end{align*}
The fourth-moment bound~\eqref{eq:profile-score-fourth-moment} implies
the conditional Lyapunov condition. Consequently,
\begin{align}
 \sqrt{NT}\{s_{NT}(\theta^0)-\mathbb E_A s_{NT}(\theta^0)\}
    \Rightarrow\mathcal N(0,V^{\rm FE}).
 \label{eq:profile-centered-clt}
\end{align}

Set $\bar\theta=\theta^0-H^{-1}B_{NT}/T$. The quantities $B_i$ and
$H_i$ are bounded. Random sampling and Lemma~\ref{lemma:hessian} give
\begin{align*}
 \frac{d^2\mathbb E_A\mathcal L_{NT}(\theta^0)}{d\theta d\theta'}
   =H+O_P(N^{-1/2}+T^{-1}).
\end{align*}
The Hessian Lipschitz bound then implies
\begin{align*}
 \mathbb E_A s_{NT}(\bar\theta)
    =O_P(T^{-2}+T^{-1}N^{-1/2})=o_P(r_{NT}).
\end{align*}
Here $\sqrt{NT}/T^2=\sqrt{N/T^3}\to0$.

Lemma~\ref{lemma:indicator}, with $d=O(T^{-1})$, gives
\begin{align*}
 \operatorname{Var}_A\big[s_{NT}(\bar\theta)-s_{NT}(\theta^0)\big]
    \leq C/(NT^2).
\end{align*}
The centered difference is $o_P(r_{NT})$.
Hence $s_{NT}(\bar\theta)=O_P(r_{NT})$.
Lemmas~\ref{lemma:expansion} and~\ref{lemma:linearize} give
\begin{align*}
 \sqrt{NT}\left(\widehat\theta^{\rm FE}-\theta^0
                  +\frac1T H^{-1}B_{NT}\right)
   =-H^{-1}\sqrt{NT}
       \{s_{NT}(\theta^0)-\mathbb E_A s_{NT}(\theta^0)\}+o_P(1).
\end{align*}
Combining with~\eqref{eq:profile-centered-clt} proves the theorem.
The conditional characteristic functions converge in probability to a
non-random limit and are bounded by one. Taking expectations gives the
unconditional convergence.
\end{proof}

\subsection{Proof of Lemma~\ref{lemma:FEweights}}
\label{app:FEweights}

\begin{proof}
Put $k=\lfloor\tau_0T\rfloor$ and $a=\langle\tau_0T\rangle$.
For fixed $\theta$, the residuals are distinct with probability one. On
each region with a fixed ordering,
\begin{align*}
 T L_{i,T}(\theta)
 &=(1-\tau_0)\sum_{q=1}^k
       \{u_{i(k+1)}(\theta)-u_{i(q)}(\theta)\}\\
 &\quad+\tau_0\sum_{q=k+2}^T
       \{u_{i(q)}(\theta)-u_{i(k+1)}(\theta)\}.
\end{align*}
Since $d u_{i(q)}(\theta)/d\theta=-z_{i[q]}(\theta)$, the coefficients
of the ordered regressors are $1-\tau_0$ for $q\leq k$,
$\tau_0T-k-\tau_0=a-\tau_0$ for $q=k+1$, and $-\tau_0$ for
$q\geq k+2$. This gives \eqref{ScoreAsWeightedOS}. If $\tau_0T$ is an
integer, every intercept between the two adjacent order statistics gives
the same profile objective and the same calculation.
\end{proof}

\subsection{Proof of Theorem~\ref{th:finiteTbias}}
\label{app:finiteTbias}

In this subsection and the next two, all expectations are conditional on
$A_i$. We suppress this conditioning and the index $i$ when there is no
ambiguity. Put $p=\tau_0$, $\xi_t=F(u_t|A_i)$, and
$d_q=w_{T,q}-w_{T,q+1}$. The variables $\xi_t$ are independent uniforms.
Write $m(v)=\mathbb E[z_t|\xi_t=v]$ and
$\Sigma(v)=\operatorname{Var}(z_t|\xi_t=v)$. For the large-$T$ results,
Assumption~\ref{ass:RC} and Bayes' rule imply that these functions are bounded
and that their required derivatives are uniformly bounded on a fixed
neighborhood of $p$. Then $m(p)=\lambda_i$ and $\Sigma(p)=V_{1,i}$.

\begin{lemma}
\label{lemma:CharacterizeBias}
Let Assumption~\ref{ass:CI} hold, let the conditional residual distributions
be continuous, assume $\mathbb E[\|z_t\||A_i]<\infty$, and suppose that
$p w_{T,1}+(1-p)w_{T,T}=0$.
If $u_{(q)}^{(-t)}$ is the $q$th order statistic after observation $t$ is
removed, then
\[
 \mathbb E\left[\frac1T\sum_{q=1}^T w_{T,q}z_{[q]}\right]
 =\sum_{q=1}^{T-1}d_q\,
   \mathbb E\left[z_t\{F(u_{(q)}^{(-t)}|z_t,A_i)-p\}\right].
\]
\end{lemma}

\begin{proof}
With probability one there are no ties. For $q<T$, the event that
observation $t$ has rank at most $q$ is exactly
$\{u_t<u_{(q)}^{(-t)}\}$. Since
$w_{T,r}=w_{T,T}+\sum_{q=r}^{T-1}d_q$,
\[
 \frac1T\sum_{r=1}^T w_{T,r}z_{[r]}
 =\frac1T\sum_{t=1}^T z_t
   \left[w_{T,T}+\sum_{q=1}^{T-1}d_q
          \mathbbm1\{u_t<u_{(q)}^{(-t)}\}\right].
\]
Conditional independence permits integration over $u_t$ while holding $z_t$
and the other observations fixed. The constant contribution after subtracting
$p$ inside each conditional probability is
$\{w_{T,T}+p(w_{T,1}-w_{T,T})\}\mathbb E[z_t]=0$.
\end{proof}

\begin{proof}[Proof of Theorem~\ref{th:finiteTbias}]
The leave-one-out quantile $F(u_{(q)}^{(-t)}|A_i)$ is independent of
$(u_t,z_t)$ and has distribution ${\cal B}(q,T-q)$.
Substituting $F(u|z_t,A_i)=G(F(u|A_i)|z_t,A_i)$ into
Lemma~\ref{lemma:CharacterizeBias} gives
\[
 \mathbb E\left[z_t\{F(u_{(q)}^{(-t)}|z_t,A_i)-p\}\right]
 =\int_0^1 g_i(v)f_{\cal B}(v,q,T-q)\,dv.
\]
Multiplication by $d_q$ and summation gives the asserted integral against
$h_T(v,w_T)$.
\end{proof}

\subsection{Proof of Theorem~\ref{th:largeTbias}}
\label{app:largeTbias}

\begin{proof}
Put $\pi_q=q/T$ and $a(v)=v(1-v)$. Summation by parts and the endpoint
conditions give
\[
 \sum_{q=1}^{T-1}d_q=1,
 \qquad \sum_{q=1}^{T-1}d_q(\pi_q-p)=0.
\]
The moment conditions imply
\begin{equation}
 \sum_q|d_q|=O(T),\qquad
 \sum_q|d_q||\pi_q-p|^{2j}=O(T^{-j}),\quad j=1,2,3.
 \label{eq:app-weight-moment-bounds}
\end{equation}
Here and below, constants are uniform over $i$.

Let $Q_q\sim{\cal B}(q,T-q)$. For a function $b$ with six bounded
derivatives near $p$, expansion about $\pi_q$ gives
\begin{align}
 \mathbb E[b(Q_q)]
 &=b(\pi_q)+\frac{a(\pi_q)}{2T}b''(\pi_q)
       +\frac{R_b(\pi_q)}{T^2}+O(T^{-3}),\label{eq:app-beta-expansion}\\
 R_b(v)&=-\frac{a(v)}2b''(v)
       +\frac{a(v)(1-2v)}3b'''(v)
       +\frac{a(v)^2}8b^{(4)}(v).\nonumber
\end{align}
The expansion is uniform for $\pi_q$ in a fixed smaller neighborhood of $p$.
To verify it, the beta moments about their mean are
\[
 \mathbb E[(Q_q-\pi_q)^2]=\frac{a(\pi_q)}{T+1},\qquad
 \mathbb E[(Q_q-\pi_q)^3]
 =\frac{2a(\pi_q)(1-2\pi_q)}{(T+1)(T+2)},
\]
and
$\mathbb E[(Q_q-\pi_q)^4]=3a(\pi_q)^2/T^2+O(T^{-3})$.
The fifth signed moment and the sixth absolute moment are $O(T^{-3})$.
These identities follow by substituting
$\mathbb E[Q_q^j]=\prod_{\ell=0}^{j-1}(q+\ell)/(T+\ell)$
into the centered moments.

If $|\pi_q-p|$ exceeds a fixed positive
constant, the sum of $|d_q|$ over these ranks is $O(T^{-3})$ by
\eqref{eq:app-weight-moment-bounds}. For the remaining ranks, the probability
that $Q_q$ leaves a fixed neighborhood of $p$ is exponentially small.
This follows from the binomial representation of a beta order statistic
and the exponential bound for sums of independent Bernoulli variables.
Boundedness of $g_i$ therefore controls all omitted tails.

Apply \eqref{eq:app-beta-expansion} to $b=g_i$. Expansion of $g_i(\pi_q)$
about $p$, using $g_i(p)=0$, gives
\[
 \sum_qd_q g_i(\pi_q)
 =\frac{M_2(w_T,p)}{2T}g_i''(p)
  +\frac{M_3(w_T,p)}{6T^{3/2}}g_i'''(p)+O(T^{-2}).
\]
The fourth-order remainder is bounded by
$C\sum_q|d_q||\pi_q-p|^4=O(T^{-2})$.
The functions $a g_i''$ and $R_{g_i}$ have two bounded derivatives locally.
Their constant and linear Taylor terms, together with
\eqref{eq:app-weight-moment-bounds}, give
\[
 \sum_qd_q a(\pi_q)g_i''(\pi_q)=a(p)g_i''(p)+O(T^{-1}),
 \qquad \sum_qd_q R_{g_i}(\pi_q)=O(1).
\]
Finally, summing the $O(T^{-3})$ error in
\eqref{eq:app-beta-expansion} against $|d_q|$ gives $O(T^{-2})$.
Theorem~\ref{th:finiteTbias} now gives the claimed expansion.
\end{proof}

\subsection{Proof of Theorem~\ref{th:VarJac}}
\label{app:VarJac}

Conditional on the ordered uniforms, the concomitants are independent, with
conditional means $m(\xi_{(q)})$ and covariances $\Sigma(\xi_{(q)})$.
Consequently, the exact variance identity is
\begin{equation}
 \operatorname{Var}\left(\frac1T\sum_qw_{T,q}z_{[q]}\right)
 =\frac1{T^2}\sum_q w_{T,q}^2\mathbb E[\Sigma(\xi_{(q)})]
  +\operatorname{Var}\left(\frac1T\sum_qw_{T,q}m(\xi_{(q)})\right).
 \label{eq:app-concomitant-variance}
\end{equation}
All matrix bounds below can be checked componentwise or after taking an
arbitrary fixed scalar projection.

\begin{proof}[Proof of Theorem~\ref{th:VarJac}]
For the FE weights, write $k=\lfloor pT\rfloor$, $r=\langle pT\rangle$, and
$a_0(u)=\mathbbm1\{u\leq p\}-p$.
The quantile restriction implies that $\mathbbm1\{\xi_t\leq p\}$ is
independent of $z_t$. Thus
\[
 \mathbb E[a_0(\xi_t)(z_t-\lambda_i)]=0,
 \qquad
 \mathbb E[a_0(\xi_t)^2(z_t-\lambda_i)(z_t-\lambda_i)']=V_{0,i}.
\]

For $B_u\sim\operatorname{Bin}(T-1,u)$, put
$\bar w(u)=\mathbb E[w^{\rm FE}_{T,1+B_u}]$.
The binomial tail bound gives, uniformly in $T$,
\begin{equation}
 \int_0^1|\bar w(u)-a_0(u)|\,du=O(T^{-1/2}),
 \qquad
 \int_0^1|u-p||\bar w(u)-a_0(u)|\,du=O(T^{-1}).
 \label{eq:app-binomial-smoothing}
\end{equation}
Also $\int_0^1(\bar w-a_0)=T^{-1}\sum_qw^{\rm FE}_{T,q}=0$.
The same bounds apply with the weights squared, except that the signed
integral is $-r(1-r)/T$, since
$\sum_q(w^{\rm FE}_{T,q})^2=Tp(1-p)-r(1-r)$.
Expanding $\Sigma(u)$ about $p$ in the signed squared-weight integral
therefore gives
\begin{equation}
 \frac1{T^2}\sum_q(w^{\rm FE}_{T,q})^2
          \mathbb E[\Sigma(\xi_{(q)})]
 =\frac1T\mathbb E[a_0(\xi_t)^2\Sigma(\xi_t)]+O(T^{-2}).
 \label{eq:app-FE-noise-variance}
\end{equation}

For a scalar component of $m$, define
\[
 L_T=T^{-1}\sum_qw^{\rm FE}_{T,q}m(\xi_{(q)}),\qquad
 \psi(u)=a_0(u)\{m(u)-m(p)\}.
\]
The mean of $\psi(\xi_1)$ is zero, because
$\int_0^p m(u)\,du=p\int_0^1m(u)\,du$.
The first-order projection
$P_T(u)=\mathbb E[L_T|\xi_1=u]-\mathbb E[L_T]$ satisfies
\[
 P_T'(u)=\frac1T m'(u)\bar w(u),
 \qquad \psi'(u)=m'(u)a_0(u)
\]
away from $p$. The second identity has no jump at $p$, since
$m(p)-m(p)=0$. By \eqref{eq:app-binomial-smoothing},
$\|P_T-\psi/T\|_\infty=O(T^{-3/2})$.
Moreover, integration in $u$ gives
\begin{align*}
 \mathbb E\left[\psi(\xi_1)\{P_T(\xi_1)-\psi(\xi_1)/T\}\right]
 &=\frac1T\int_0^1m'(v)\{\bar w(v)-a_0(v)\}
           \left\{\int_v^1\psi(u)\,du\right\}dv\\
 &=O(T^{-2}).
\end{align*}
For this bound, subtract the value at $p$ of
$m'(v)\int_v^1\psi(u)\,du$ and use its local Lipschitz bound, the zero
signed integral, and \eqref{eq:app-binomial-smoothing}.
It follows that
$T\operatorname{Var}(P_T(\xi_1))=T^{-1}\operatorname{Var}(\psi(\xi_1))+O(T^{-2})$.

The remaining projection terms have total variance $O(T^{-2})$.
To check this bound, replace observations 1 and 2 by independent copies and
write $\Delta_{12}L_T$ for the resulting mixed difference.
Inserting the two observations into the ordered sample of the other $T-2$
uniforms shows that
\[
 |\Delta_{12}L_T|
 \leq \frac{C}{T}\{\xi^{(-1,-2)}_{(k+1)}-\xi^{(-1,-2)}_{(k-2)}\}.
\]
The bound applies on a fixed neighborhood of $p$, where $m$ is Lipschitz.
Only ranks within three places of $k$ can contribute to the mixed
difference. The squared spacing has expectation $O(T^{-2})$.
The orthogonal projection decomposition for independent observations gives
\[
 \operatorname{Var}\left(L_T-\mathbb E L_T-\sum_{t=1}^T P_T(\xi_t)\right)
 \leq \frac14\sum_{s<t}\mathbb E[(\Delta_{st}L_T)^2]=O(T^{-2}).
\]
Indeed, a projection involving $j\geq2$ observations appears
$\binom j2$ times in the sum of squared mixed differences, with coefficient
four. This also proves the displayed inequality.

Extend $m$
smoothly outside a neighborhood of $p$. The difference between the original
and extended functions contributes the additive statistic
$T^{-1}\sum_ta_0(\xi_t)\{m(\xi_t)-m_{\rm ext}(\xi_t)\}$ except on an event
with exponentially small probability. It has no higher-order projections.
The bounds for the first projection use smoothness only near $p$.
We conclude that
\[
 \operatorname{Var}(L_T)
 =\frac1T\mathbb E[\psi(\xi_t)\psi(\xi_t)']+O(T^{-2}).
\]
Combining this equality with \eqref{eq:app-FE-noise-variance} and
\eqref{eq:app-concomitant-variance} gives $V_{0,i}/T+O_{\rm u}(T^{-2})$.

For general weights, put $\delta_q=w_{T,q}-w^{\rm FE}_{T,q}$,
$D_T^2=\sum_q\delta_q^2$, and $A_q=\sum_{j=1}^q\delta_j$.
The weight conditions imply $\sum_q\delta_q=0$ and
\begin{equation}
 \sum_{q=1}^{T-1}|A_q|=O(T).
 \label{eq:app-cumulative-weight-bound}
\end{equation}
For a direct verification, set
$e_q=\delta_q-\delta_{q+1}$.
Then $\sum_qe_q=\sum_qqe_q=0$ and
\[
 A_q=-\sum_re_r(q-r)_+=-\sum_re_r(r-q)_+.
\]
Use the first expression for $q\leq pT$ and the second for $q>pT$.
Summation over $q$ bounds $\sum_q|A_q|$ by
$C\sum_r|e_r|\{1+(r-pT)^2\}=O(T)$.

Since $\sum_q\delta_q=0$, summation by parts gives
\[
 \frac1T\sum_q\delta_qm(\xi_{(q)})
 =-\frac1T\sum_{q=1}^{T-1}A_q
                    \{m(\xi_{(q+1)})-m(\xi_{(q)})\}.
\]
The $L^2$ norm of a uniform spacing is
$\{2/[(T+1)(T+2)]\}^{1/2}$.
The local Lipschitz bound and \eqref{eq:app-cumulative-weight-bound}
therefore imply
\begin{equation}
 \left\|\frac1T\sum_q\delta_qm(\xi_{(q)})\right\|_2=O(T^{-1}).
 \label{eq:app-correction-mean-bound}
\end{equation}
To control the tails, for every fixed $\eta>0$ the sixth moment condition
implies $|\delta_q|=O(T^{-3})$ uniformly over $|q/T-p|>\eta$.
For the other ranks, leaving a fixed neighborhood of $p$ has exponentially
small probability. The same argument proves
\eqref{eq:app-correction-mean-bound} with the $L^4$ norm.

Uniform continuity of $\Sigma$ near $p$ and rank concentration now give
\[
 \sum_q\delta_q^2\mathbb E[\Sigma(\xi_{(q)})]
 =V_{1,i}D_T^2+o_{\rm u}(T+D_T^2).
\]
The cross term with the FE weights is $o_{\rm u}(T+D_T^2)$.
For $|q/T-p|\leq\eta$, its absolute value is bounded by
$C\sqrt{\eta T}\,D_T$. Outside these ranks,
$\sum_q|\delta_q|=O(T^{-2})$.
Divide by $T+D_T^2$, first let $T$ increase, and then let $\eta$ decrease
to zero. The covariance between the conditional-mean FE statistic and the
correction is $O(T^{-3/2})$ by \eqref{eq:app-correction-mean-bound}.
Using \eqref{eq:app-concomitant-variance}, we have
\[
 \operatorname{Var}\left(\frac1T\sum_qw_{T,q}z_{[q]}\right)
 =\frac{V_{0,i}}T+\frac{D_T^2}{T^2}V_{1,i}
       +o_{\rm u}\left(\frac1T+\frac{D_T^2}{T^2}\right).
\]

To calculate the Jacobian in the endogenous case, write $u_t(\theta)=y_t-x_t'\theta$.
Let $F_\theta$ be its marginal distribution given $A_i$,
$c_\theta(v)=F_\theta^{-1}(v)$, and
$\lambda_{x,\theta}(v)=\mathbb E[x_t|u_t(\theta)=c_\theta(v),A_i]$.
At the exogenous model, replace $x_t$ by $z_t$.
Define
\[
 g_i(v,\theta)=\mathbb E\left[
 z_t\{\mathbbm1(u_t(\theta)\leq c_\theta(v))-p\}\right].
\]
This agrees with~\eqref{Defg} at $\theta=\theta^0$.
The leave-one-out argument in Lemma~\ref{lemma:CharacterizeBias} does not
use the quantile restriction. Applied at $\theta$, it gives
\[
 \mathbb E\left[\frac1T\sum_qw_{T,q}z_{[q]}(\theta)\right]
 =\int_0^1g_i(v,\theta)h_T(v,w_T)\,dv.
\]
Differentiation of the residual distribution and its inverse gives
\[
 \partial_\theta F_\theta(u)
  =\mathbb E[f_y(u+x_t'\theta|x_t,z_t,A_i)x_t'],
 \qquad \partial_\theta c_\theta(v)=-\lambda_{x,\theta}(v)'.
\]
With $f_\theta$ the marginal residual density, differentiation of $g_i$ gives
\begin{align}
 \partial_\theta g_i(v,\theta)
 &=\mathbb E\left[z_t f_y(c_\theta(v)+x_t'\theta|x_t,z_t,A_i)
                 \{x_t-\lambda_{x,\theta}(v)\}'\right]\nonumber\\
 &=K_i(v;\theta),\nonumber\\
 K_i(v;\theta)&:=f_\theta(c_\theta(v))
       \operatorname{Cov}(z_t,x_t|u_t(\theta)=c_\theta(v),A_i),\nonumber\\
 \partial_\theta\mathbb E\left[\frac1T\sum_qw_{T,q}z_{[q]}(\theta)\right]
 &=\int_0^1 K_i(v;\theta)h_T(v,w_T)\,dv.
 \label{eq:app-exact-rank-jacobian}
\end{align}
No global positive lower bound on the residual density is required here.
If $\|x_t\|\leq M$, then
$|c_{\theta+h}(v)-c_\theta(v)|\leq M\|h\|$ by the ordering of the two
residual distributions. The inverse derivative exists at almost every $v$
whose residual quantile has positive density. The remaining $v$ have
Lebesgue measure zero. Bounded conditional densities and the quantile
Lipschitz bound dominate the difference quotients in the expectation and
the integral. These facts justify \eqref{eq:app-exact-rank-jacobian}.

The function $K_i$ is bounded on $(0,1)$ and has four uniformly bounded
derivatives in $v$ near $p$ and $\theta^0$.
Using the beta expansion through its $T^{-1}$ term and
\eqref{eq:app-weight-moment-bounds} gives
\[
 \int_0^1 K_i(v;\theta)h_T(v,w_T)\,dv
 =K_i(p;\theta)+O_{\rm u}(T^{-1}).
\]
The constant and linear terms of $K_i(\pi_q;\theta)$ sum to
$K_i(p;\theta)$, and the remaining term is bounded by
$C\sum_q|d_q||\pi_q-p|^2=O(T^{-1})$.
The $T^{-1}$ beta correction has a bounded signed sum, and its remaining
$O(T^{-2})$ error times $\sum_q|d_q|$ is $O(T^{-1})$.
At $(p,\theta^0)$, $K_i=J_i'$ in the endogenous case and $K_i=H_i$
in the exogenous case. This proves the Jacobian conclusion, with an
$O_{\rm u}(T^{-1})$ remainder.
\end{proof}

\subsection{Proof of Theorem~\ref{th:BCAsy}}
\label{app:BiasCorrection}

Write $m=T-S$, $v_\tau=\tau_0(1-\tau_0)$, and
$a_n=\langle\tau_0n\rangle$. We condition on
$A_1,\ldots,A_N$, except where stated otherwise. Constants in the bounds
below are uniform in these conditioning variables. The letter $C$ denotes
a finite constant whose value can change between displays.

\subsubsection{Weight regularity}
\label{app:WeightRegularity}

Set $p_{n,q}=w^{\rm FE}_{n,q}+\tau_0$ and define
\[
 v_q=\mathbb E[p_{S,K_q}],\qquad
 K_q=1+X_q,\qquad
 X_q\sim\operatorname{Hypergeom}(T-1,q-1,S-1).
\]
The probability that a uniform size-$S$ subsample contains rank $q$ in
position $k$ is
\[
 \binom{T}{S}^{-1}\binom{q-1}{k-1}\binom{T-q}{S-k}
 =\frac ST\Pr(K_q=k).
\]
Substitution in~\eqref{HowToComputeWJK} gives
\begin{align}
 \delta_q:=w^{\rm MJ}_{T,q}-w^{\rm FE}_{T,q}
 =\kappa\frac{S}{m}(p_{T,q}-v_q).
 \label{eq:BC-hypergeom}
\end{align}
The second-moment formulas in~\eqref{DefKappa} imply
$\kappa=1+O(m^{-1})$ when $S/T$ is bounded away from zero and
$m\to\infty$. Hence $\kappa$ is bounded and positive eventually.

Select $S-1$ positions uniformly from $1,\ldots,T-1$, and let $Z_j$ be
their $j$th order statistic. If $j=\lfloor\tau_0S\rfloor$, define $Z$ by
mixing $Z_j$ and $Z_{j+1}$ with probabilities $1-a_S$ and $a_S$.
The second component is used only if $a_S>0$.
Counting the selected positions below $q$ gives
\[
 v_q-v_{q+1}=\Pr(Z=q),\qquad
 \mathbb E Z=\tau_0T.
\]
The distribution of $Z_j-j$ is beta-binomial. More explicitly,
$Z_j=j+B_j$, where, conditional on
$P_j\sim{\cal B}(j,S-j)$, $B_j\sim\operatorname{Bin}(m,P_j)$.
The binomial and beta moment formulas give, for $k=1,2,3$,
\begin{align}
 \mathbb E|Z-\tau_0T|^{2k}&\leq C m^k,&
 |\mathbb E(Z-\tau_0T)^3|&\leq C m.
 \label{eq:BC-position-moments}
\end{align}
The first bound follows from $\mathbb E|B_j-mP_j|^{2k}\leq C(m^k+m)$ and
$m^{2k}\mathbb E|P_j-j/S|^{2k}\leq Cm^{2k}/S^k\leq Cm^k$.
The third signed moment is bounded by
$C(m+m^2/S+m^3/S^2)$. Mixing the adjacent positions changes these bounds
only by a constant factor, since $T/S$ is bounded.

For finite admissible $S<T$, differencing the survival probabilities gives
\begin{align*}
 M_2(w^{\rm FE}_T,\tau_0)&=\frac{a_T(1-a_T)}T,\\
 M_2(w^{\rm JK}_T,\tau_0)-M_2(w^{\rm FE}_T,\tau_0)
 &=\frac{S}{mT}\{a_T(1-a_T)-\operatorname{Var}(Z)\}.
\end{align*}
The beta-binomial moments give
$\operatorname{Var}(Z)=Tmv_\tau/(S+1)+T(T+1)a_S(1-a_S)/\{S(S+1)\}$.
These identities give the closed forms in~\eqref{DefKappa}.
The denominator in the explicit expression for $\kappa$ is
\[
 S(S+1)\{\operatorname{Var}(Z)-a_T(1-a_T)\}.
\]
With $k=\lfloor\tau_0T\rfloor$, integer-valuedness and
$\mathbb EZ=\tau_0T$ imply
\[
 \operatorname{Var}(Z)-a_T(1-a_T)
 =\mathbb E[(Z-k)(Z-k-1)]\geq0.
\]
Equality holds exactly when $Z$ is supported on $\{k,k+1\}$.
The variable $Z_j$ has positive probability at every integer from $j$ to
$j+m$. Consequently, the denominator is positive for $m\geq2$.
For $m=1$, it vanishes exactly when $a_S=0$.
At these exceptional points $v_q=p_{T,q}$, so the standard Jackknife
weights equal the FE weights and the rescaling cannot correct the bias.
Under the stronger conditions $\tau_0T\geq2$ and $(1-\tau_0)T\geq2$,
the explicit modified delete-one weights in the main text also define a
bias-correcting sequence at these exceptional points. Their branch with
$a_T\geq\tau_0$ specifies which adjacent pair receives the correction.
These conditions keep both affected ranks strictly between $1$ and $T$.

Under these interior conditions, let $c_T$ be the correction at the first of
the two affected ranks. It equals $a$ or $b$ for the standard weights and
$\{Tv_\tau+a_T(1-a_T)\}/2$ for the modified weights. The two corrections are
$c_T$ and $-c_T$, so their squared sum is $2c_T^2$.
For each affected rank,
$\mathbb E[\Sigma(\xi_{(q)})]=\Sigma(\tau_0)+O(T^{-1})$.
The exact conditional variance decomposition
\eqref{eq:app-concomitant-variance}, the bound $|c_T|\leq CT$, and
\eqref{eq:app-correction-mean-bound} therefore give
\[
 \operatorname{Var}\left(\frac1T\sum_q w_{T,q}z_{i[q]}\right)
 =\frac{2c_T^2}{T^2}V_{1,i}+O_{\rm u}(T^{-1}).
\]
For the modified weights this coefficient is $v_\tau^2/2+O(T^{-1})$.
For the standard weights it depends on the fractional part of $\tau_0T$
and can vanish.

Both $p_{T,q}$ and $v_q$ decrease from one to zero. They are the survival
probabilities of integer-valued variables with mean $\tau_0T$.
The variable corresponding to $p_{T,q}$ is supported on
$\lfloor\tau_0T\rfloor$ and $\lfloor\tau_0T\rfloor+1$.
It follows from~\eqref{eq:BC-hypergeom}--\eqref{eq:BC-position-moments} that
\begin{align}
 \delta_1=\delta_T=0,\qquad &\sum_q\delta_q=0,\qquad
 \max_q|\delta_q|\leq C\beta_T,\qquad \beta_T:=\frac{T}{m},\nonumber\\
 \sum_{q=1}^{T-1}|\delta_q-\delta_{q+1}|&\leq C\frac{T}{m},&
 \sum_q|\delta_q|&\leq C\frac{T}{\sqrt m}.
 \label{eq:BC-weight-bounds}
\end{align}
The last bound follows by summing the survival probabilities and using
$\mathbb E|Z-\tau_0T|\leq C\sqrt m$.
Moreover,
\[
 T^{-k}\sum_{q=1}^{T-1}|w^{\rm MJ}_{T,q}-w^{\rm MJ}_{T,q+1}|
       |q-\tau_0T|^{2k}\leq C,
 \qquad k=1,2,3.
\]
Thus the moment conditions on the weights in
Theorem~\ref{th:largeTbias} hold. The signed third-moment bound in
\eqref{eq:BC-position-moments} also gives
$M_3(w^{\rm MJ}_T,\tau_0)=O(T^{-1/2})$.

For the order of $b_T$, put $\omega_{TS}=\sqrt{Tm/S}$ and
$x_q=(q-\tau_0T)/\omega_{TS}$. The beta central limit theorem gives
$\sqrt S(P_j-j/S)\Rightarrow{\cal N}(0,v_\tau)$.
Conditional on $P_j$, the characteristic function of
$(B_j-mP_j)/\sqrt m$ converges to that of
${\cal N}(0,v_\tau)$, uniformly for $P_j$ near $\tau_0$.
The two limiting normal variables are therefore independent. Since
\[
 \frac{Z_j-jT/S}{\omega_{TS}}
 =\sqrt{\frac ST}\frac{B_j-mP_j}{\sqrt m}
  +\sqrt{\frac mT}\sqrt S(P_j-j/S),
\]
their weighted sum converges to ${\cal N}(0,v_\tau)$.
Both mixture components have this limit, and their means differ by
$T/S=O(1)$.
It follows that $v_q\to\Phi(-x/\sqrt{v_\tau})$ for
$q=\tau_0T+x\omega_{TS}+O(1)$, uniformly on bounded intervals of $x$.
Chebyshev's inequality gives
\[
 |p_{T,q}-v_q|^2\leq C\min\{1,|x_q|^{-4}\}
\]
away from the bounded number of ranks at the cutoff.
This bound permits summing the limit over the ranks. Consequently,
\begin{align}
 b_T:=\frac1T\sum_q\delta_q^2
 &\sim \frac{C_{\tau_0}}{\sqrt T}
       \left(\frac{S}{m}\right)^{3/2},&
 C_{\tau_0}&=\sqrt{v_\tau}\frac{\sqrt2-1}{\sqrt\pi}.
 \label{DefbTbetaT}
\end{align}
Here we used
\[
 \int_{\mathbb R}
 \{\mathbbm{1}(x<0)-\Phi(-x/\sqrt{v_\tau})\}^2\,dx
 =2\sqrt{v_\tau}\int_0^\infty\{1-\Phi(u)\}^2du
 =C_{\tau_0}.
\]
It follows that $b_T=O(T/m^{3/2})$ and
$b_T\to C_{\tau_0}c_2^{3/2}$ under the rates in the theorems.
This limit is zero if $S/T$ converges to a number below one.

\subsubsection{Evaluation at the preliminary estimator}

For either the exogenous or endogenous model, write
\[
 D_i(\theta)=\frac1T\sum_q\delta_qz_{i[q]}(\theta),\qquad
 D_{NT}(\theta)=\frac1N\sum_iD_i(\theta),\qquad
 a_{NT}=T^{-1}+(NT)^{-1/2}.
\]
In the endogenous model the ordering uses $y_{it}-x_{it}'\theta$.
In the exogenous model it uses $y_{it}-z_{it}'\theta$.
Ties are ordered by the observation index.

\begin{lemma}
\label{lemma:BiasCorrection}
Suppose the assumptions of Theorem~\ref{th:BCAsy} or
Theorem~\ref{th:BCAsy:End} hold, including
\begin{align}
 \frac{T^2\log^2(NT)}{N(T-S)^2}\longrightarrow0.
 \label{eq:BC-plugin-rate}
\end{align}
If $\widetilde\theta-\theta^0=O_P(a_{NT})$, then
\[
 \sqrt{NT}\{D_{NT}(\widetilde\theta)-D_{NT}(\theta^0)\}
 =o_P(1).
\]
\end{lemma}

\begin{proof}
For a deterministic radius $a$, let
$F_i(\theta)=D_i(\theta)-D_i(\theta^0)$, with
$\|\theta-\theta^0\|\leq a$.
Summation by parts expresses $F_i(\theta)$ as a sum of changes in the
regressor partial sums below the $k$th residual, multiplied by
$(\delta_k-\delta_{k+1})/T$.
All the indices are central, since the differences in
\eqref{eq:BC-hypergeom} are supported between
$\lfloor\tau_0S\rfloor-1$ and $T-S+\lfloor\tau_0S\rfloor+1$.
Use~\eqref{eq:central-cutoff-count-bound}, the triangle inequality in
$L^2$, and~\eqref{eq:BC-weight-bounds} to get
\begin{align}
 \sup_{\|\theta-\theta^0\|\leq a}
   \mathbb E[\|F_i(\theta)\|^2\mid A_i]
 &\leq C\beta_T^2(a^2+a/T),&
 \sup_\theta\|F_i(\theta)\|&\leq C m^{-1/2}.
 \label{eq:BC-row-increment}
\end{align}

We use a uniform bound across individuals. For any fixed data set with
$n$ individuals, the orderings change only across the pairwise residual
equalities. There are at most $nT(T-1)/2$ such hyperplanes in the
fixed-dimensional parameter space. They partition the parameter space
into at most $C(nT^2)^{d_\theta}$ cells, including their faces, where
$d_\theta=\dim\theta$.
Each coordinate of $F_i(\theta)$ is constant on every cell.
The conditional means are continuous in $\theta$, by the same increment
bound as in~\eqref{eq:BC-row-increment} applied at two parameter values.
Measurable countable dense selections on the faces therefore give
measurable suprema, including after centering, as in
Appendix~\ref{app:measurability}.

Here is a symmetrization argument that allows the cells to depend on the
observations. For one coordinate, put
$\sigma^2=C\beta_T^2(a^2+a/T)$, $K=Cm^{-1/2}$, and
$\ell=\log(NT)$.
Let $R$ be the expected supremum of
$|N^{-1}\sum_i\epsilon_iF_i(\theta)|$, where the $\epsilon_i$ are
independent random signs, independent of the data.
Conditional on the data, the exponential bound for sums of random signs
and the number of cells give
\[
 R\leq C\sqrt{\ell/N}
       \left\{\mathbb E\sup_\theta
                    N^{-1}\sum_iF_i(\theta)^2\right\}^{1/2}.
\]
Symmetrization with an independent copy of each individual's observations
and the contraction inequality for $u\mapsto u^2$ on $[-K,K]$ give
\[
 \mathbb E\sup_\theta N^{-1}\sum_iF_i(\theta)^2
\leq\sigma^2+CKR.
\]
Solving the resulting quadratic inequality for $R$, and symmetrizing
the centered average once more, proves
\begin{align}
 \mathbb E\sup_{\|\theta-\theta^0\|\leq a}
 \left\|\frac1N\sum_i
   \{F_i(\theta)-\mathbb E[F_i(\theta)\mid A_i]\}\right\|
 \leq C\left\{
   \beta_T\sqrt{\frac{(a^2+a/T)\ell}{N}}
   +\frac{\ell}{N\sqrt m}\right\}.
 \label{eq:BC-uniform-increment}
\end{align}
The same proof works for independent, non-identically distributed
individuals conditional on $A_1,\ldots,A_N$.

For the mean, let $c_i(v,\theta)$ be the conditional $v$-quantile of
$u_{it}(\theta)$ given $A_i$, and use $g_i(v,\theta)$ defined in the
Jacobian proof in Appendix~\ref{app:VarJac}.
The subsample definition of the weights gives the exact
identity
\[
 \mathbb E[D_i(\theta)\mid A_i]
 =\kappa\frac{S}{m}
  \{\mathbb E[m^{\rm FE}_{i,T}(\theta)\mid A_i]
       -\mathbb E[m^{\rm FE}_{i,S}(\theta)\mid A_i]\}.
\]
Here $m^{\rm FE}_{i,n}(\theta)$ is the FE score from $n$ observations.
The beta-density representation of its expectation, expanded about
$\tau_0$, gives uniformly near $\theta^0$
\[
 \mathbb E[m^{\rm FE}_{i,n}(\theta)\mid A_i]
 =g_i(\tau_0,\theta)+\frac{v_\tau}{2n}
       \partial_\tau^2g_i(\tau_0,\theta)+O(n^{-2}).
\]
The density is bounded away from zero near the relevant quantile, so
implicit differentiation of that quantile is valid. The six bounded
density derivatives and bounded regressors bound the required derivatives
of $g_i$ locally, also after one derivative in $\theta$.
For all $v\in(0,1)$, bounded regressors give
$|c_i(v,\theta)-c_i(v,\theta^0)|\leq C\|\theta-\theta^0\|$.
The bounded conditional density then gives
$\|g_i(v,\theta)-g_i(v,\theta^0)\|\leq C\|\theta-\theta^0\|$.
The beta probability outside the fixed quantile neighborhood is
exponentially small. Hence the difference of the $O(n^{-2})$ remainders
at $\theta$ and $\theta^0$ is bounded by
$Cn^{-2}\|\theta-\theta^0\|$, including the beta tails.
The first term cancels between sample sizes, and
$|T^{-1}-S^{-1}|=m/(ST)$. The remaining bound is
$Ca\{T^{-1}+(mS)^{-1}\}\leq Ca/T$, since $T/S$ is bounded and $m\geq1$.
Therefore
\begin{align}
 \sup_{\|\theta-\theta^0\|\leq a}
 \|\mathbb E[F_i(\theta)\mid A_i]\|
 \leq Ca/T.
 \label{eq:BC-mean-increment}
\end{align}

Apply~\eqref{eq:BC-uniform-increment}--\eqref{eq:BC-mean-increment}
with $a=M a_{NT}$ for fixed $M$.
After multiplication by $\sqrt{NT}$, the stochastic bound is at most
\[
 C_M\left\{
   \frac{T}{m}\sqrt{\ell(T^{-1}+N^{-1})}
       +\sqrt{\frac{T}{Nm}}\,\ell\right\}.
\]
It tends to zero by~\eqref{eq:BC-plugin-rate},
$m\geq cT^{2/3}$, and $N/T^3\to0$.
The mean bound tends to zero because
$\sqrt{NT}a_{NT}/T=\sqrt{N/T^3}+T^{-1}\to0$.
Finally, $\widetilde\theta-\theta^0=O_P(a_{NT})$ permits $M$ to be
chosen arbitrarily large after taking the limit.
\end{proof}

\subsubsection{Moments and the central limit theorem}

Let
\[
 Q^*_{i,T}=\frac1T\sum_qw^{\rm MJ}_{T,q}z_{i[q]}(\theta^0),\qquad
 \mu^*_{i,T}=\mathbb E[Q^*_{i,T}\mid A_i].
\]
The exact second-moment normalization and the bound on $M_3$ give
\begin{align}
 \|\mu^*_{i,T}\|&\leq CT^{-2},&
 T\operatorname{Var}(Q^*_{i,T}\mid A_i)
 &=V_{0,i}+b_TV_{1,i}+o_{\rm u}(1).
 \label{eq:BC-score-moments}
\end{align}
The mean follows from Theorem~\ref{th:largeTbias}, and the variance from
Theorem~\ref{th:VarJac}, since $b_T$ is bounded.

For the central limit theorem, the centered vectors
$M^*_{i,T}=\sqrt T(Q^*_{i,T}-\mu^*_{i,T})$ have uniformly bounded
fourth moments. To verify this, let
$\xi_{it}=F(u_{it}(\theta^0)\mid A_i)$ and
$m_i(u)=\mathbb E[z_{it}\mid \xi_{it}=u,A_i]$.
Conditional on the ordered $\xi$'s, the concomitant errors
$z_{i[q]}-m_i(\xi_{i(q)})$ are independent, centered, and bounded.
Expansion of the fourth power of their weighted sum gives
\[
 \mathbb E\left[
 \left\|T^{-1/2}\sum_qw^{\rm MJ}_{T,q}
       \{z_{i[q]}-m_i(\xi_{i(q)})\}\right\|^4
       \Bigm|A_i\right]
 \leq C\left(T^{-1}\sum_q(w^{\rm MJ}_{T,q})^2\right)^2\leq C.
\]

The FE part of the conditional mean has bounded fourth moment by the
fixed-cutoff comparison used in~\eqref{eq:profile-score-fourth-moment}.
For the correction part, the $L^4$ version of
\eqref{eq:app-correction-mean-bound} gives
\[
 \left\|T^{-1/2}\sum_q\delta_qm_i(\xi_{i(q)})\right\|_4
 \leq C T^{-1/2}.
\]
Thus the conditional-mean contribution has bounded fourth moment as well.

The conditional Lyapunov condition now follows from
$N^{-2}\sum_i\mathbb E[\|M^*_{i,T}\|^4\mid A_i]\leq C/N$.
Using~\eqref{eq:BC-score-moments} and the definition of $V^*$, we have
\begin{align}
 \sqrt{NT}\,\frac1N\sum_iQ^*_{i,T}
 \ \Rightarrow\ {\cal N}(0,V^*).
 \label{eq:BC-score-CLT}
\end{align}
The conditional mean is negligible because
$\sqrt{NT}\,\|N^{-1}\sum_i\mu^*_{i,T}\|
\leq C\sqrt{N/T^3}\to0$.

\subsubsection{Linearization}

\begin{proof}[Proof of Theorem~\ref{th:BCAsy}]
The preliminary FE estimator has error $O_P(a_{NT})$ by the
exogenous expansion. Lemma~\ref{lemma:BiasCorrection} therefore permits replacing the
preliminary estimator in the correction by $\theta^0$.
At $\theta^0$, the gradient of the exogenous corrected objective is
\begin{align}
 \nabla_\theta{\cal L}^*_{NT}(\theta^0,S)
 =\frac1N\sum_iQ^*_{i,T}+o_P((NT)^{-1/2}).
 \label{GradLstar}
\end{align}
The correction is affine in $\theta$. Its addition changes the linear
term in the local convex objective expansion, and leaves the quadratic
term and remainder unchanged. By~\eqref{eq:BC-score-CLT}, the new
gradient is $O_P((NT)^{-1/2})$. Lemma~\ref{lemma:linearize} applied to
this objective gives
\[
 \sqrt{NT}(\widehat\theta^*(S)-\theta^0)
 =-H^{-1}\sqrt{NT}\,\frac1N\sum_iQ^*_{i,T}+o_P(1).
\]
Combining this expansion with~\eqref{eq:BC-score-CLT} gives the covariance
$H^{-1}V^*H^{-1}$.
The conditional limiting distribution has a nonrandom covariance, so
the same convergence holds unconditionally.
\end{proof}

\subsection{Proof of Theorem~\ref{th:EndAsy}}
\label{app:Endog}

Let $\mu=(\theta',\gamma')'$, $\mu^0=(\theta^{0\prime},0')'$ and
$v_{it}=(x_{it}',z_{it}')'$. The joint profile objective is convex in
$\mu$. The objective, Hessian, and probability bounds in
Appendix~\ref{app:Exog} apply with $v_{it}$ as the regressor vector.
They do not require the quantile restriction conditional on $x_{it}$.
The limiting expected Hessian is
\begin{align*}
 \mathcal H=\begin{pmatrix}G&J\\J'&H\end{pmatrix}.
\end{align*}
Only $H$ needs to be positive definite. We continue to condition on $A$
and write $r_{NT}=(NT)^{-1/2}$.

\subsubsection{Uniform expansion and the inner minimizer}

Let $q_{NT}=\partial_\gamma\mathcal L_{NT}(\mu^0)$ and let
$a_{NT}\to0$ be deterministic with $a_{NT}\geq r_{NT}$. Write
\begin{align*}
 R_{NT}(h)=\mathcal L_{NT}(\mu^0+h)-\mathcal L_{NT}(\mu^0)
          -\partial_\mu\mathcal L_{NT}(\mu^0)'h
          -\tfrac12h'\mathcal Hh.
\end{align*}
For every deterministic $\epsilon_{NT}\to0$,
\begin{align}
 \sup_{\|h\|\leq\epsilon_{NT}}
       \frac{|R_{NT}(h)|}{a_{NT}^2+\|h\|^2}=o_P(1).
 \label{JointExpansion}
\end{align}

To verify the uniform bound, let $R_{i,T}(h)$ be the individual remainder
after subtracting the sample linear term, as in the proof of
Lemma~\ref{lemma:expansion}. Replacing one observation changes a profile
score by at most $C/T$. Integrating the difference of the scores along
the segment therefore changes $R_{i,T}(h)$ by at most $C\|h\|/T$.
The variance inequality for independent replacements gives
\begin{align*}
 \operatorname{Var}_A\left[\frac1N\sum_iR_{i,T}(h)\right]
      \leq \frac{C\|h\|^2}{NT}.
\end{align*}
Together with~\eqref{eq:profile-remainder-variance}, this bounds the
variance divided by $d^4$, where $d=\|h\|$, by
\begin{align*}
 C\min\left\{\frac1N+\frac1{NTd}+\frac1{NT^2d^2},
                       \frac1{NTd^2}\right\}.
\end{align*}
Apply this bound at a fixed finite grid on each rescaled ball of radius
$d_j=2^j a_{NT}$ up to $2\epsilon_{NT}$. Convexity extends the bounds
from a sufficiently fine grid to the interior ball. More precisely,
for each fixed error tolerance a finite grid suffices, because the
quadratic limit is uniformly continuous on the ball and a convex
function is bounded above and below on the interior by its values on
that grid. This is the finite-grid argument in the convexity lemma of
\cite{Pollard1991}.
The sum of the probability bounds tends to zero. Indeed,
\begin{align*}
 \sum_{j\geq0}\min\{N^{-1},(NTd_j^2)^{-1}\}
       &\leq C(1+\log N)/N,\\
 \sum_{j\geq0}\left\{(NTd_j)^{-1}+(NT^2d_j^2)^{-1}\right\}
       &\leq C(r_{NT}+T^{-1}).
\end{align*}
The conditional mean remainder divided by $d^2$ is bounded by
$O_P(N^{-1/2}+T^{-1})+C\epsilon_{NT}$, by
Lemma~\ref{lemma:hessian}. The innermost ball is covered by the same
argument at scale $a_{NT}$. This proves~\eqref{JointExpansion}.

\begin{lemma}[Inner and outer minimizers]
\label{lemma:inner}
Suppose that $q_{NT}=O_P(a_{NT})$, $H>0$, $J$ has full row rank, and
$\mathcal W_{NT}\to_P\mathcal W>0$.
Let $\Theta$ be a compact parameter space containing $\theta^0$ in its
interior, and let $\widehat\theta$ be a consistent global minimizer of
the outer objective over $\Theta$.
For every fixed $M<\infty$,
\begin{align}
 \sup_{\|\theta-\theta^0\|\leq M a_{NT}}
 \left\|\widehat\gamma(\theta)
       +H^{-1}\{q_{NT}+J'(\theta-\theta^0)\}\right\|
       =o_P(a_{NT}).
 \label{InnerApprox}
\end{align}
Then $\widehat\theta-\theta^0=O_P(a_{NT})$.
If $q_{NT}=O_P(r_{NT})$, then
\begin{align}
 \widehat\theta-\theta^0
    =-(JWJ')^{-1}JWq_{NT}+o_P(r_{NT}),
 \qquad W=H^{-1}\mathcal WH^{-1}.
 \label{eq:QIV-centered-score-linearization}
\end{align}
The conclusions also apply after adding a possibly random term
$\gamma'C_{NT}$ to the inner objective, with $q_{NT}$ replaced by
$q_{NT}+C_{NT}$ in the score-order conditions and conclusions.
\end{lemma}

\begin{proof}
For fixed $\theta$, terms involving only $\theta$ do not affect the
inner minimizer. The remaining quadratic in~\eqref{JointExpansion} is
\begin{align*}
 \gamma'\{q_{NT}+J'(\theta-\theta^0)\}+\tfrac12\gamma'H\gamma.
\end{align*}
Its minimizer is
$g_{NT}(\theta)=-H^{-1}\{q_{NT}+J'(\theta-\theta^0)\}$.
Since $H>0$, comparison on a sufficiently large ball in $\gamma$ and
convexity confine every inner minimizer to that ball. Completing the
square in the uniform expansion gives~\eqref{InnerApprox}.
The weighted form~\eqref{JointExpansion} also gives, uniformly on the
shrinking parameter neighborhood,
\begin{align*}
 \|\widehat\gamma(\theta)-g_{NT}(\theta)\|
     =o_P(a_{NT}+\|\theta-\theta^0\|).
\end{align*}

By consistency, choose $\epsilon_{NT}\to0$ slowly enough that
$\epsilon_{NT}/a_{NT}\to\infty$ and
$\mathbb P(\|\widehat\theta-\theta^0\|\leq\epsilon_{NT})\to1$.
Full row rank of $J$ gives a constant $c>0$ such that
$\|H^{-1}J'd\|\geq c\|d\|$.
Since $q_{NT}=O_P(a_{NT})$, the outer objective at $\theta^0$ is
$O_P(a_{NT}^2)$. For a sufficiently large $M$, its value at every
$\theta$ with $\|\theta-\theta^0\|>Ma_{NT}$ in the shrinking
neighborhood exceeds its value at $\theta^0$, with arbitrarily high
probability. This uses the preceding uniform bound and the eigenvalue
bounds for $\mathcal W_{NT}$. Global minimization on $\Theta$
therefore gives the stated rate.

If $q_{NT}=O_P(r_{NT})$, take $a_{NT}=r_{NT}$.
Uniformly on bounded $r_{NT}$-balls, the outer objective equals
\begin{align*}
 \{q_{NT}+J'(\theta-\theta^0)\}'W_{NT}
       \{q_{NT}+J'(\theta-\theta^0)\}+o_P(r_{NT}^2),
 \qquad W_{NT}=H^{-1}\mathcal W_{NT}H^{-1}.
\end{align*}
The quadratic has minimizer
$\theta^0-(JW_{NT}J')^{-1}JW_{NT}q_{NT}$.
Both this point and the outer minimizer lie in a bounded
$r_{NT}$-ball with arbitrarily high probability. Comparing their
objective values and using $JWJ'>0$ makes their difference
$o_P(r_{NT})$. Since $W_{NT}\to_P W$ and $q_{NT}=O_P(r_{NT})$,
this proves~\eqref{eq:QIV-centered-score-linearization}.
An affine addition changes only $q_{NT}$ and leaves all remainder
bounds unchanged.
\end{proof}

\subsubsection{Stochastic increments of the inner score}

Write $\ell=\log(NT)$ and
\begin{align*}
 \xi_{NT}(\mu)
   =\partial_\gamma\mathcal L_{NT}(\mu)
           -\mathbb E_A\partial_\gamma\mathcal L_{NT}(\mu).
\end{align*}
At ties, define the score by ordering tied residuals in a fixed order.
For $N\geq T$, any fixed $C<\infty$, and $0<d\leq C/T$,
\begin{align}
 \sup_{\substack{\|\mu-\mu^0\|,\|\mu'-\mu^0\|\leq C/T\\
                         \|\mu'-\mu\|\leq d}}
   \|\xi_{NT}(\mu')-\xi_{NT}(\mu)\|
   =O_P\big(r_{NT}\sqrt{d\ell}+r_{NT}^2\ell^2\big).
 \label{eq:QIV-score-modulus}
\end{align}
The bound also holds in conditional expectation after truncation on an
event of probability tending to zero.

Apply~\eqref{eq:FE-score-increment} to each coordinate of
the score increment. Its conditional second moment is at most $Cd/T$.
Throughout the $C/T$-ball, a rank-weight change involves only observations
within $C'/T$ of an original central order statistic.
Conditional on that order statistic, the count is a sum of two binomial
counts with bounded means, except on a binomial tail event of probability
$Ce^{-cT}$. Thus, outside an event with probability at most
$Ce^{-cT}+(NT)^{-K}$, the count is at most $C_K\ell$.
Set each individual score increment to zero on the exceptional event. Its envelope is
then $K_T=C\ell/T$, uniformly over the parameter pairs. The truncation
changes the empirical process with probability tending to zero and its
mean by $o((NT)^{-1})$. Here $N$ grows at most polynomially in $T$ under
the theorem's growth condition.

The centered suprema are measurable by the face-selection argument in
Appendix~\ref{app:measurability}. On any collection of $N$ individual
samples, the uncentered score is constant
between the residual-ordering hyperplanes. There are at most $NT^2$
such hyperplanes in the fixed-dimensional joint parameter space.
For parameter pairs, the number of distinct evaluation vectors is
therefore at most $(C NT^2)^{2\dim\mu}$.
Let $R$ be the expected supremum of the Rademacher average of one
coordinate of the truncated increments, and let $\sigma^2=Cd/T$.
The finite-class Rademacher bound and the contraction inequality for
the squared functions give
\begin{align*}
 R^2\leq\frac{C\ell}{N}\{\sigma^2+4K_TR\}.
\end{align*}
Indeed, the square of the conditional Rademacher bound is at most the
largest empirical second moment times $C\ell/N$. The expected excess of that
second moment over $\sigma^2$ is at most $4K_TR$, by symmetrization
and contraction. Solving the quadratic inequality and symmetrizing the
centered empirical average gives
\begin{align*}
 C\left\{\sqrt{\frac{d\ell}{NT}}+\frac{\ell^2}{NT}\right\},
\end{align*}
which proves~\eqref{eq:QIV-score-modulus}.

The selected score at an inner minimizer has norm $O((NT)^{-1})$,
uniformly in the joint parameter neighborhood. To check this, condition
on all regressors. The outcomes have a joint density. If tied residual
groups have sizes $m_1,m_2,\ldots$, their independent equality
constraints number $\sum_j(m_j-1)$. With probability one, this number
cannot exceed $\dim\mu$, because the joint parameter has that fixed
dimension. Therefore at most $2\dim\mu$ observations belong to
nontrivial tied groups. Reordering them changes the full-sample score
by at most $C\dim\mu/(NT)$. The zero vector belongs to the inner
subdifferential, giving the bound for the selected score.

\subsubsection{Asymptotic distribution}

\begin{proof}[Proof of Theorem~\ref{th:EndAsy}]
The moment results and~\eqref{eq:profile-score-fourth-moment} give
\begin{align}
 q_{NT}=B_{NT}/T+r_{NT}Z_{NT}+O_P(T^{-2}),
 \qquad Z_{NT}\Rightarrow\mathcal N(0,V^{\rm FE}).
 \label{eq:QIV-score-decomposition}
\end{align}
Put $W_{NT}=H^{-1}\mathcal W_{NT}H^{-1}$ and
$A_{NT}=(JW_{NT}J')^{-1}JW_{NT}$.
When $N\leq T$, the score is $O_P(r_{NT})$.
Lemma~\ref{lemma:inner} then gives
\begin{align*}
 \widehat\theta^{\rm FE\text{-}IV}-\theta^0
       =-A_{NT}q_{NT}+o_P(r_{NT}),
\end{align*}
and the result follows directly.

Consider $N\geq T$. Let $\gamma_T(\theta)$ minimize the conditional
expected inner objective near zero. The implicit function theorem,
$H>0$, and Lemma~\ref{lemma:hessian} give a twice differentiable map
with bounded second derivative. Its derivative converges locally to
$-H^{-1}J'$. Let $\theta_T$ minimize
$\gamma_T(\theta)'\mathcal W_{NT}\gamma_T(\theta)$ near $\theta^0$.
The Hessian of this population criterion is positive definite there
for all sufficiently large $N,T$, with probability tending to one.
Although $\mathcal W_{NT}$ can be data dependent, all subsequent bounds
are uniform over matrices in a fixed compact set of positive definite
matrices. Taylor expansion of the population first-order conditions gives
\begin{align}
 \theta_T-\theta^0
    &=-A_{NT}B_{NT}/T
           +O_P(T^{-2}+T^{-1}N^{-1/2}),
 \label{eq:QIV-population-center}\\
 b_T^\gamma:=\gamma_T(\theta_T)&=O_P(T^{-1}),
 \qquad
 D_T:=\frac{\partial\gamma_T(\theta_T)}{\partial\theta'}
        =-H^{-1}J'+O_P(T^{-1}+N^{-1/2}),
 \nonumber\\
 D_T'\mathcal W_{NT}b_T^\gamma&=0.
 \nonumber
\end{align}
For the first equality, the population inner equation is
$B_{NT}/T+J'(\theta-\theta^0)+H\gamma$
up to $O_P(T^{-2}+T^{-1}N^{-1/2})$ on an $O(T^{-1})$-ball.
Substituting its solution into the outer first-order condition gives
$JW_{NT}\{B_{NT}/T+J'(\theta_T-\theta^0)\}=O_P(T^{-2}+T^{-1}N^{-1/2})$.

The rate part of Lemma~\ref{lemma:inner} first gives
$\widehat\theta^{\rm FE\text{-}IV}-\theta^0=O_P(T^{-1})$.
Uniformly on such a neighborhood, the inner first-order condition,
the subgradient bound, and~\eqref{eq:QIV-score-modulus} give
$\widehat\gamma(\theta)-\gamma_T(\theta)=O_P(r_{NT})$.
To obtain this bound, note that
$\xi_{NT}(\mu^0)=O_P(r_{NT})$ and the modulus at $d=C/T$ is
$o_P(r_{NT})$. The expected inner Hessian is uniformly positive definite.

Let $H_T^\gamma$ be this expected Hessian at
$\mu_T=(\theta_T',\gamma_T(\theta_T)')'$ and set
$\eta_T=-(H_T^\gamma)^{-1}\xi_{NT}(\mu_T)$.
For $\|h\|\leq d$, where $r_{NT}\leq d\leq C/T$, expanding the
inner first-order condition gives
\begin{align}
 \widehat\gamma(\theta_T+h)
    &=b_T^\gamma+D_Th+\eta_T+e_{NT}(h),
 \label{eq:QIV-refined-inner}\\
 \sup_{\|h\|\leq d}\|e_{NT}(h)\|
    &=O_P\big(d^2+r_{NT}\sqrt{d\ell}+r_{NT}^2\ell^2\big).
 \nonumber
\end{align}
The deterministic Taylor remainder is $O(d^2)$.
The inner displacement from its population value is $O_P(r_{NT})$,
so the stochastic difference is bounded by the modulus at $d+O_P(r_{NT})$.
The subgradient remainder is $O(r_{NT}^2)$.
The expectation form of~\eqref{eq:QIV-score-modulus} makes these bounds
available simultaneously on dyadic balls.

The quadratic outer approximation from~\eqref{eq:QIV-refined-inner}
has minimizer
\begin{align*}
 h_T^*=-(D_T'\mathcal W_{NT}D_T)^{-1}D_T'\mathcal W_{NT}\eta_T
       =O_P(r_{NT}).
\end{align*}
Write $\bar\gamma_T(h)=b_T^\gamma+D_Th+\eta_T$ and define
$Q_{NT}(h)=\widehat\gamma(\theta_T+h)'\mathcal W_{NT}
\widehat\gamma(\theta_T+h)$ and
$Q_{NT}^{\rm quad}(h)=\bar\gamma_T(h)'\mathcal W_{NT}\bar\gamma_T(h)$.
Substitution of~\eqref{eq:QIV-refined-inner} gives the exact difference
\begin{align*}
 Q_{NT}(h)-Q_{NT}^{\rm quad}(h)
 =2e_{NT}(h)'\mathcal W_{NT}\bar\gamma_T(h)
       +e_{NT}(h)'\mathcal W_{NT}e_{NT}(h).
\end{align*}
Uniformly for $\|h\|\leq d$,
$\|\bar\gamma_T(h)\|=O_P(T^{-1}+d+r_{NT})$.
The eigenvalues of $\mathcal W_{NT}$ are bounded in probability.
The outer approximation error is consequently
\begin{align*}
 &\sup_{\|h\|\leq d}|Q_{NT}(h)-Q_{NT}^{\rm quad}(h)|\\
 &\quad=O_P\!\bigg[(T^{-1}+d+r_{NT})
       \big\{d^2+r_{NT}\sqrt{d\ell}+r_{NT}^2\ell^2\big\}\\
 &\hspace{43mm}+\big\{d^2+r_{NT}\sqrt{d\ell}+r_{NT}^2\ell^2\big\}^2\bigg].
\end{align*}
At $d=r_{NT}$ this is $o_P(r_{NT}^2)$ under
$N\ell^2/T^3\to0$. The potentially largest ratios are
\begin{align*}
 \frac{\sqrt\ell}{T\sqrt{r_{NT}}}
       =\left(\frac{N\ell^2}{T^3}\right)^{1/4},
 \qquad \frac{\ell^2}{T},
 \qquad \sqrt{r_{NT}\ell},
 \qquad r_{NT}\ell^2.
\end{align*}
They all converge to zero. For radii $d=2^jMr_{NT}$, the stochastic
terms in the first product, divided by $d^2$, decrease at least as
$2^{-j/2}$. Their sums are finite.
The inner error is uniformly $o_P(T^{-1})$ on the $C/T$-ball, so its
squared contribution is bounded by $o_P(T^{-1})$ times its norm.
Since $D_T'\mathcal W_{NT}b_T^\gamma=0$, the quadratic difference is
\[
 Q_{NT}^{\rm quad}(h)-Q_{NT}^{\rm quad}(0)
 =h'D_T'\mathcal W_{NT}D_Th+2h'D_T'\mathcal W_{NT}\eta_T.
\]
The smallest eigenvalue of $D_T'\mathcal W_{NT}D_T$ is bounded away
from zero. Thus this difference is at least
$c\|h\|^2-O_P(r_{NT})\|h\|$.
For sufficiently large $M$, it is positive on
$Mr_{NT}\leq\|h\|\leq C/T$ with arbitrarily high probability.
The summable error bounds on the dyadic balls preserve this comparison
for $Q_{NT}$. The outer minimizer already lies in a $C/T$-ball with
arbitrarily high probability, so global minimization gives
$\widehat\theta^{\rm FE\text{-}IV}-\theta_T=O_P(r_{NT})$.
On a fixed multiple of this smaller ball, the approximation error is
$o_P(r_{NT}^2)$ and
\[
 Q_{NT}^{\rm quad}(h)-Q_{NT}^{\rm quad}(h_T^*)
 =(h-h_T^*)'D_T'\mathcal W_{NT}D_T(h-h_T^*).
\]
Comparing $Q_{NT}$ at its minimizer and at $h_T^*$ therefore gives
\begin{align*}
 \widehat\theta^{\rm FE\text{-}IV}-\theta_T=h_T^*+o_P(r_{NT}).
\end{align*}

By~\eqref{eq:QIV-score-modulus},
$\xi_{NT}(\mu_T)-\xi_{NT}(\mu^0)=o_P(r_{NT})$.
Also $D_T\to_P-H^{-1}J'$ and $H_T^\gamma\to_P H$.
Combining~\eqref{eq:QIV-population-center}--\eqref{eq:QIV-refined-inner}
therefore gives
\begin{align*}
 \sqrt{NT}\left(\widehat\theta^{\rm FE\text{-}IV}-\theta^0
                   +A_{NT}B_{NT}/T\right)
    =-(JWJ')^{-1}JWZ_{NT}+o_P(1).
\end{align*}
The error in the population center is negligible because
$\sqrt{NT}(T^{-2}+T^{-1}N^{-1/2})\to0$.
This proves the stated limit. Conditional convergence of the score
implies its unconditional convergence as in the exogenous proof.
\end{proof}

The pilot rate
$\widehat\theta^{\rm FE\text{-}IV}-\theta^0
=O_P(T^{-1}+(NT)^{-1/2})$
uses only the rate conclusion of Lemma~\ref{lemma:inner}.
It does not require the logarithmic growth restriction used for the
uncorrected centered limit.

\subsection{Proof of Theorem~\ref{th:BCAsy:End}}
\label{app:BiasCorrectionEnd}

\begin{proof}
The weight calculations and the score limit
\eqref{eq:BC-score-CLT} continue to hold when the residual ordering is
based on $y_{it}-x_{it}'\theta^0$. The preliminary estimator has error
$O_P(a_{NT})$ by Lemma~\ref{lemma:inner} and the full row rank of $J$.
Lemma~\ref{lemma:BiasCorrection} therefore permits replacing the
preliminary estimator in the correction by $\theta^0$.

The correction is affine in $\gamma$. At $\theta^0$,
\[
 q^*_{NT}:=\partial_\gamma
    {\cal L}^*_{NT}(\theta^0,0,S)
 =\frac1N\sum_iQ^*_{i,T}+o_P((NT)^{-1/2}).
\]
Since $q^*_{NT}=O_P(r_{NT})$, Lemma~\ref{lemma:inner} applies to the
corrected objective with $a_{NT}=r_{NT}$. Uniformly for bounded
$\sqrt{NT}\,h$,
\[
 \widehat\gamma^*(\theta^0+h,S)
 =-H^{-1}(q^*_{NT}+J'h)+o_P((NT)^{-1/2}).
\]
The consistency and global minimization conditions in
Theorem~\ref{th:BCAsy:End} and the positive definiteness of $JWJ'$ imply
$\widehat\theta^*(S)-\theta^0=O_P((NT)^{-1/2})$.
After multiplication by $NT$, the criterion has the uniform quadratic
approximation
\[
 (\sqrt{NT}\,q^*_{NT}+J'v)'W
       (\sqrt{NT}\,q^*_{NT}+J'v)
\]
on bounded sets of $v=\sqrt{NT}(\theta-\theta^0)$.
Its minimizer is
$-(JWJ')^{-1}JW\sqrt{NT}\,q^*_{NT}$.
The positive definite quadratic term and the uniform remainder therefore
give
\[
 \sqrt{NT}(\widehat\theta^*(S)-\theta^0)
 =-(JWJ')^{-1}JW\sqrt{NT}\,\frac1N\sum_iQ^*_{i,T}+o_P(1).
\]
Combining this expansion with~\eqref{eq:BC-score-CLT} gives the covariance
$(JWJ')^{-1}JWV^*WJ'(JWJ')^{-1}$. The conditional limiting distribution
has a nonrandom covariance, so the same convergence holds unconditionally.
\end{proof}

\subsection{Proof of Consistency of the Asymptotic Variance Estimators}
\label{app:VarianceConsistency}

Write $d_{NT}=\sqrt{\log(N+1)/T}+T^{-1}+(NT)^{-1/2}$ and
$\varepsilon_{it}=y_{it}-z_{it}'\theta^0-\alpha_i^0$ in the exogenous case.
The conditional density lower bound and the exponential bound for a binomial
tail give
\[
 \max_{i\leq N}|\widehat\alpha_i-\alpha_i^0|
 =O_P(d_{NT}),\qquad
 \max_{i,t}|\hat\varepsilon_{it}-\varepsilon_{it}|=O_P(d_{NT}).
\]
Indeed, at either endpoint $\alpha_i^0\pm v$ the population cdf differs
from $\tau_0$ by at least $cv$ for small $v$. The probability that the
sample quantile lies outside this interval is at most $2\exp(-c'Tv^2)$.
Take $v=C\sqrt{\log(N+1)/T}$, sum over $i$, and use the bounded regressors
and $\widehat\theta^{\rm FE}-\theta^0=O_P(T^{-1}+(NT)^{-1/2})$.
The empirical quantile changes by at most
$\max_t|z_{it}'(\widehat\theta^{\rm FE}-\theta^0)|$ when the residuals
are evaluated at $\widehat\theta^{\rm FE}$.

Let $P_{it}$ denote any entry of $1$, $z_{it}$, or $z_{it}z_{it}'$.
Conditional on $A_i$, the variables $K_h(\varepsilon_{it})P_{it}$ are
independent and bounded by $C/h$. An exponential bound and a union bound give
\[
 \max_{i\leq N}\left|\frac1T\sum_t K_h(\varepsilon_{it})P_{it}
 -\mathbb E[K_h(\varepsilon_{it})P_{it}\mid A_i]\right|
 =O_P\!\left(\sqrt{\frac{\log(N+1)}{Th^2}}\right)=o_P(1).
\]
Symmetry of $K$ and the uniform density derivative bounds imply
\[
 \mathbb E[K_h(\varepsilon_{it})P_{it}\mid A_i]
 =\mathbb E[P_{it}f(\alpha_i^0\mid z_{it},A_i)\mid A_i]+O(h^2)
\]
uniformly in $i$. The Lipschitz bound for $K_h$ also gives
\[
 \max_{i\leq N}\left|\frac1T\sum_t
 \{K_h(\hat\varepsilon_{it})-K_h(\varepsilon_{it})\}P_{it}\right|
 =O_P(d_{NT}/h^2)=o_P(1).
\]
Consequently, $\max_i|\hat f_i-f(\alpha_i^0\mid A_i)|=o_P(1)$,
$\max_i\|\hat\lambda_i-\lambda_i\|=o_P(1)$, and
$\max_i\|\widehat V_{1,i}-V_{1,i}\|=o_P(1)$.
The same bounds for the unweighted first and second moments of $z_{it}$
give $\widehat H\to_P H$ and $\widehat V^{\rm FE}\to_P V^{\rm FE}$.
Since $b_T=O(1)$ under Theorem~\ref{th:BCAsy}, we also have
$\widehat V^*\to_P V^*$.

For the endogenous case, replace $z_{it}'\theta^0$ by $x_{it}'\theta^0$
in the residual and include entries of $x_{it}$ and $x_{it}z_{it}'$ among the $P_{it}$.
The conditional density calculation then gives
\[
 \mathbb E[K_h(\varepsilon_{it})P_{it}\mid A_i]
 =\mathbb E[P_{it}f(\alpha_i^0\mid x_{it},z_{it},A_i)\mid A_i]
       +O(h^2),
\]
where $f$ is the density of $y_{it}-x_{it}'\theta^0$ conditional on
$(x_{it},z_{it},A_i)$. The residual substitution bounds use
Assumption~\ref{ass:End} and the pilot rate from
Lemma~\ref{lemma:inner}.
They give $\widehat J\to_P J$ as well.
The asserted covariance consistency follows by continuity of matrix inversion
at the nonsingular limiting matrices.

\section{Supplementary Appendix}\label{app:Supplement}\label{app:MC}

This supplementary appendix contains plots of the empirical bias and standard deviation of the JK and MJ bias-corrected estimators as functions of $S$, for every $(T, \tau_0)$ combination considered in Section~\ref{sec:MC}. Results are reported separately for the exogenous design ($\rho = 0$) of Subsection~\ref{subsec:MC-exog} and for the endogenous design ($\rho = 0.5$) of Subsection~\ref{subsec:MC-endog}.

\Needspace{9\baselineskip}
\subsection{Exogenous Design: Empirical Bias}\label{exog:bias}

\begin{figure}[!htbp]
	\centering
	\begin{tabular}{cc}
		\includegraphics[width=0.40\textwidth]{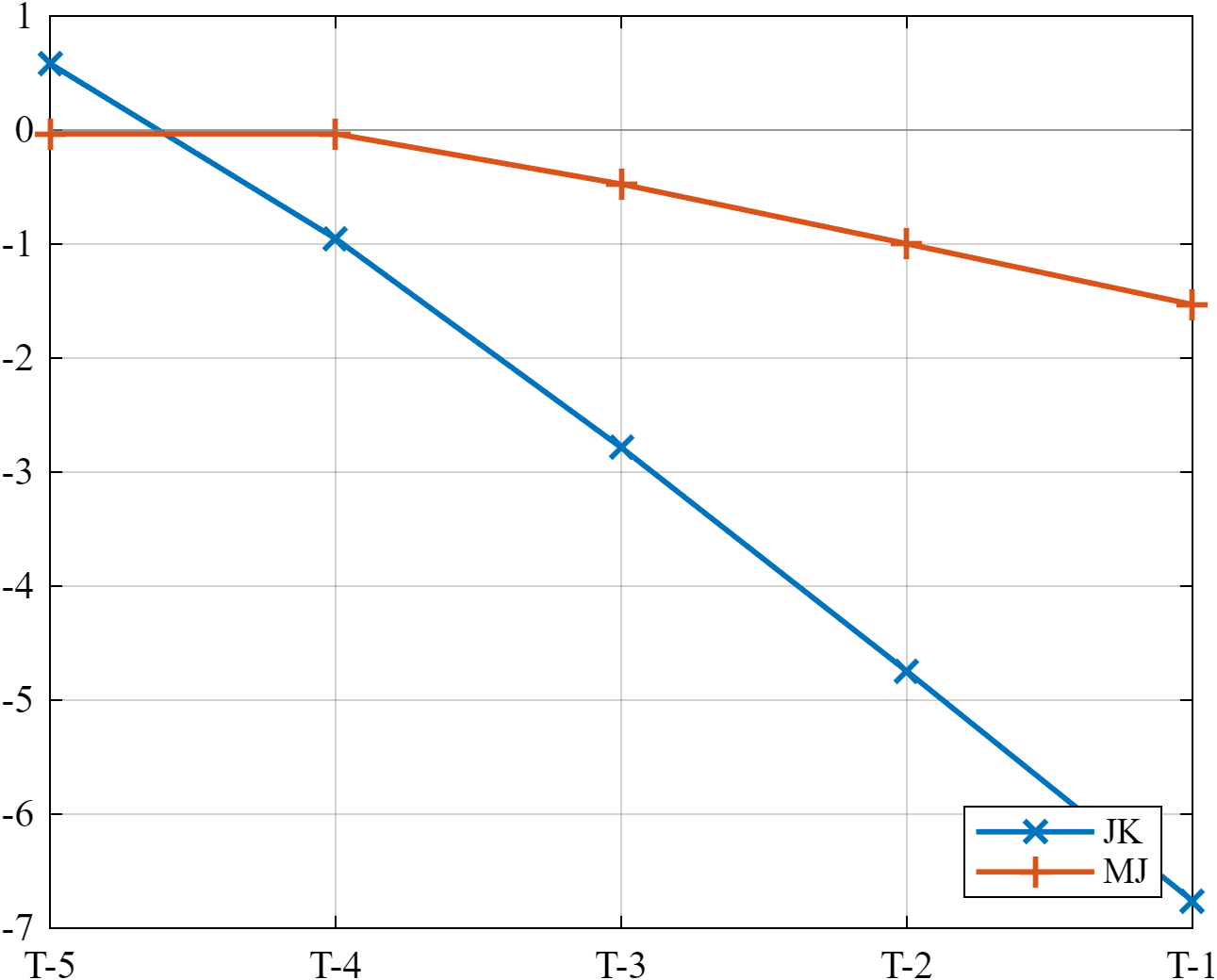} &
		\includegraphics[width=0.40\textwidth]{./Graphs/fig_T20_N1000_MC1000_tau20_bias.png} \\
		\footnotesize $T = 10$ & \footnotesize $T = 20$ \\[6pt]
		\includegraphics[width=0.40\textwidth]{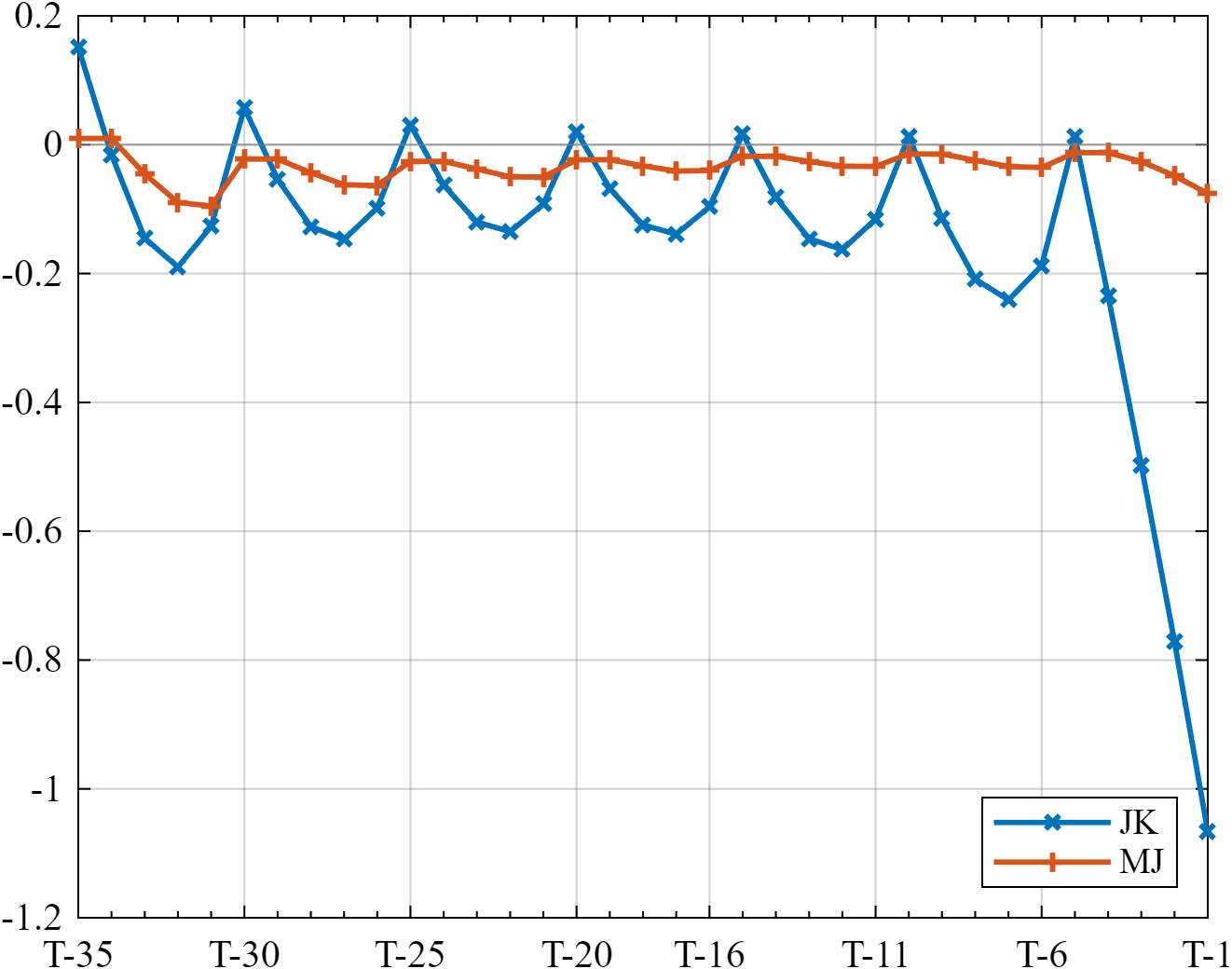} &
		\includegraphics[width=0.40\textwidth]{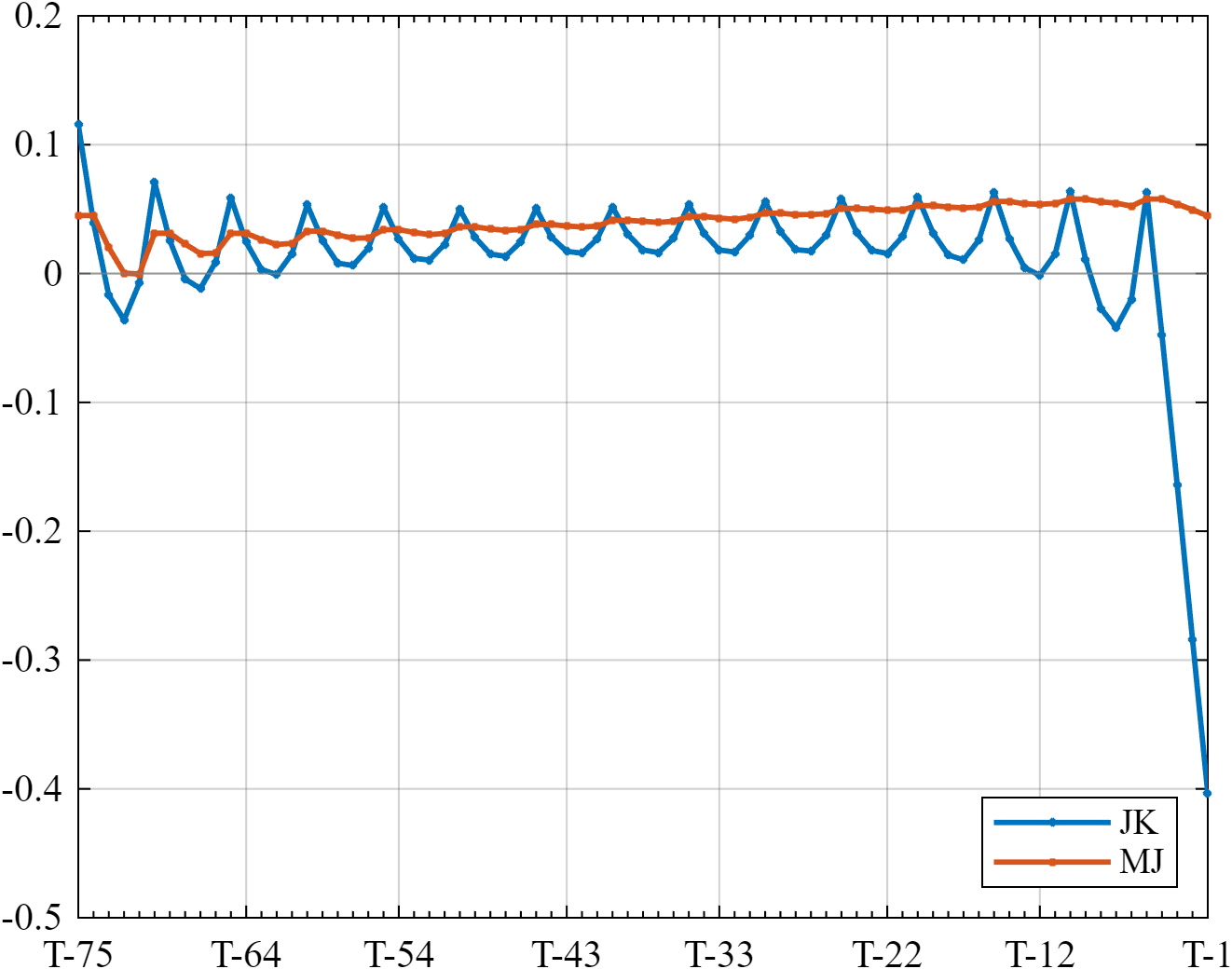} \\
		\footnotesize $T = 40$ & \footnotesize $T = 80$
	\end{tabular}
	\caption{Empirical bias of the JK (blue) and MJ (red) bias-corrected estimators as a function of $S$ at $\tau_0 = 0.2$ with $\rho = 0$.}
	\label{fig:appbias-tau20}
\end{figure}

\begin{figure}[!htbp]
	\centering
	\begin{tabular}{cc}
		\includegraphics[width=0.40\textwidth]{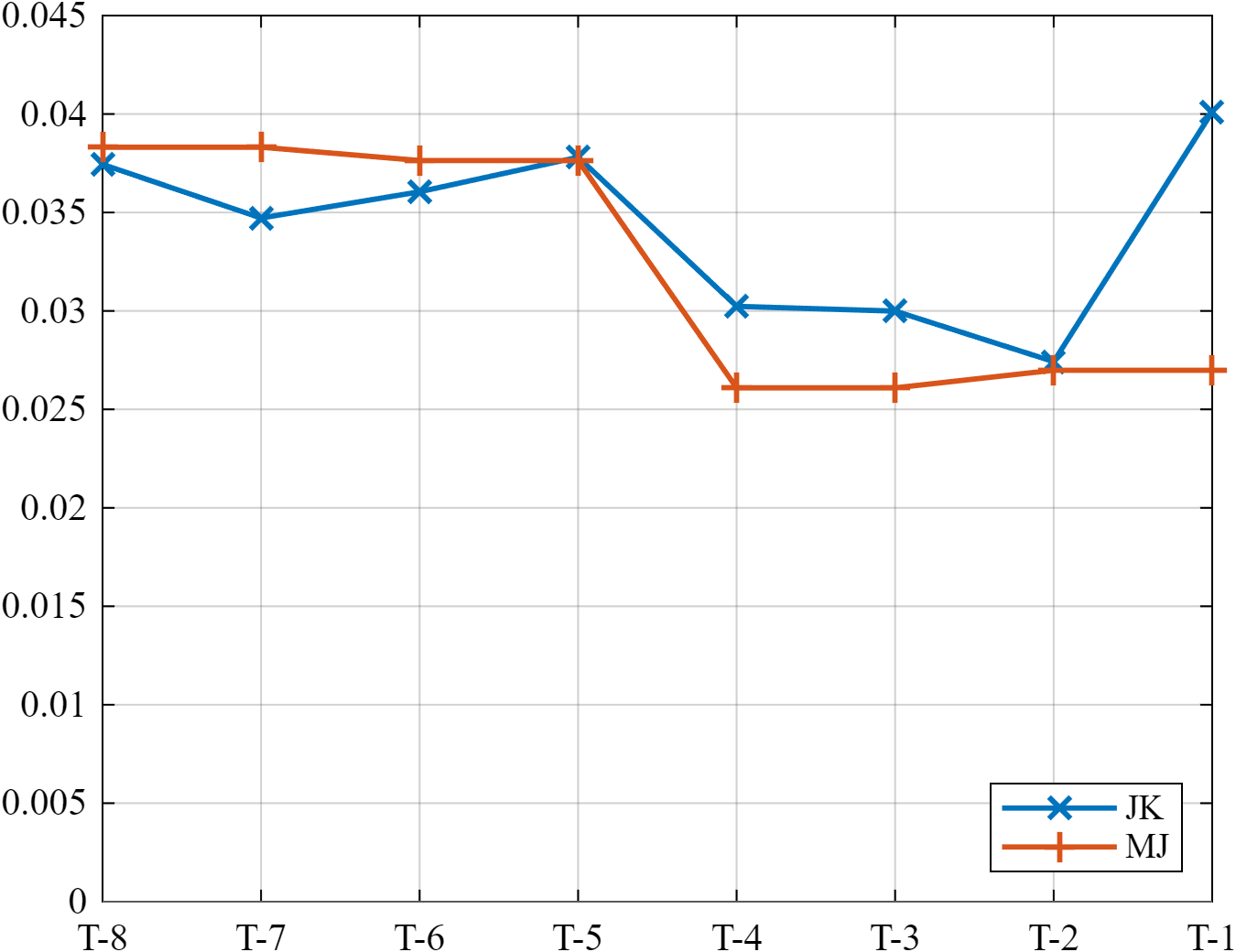} &
		\includegraphics[width=0.40\textwidth]{./Graphs/fig_T20_N1000_MC1000_tau50_bias.png} \\
		\footnotesize $T = 10$ & \footnotesize $T = 20$ \\[6pt]
		\includegraphics[width=0.40\textwidth]{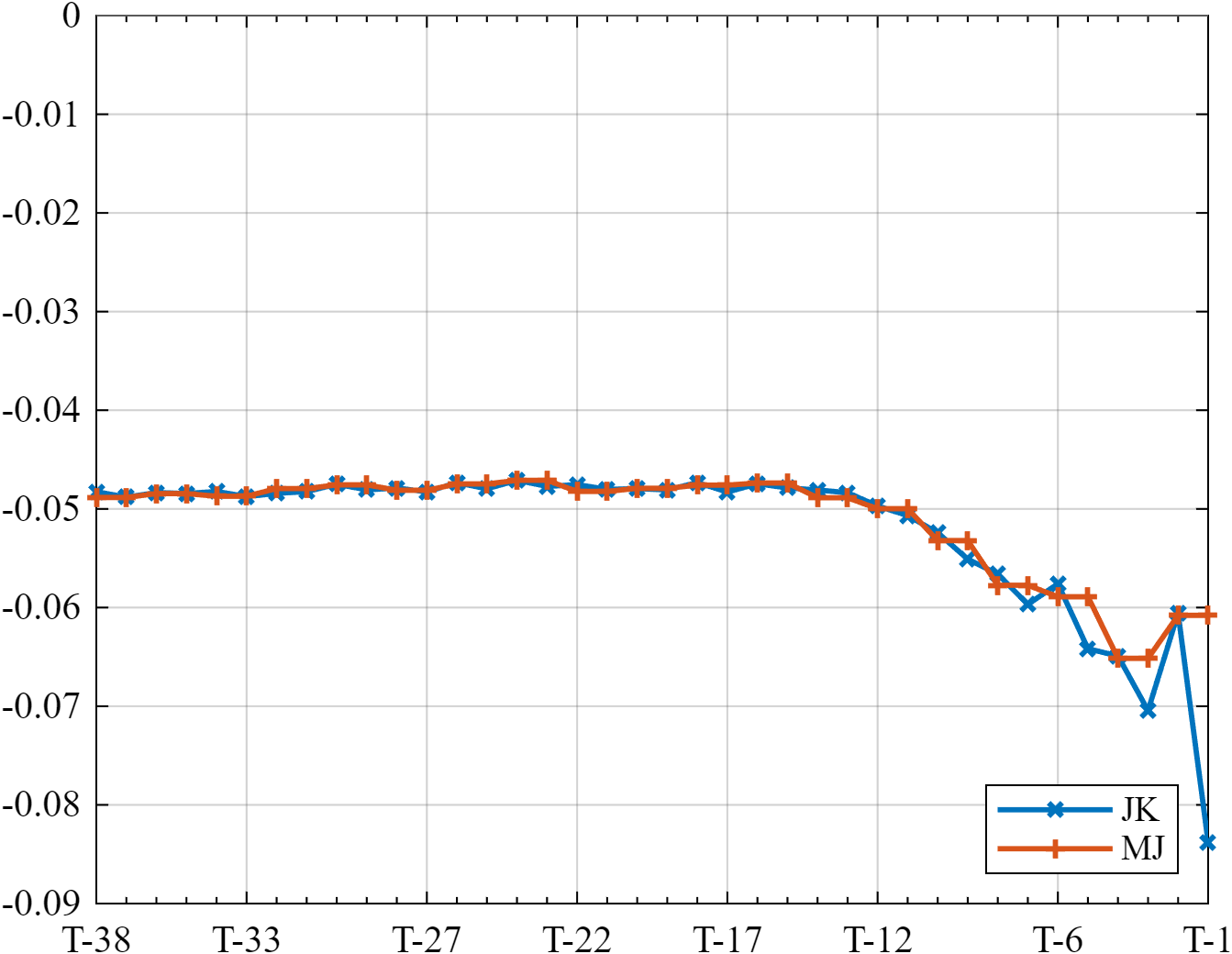} &
		\includegraphics[width=0.40\textwidth]{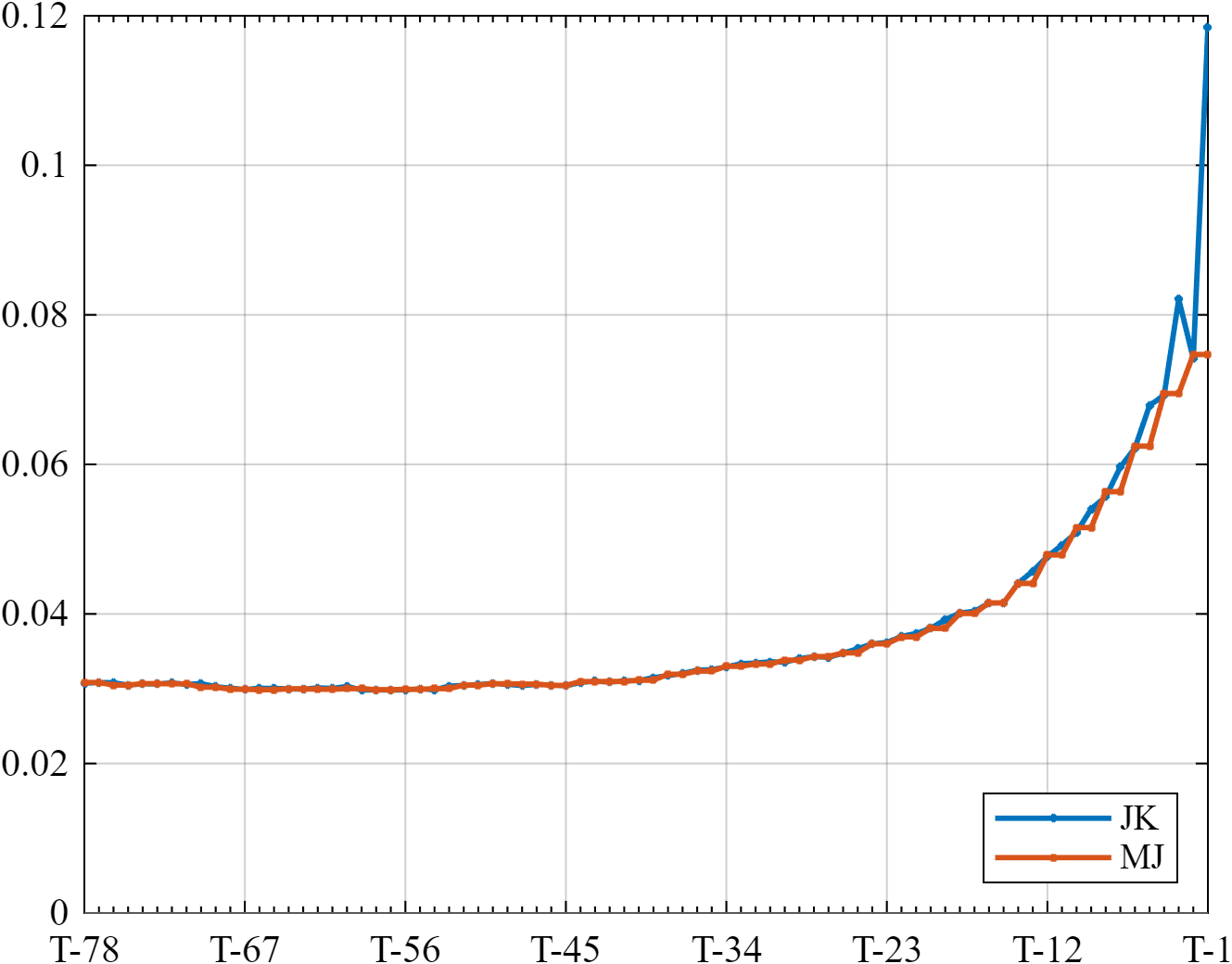} \\
		\footnotesize $T = 40$ & \footnotesize $T = 80$
	\end{tabular}
	\caption{Empirical bias of the JK (blue) and MJ (red) estimators as a function of $S$ at $\tau_0 = 0.5$ with $\rho = 0$.}
	\label{fig:appbias-tau50}
\end{figure}

\begin{figure}[!htbp]
	\centering
	\begin{tabular}{cc}
		\includegraphics[width=0.40\textwidth]{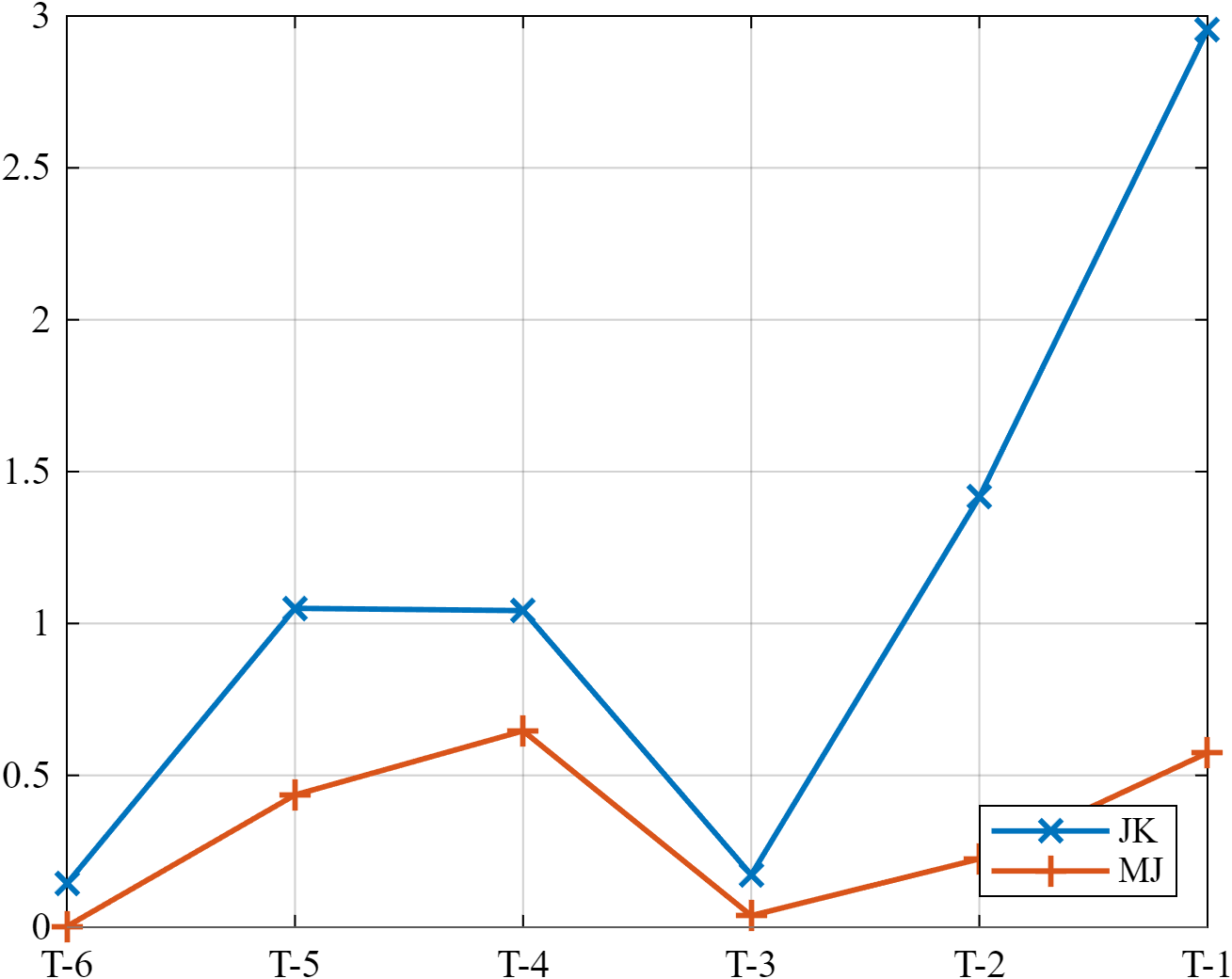} &
		\includegraphics[width=0.40\textwidth]{./Graphs/fig_T20_N1000_MC1000_tau70_bias.png} \\
		\footnotesize $T = 10$ & \footnotesize $T = 20$ \\[6pt]
		\includegraphics[width=0.40\textwidth]{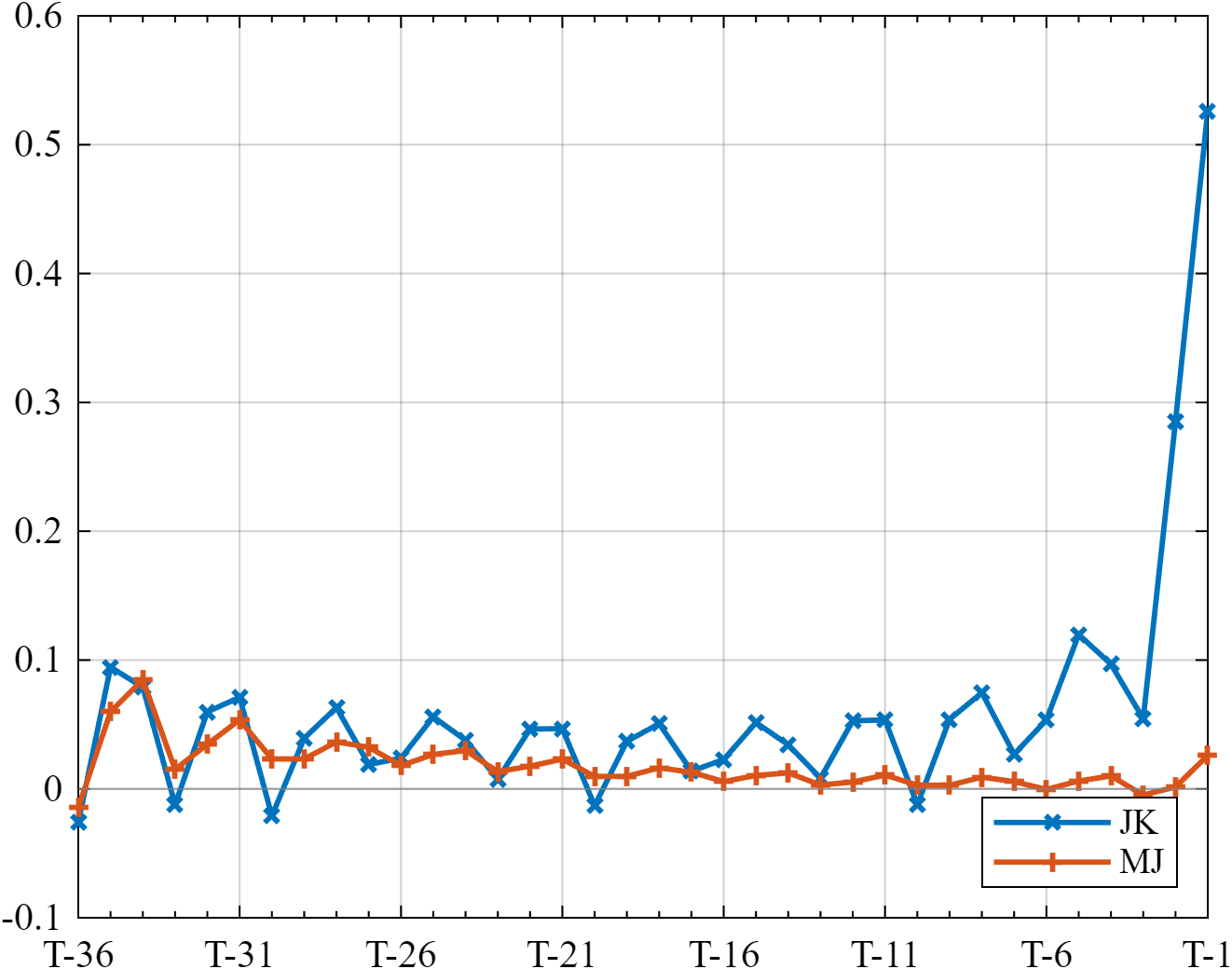} &
		\includegraphics[width=0.40\textwidth]{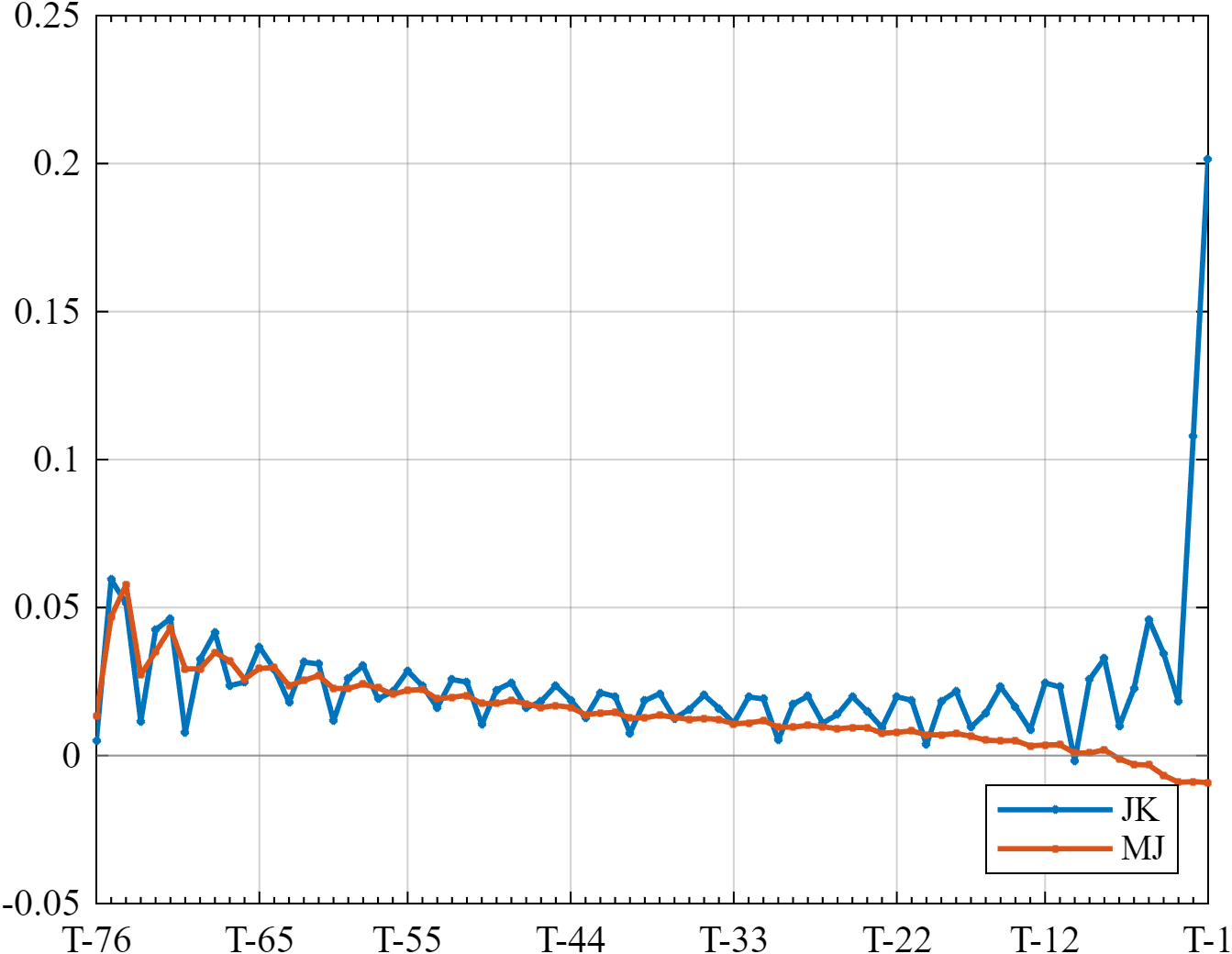} \\
		\footnotesize $T = 40$ & \footnotesize $T = 80$
	\end{tabular}
	\caption{Empirical bias of the JK (blue) and MJ (red) estimators as a function of $S$ at $\tau_0 = 0.7$ with $\rho = 0$.}
	\label{fig:appbias-tau70}
\end{figure}

\FloatBarrier
\subsection{Exogenous Design: Empirical Standard Deviation}\label{exog:std}

\begin{figure}[!htbp]
	\centering
	\begin{tabular}{cc}
		\includegraphics[width=0.40\textwidth]{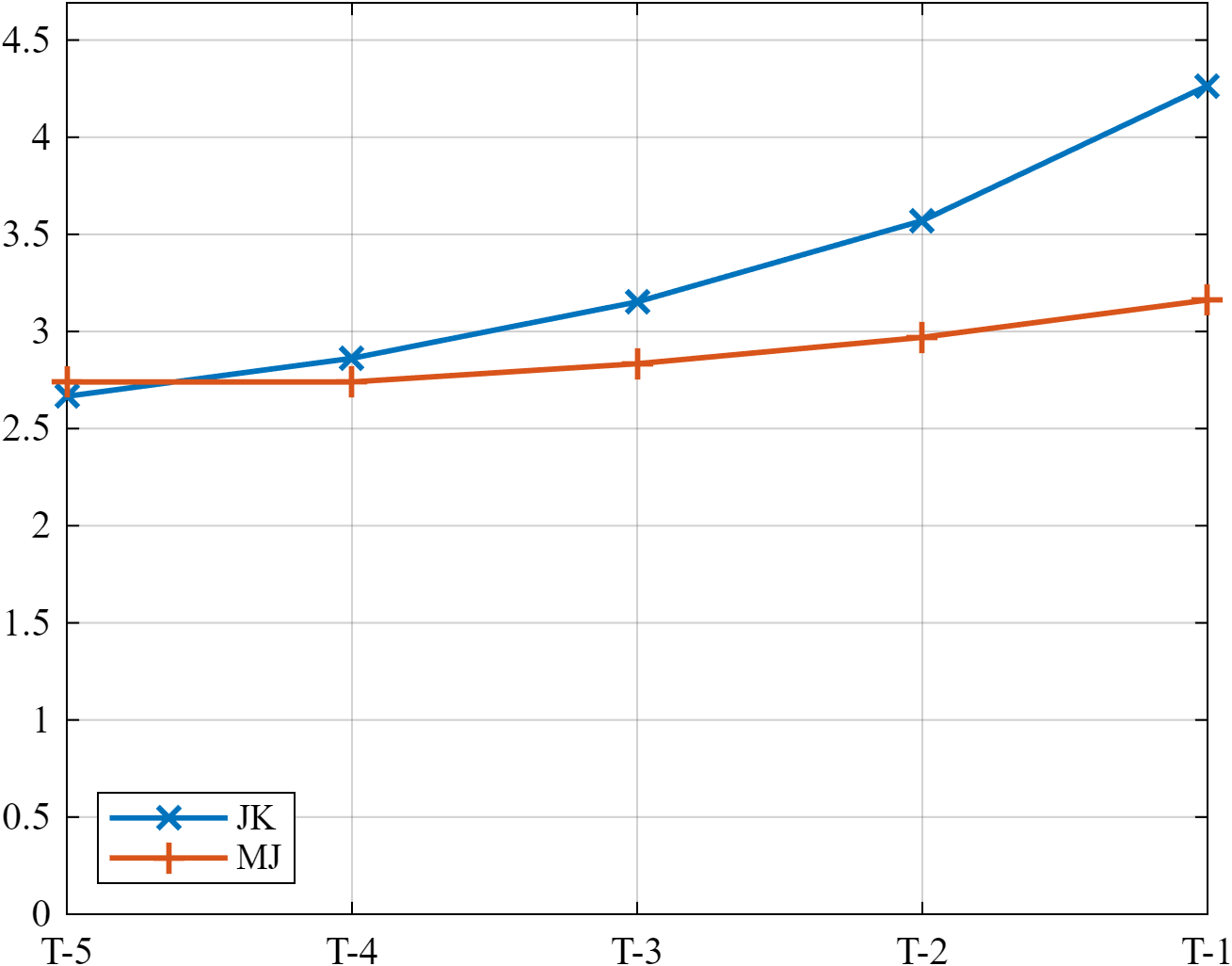} &
		\includegraphics[width=0.40\textwidth]{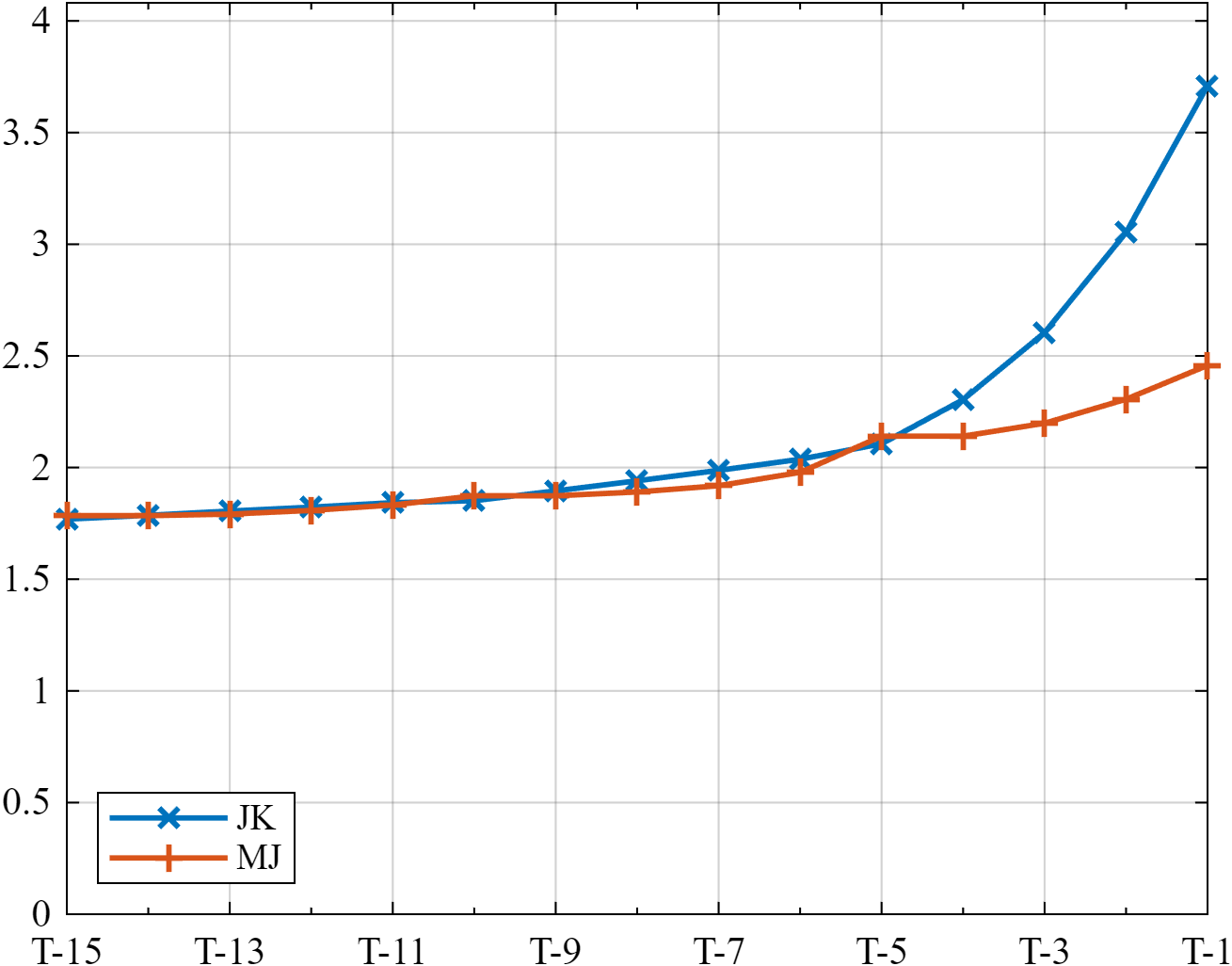} \\
		\footnotesize $T = 10$ & \footnotesize $T = 20$ \\[6pt]
		\includegraphics[width=0.40\textwidth]{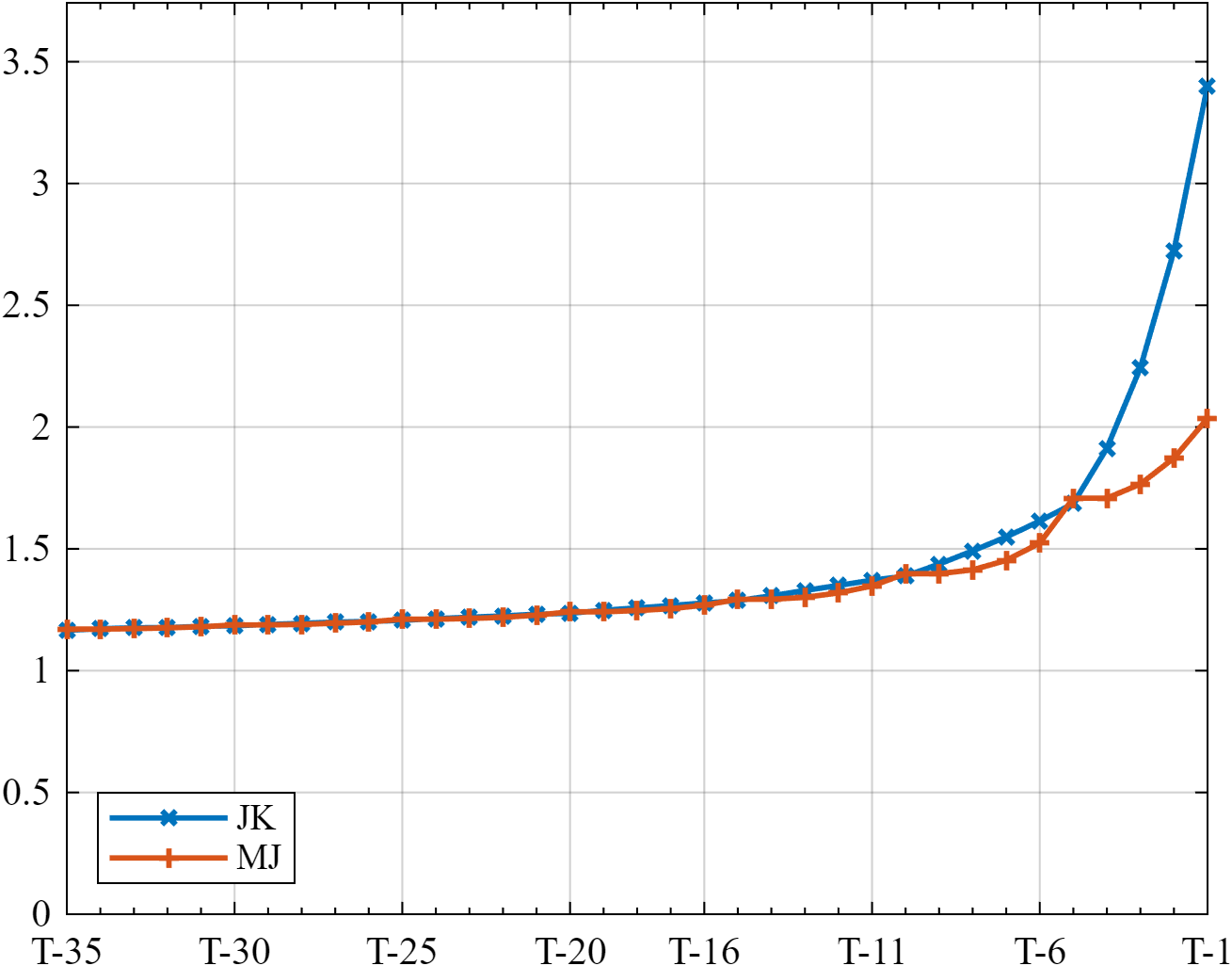} &
		\includegraphics[width=0.40\textwidth]{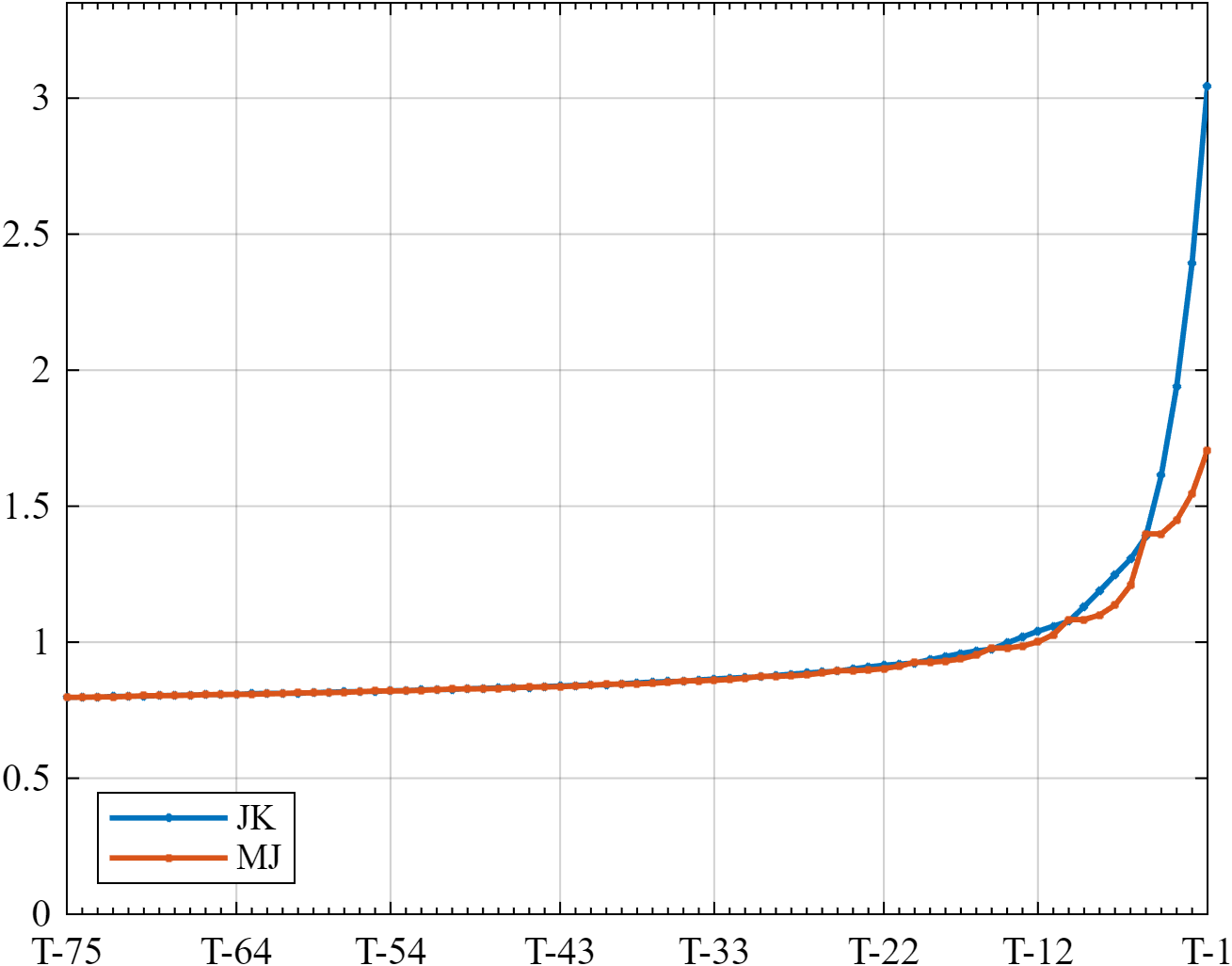} \\
		\footnotesize $T = 40$ & \footnotesize $T = 80$
	\end{tabular}
	\caption{Empirical standard deviation of the JK (blue) and MJ (red) estimators as a function of $S$ at $\tau_0 = 0.2$ with $\rho = 0$.}
	\label{fig:appstd-tau20}
\end{figure}

\begin{figure}[!htbp]
	\centering
	\begin{tabular}{cc}
		\includegraphics[width=0.40\textwidth]{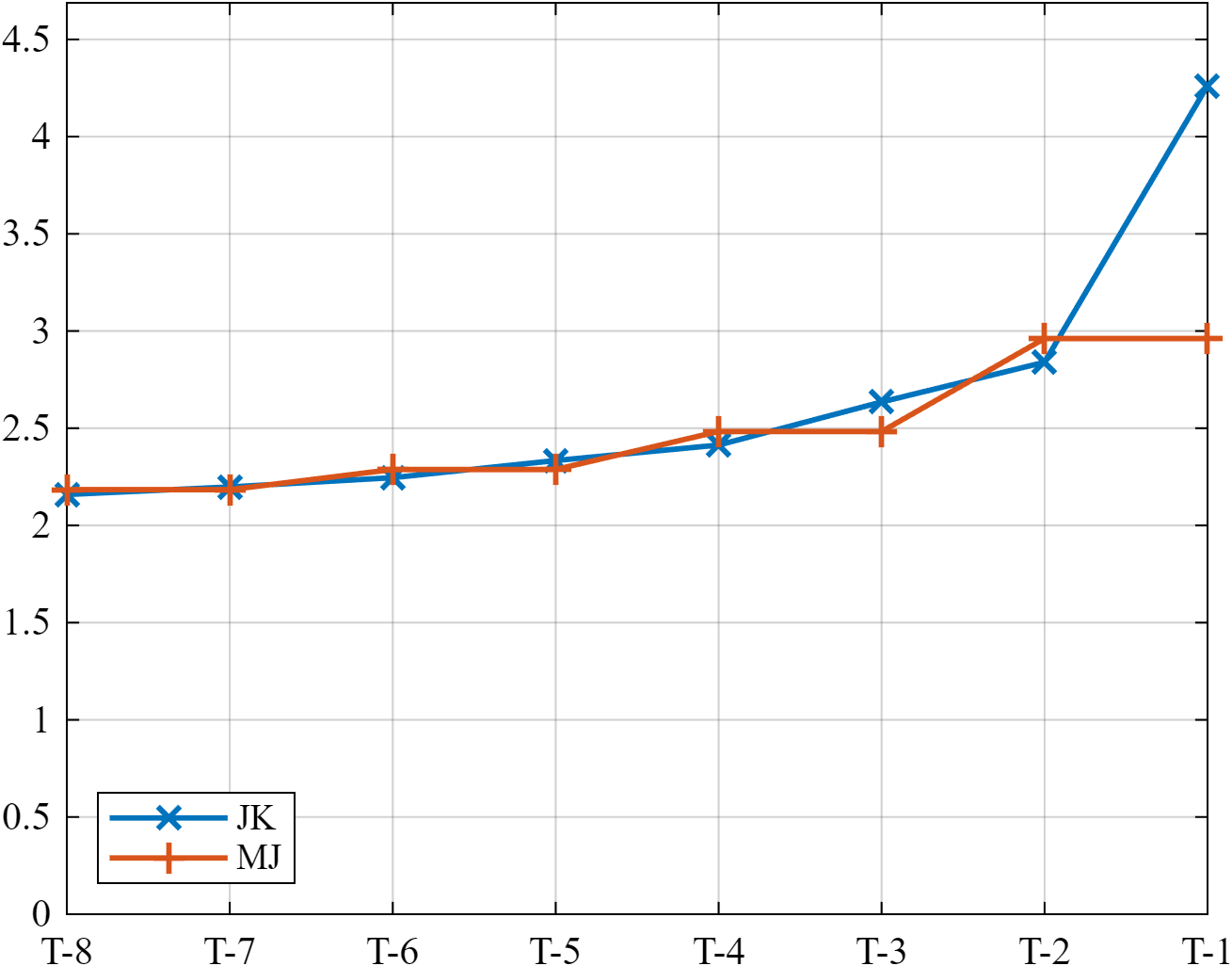} &
		\includegraphics[width=0.40\textwidth]{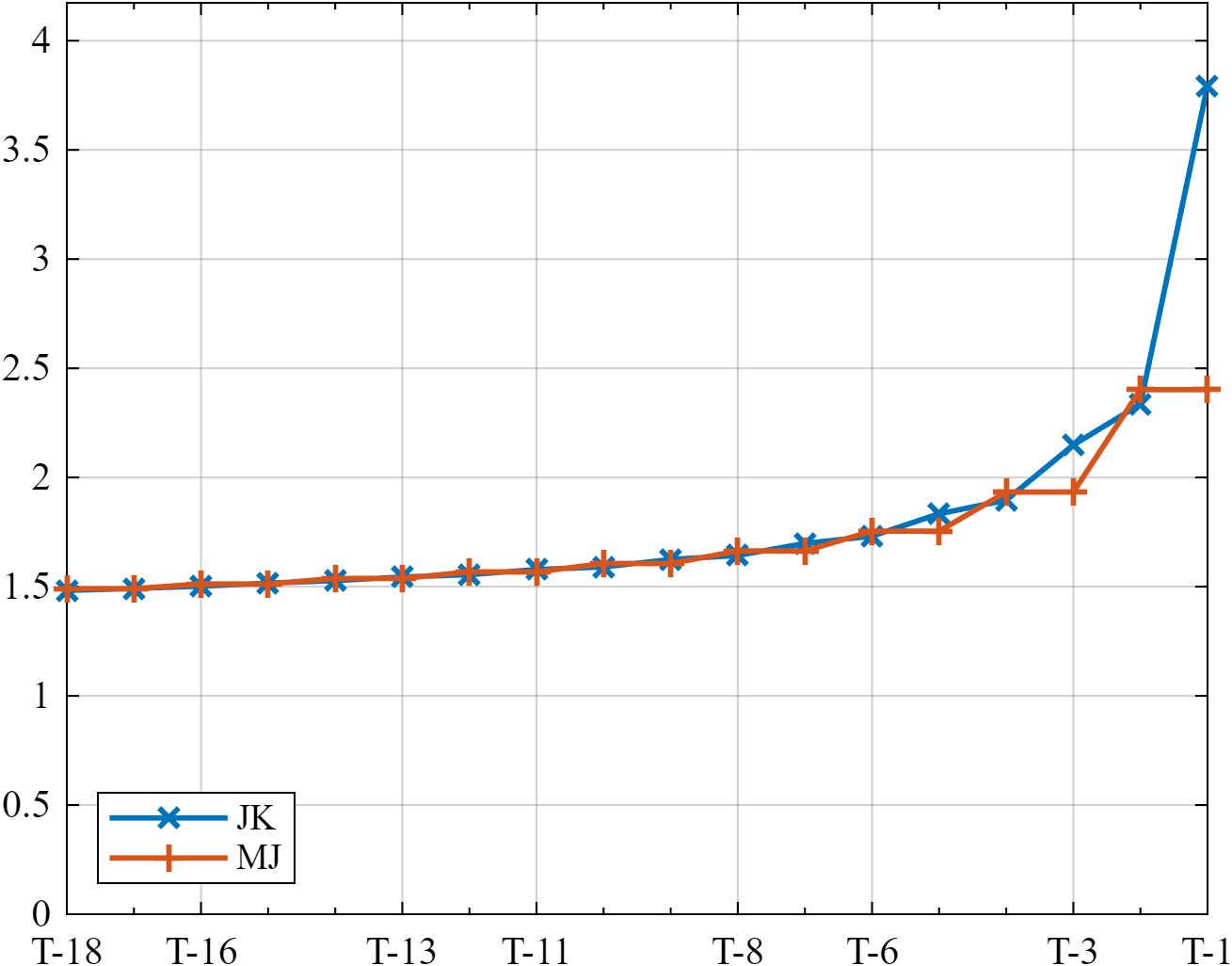} \\
		\footnotesize $T = 10$ & \footnotesize $T = 20$ \\[6pt]
		\includegraphics[width=0.40\textwidth]{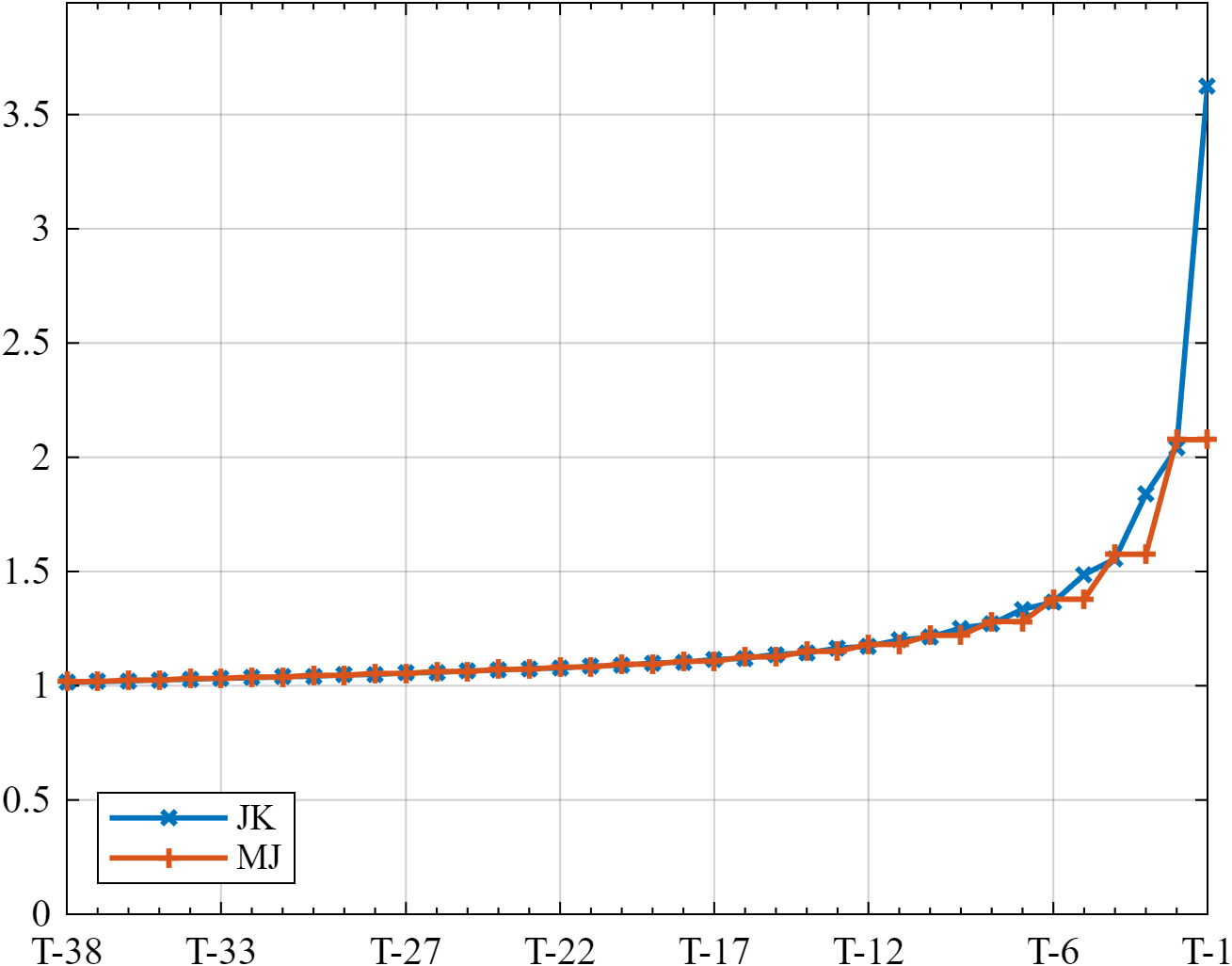} &
		\includegraphics[width=0.40\textwidth]{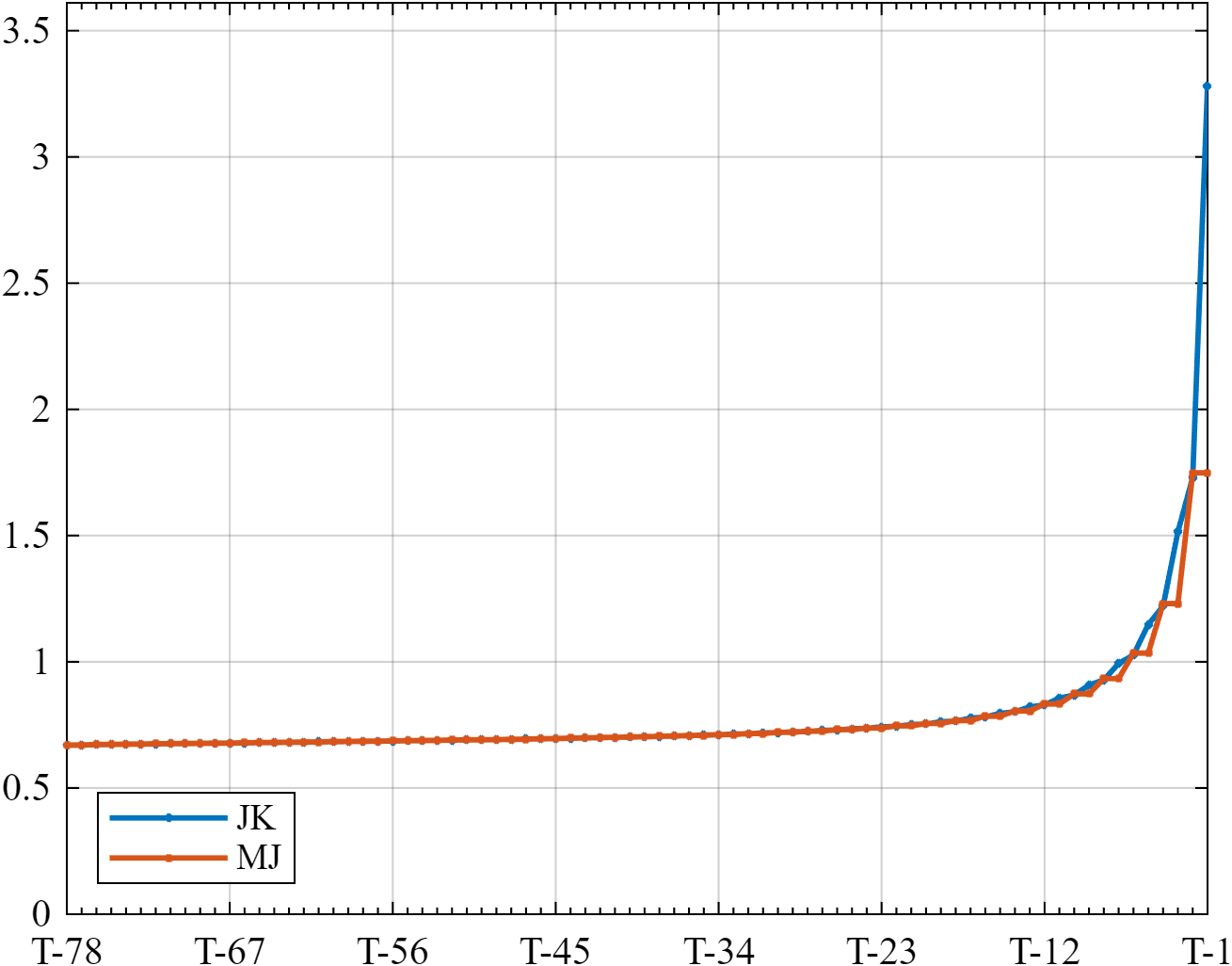} \\
		\footnotesize $T = 40$ & \footnotesize $T = 80$
	\end{tabular}
	\caption{Empirical standard deviation of the JK (blue) and MJ (red) estimators as a function of $S$ at $\tau_0 = 0.5$ with $\rho = 0$.}
	\label{fig:appstd-tau50}
\end{figure}

\begin{figure}[!htbp]
	\centering
	\begin{tabular}{cc}
		\includegraphics[width=0.40\textwidth]{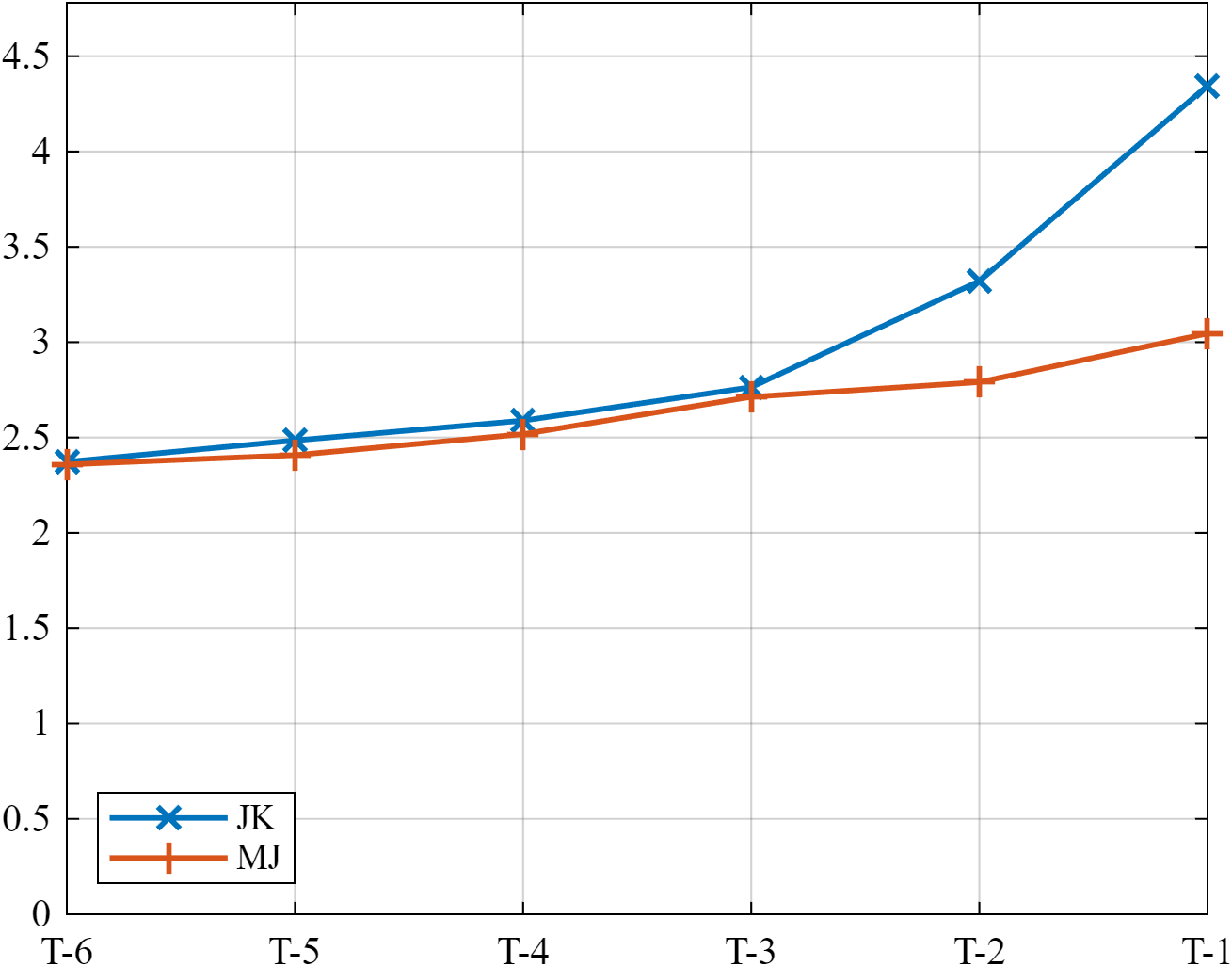} &
		\includegraphics[width=0.40\textwidth]{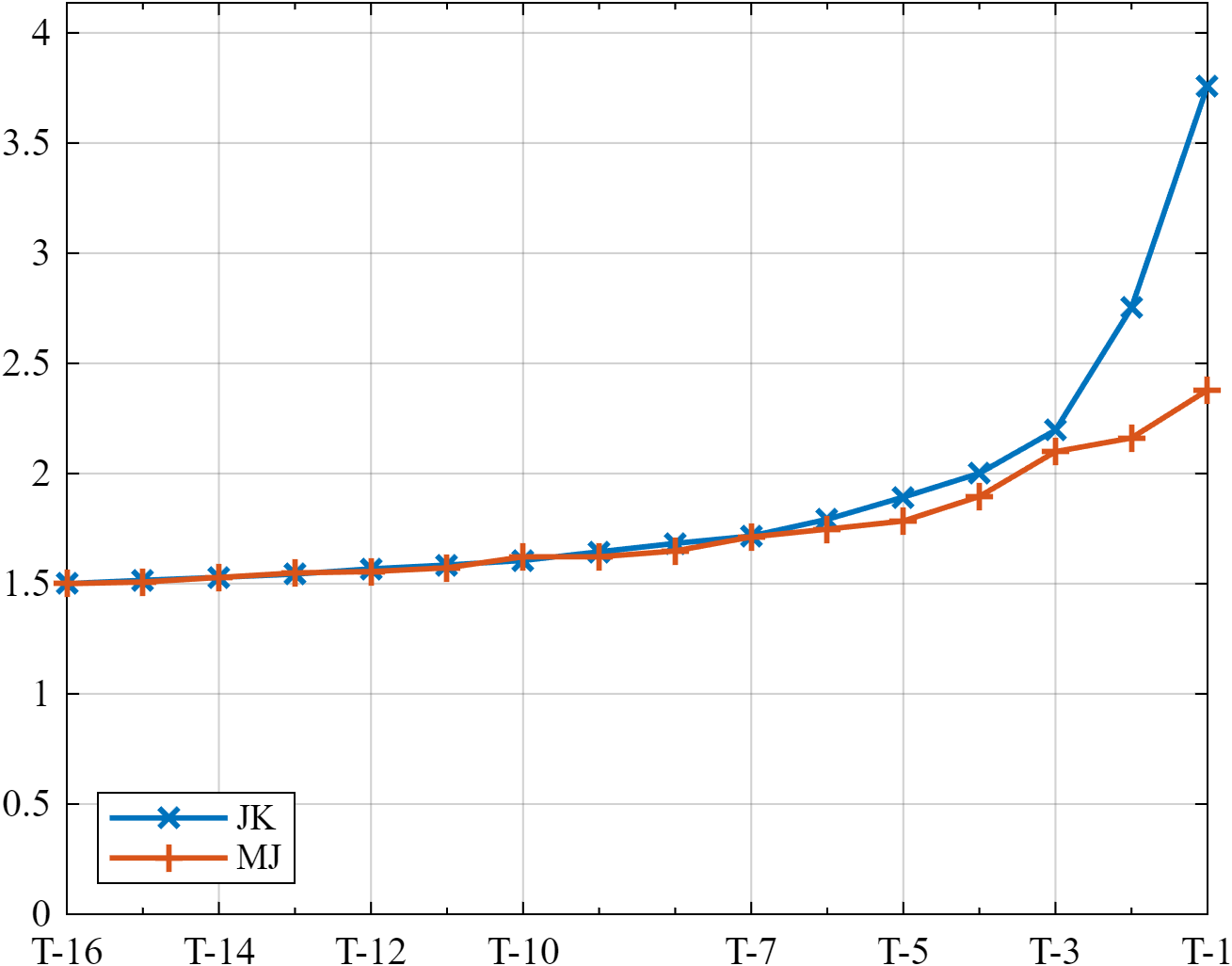} \\
		\footnotesize $T = 10$ & \footnotesize $T = 20$ \\[6pt]
		\includegraphics[width=0.40\textwidth]{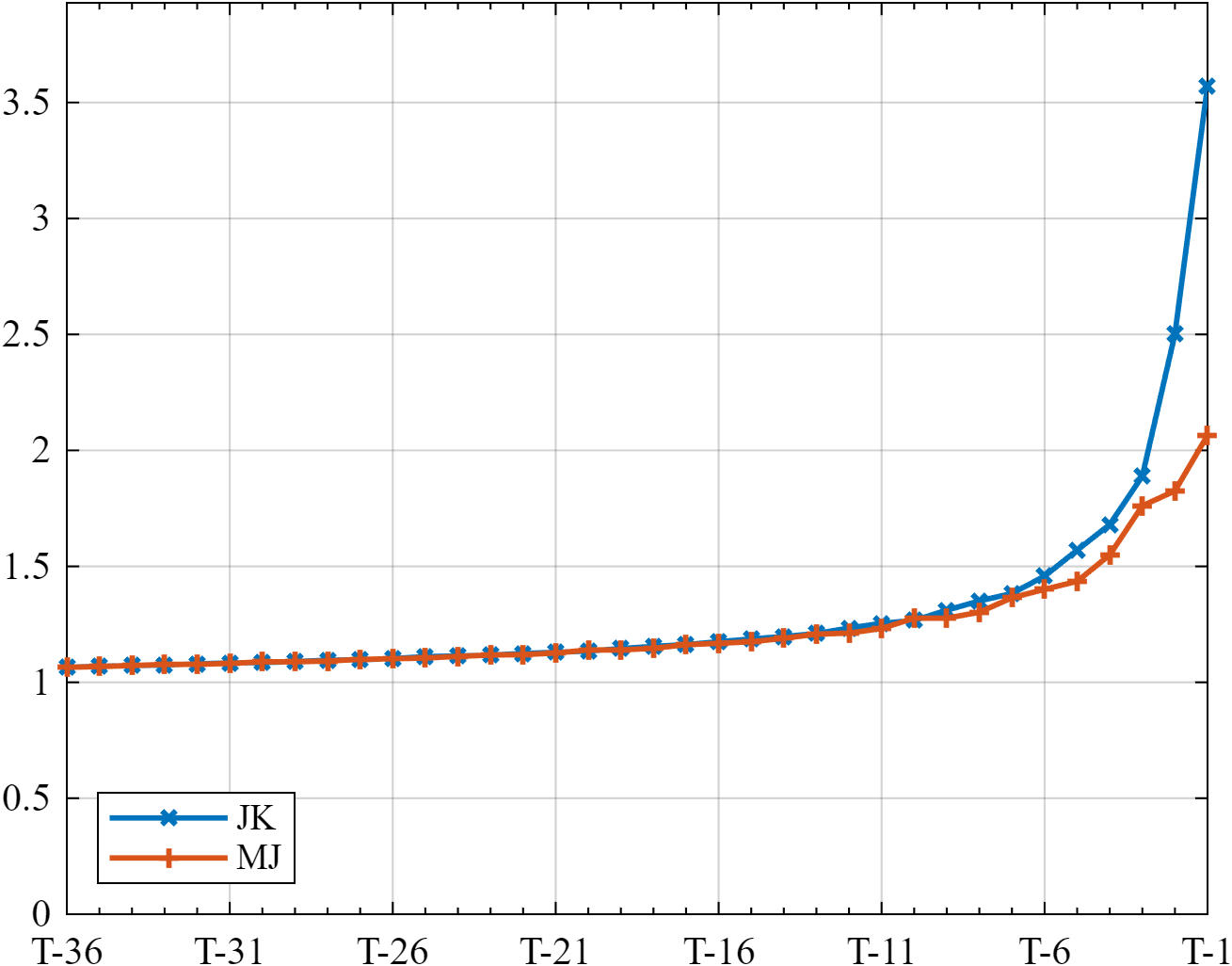} &
		\includegraphics[width=0.40\textwidth]{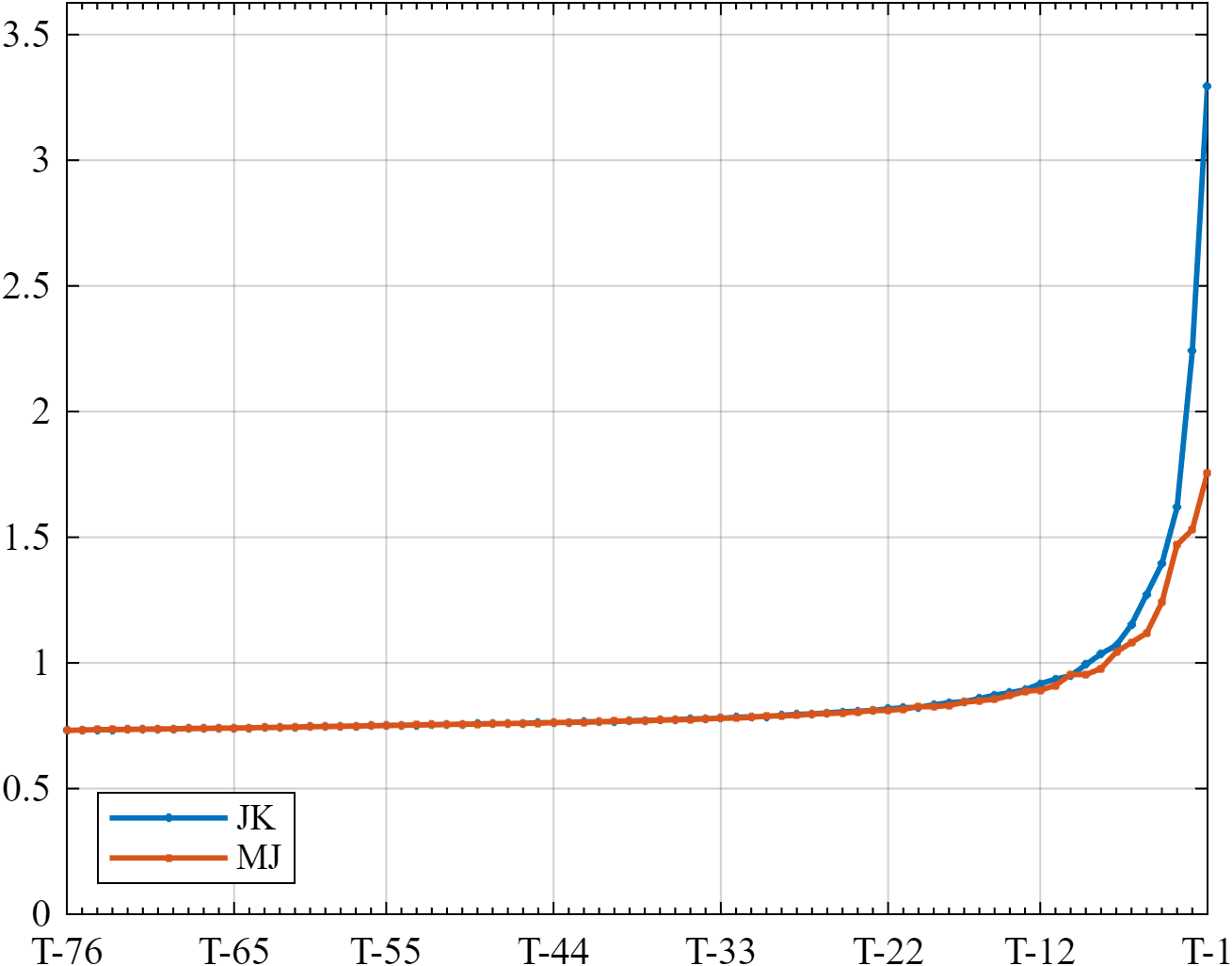} \\
		\footnotesize $T = 40$ & \footnotesize $T = 80$
	\end{tabular}
	\caption{Empirical standard deviation of the JK (blue) and MJ (red) estimators as a function of $S$ at $\tau_0 = 0.7$ with $\rho = 0$.}
	\label{fig:appstd-tau70}
\end{figure}

\FloatBarrier

\subsection{Endogenous Design: Empirical Bias}\label{endog:bias}

\begin{figure}[!htbp]
	\centering
	\begin{tabular}{cc}
		\includegraphics[width=0.40\textwidth]{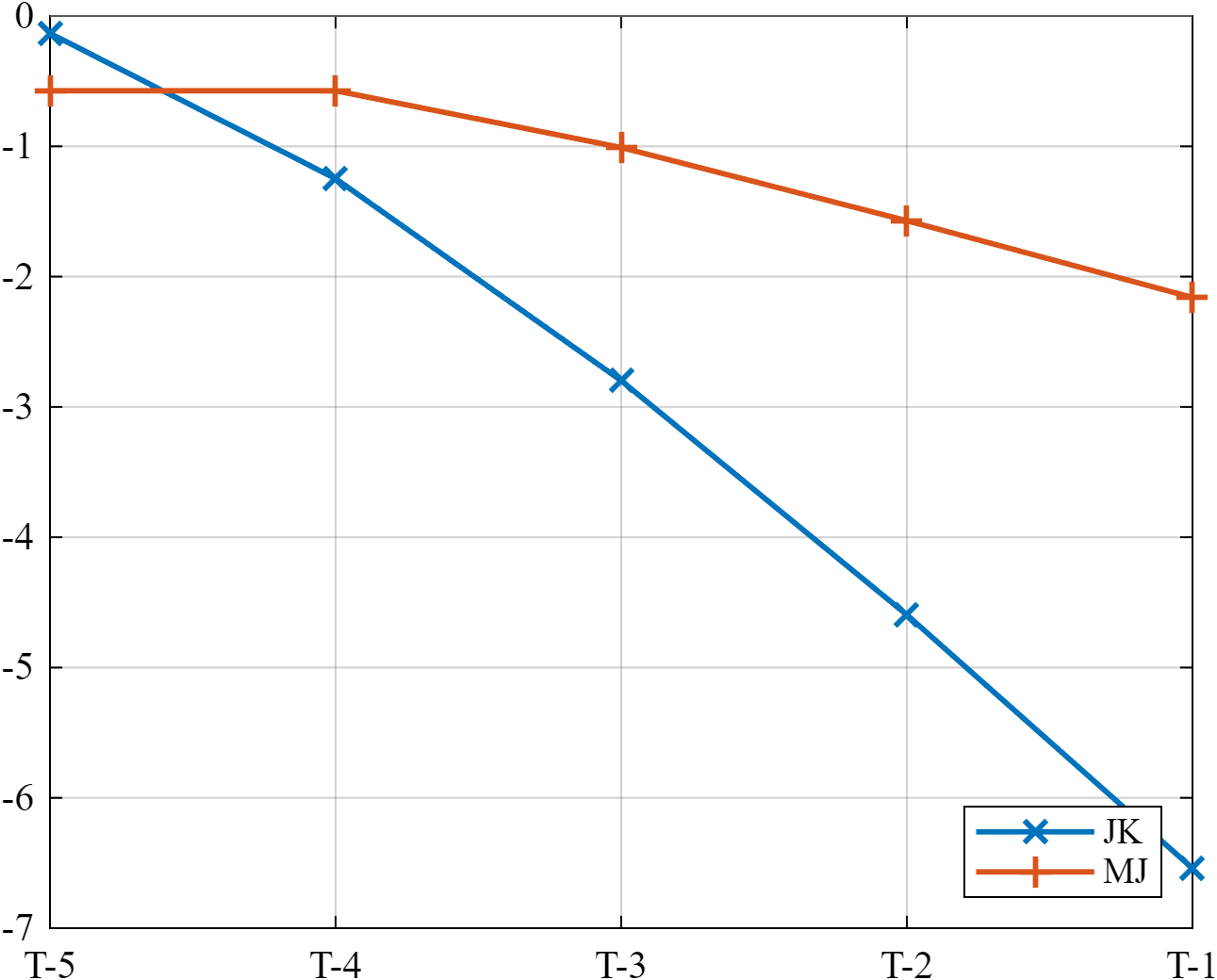} &
		\includegraphics[width=0.40\textwidth]{./Graphs/fig_endog_T20_N1000_MC1000_tau20_bias.png} \\
		\footnotesize $T = 10$ & \footnotesize $T = 20$ \\[6pt]
		\includegraphics[width=0.40\textwidth]{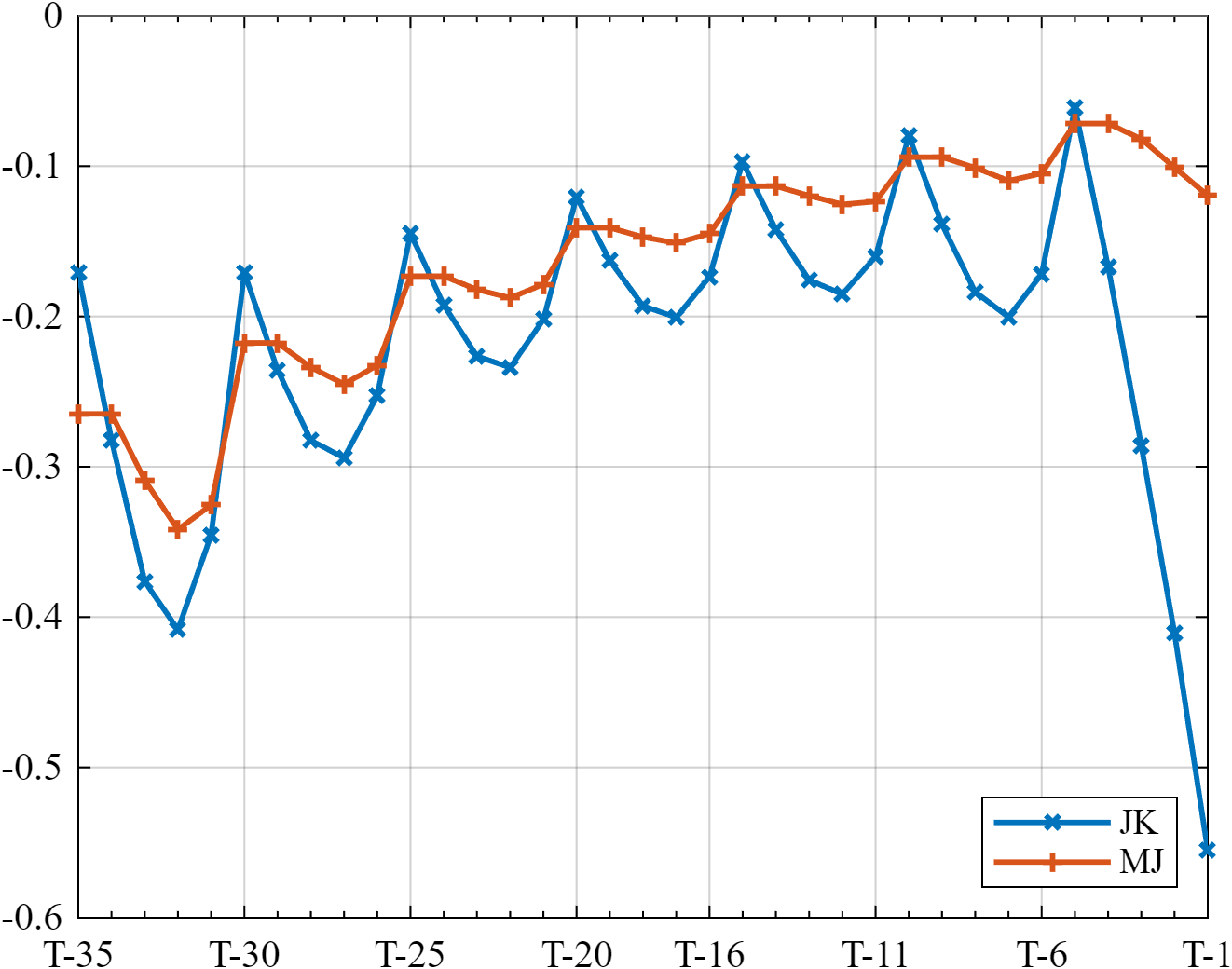} &
		\includegraphics[width=0.40\textwidth]{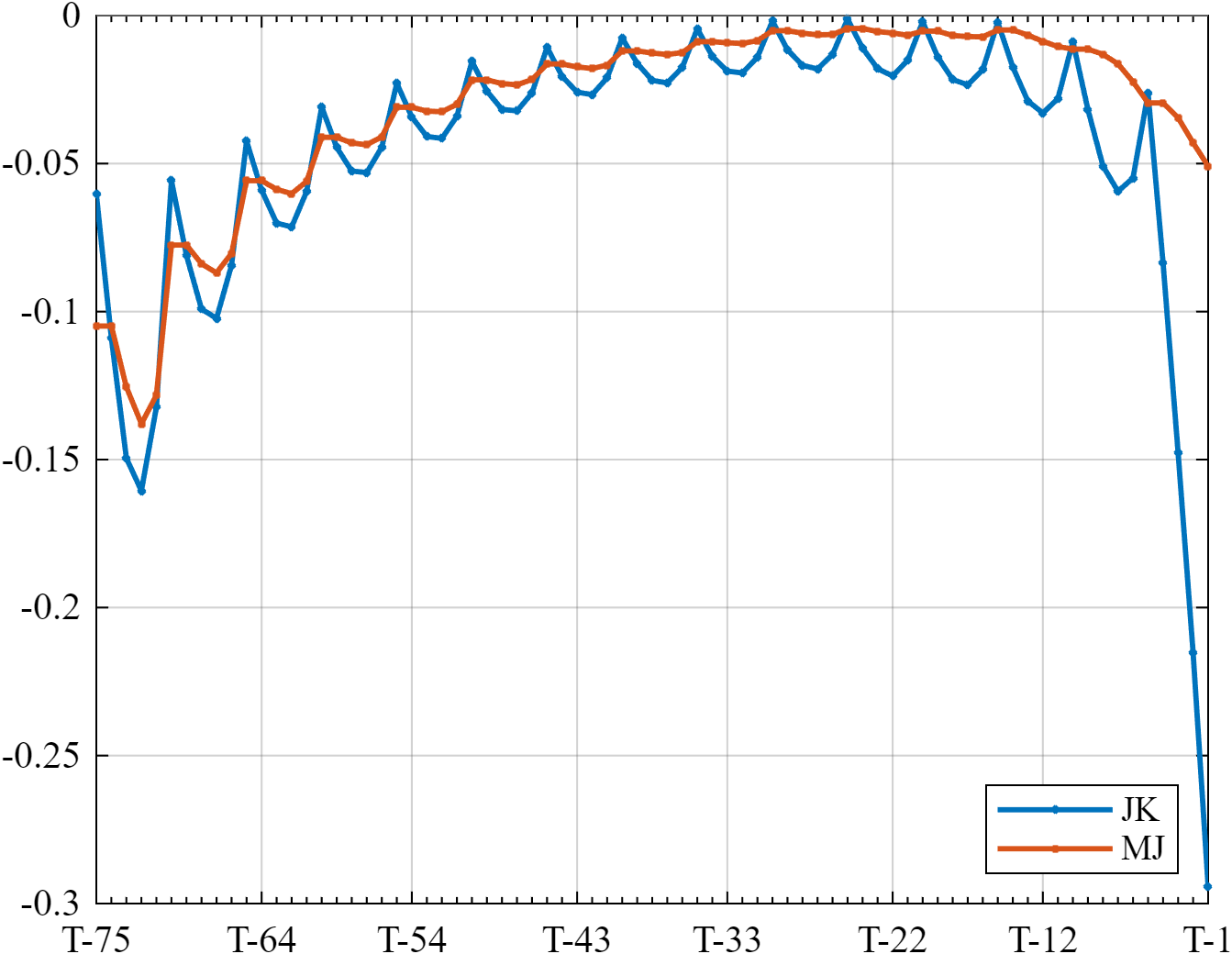} \\
		\footnotesize $T = 40$ & \footnotesize $T = 80$
	\end{tabular}
	\caption{Empirical bias of the JK (blue) and MJ (red) estimators as a function of $S$ at $\tau_0 = 0.2$ with $\rho = 0.5$.}
	\label{fig:appbias-endog-tau20}
\end{figure}

\begin{figure}[!htbp]
	\centering
	\begin{tabular}{cc}
		\includegraphics[width=0.40\textwidth]{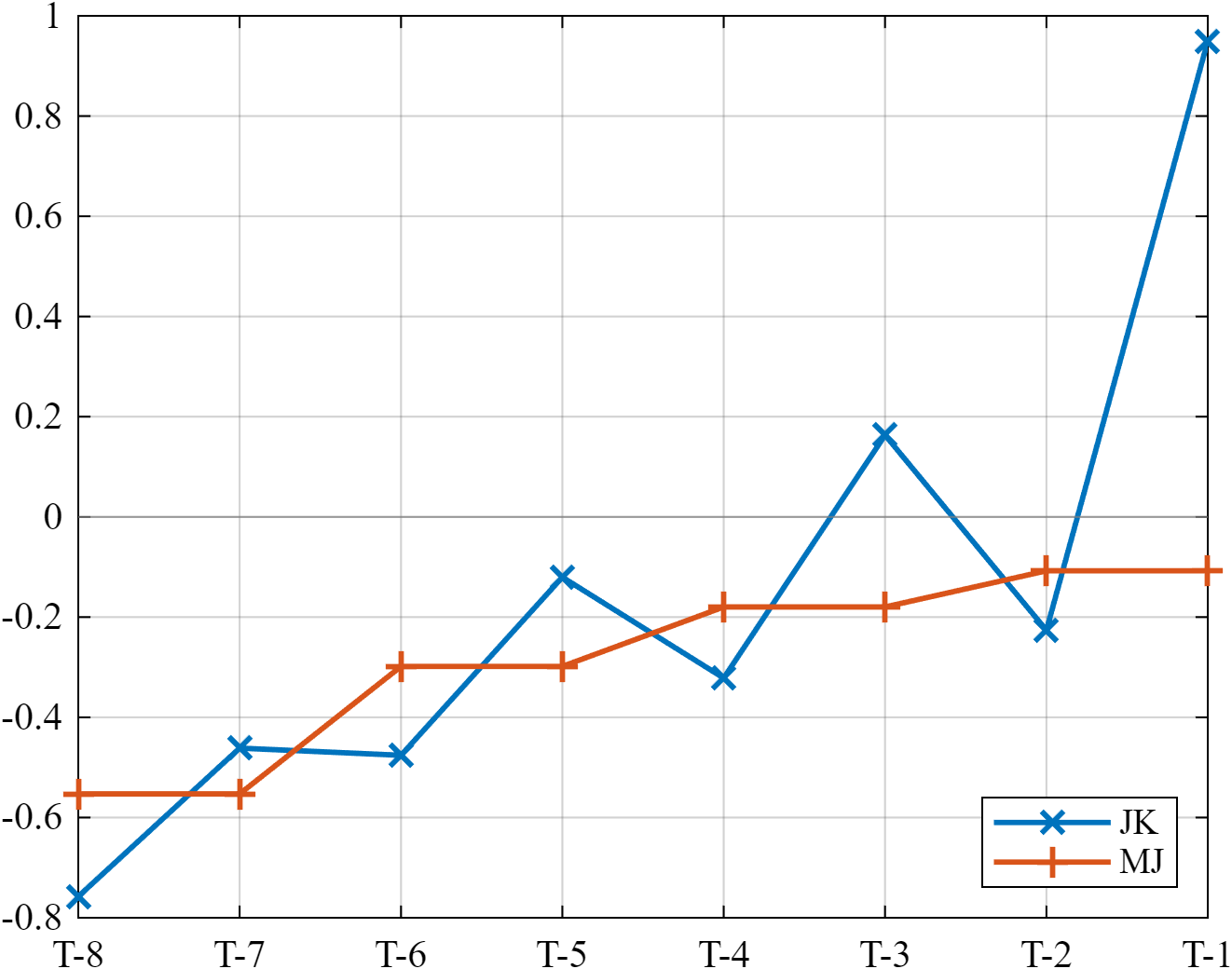} &
		\includegraphics[width=0.40\textwidth]{./Graphs/fig_endog_T20_N1000_MC1000_tau50_bias.png} \\
		\footnotesize $T = 10$ & \footnotesize $T = 20$ \\[6pt]
		\includegraphics[width=0.40\textwidth]{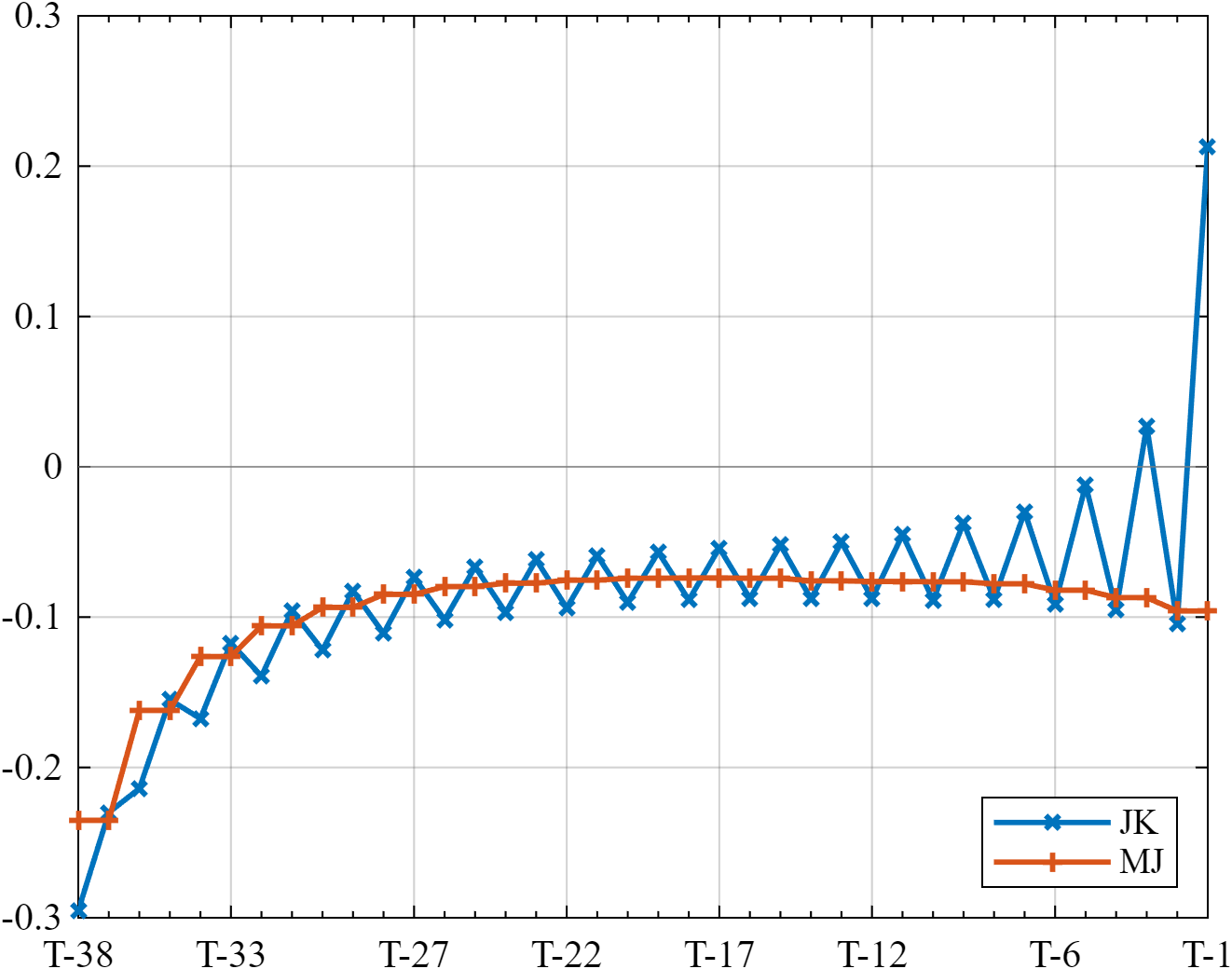} &
		\includegraphics[width=0.40\textwidth]{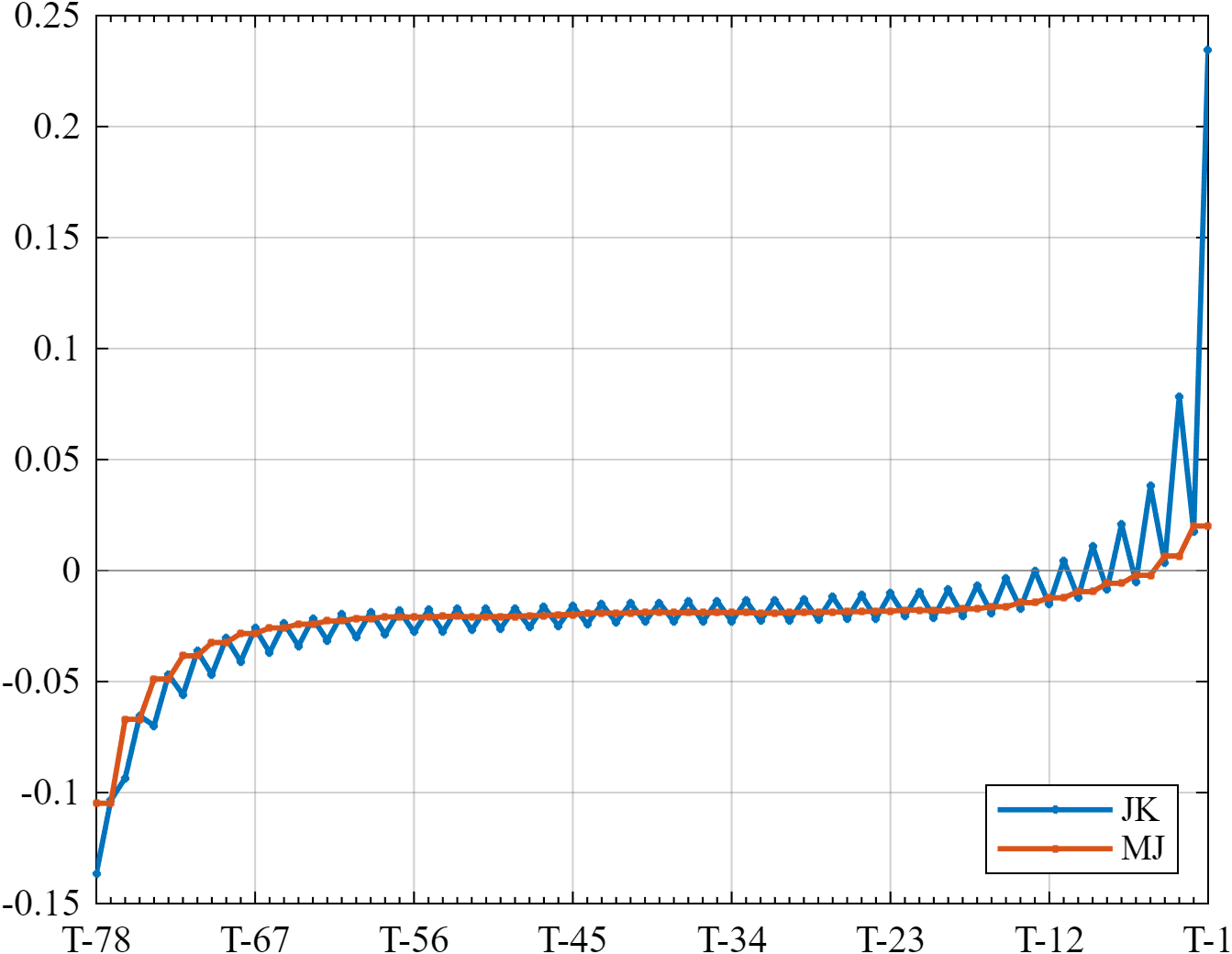} \\
		\footnotesize $T = 40$ & \footnotesize $T = 80$
	\end{tabular}
	\caption{Empirical bias of the JK (blue) and MJ (red) estimators as a function of $S$ at $\tau_0 = 0.5$ with $\rho = 0.5$.}
	\label{fig:appbias-endog-tau50}
\end{figure}

\begin{figure}[!htbp]
	\centering
	\begin{tabular}{cc}
		\includegraphics[width=0.40\textwidth]{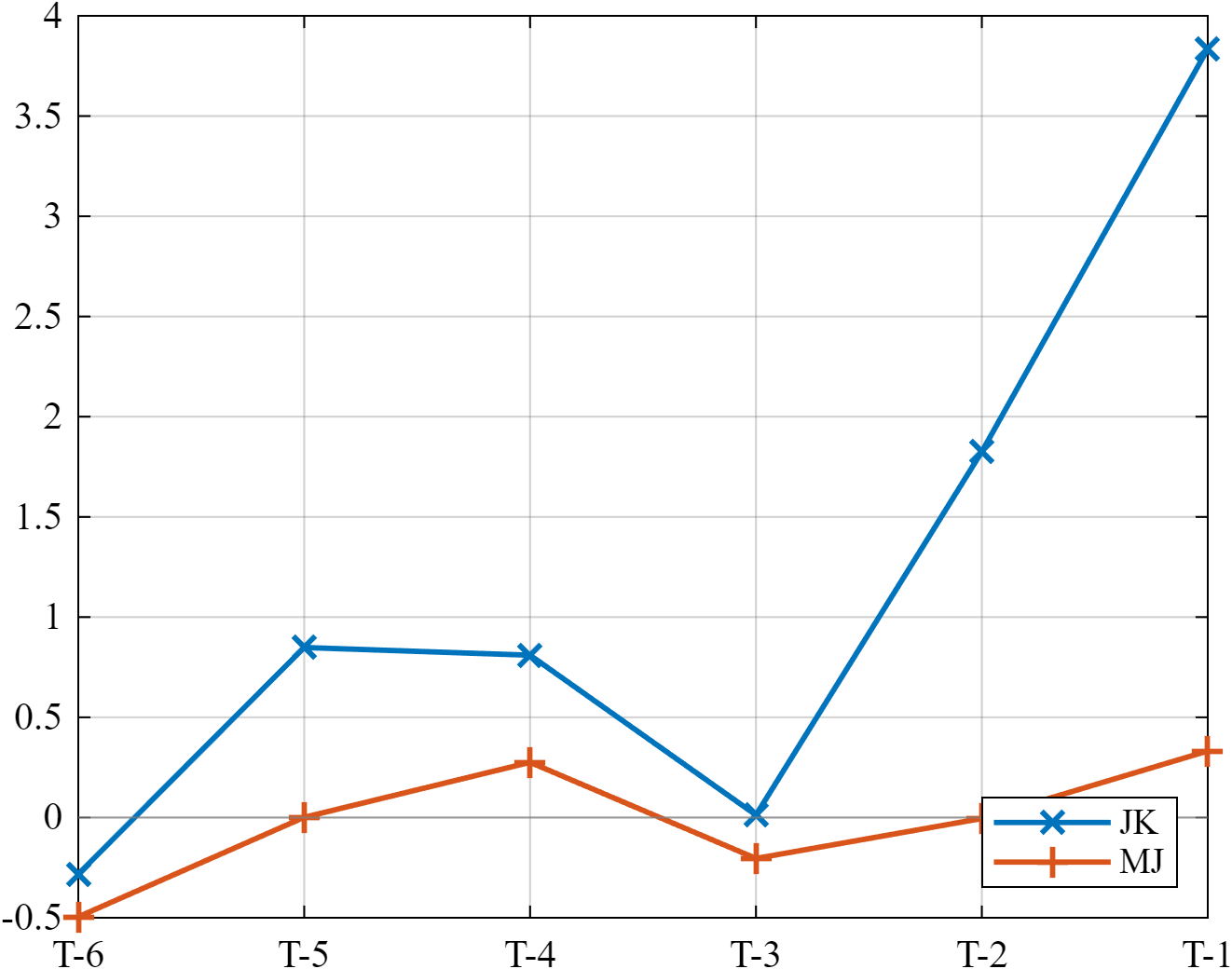} &
		\includegraphics[width=0.40\textwidth]{./Graphs/fig_endog_T20_N1000_MC1000_tau70_bias.png} \\
		\footnotesize $T = 10$ & \footnotesize $T = 20$ \\[6pt]
		\includegraphics[width=0.40\textwidth]{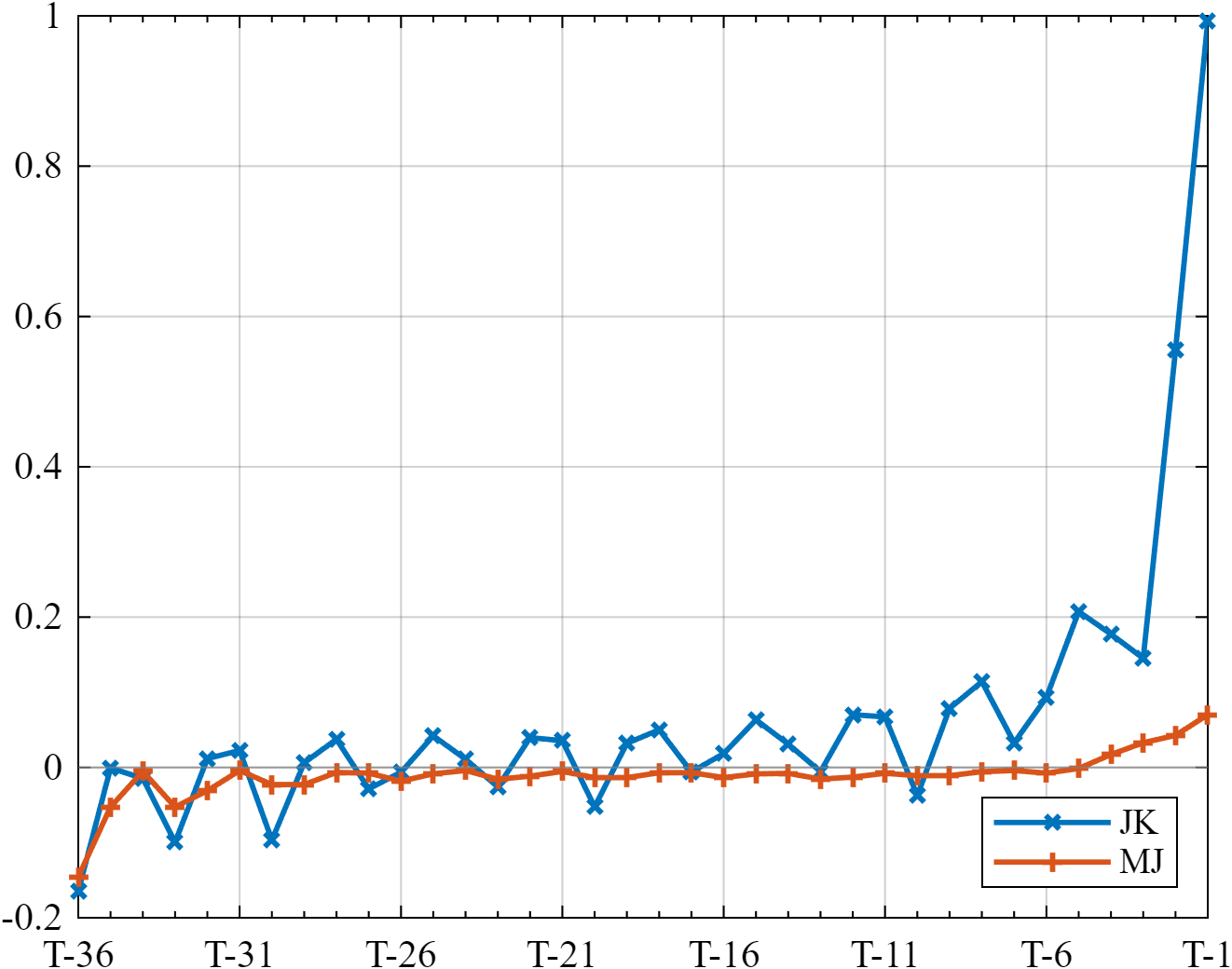} &
		\includegraphics[width=0.40\textwidth]{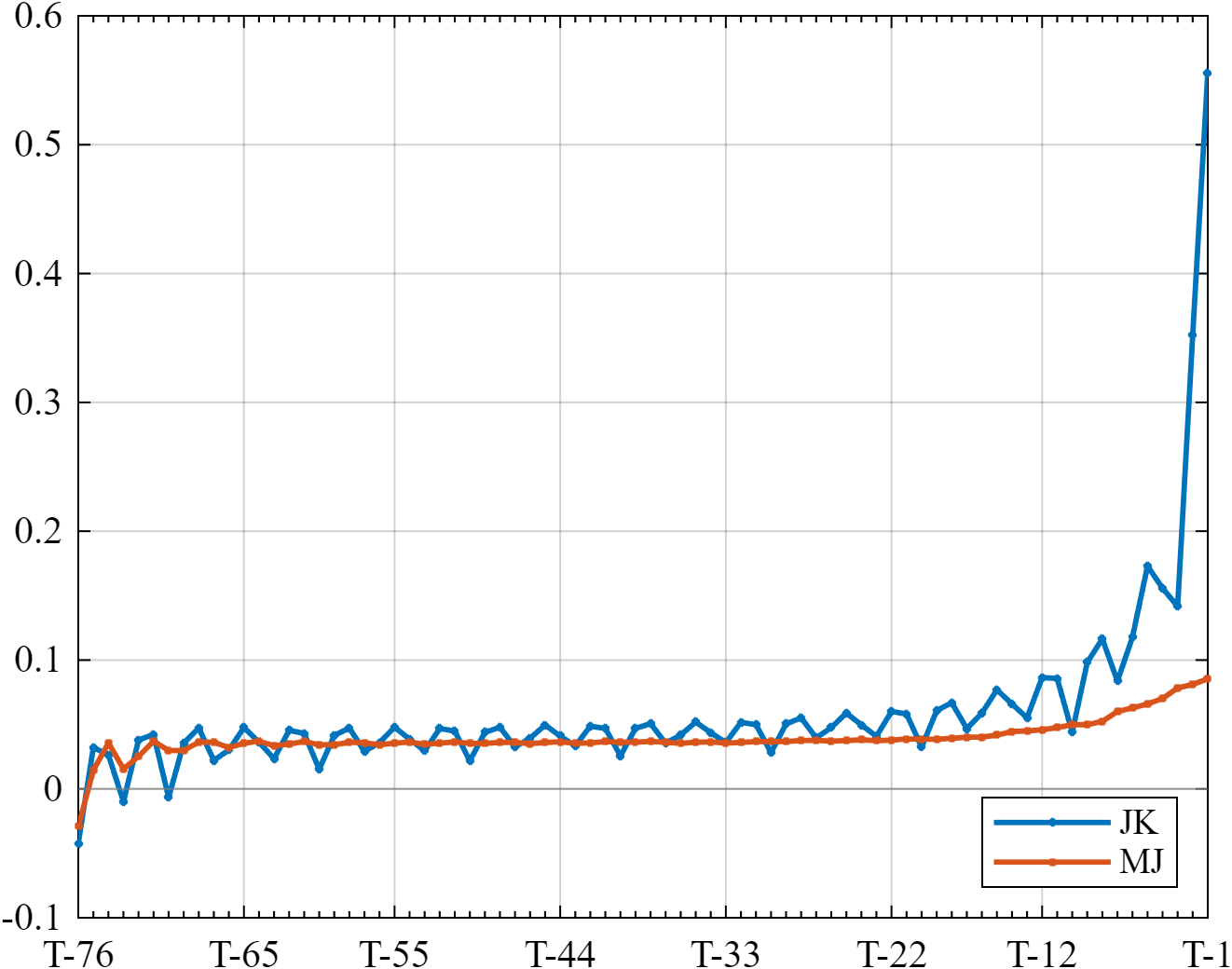} \\
		\footnotesize $T = 40$ & \footnotesize $T = 80$
	\end{tabular}
	\caption{Empirical bias of the JK (blue) and MJ (red) estimators as a function of $S$ at $\tau_0 = 0.7$ with $\rho = 0.5$.}
	\label{fig:appbias-endog-tau70}
\end{figure}

\FloatBarrier

\subsection{Endogenous Design: Empirical Standard Deviation}\label{endog:std}

\begin{figure}[!htbp]
	\centering
	\begin{tabular}{cc}
		\includegraphics[width=0.40\textwidth]{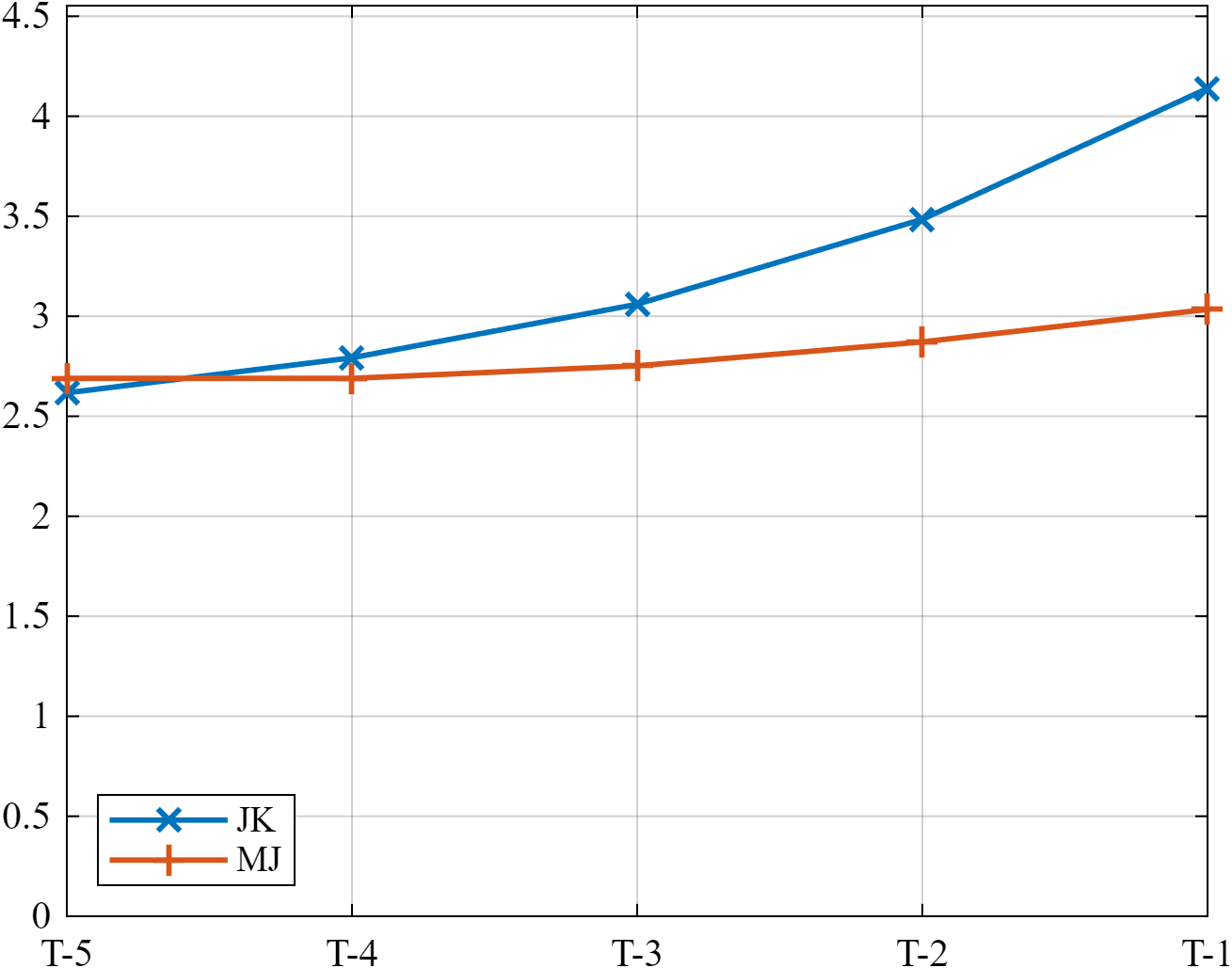} &
		\includegraphics[width=0.40\textwidth]{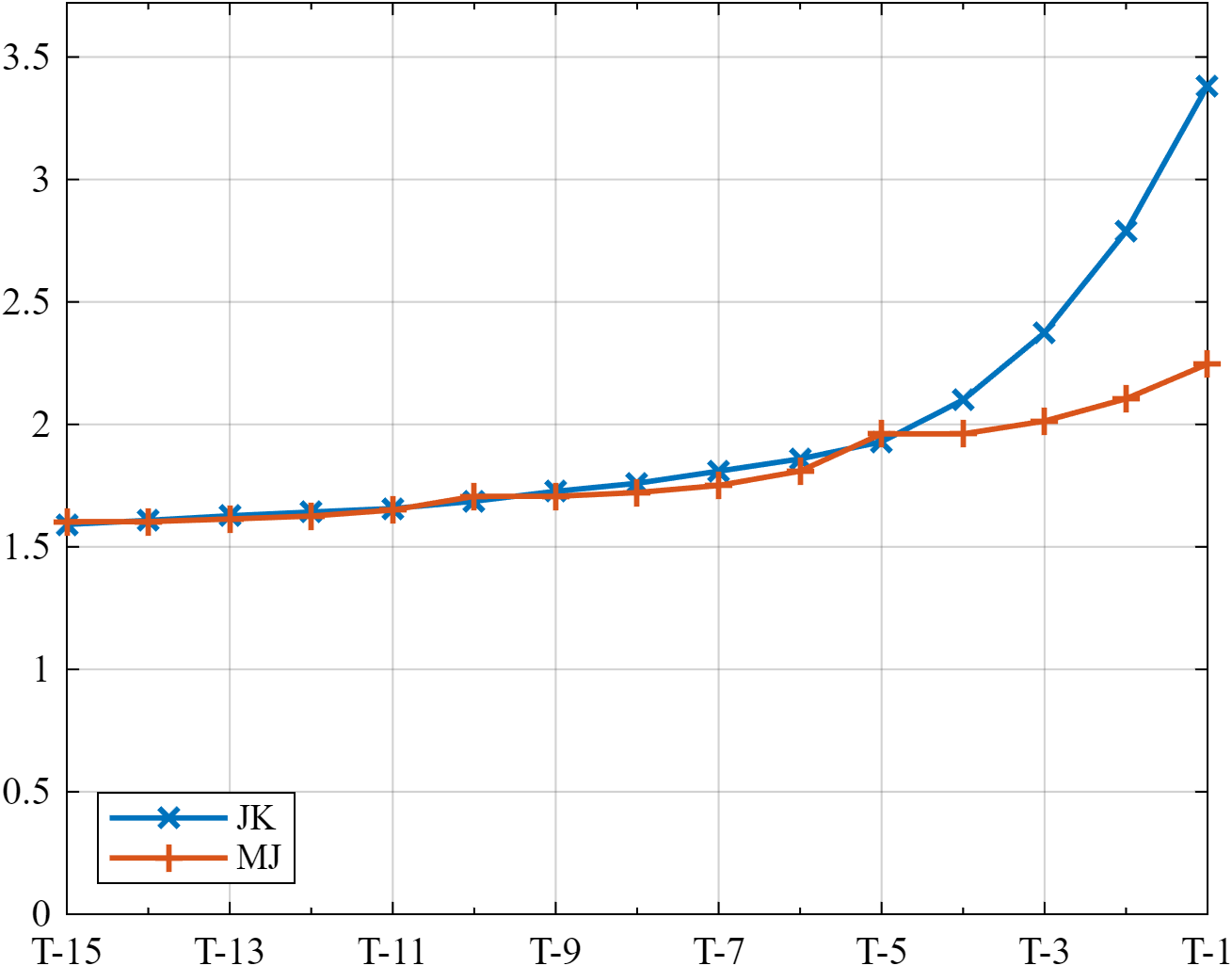} \\
		\footnotesize $T = 10$ & \footnotesize $T = 20$ \\[6pt]
		\includegraphics[width=0.40\textwidth]{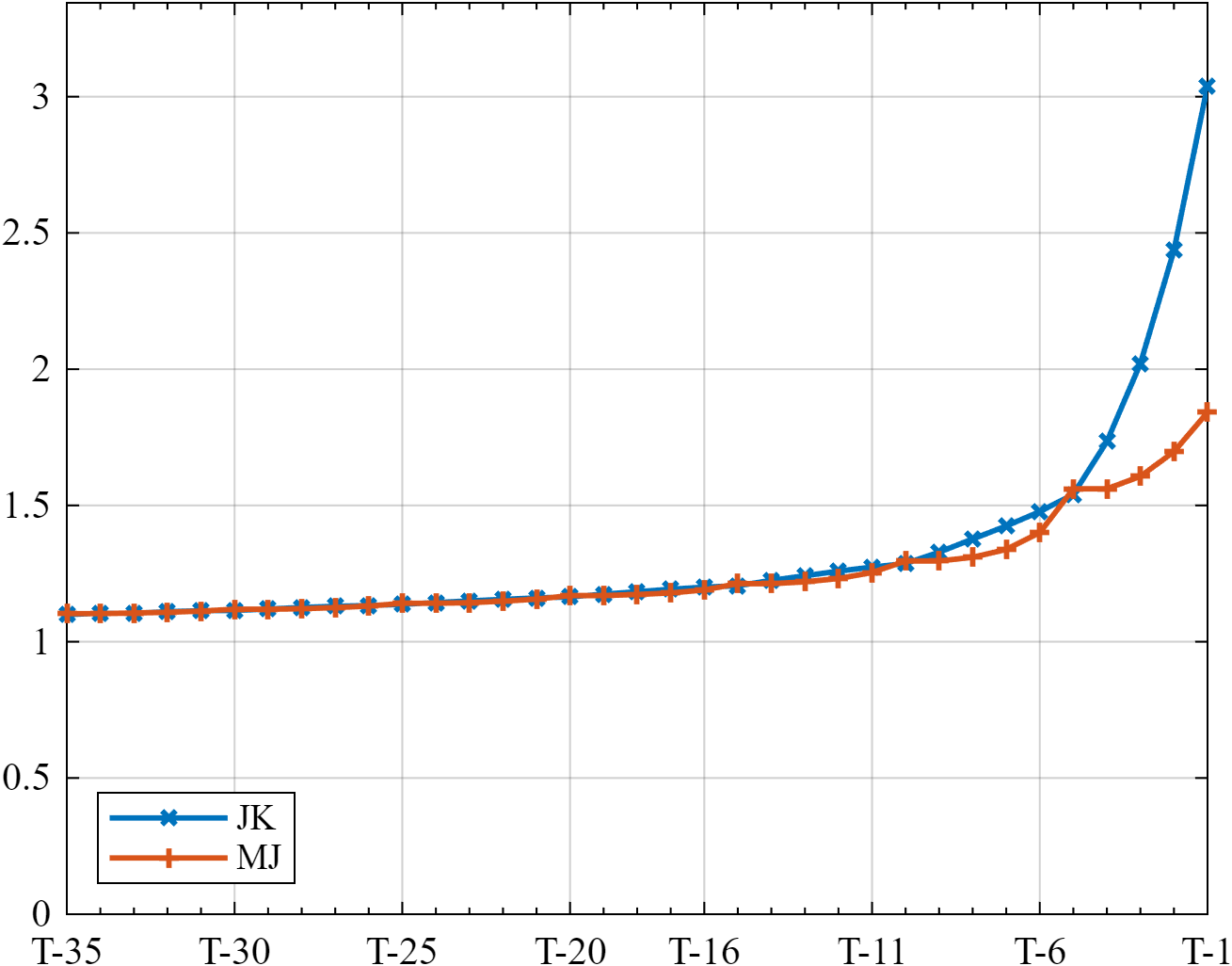} &
		\includegraphics[width=0.40\textwidth]{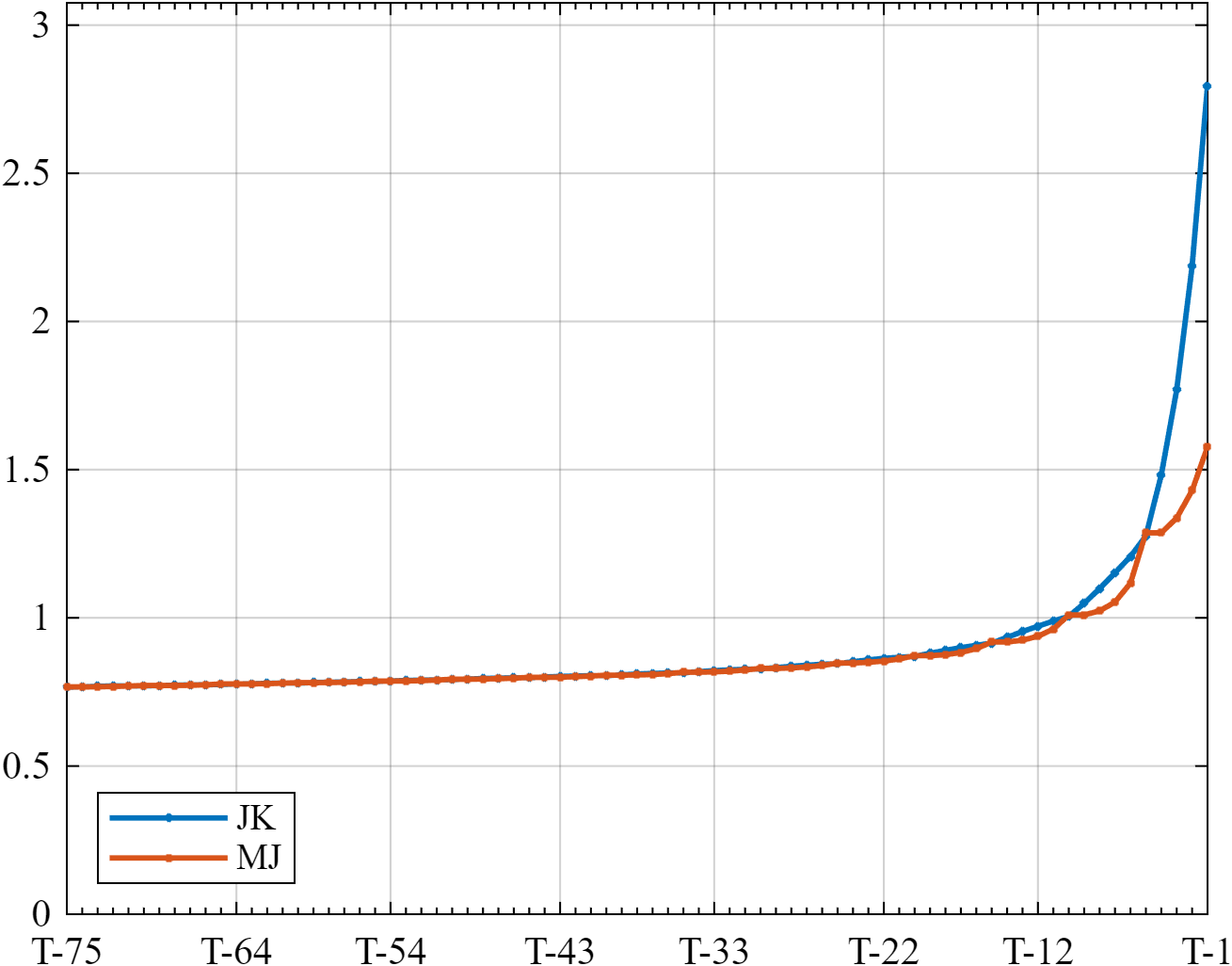} \\
		\footnotesize $T = 40$ & \footnotesize $T = 80$
	\end{tabular}
	\caption{Empirical standard deviation of the JK (blue) and MJ (red) estimators as a function of $S$ at $\tau_0 = 0.2$ with $\rho = 0.5$.}
	\label{fig:appstd-endog-tau20}
\end{figure}

\begin{figure}[!htbp]
	\centering
	\begin{tabular}{cc}
		\includegraphics[width=0.40\textwidth]{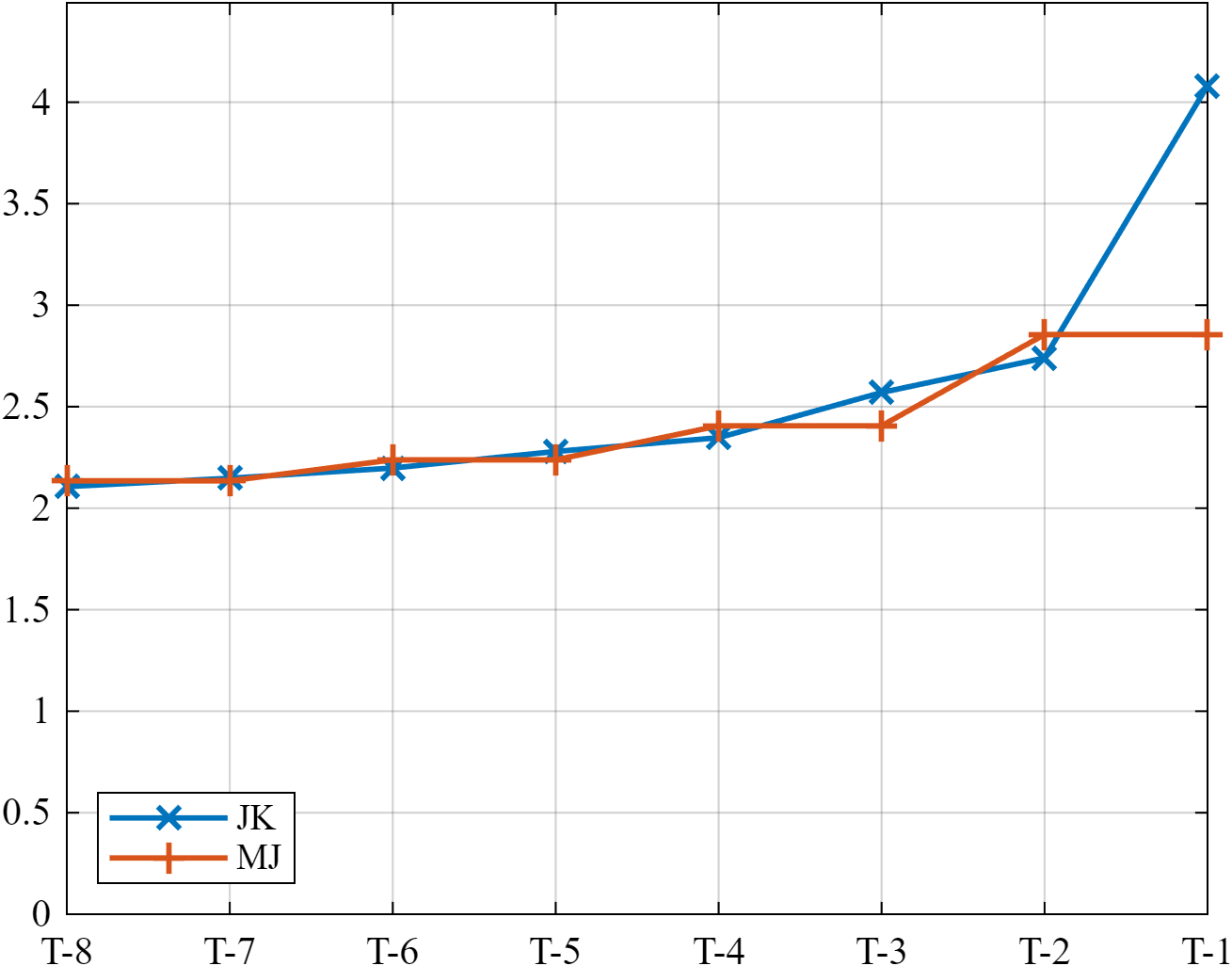} &
		\includegraphics[width=0.40\textwidth]{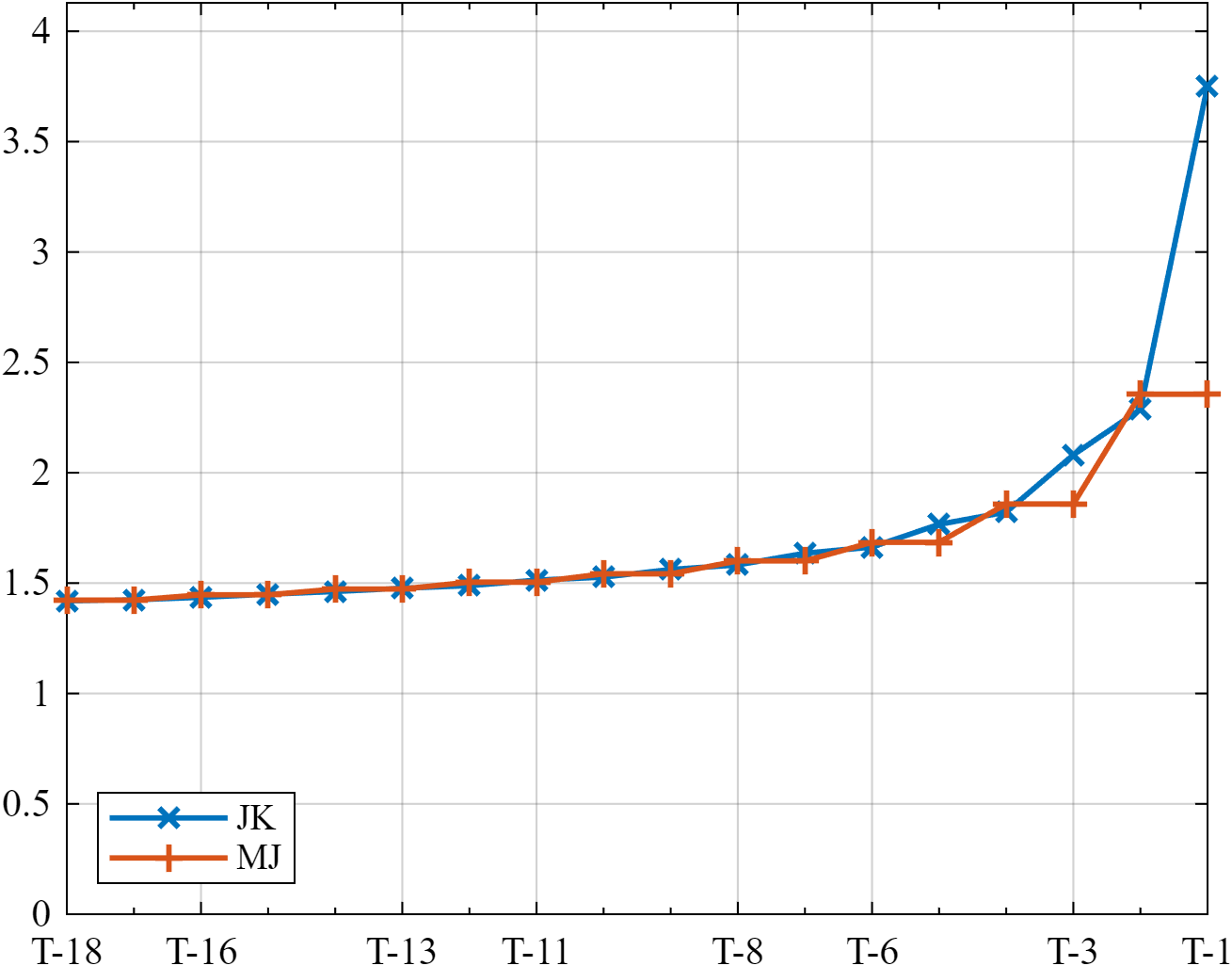} \\
		\footnotesize $T = 10$ & \footnotesize $T = 20$ \\[6pt]
		\includegraphics[width=0.40\textwidth]{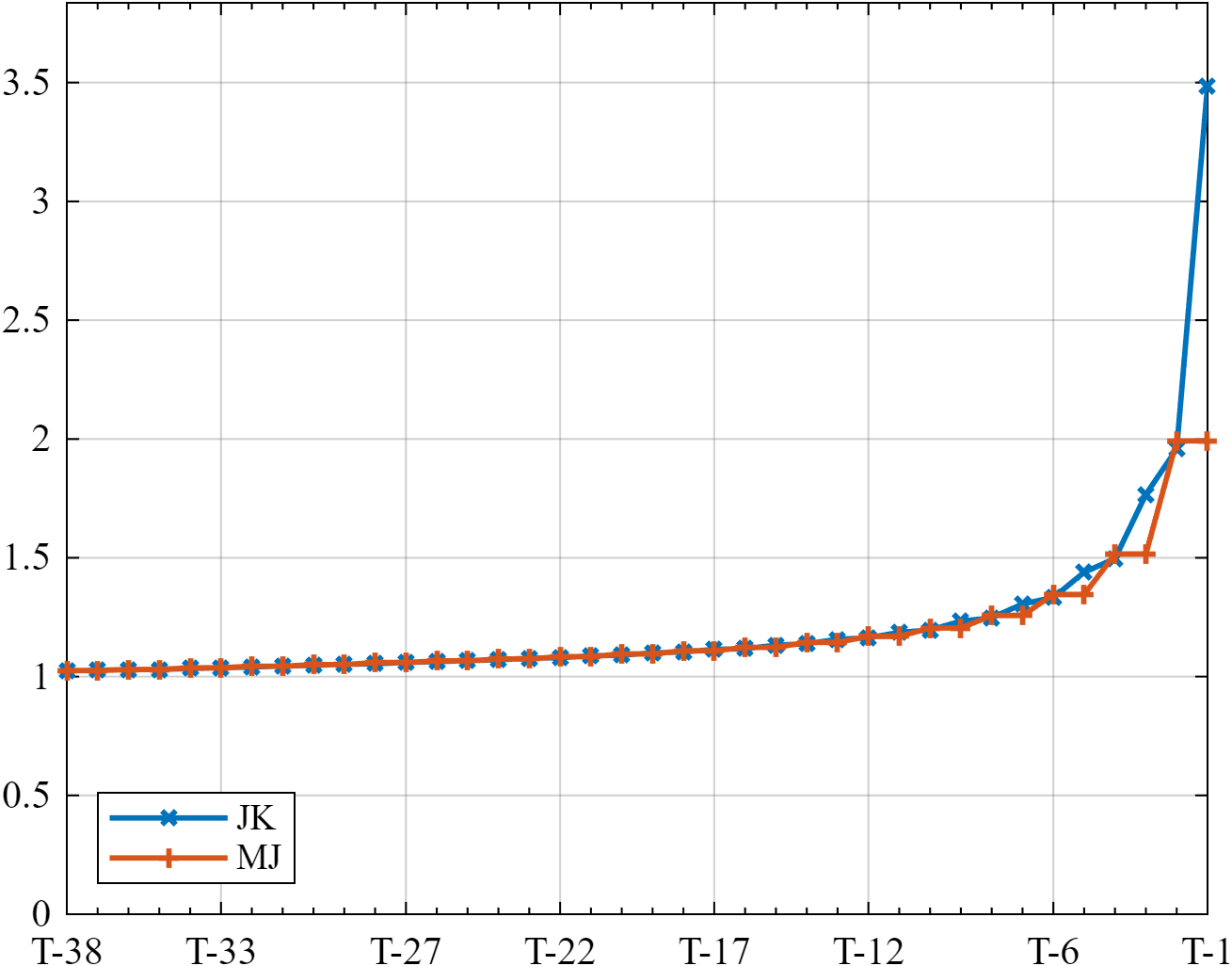} &
		\includegraphics[width=0.40\textwidth]{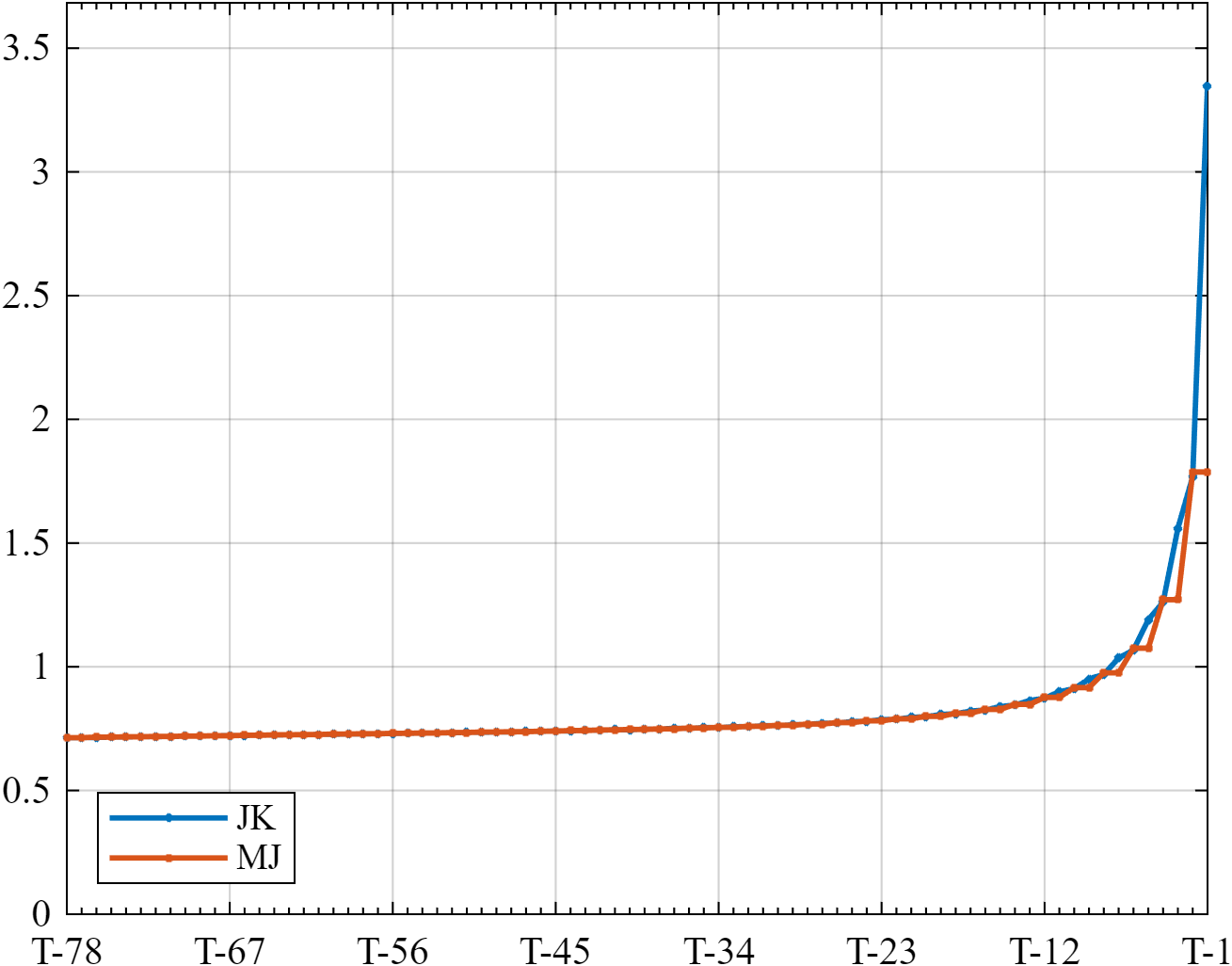} \\
		\footnotesize $T = 40$ & \footnotesize $T = 80$
	\end{tabular}
	\caption{Empirical standard deviation of the JK (blue) and MJ (red) estimators as a function of $S$ at $\tau_0 = 0.5$ with $\rho = 0.5$.}
	\label{fig:appstd-endog-tau50}
\end{figure}

\begin{figure}[!htbp]
	\centering
	\begin{tabular}{cc}
		\includegraphics[width=0.40\textwidth]{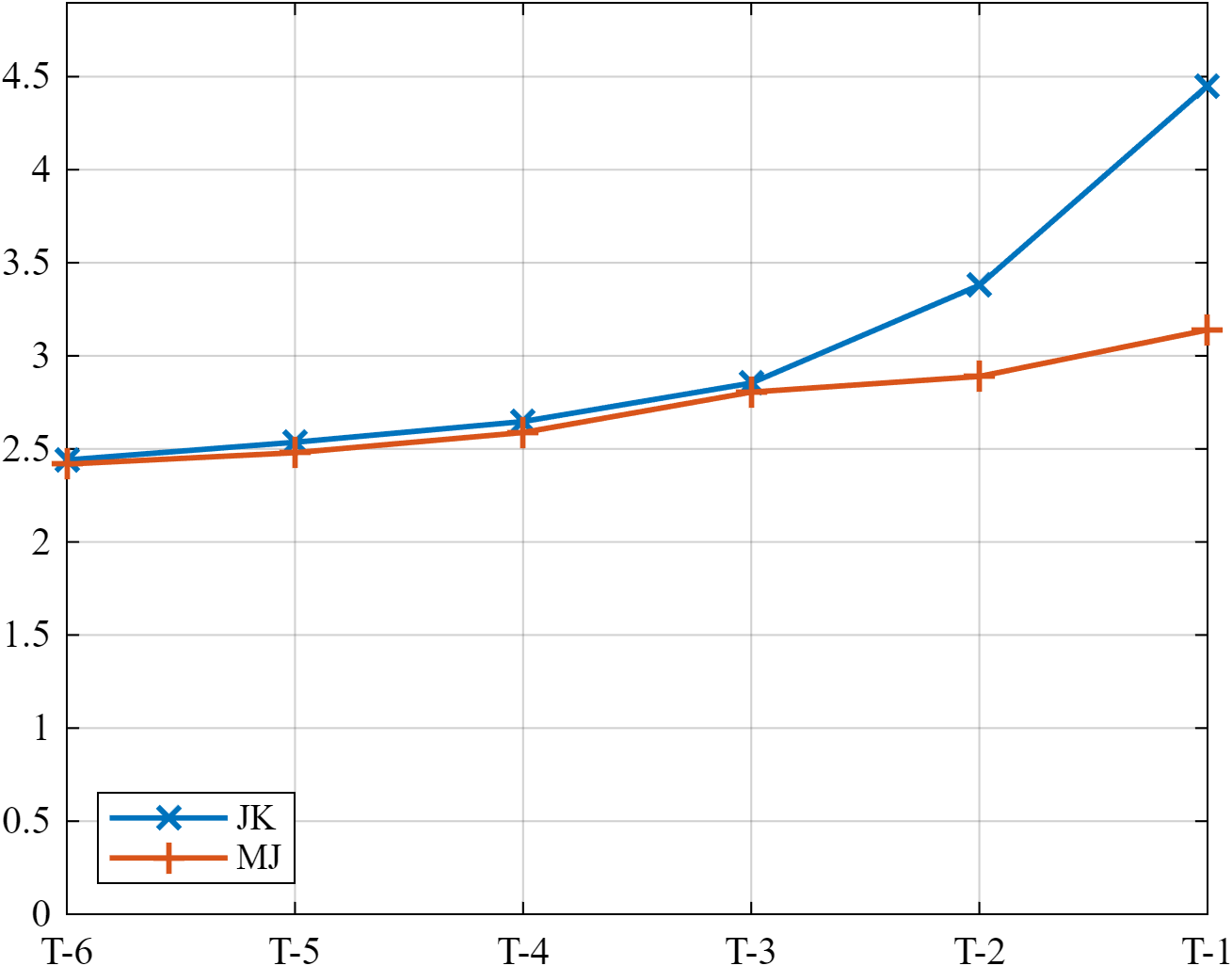} &
		\includegraphics[width=0.40\textwidth]{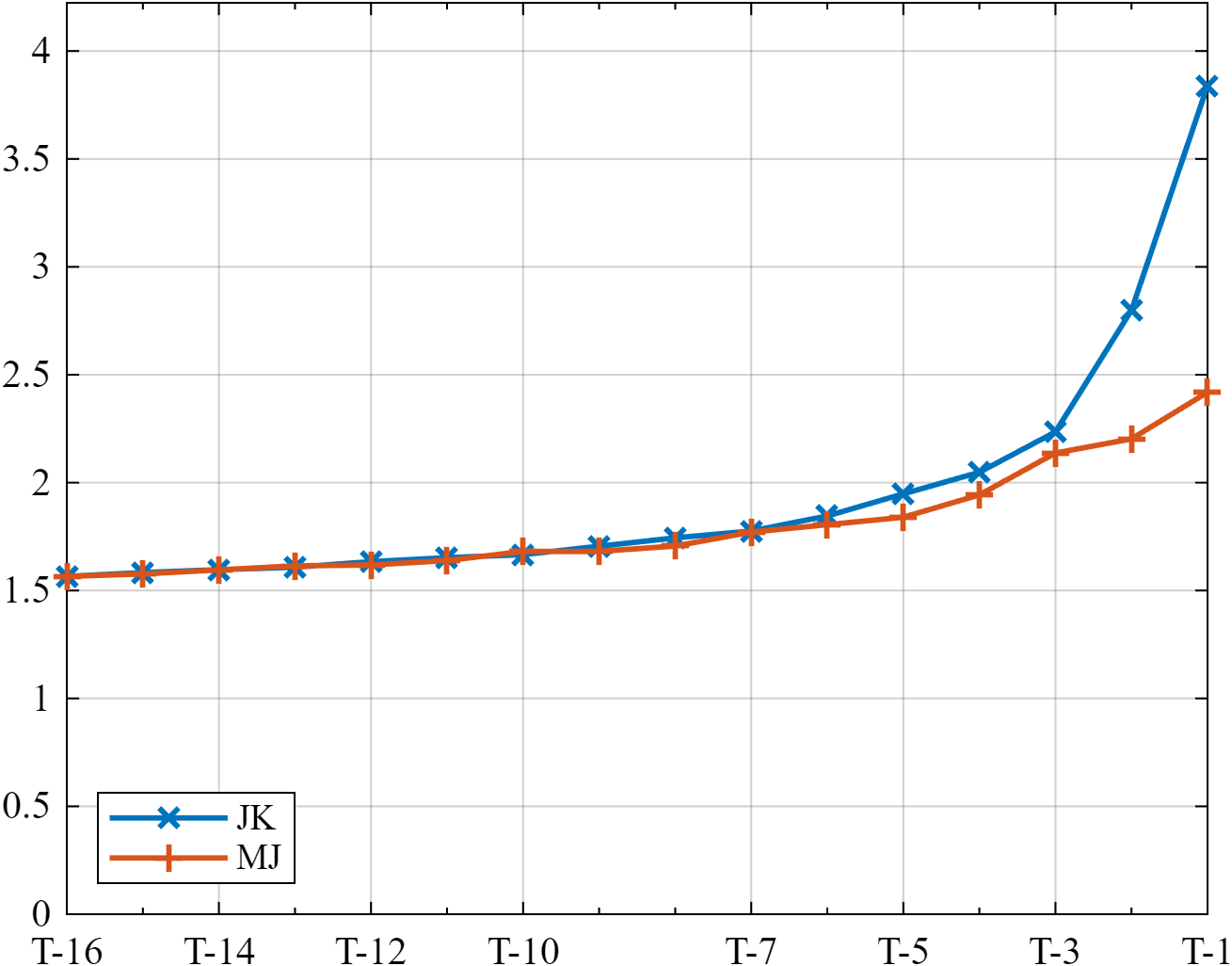} \\
		\footnotesize $T = 10$ & \footnotesize $T = 20$ \\[6pt]
		\includegraphics[width=0.40\textwidth]{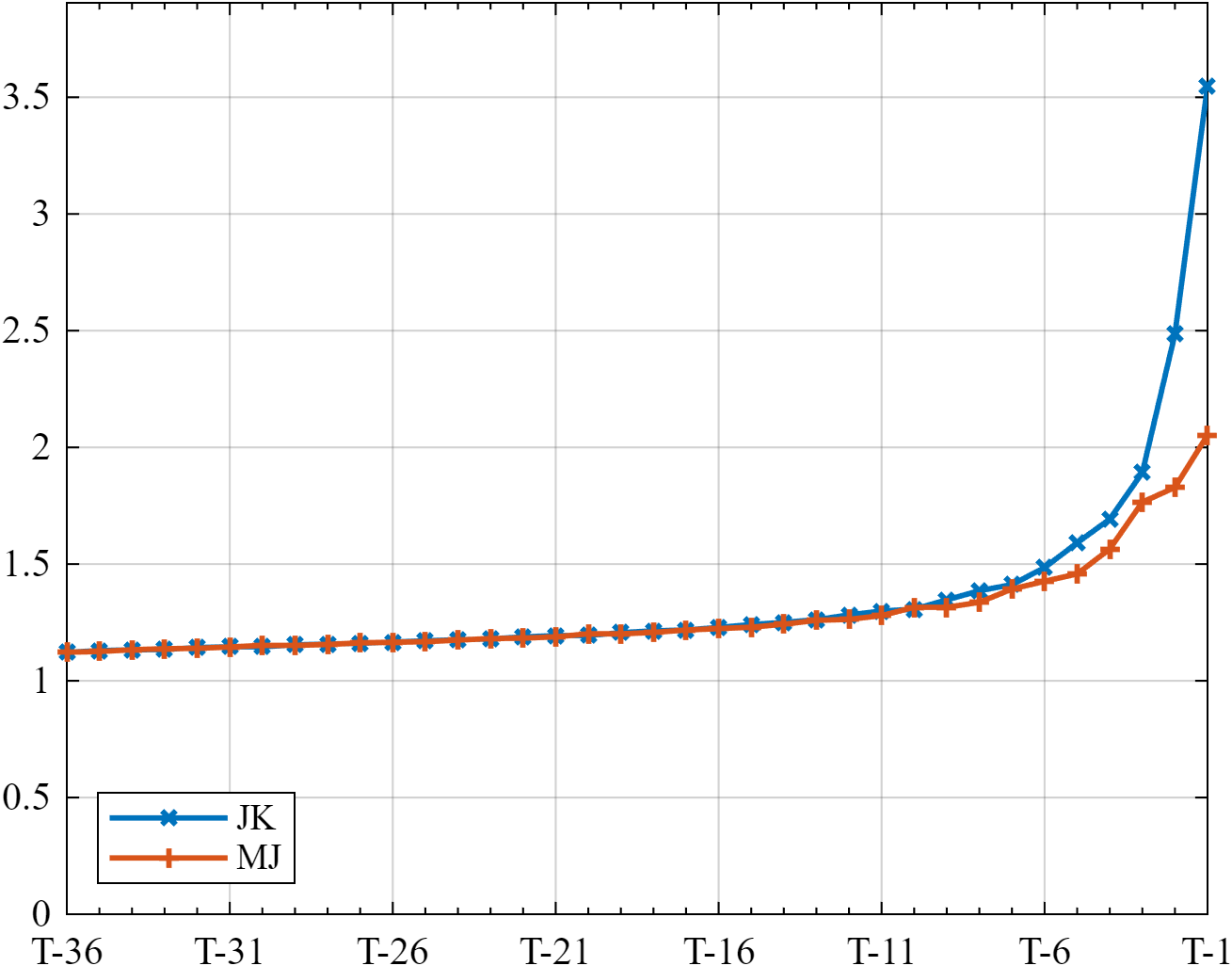} &
		\includegraphics[width=0.40\textwidth]{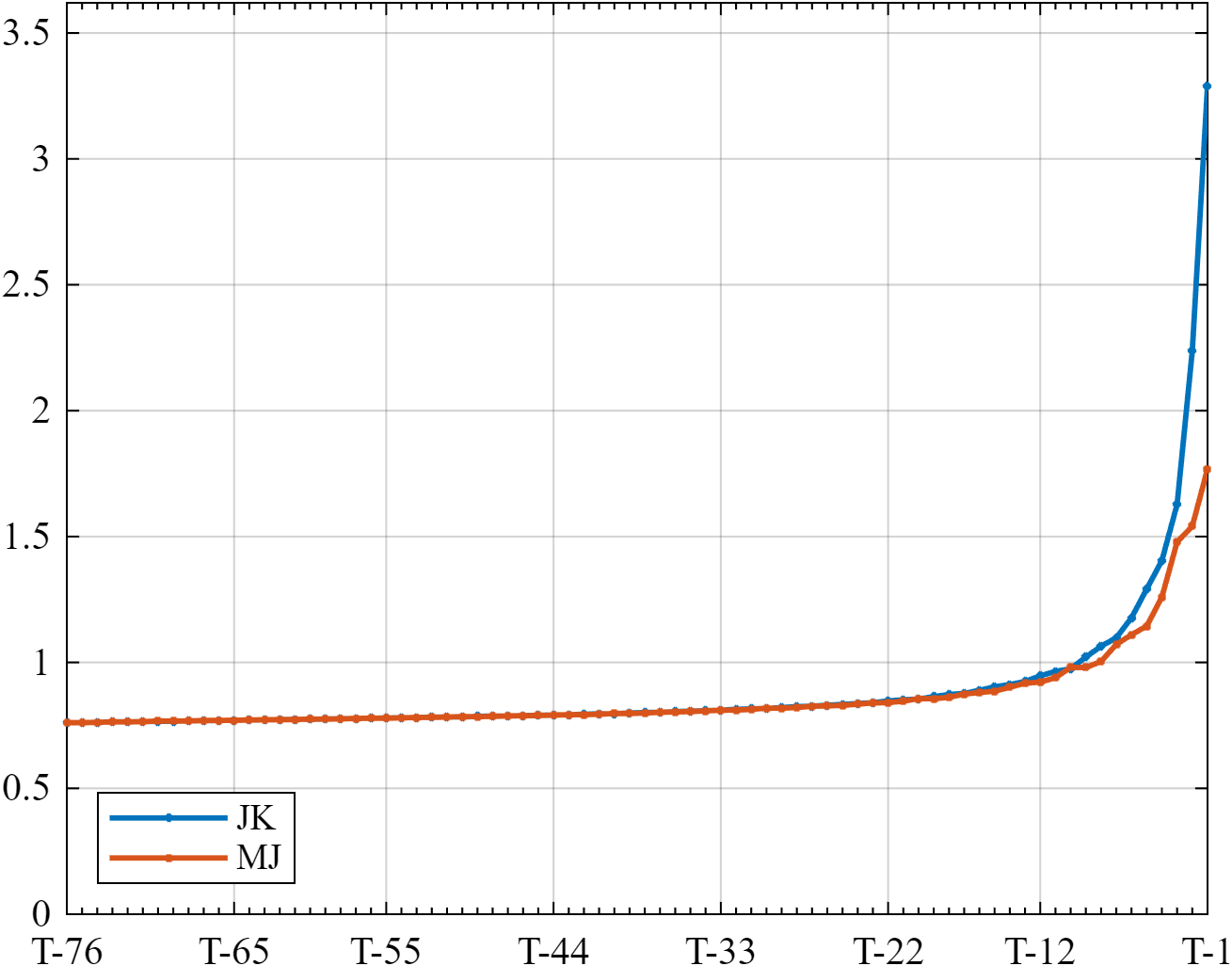} \\
		\footnotesize $T = 40$ & \footnotesize $T = 80$
	\end{tabular}
	\caption{Empirical standard deviation of the JK (blue) and MJ (red) {estimators} as a function of $S$ at $\tau_0 = 0.7$ with $\rho = 0.5$.}
	\label{fig:appstd-endog-tau70}
\end{figure}

\end{document}